\documentclass[aps,prl,superscriptaddress,reprint]{revtex4-2}
\usepackage{amsmath, amssymb}
\usepackage[export]{adjustbox}
\usepackage[colorlinks=true,breaklinks=true,linkcolor=blue,citecolor=blue,urlcolor=blue]{hyperref}
\usepackage{multirow}
\usepackage{xcolor}

\definecolor{tab:red}{rgb}{0.8392156862745098, 0.15294117647058825, 0.1568627450980392}
\definecolor{royalblue}{rgb}{0.2549019607843137, 0.4117647058823529, 0.8823529411764706}
\definecolor{bronze}{RGB}{184, 145, 55}
\definecolor{mpl_gray}{rgb}{0.5019607843137255, 0.5019607843137255, 0.5019607843137255}

\graphicspath{ {./figures/} }

\newcommand{\ket}[1]{{| {#1}  \rangle}}
\newcommand{\bra}[1]{{\langle {#1} |}}
\DeclareMathOperator{\tr}{tr}

\begin{document}

\title{Quantum-enhanced Markov chain Monte Carlo}

\author{David Layden}
\email{david.layden@ibm.com}
\affiliation{IBM Quantum, Almaden Research Center, San Jose, California 95120, USA}

\author{Guglielmo Mazzola}
\affiliation{IBM Quantum, IBM Research -- Zurich, 8803 R\"uschlikon, Switzerland}

\author{Ryan V. Mishmash}
\affiliation{IBM Quantum, Almaden Research Center, San Jose, California 95120, USA}

\author{Mario Motta} 
\affiliation{IBM Quantum, Almaden Research Center, San Jose, California 95120, USA}

\author{Pawel Wocjan}
\affiliation{IBM Quantum, T. J. Watson Research Center, Yorktown Heights, NY 10598, USA}

\author{Jin-Sung Kim}
\altaffiliation[Current affiliation: ]{NVIDIA}
\affiliation{IBM Quantum, Almaden Research Center, San Jose, California 95120, USA}

\author{Sarah Sheldon}
\affiliation{IBM Quantum, Almaden Research Center, San Jose, California 95120, USA}

\begin{abstract}
Sampling from complicated probability distributions is a hard computational problem arising in many fields, including statistical physics, optimization, and machine learning. Quantum computers have recently been used to sample from complicated distributions that are hard to sample from classically, but which seldom arise in applications. Here we introduce a quantum algorithm to sample from distributions that pose a bottleneck in several applications, which we implement on a superconducting quantum processor. The algorithm performs Markov chain Monte Carlo (MCMC), a popular iterative sampling technique, to sample from the Boltzmann distribution of classical Ising models. In each step, the quantum processor explores the model in superposition to propose a random move, which is then accepted or rejected by a classical computer and returned to the quantum processor, ensuring convergence to the desired Boltzmann distribution. We find that this quantum algorithm converges in fewer iterations than common classical MCMC alternatives on relevant problem instances, both in simulations and experiments. It therefore opens a new path for quantum computers to solve useful---not merely difficult---problems in the near term.
\end{abstract}

\maketitle

Quantum computers promise to solve certain computational problems much faster than classical computers. However, current quantum processors are limited by their modest size and appreciable error rates. Recent efforts to demonstrate quantum speedups have therefore focused on problems that are both classically hard and naturally suited to current quantum devices, like sampling from complicated---though not explicitly useful---probability distributions \cite{arute:2019, wu:2021, zhong:2021}. Here we introduce and experimentally demonstrate a quantum algorithm that is similarly well-suited to current devices, which samples from distributions that can be both complicated and useful: the Boltzmann distribution of classical Ising models.

\begin{figure}
\centering
\includegraphics[width=3.41in]{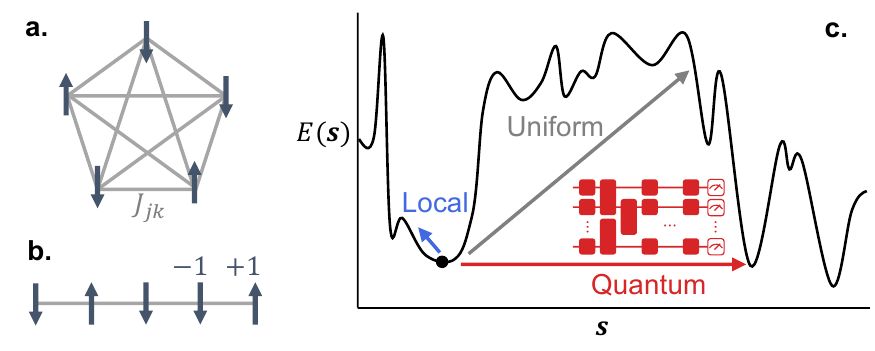}
\caption{\textbf{Ising model representations.} \textbf{a.} Graph depicting an $n=5$ model instance where arrows (vertices) represent spins and edges represent the $\binom{n}{2}$ non-zero couplings $J_{jk}$. Fields $h_j$ are not shown. \textbf{b.} An $n=5$ model instance with only $n-1$ non-zero couplings. \textbf{c.} A rugged energy landscape typical of spin glasses, with the configurations $\boldsymbol{s} \in \{-1, 1\}^n$ depicted in 1D. Typical proposed jumps for three MCMC algorithms, from a local minimum, are shown for illustration.}
\label{fig:overview}
\end{figure}

A classical Ising model consists of $n$ variables $(s_1, \dots, s_n)=\boldsymbol{s}$ called spins that can take values $s_j=\pm 1$ independently \cite{ising:1925}. A model instance is defined by coefficients $\{J_{jk}\}_{j>k=1}^n$ and $\{h_j\}_{j=1}^n$ called couplings and fields respectively, together with a temperature $T>0$. It is often represented as a graph, as in Figs.~\ref{fig:overview}a-b. Each spin configuration $\boldsymbol{s} \in \{1, -1\}^n$ is assigned an energy 
\begin{equation}
E(\boldsymbol{s}) = -\sum_{j>k=1}^n J_{jk} s_j s_k - \sum_{j=1}^n h_j s_j
\label{eq:E_function}
\end{equation}
and a corresponding Boltzmann probability $\mu(\boldsymbol{s}) = \frac{1}{\mathcal{Z}} e^{-E(\boldsymbol{s})/T}$, where the partition function $\mathcal{Z}=\sum_{\boldsymbol{s}} e^{-E(\boldsymbol{s})/T}$ ensures normalization. Sampling from $\mu$ is a common subroutine in many disparate applications, including in statistical physics, where it is used to compute thermal averages \cite{huang:2008}; in machine learning, to train Boltzmann machines \cite{ackley:1985}; and in combinatorial optimization, as part of the famous simulated annealing algorithm \cite{kirkpatrick:1983}. Indeed, this sampling is often a computational bottleneck when $J_{jk}$ and $h_{j}$ have varying signs and follow no particular pattern. Eq.~\eqref{eq:E_function} typically defines a rugged energy landscape for such instances, informally termed \emph{spin glasses} \cite{lucas:2014}, with many local minima that can be far from one another in Hamming distance, as depicted in Fig.~\ref{fig:overview}c. In the $T\rightarrow0$ limit, sampling from $\mu$ amounts to minimizing $E(\boldsymbol{s})$, which is NP-hard for spin glasses \cite{barahona:1982}. For small but non-zero $T$, sampling from $\mu$ requires finding several of the lowest-energy configurations, which may be far apart, not just the ground configuration(s). This hard regime will be our main focus.

\section*{Markov chain Monte Carlo} 
Markov chain Monte Carlo (MCMC) is the most popular algorithmic technique for sampling from the Boltzmann distribution $\mu$ of Ising models in all of the aforementioned applications. In fact, it was named one of the ten most influential algorithms of the 20\textsuperscript{th} century in science and engineering \cite{dongarra:2000}. It approaches the problem indirectly, without ever computing $\mu(\boldsymbol{s})$ and in turn the partition function $\mathcal{Z}$, which is believed to take $\Omega(2^n)$ time. Rather, MCMC performs a sequence of random jumps between spin configurations, starting from a generic one and jumping from $\boldsymbol{s}$ to $\boldsymbol{s'}$ with fixed transition probability $P(\boldsymbol{s'} |\, \boldsymbol{s})$ in any iteration. Such a process is called a Markov chain. For it to form a useful algorithm, the transitions probabilities must be carefully chosen to make the process converge to $\mu$; that is, for the probability of being in $\boldsymbol{s}$ after many iterations to approach $\mu(\boldsymbol{s})$. Sufficient conditions for such convergence are that the Markov chain be irreducible and aperiodic (technical requirements that will always be met here \cite{SM}), and satisfy the detailed balance condition:
\begin{equation}
P(\boldsymbol{s'} |\,\boldsymbol{s}) \, \mu(\boldsymbol{s})
=
P(\boldsymbol{s} \, | \boldsymbol{s'}) \, \mu(\boldsymbol{s'})
\label{eq:detailed_balance}
\end{equation}
for all $\boldsymbol{s} \neq \boldsymbol{s'}$ \cite{levin:2017}. 

How to efficiently realize a $P$ satisfying these conditions? A powerful approach is to decompose each jump into two steps: \textit{Step~1 (proposal)-} If the current configuration is $\boldsymbol{s}$, propose $\boldsymbol{s'}$ with some probability $Q(\boldsymbol{s'} |\, \boldsymbol{s})$. \textit{Step~2 (accept/reject)-} Compute an appropriate acceptance probability $A(\boldsymbol{s'} |\, \boldsymbol{s})$ based on $Q$ and $\mu$, then move to $\boldsymbol{s'}$ with that probability (i.e., accept the proposal), otherwise remain at $\boldsymbol{s}$. This gives $P(\boldsymbol{s'} |\,\boldsymbol{s}) = A(\boldsymbol{s'} |\,\boldsymbol{s}) \, Q(\boldsymbol{s'} |\,\boldsymbol{s})$ for all $\boldsymbol{s'}\neq \boldsymbol{s}$. For a given $Q$ there are infinitely many choices of $A$ that satisfy detailed balance \eqref{eq:detailed_balance}. One of the most popular is the Metropolis-Hastings (M-H) acceptance probability \cite{metropolis:1953, hastings:1970}:
\begin{equation}
A(\boldsymbol{s'} |\,\boldsymbol{s})
=
\min \left(1, \;
\frac{\mu(\boldsymbol{s'})}{\mu(\boldsymbol{s})}
\frac{Q(\boldsymbol{s}\,|\boldsymbol{s'})}{Q(\boldsymbol{s'}|\,\boldsymbol{s})}
\right).
\label{eq:MH_acceptance}
\end{equation}
Although $\mu(\boldsymbol{s})$ and $\mu(\boldsymbol{s'})$ cannot be efficiently computed, Eq.~\eqref{eq:MH_acceptance} depends only on their ratio, which can be evaluated in $O(n^2)$ time since $\mathcal{Z}$ cancels out. This cancellation underpins MCMC, and ensures that $A(\boldsymbol{s'} |\,\boldsymbol{s})$ can be efficiently computed provided $Q(\boldsymbol{s}\,|\boldsymbol{s'})/Q(\boldsymbol{s'}|\,\boldsymbol{s})$ can too. The same is true for the other popular choice of $A$, called the Glauber or Gibbs sampler acceptance probability \cite{SM}. 

The most common way to propose a candidate $\boldsymbol{s'}$ is by flipping ($s_j \! \mapsto \! -s_j$) a uniformly random spin $j\in[1,n]$ of the current configuration $\boldsymbol{s}$. We call this the \textit{local proposal strategy} since $\boldsymbol{s}$ and $\boldsymbol{s'}$ are always neighbors in terms of Hamming distance. The resulting acceptance probability can be computed efficiently, and the overall Markov chain converges quickly to $\mu$ for simple model instances at moderate $T$ \cite{levin:2017}. Various non-local strategies can also be used (sometimes in combination \cite{andrieu:2003}), including the \textit{uniform proposal strategy} where $\boldsymbol{s'}$ is picked uniformly at random from $\{-1, 1\}^n$, and more complex ones which flip clusters of spins \cite{swendsen:1987, wolff:1989, houdayer:2001, zhu:2015}. Broadly, spin glasses at low $T$ present a formidable challenge for all of these approaches, manifesting in long autocorrelation, slow convergence, and ultimately long MCMC running times. This challenge is an active area of research due to the many applications in which it arises \cite{goodfellow:2016}. The problem: $\mu(\boldsymbol{s'})/\mu(\boldsymbol{s})$ in Eq.~\eqref{eq:MH_acceptance}, which comes from demanding detailed balance \eqref{eq:detailed_balance}, is exponentially small in $\Delta E/T = [E(\boldsymbol{s'}) - E(\boldsymbol{s})]/T$ for energy-increasing proposals $\boldsymbol{s} \rightarrow \boldsymbol{s'}$. Consequently, such proposals are frequently rejected. (And a rejected proposal still counts as an MCMC iteration.) This makes it difficult for Markov chains to explore rugged energy landscapes at low $T$, as they tend to get stuck in local minima for long stretches and to rarely cross large barriers in the landscape.

\section*{Quantum algorithm}

To alleviate this issue, we introduce an MCMC algorithm which uses a quantum computer to propose moves and a classical computer to accept/reject them. It alternates between two steps: \textit{Step~1 (quantum proposal)}- If the current configuration is $\boldsymbol{s}$, prepare the computational basis state $\ket{\boldsymbol{s}}$ on the quantum processor, where $s_j=\pm 1$ refers to an eigenvalue of $Z_j$. (E.g., if $\boldsymbol{s}=(1,1,-1)$, prepare $\ket{\boldsymbol{s}} = \ket{001}$. We use $X_j$, $Y_j$ and $Z_j$ to denote $\sigma_x$, $\sigma_y$ and $\sigma_z$ on qubit $j$ respectively.) Then apply a unitary $U$ satisfying the symmetry constraint:
\begin{equation}
\label{eq:symmetry}
\big| \bra{\boldsymbol{s'}} U \ket{\boldsymbol{s}} \big|
=
\big| \bra{\boldsymbol{s}} U \ket{\boldsymbol{s'}} \big|
\;\; \text{for all } \boldsymbol{s}, \boldsymbol{s'} \in \{-1, 1\}^n.
\end{equation}
Finally, measure each qubit in the $Z$ eigenbasis, i.e., the computational basis, denoting the outcome $\boldsymbol{s'}$. \textit{Step~2 (classical accept/reject)}- Compute $A(\boldsymbol{s'}|\,\boldsymbol{s})$ from Eq.~\eqref{eq:MH_acceptance} on a classical computer and jump to $\boldsymbol{s'}$ with this probability, otherwise stay at $\boldsymbol{s}$. While computing $Q(\boldsymbol{s'} | \, \boldsymbol{s})$ and $Q(\boldsymbol{s} \,| \boldsymbol{s'})$ may take exponential (in $n$) time in general, there is no need to do so: Eq.~\eqref{eq:MH_acceptance} depends only on their ratio, which equals 1 since $Q(\boldsymbol{s'} | \, \boldsymbol{s}) = |\bra{\boldsymbol{s'}} U \ket{ \boldsymbol{s}}|^2 = Q(\boldsymbol{s} \,| \boldsymbol{s'})$ due to \eqref{eq:symmetry}. This cancellation underpins our algorithm, and mirrors that between $\mu(\boldsymbol{s})$ and $\mu(\boldsymbol{s'})$. The resulting Markov chain provably converges to the Boltzmann distribution $\mu$, but is hard to mimic classically, provided it is classically hard to sample the measurement outcomes of $U \ket{\boldsymbol{s}}$. This combination opens the possibility of a useful quantum advantage. 

The symmetry requirement \eqref{eq:symmetry} ensures convergence to $\mu$, but does not uniquely specify $U$. Rather, we pick the quantum step of our algorithm heuristically with the aim of accelerating MCMC convergence in spin glasses at low $T$, while still satisfying condition \eqref{eq:symmetry}. We then evaluate the resulting Markov chains through simulations and experiments. Several choices of $U$ are promising, including ones arising from quantum phase estimation and quantum annealing \cite{SM}. However, we focus here on evolution $U =e^{-iHt}$ under a \textit{time-independent} Hamiltonian 
\begin{equation}
H = (1-\gamma) \alpha \, H_\text{prob} + \gamma H_\text{mix}
\label{eq:H}
\end{equation}
for a time $t$, where
\begin{equation}
H_\text{prob} 
=
 -\!\! \sum_{j>k=1}^n J_{jk} Z_j Z_k - \sum_{j=1}^n h_j Z_j
 =
\sum_{\boldsymbol{s}}
E(\boldsymbol{s}) \ket{\boldsymbol{s}} \! \bra{\boldsymbol{s}}
\label{eq:H_prob}
\end{equation}
encodes the classical model instance, $H_\text{mix}$ (discussed below) generates quantum transitions, and $\gamma \in [0,1]$ is a parameter controlling the relative weights of both terms. It is convenient to include a normalizing factor $\alpha = \|H_\text{mix}\|_\textsc{f}  / \|H_\text{prob}\|_\textsc{f}$ in Eq.~\eqref{eq:H} so that both terms of $H$ share a common scale regardless of  $\{J_{jk}, h_j\}$, which can be arbitrary. ($\|M\|_\textsc{f} = \tr(M^\dagger M)^{1/2}$ is the Frobenius norm of a matrix $M$.) 

In principle, $H_\text{mix}$ could comprise arbitrary weighted sums and products of $X_j$ and $Y_j Y_k$ terms. These produce a symmetric $H$ and therefore $U=U^T$ (although $U \neq U^\dagger$ generically), thus satisfying condition~\eqref{eq:symmetry}. If $H_\text{mix}$ were a dense matrix and $\gamma \ll 1$, a perturbative analysis shows that non-trivial quantum proposals $\boldsymbol{s} \rightarrow \boldsymbol{s'}$ would exhibit a remarkable combination of features that suggest fast MCMC convergence \cite{SM}: Like local proposals, their absolute energy change $|\Delta E| = |E(\boldsymbol{s'}) - E(\boldsymbol{s})|$ is typically small, meaning they are likely to be accepted even if $\Delta E > 0$. However, like uniform or cluster proposals, $\boldsymbol{s}$ and $\boldsymbol{s'}$ are typically far in Hamming distance, so the resulting Markov chain can move rapidly between distant local energy minima. While a dense $H_\text{mix}$ would pose experimental difficulties, similar behavior is reported for various spin glasses in Refs.~\cite{baldwin:2018, smelyanskiy:2020, smelyanskiy:2019} using 
\begin{equation}
H_\text{mix} = \sum_{j=1}^n X_j
\label{eq:H_mix}
\end{equation} 
and larger $\gamma$. Inspired by these results, we take Eq.~\eqref{eq:H_mix} as the definition of $H_\text{mix}$ here. The normalizing factor $\alpha$ in Eq.~\eqref{eq:H} then takes the simple form
\begin{equation}
\alpha = \frac{\|H_\text{mix}\|_\textsc{f} }{\|H_\text{prob}\|_\textsc{f}} = \frac{\sqrt{n}}{\sqrt{\sum_{j>k=1}^n J_{jk}^2 + \sum_{j=1}^n h_j^2}}
\label{eq:scaling}
\end{equation}
which---crucially---can be computed in $O(n^2)$ time. Step~1 of our algorithm therefore consists of realizing quenched dynamics of a transverse-field quantum Ising model encoding the classical model instance, which can be efficiently simulated on a quantum computer \cite{lloyd:1996}. Note, however, that our algorithm samples from a classical Boltzmann distribution $\mu$, not from a quantum Gibbs state as in Refs.~\cite{temme:2011, yung:2012, moussa:2019, wocjan:2021}. 

As an initial state $\ket{\boldsymbol{s}}$ evolves under $H$, it becomes delocalized due to $H_\text{mix}$ and effectively queries the classical energy function $E$ from Eq.~\eqref{eq:E_function} in quantum superposition through $H_\text{prob}$. The measurement outcome $\boldsymbol{s'}$ is therefore influenced by the entire energy landscape. The details of this quantum evolution depend on the free parameters $\gamma$ and $t$. Rather than try to optimize these, we rely on a simple observation: instead of applying the same $U$ in every iteration of our algorithm, one could equally well pick $U$ at random in each iteration from an ensemble $\{U_1, U_2, \dots\}$ of unitaries, each satisfying condition~\eqref{eq:symmetry}. This amounts to sampling outputs from random circuits, initialized according to the state of a Markov chain. It is this randomized approach that we implement, by picking $\gamma$ and $t$ uniformly at random in each iteration \cite{SM}, thus obviating the need to optimize these parameters. Note that we do not invoke adiabatic quantum evolution in any way, despite the familiar form of Eq.~\eqref{eq:H}. Rather, the relative weights of $H_\text{prob}$ and $H_\text{mix}$ are held fixed throughout each quantum step.

\section*{Performance}

\begin{figure*}
\includegraphics{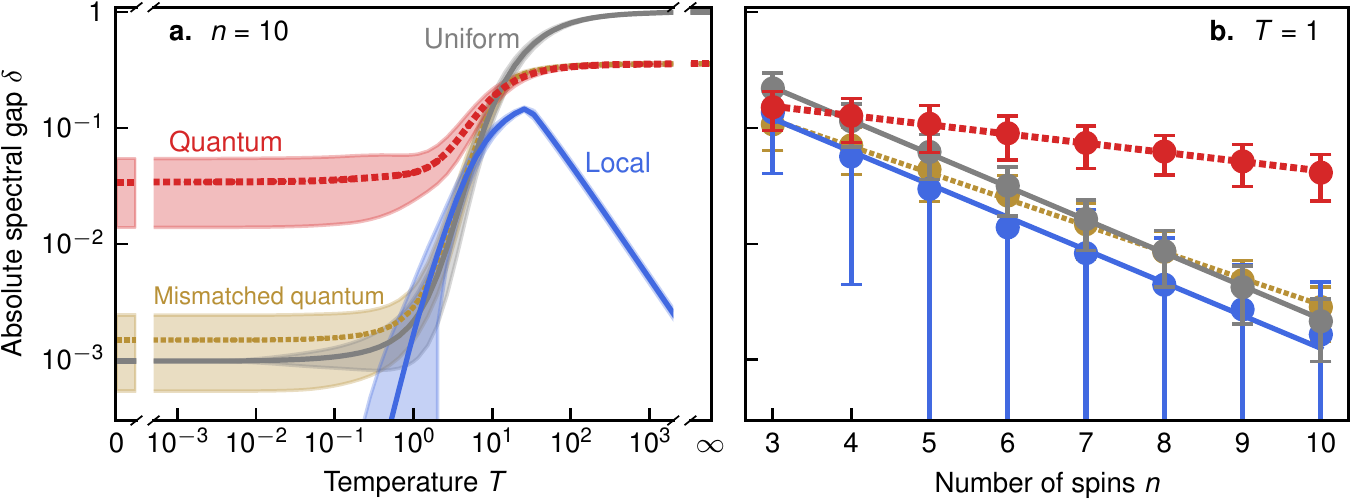}
\hfill
\raisebox{8.5em}{
\begin{tabular}{ c c }
\multicolumn{2}{c}{
\textbf{c.} $\boldsymbol{\;\langle \delta \rangle \propto 2^{-kn}}$ \textbf{fits} } \\
\hline
\color{tab:red} Quantum & $\color{tab:red}k=0.264(4)$\\
\color{bronze} Mismatched & \multirow{2}{*}{$\color{bronze} k=0.76(1)$} \\[-1ex]
\color{bronze} quantum & \\
\color{royalblue} Local & $ \color{royalblue} k=0.94(4)$ \\ 
\color{mpl_gray} Uniform & $\color{mpl_gray}  k=0.948(7)$\\
\hline 
\multicolumn{2}{c}{
Quantum enhancement}\\[-1ex]
\multicolumn{2}{c}{
factor in $k$: 3.6(1)}
\end{tabular}
}
\caption{\textbf{Average-case convergence rate simulations.} The absolute spectral gap $\delta$, a measure of MCMC convergence rate, using the M-H acceptance probability \eqref{eq:MH_acceptance} with different proposal strategies. All strategies were simulated classically. Lines/markers show the average $\delta$ over 500 random fully-connected Ising model instances for each $n$; error bands/bars show the standard deviation in $\delta$ over these instances. Dotted lines are for visibility. \textbf{a.} The slow-down of each strategy at low $T$. For the local proposal strategy, $\delta \rightarrow 0$ also at high $T$ because an eigenvalue of its transition matrix approaches $-1$. This artifact can be easily be remedied by using a lazy chain or the Gibbs sampler acceptance probability; the same is not true at low $T$, however \cite{SM}. \textbf{b.} Problem size dependence, with least squares exponential fits to the average $\delta$, weighted by the standard error of the mean. \textbf{c.} The resulting fit parameters and the average quantum enhancement exponent, which is the ratio of $k$ for the quantum algorithm and the smallest $k$ among classical proposal strategies (the local strategy, here). Uncertainties are from the fit covariance matrices. Similar data for different $n$ and $T$, model connectivity, and acceptance probabilities is shown in \cite{SM}.
}
\label{fig:bulk_numerics}
\end{figure*}

We now analyze the running time of our quantum algorithm through its convergence rate. MCMC convergence is inherently multifaceted. However, a Markov chain's convergence rate is often summarized by its absolute spectral gap, $\delta \in [0,1]$, where the extremes of $\delta=0$ and $1$ describe the slowest and fastest possible convergence, respectively. This quantity is found by forming the transition probabilities $P(\boldsymbol{s'}|\, \boldsymbol{s})$ into a $2^n \times 2^n$ matrix and computing its eigenvalues $\{\lambda \}$, all of which satisfy $|\lambda|\le 1$ and describe a facet of the chain's convergence. The absolute spectral gap, $\delta = 1-\max_{\lambda\neq 1}|\lambda|$, describes the slowest facet. More concretely, it bounds the mixing time $\tau_\varepsilon$, defined as the minimum number of iterations required for the Markov chain's distribution to get within any $\varepsilon>0$ of $\mu$ in total variation distance, for any initial distribution, as \cite{levin:2017}

\begin{equation}
\left(\delta^{-1}-1 \right) \ln \left( \frac{1}{2\varepsilon} \right)
\; \le \; 
\tau_\varepsilon 
\; \le \; 
\delta^{-1}
\ln \left( \frac{1}{\varepsilon \, \min_{\boldsymbol{s}}  \mu(\boldsymbol{s})} \right).
\label{eq:mixing_time}
\end{equation}
Since $\delta$ is the only quantity on either side of \eqref{eq:mixing_time} that depends on the proposal strategy $Q$, it is a particularly good metric for comparing the convergence rate of our quantum algorithm with that of common classical alternatives. Because $\delta$ depends on the model instance and on $T$, not just on $Q$, we analyzed our algorithm's convergence in two complementary ways: First, we simulated it on a classical computer for many instances to elucidate the average-case $\delta$. Second, we implemented it experimentally for illustrative instances and analyzed both the convergence rate and the mechanism underlying the speedup observed in simulations. (Note that computing $\delta$ is different---and much more demanding---than simply running MCMC.) We used the M-H acceptance probability \eqref{eq:MH_acceptance} throughout, although this choice has little impact on the results \cite{SM}. Unlike previously proposed algorithms which prepare a quantum state encoding $\mu$, ours uses simple, shallow quantum circuits \cite{szegedy:2004, richter:2007, somma:2008, wocjan:2008, harrow:2020, lemieux:2020, arunachalam:2021}. Its alternating quantum/classical structure means that quantum coherence need only be maintained in each repetition of Step~1, rather than over the whole algorithm \cite{wild:2021prl, wild:2021pra}. Moreover, it generates numerous samples per run, like a fully classical MCMC algorithm, rather than a single one. These features make it sufficiently well-suited to current quantum processors that we observed a quantum speedup experimentally on up to $n=10$ qubits.

\subsection*{Average-Case Performance}

For the first part of the analysis, we generated 500 random spin glass instances on $n$ spins by drawing each $J_{jk}$ and $h_j$ independently from standard normal distributions. We did not explicitly fix any couplings $J_{jk}$ to zero; the random instances are therefore fully connected as in Fig.~\ref{fig:overview}a. This ensemble is the archetypal Sherrington-Kirkpatrick model \cite{sherrington:1975} (up to a scale factor) with random local fields, where the fields serve to break inversion symmetry and thus increase the complexity. For each instance, we explicitly computed all the transition probabilities $\{P(\boldsymbol{s'}|\, \boldsymbol{s}) \}$ and then $\delta$ as a function of $T$ for different proposal strategies $Q$. We then averaged $\delta$ over the model instances, and repeated this process for $3 \le n \le 10$. The results describe the average MCMC convergence rate as a function of $n$ and $T$. Two illustrative slices are shown in Fig.~\ref{fig:bulk_numerics}, where $n$ and $T$ are held fixed in turn. At high $T$, where $\mu$ is nearly uniform, the uniform proposal produces a fast-converging Markov chain with $\delta$ near $1$, as shown in Fig.~\ref{fig:bulk_numerics}a. However, both the uniform and local proposals suffer a sharp decrease in $\delta$ at lower $T$. This slow-down is much less pronounced for our quantum algorithm, which gives a substantially better $\delta$ on average than either classical alternative for $T \lesssim 1$. 

For all three proposal strategies, the average scaling of $\delta$ with $n$ at $T=1$ fits well to $\langle \delta \rangle \propto 2^{-kn}$, as shown in Fig.~\ref{fig:bulk_numerics}b. Our algorithm appears to give an average-case polynomial enhancement in $\delta$ over the local and uniform proposals based on the fitted values of $k$, shown in Fig.~\ref{fig:bulk_numerics}c. These values depend on $T$, but their ratios suggest a roughly cubic/quartic enhancement at low temperatures regardless of the exact $T$ \cite{SM}. Finally, to elucidate the source of this average-case quantum enhancement, we also computed $\delta$ for a \textit{mismatched quantum proposal strategy}, where moves are proposed like in our algorithm, but based on the wrong energy landscape. That is, rather than explore the relevant $E$ defined by coefficients $\{J_{jk}\}$ and $\{h_j\}$ as in Eqs.~\eqref{eq:E_function} and \eqref{eq:H_prob}, the mismatched quantum strategy uses the wrong coefficients $\{\tilde{J}_{jk}\}$ and $\{\tilde{h}_j\}$ (drawn randomly from the same distribution as $J_{jk}$ and $h_j$) in its quantum Hamiltonian and therefore explores the wrong energy landscape $\tilde{E}$ in Step~1. The resulting $\delta$ is comparable with that of the uniform proposal in Fig.~\ref{fig:bulk_numerics}. This suggests that the $\delta$ enhancement in our algorithm indeed arises from exploring the classical energy landscape $E$ in quantum superposition to propose moves.

\subsection*{Experimental Implementation}

For the second part of the analysis we focus in on individual model instances, for which we implemented our quantum algorithm experimentally and analyzed it in more depth than is feasible for a large number of instances. We generated random model instances and picked illustrative ones whose lowest-$E$ configurations include several near-degenerate local minima---a central feature of spin glasses at larger $n$ which hampers MCMC at low $T$. We then implemented our quantum algorithm experimentally for these instances on \textit{ibmq\_mumbai}, a 27-qubit superconducting quantum processor, using Qiskit \cite{qiskit}, an open-source quantum software development platform. (The Ising model $T$ has no relation to the processor's physical temperature.) We approximated $U = e^{-iHt}$ on this device through a randomly-compiled second-order Trotter-Suzuki product formula \cite{suzuki:1985, lloyd:1996, wallman:2016} with up to 48 layers of pulse-efficient 2-qubit gates \cite{earnest:2021} acting on up to 5 pairs of qubits in parallel \cite{SM}, and set $Q(\boldsymbol{s}\,|\boldsymbol{s'})/Q(\boldsymbol{s'}|\,\boldsymbol{s})=1$ in the acceptance probability. Unlike in the first part of the analysis, we restricted our focus to model instances where $J_{jk}=0$ for $|j-k| \neq 1$ as in Fig.~\ref{fig:overview}b, in order to match the connectivity of qubits in the quantum processor for this initial demonstration. Simulations show that the average-case $\delta$ for such instances is qualitatively similar to Fig.~\ref{fig:bulk_numerics} \cite{SM}.

\begin{figure}[t]
\centering
\includegraphics[width=3.41in]{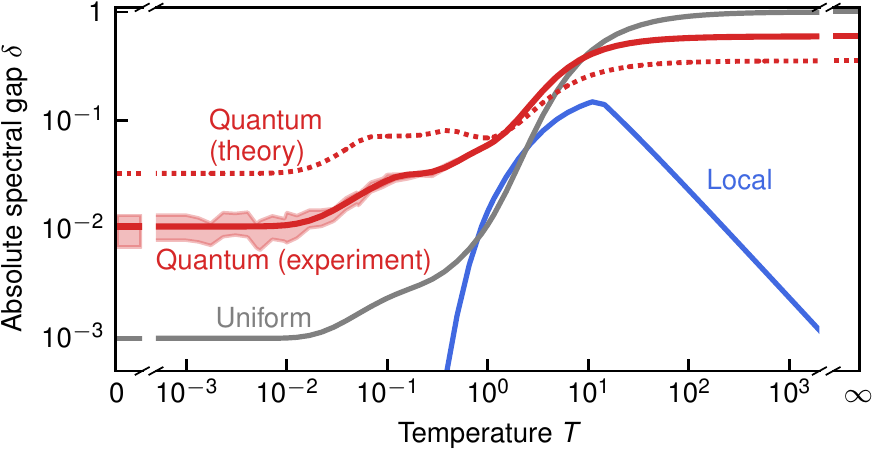}
\caption{\textbf{Convergence rate experiment.} The absolute spectral gap $\delta$, a measure of MCMC convergence rate, for an illustrative model instance on $n=10$ spins with 1D connectivity. Each proposal strategy is combined with the M-H acceptance probability \eqref{eq:MH_acceptance}. To infer $\delta$ for the experimental realization of our algorithm, we recorded $5.76\times 10^7$ quantum transitions $\ket{\boldsymbol{s}} \rightarrow \ket{\boldsymbol{s'}}$ between computational states with $\ket{\boldsymbol{s}}$ uniformly distributed, to estimate each $Q(\boldsymbol{s'} |\, \boldsymbol{s})$. For each $T$, we used the full data set to estimate the MCMC transition matrix $P$ and in turn form a point estimate of $\delta$ (solid red line), then a random subsample of the data to compute 99\% bootstrap confidence intervals (red error bands).
}
\label{fig:n=10_gap}
\end{figure}

We focus on an $n=10$ model instance here in which the six lowest-$E$ configurations are all local minima, and the two lowest have an energy difference of just $|\Delta E| = 0.05$. Similar instances with $n=8$ and $9$ are analyzed in \cite{SM}. For the present $n=10$ instance, $\delta$ closely follows the average in Fig.~\ref{fig:bulk_numerics}a for the local and uniform proposal strategies, as well as for our quantum algorithm in theory, as shown in Fig.~\ref{fig:n=10_gap}. We estimated $\delta$ as a function of $T$ for the experimental realization of our quantum algorithm by counting quantum transitions $\ket{\boldsymbol{s}} \rightarrow \ket{\boldsymbol{s'}}$ between all computational states $\ket{\boldsymbol{s}}$ and $\ket{\boldsymbol{s'}}$ to estimate the MCMC transition matrix, which we then diagonalized. At low $T$, the inferred $\delta$ is smaller than the theoretical value due to experimental imperfections, but still significantly larger than that of either the local or uniform alternative. This constitutes an experimental quantum enhancement in MCMC convergence on current quantum hardware. At high $T$, the experimental $\delta$ is larger than the theoretical value for our quantum algorithm, which we attribute to noise in the quantum processor mimicking the uniform proposal to a degree. We also plot $\delta$ for common MCMC cluster algorithms (those of Swendsen-Wang, Wolff and Houdayer) in \cite{SM}. These are substantially more complicated than the local and uniform proposals, both conceptually and practically, but offer almost no $\delta$ enhancement in this setting compared to the simpler classical alternatives, so we do not focus on them here. 

To further illustrate the increased convergence rate of our quantum-enhanced MCMC algorithm compared to these classical alternatives, we use it to estimate the average magnetization (with respect to the Boltzmann distribution $\mu$) of this same $n=10$ instance. The magnetization of a spin configuration $\boldsymbol{s}$ is $m(\boldsymbol{s}) = \frac{1}{n} \sum_{j=1}^n s_j$, and the Boltzmann average magnetization is
\begin{equation}
\langle m \rangle_{\mu}
=
\sum_{\boldsymbol{s}}
\mu(\boldsymbol{s}) \, m(\boldsymbol{s}).
\label{eq:avg_magnetization}
\end{equation}
Eq.~\eqref{eq:avg_magnetization} involves a sum over all $2^n$ configurations, but given $N$ samples $\{\boldsymbol{s}\} = \mathcal{S}$ from $\mu$, the approximation $\langle m \rangle_{\mu} \approx N^{-1} \sum_{\boldsymbol{s} \in \mathcal{S}} m(\boldsymbol{s})$ can be accurate with high probability even if $N \ll 2^n$ \cite{ambegaokar:2010}. While sampling from $\mu$ exactly may be infeasible, it is common to approximate $\langle m \rangle_{\mu}$ by the running average of $m(\boldsymbol{s})$ over MCMC trajectories (of one or several independent chains), and likewise for other average quantities. The quality of this approximation after a fixed number of MCMC iterations reflects the Markov chains' convergence rate \cite{levin:2017}.

\begin{figure}
\centering
\includegraphics[width=3.41in]{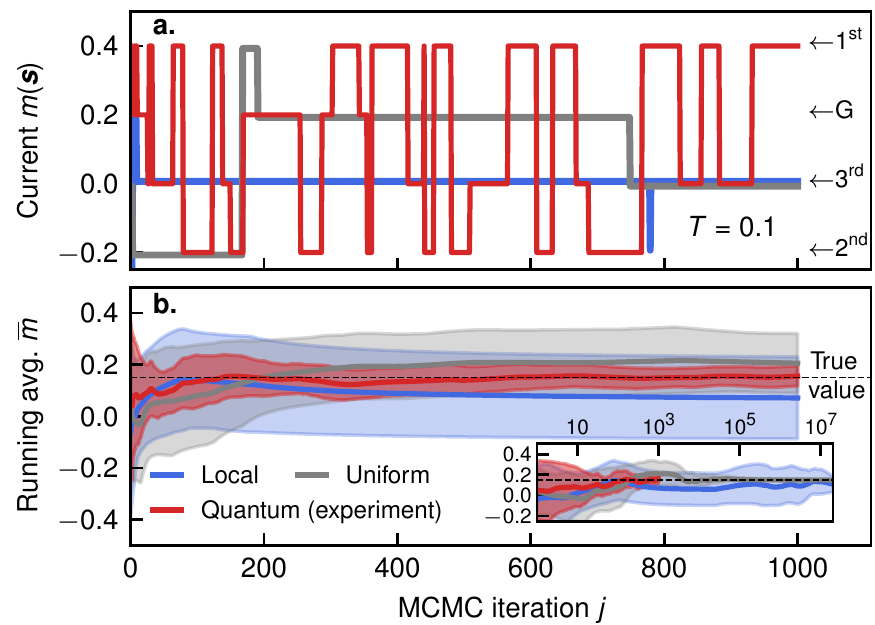}
\caption{\textbf{Magnetization estimate experiment.} \textbf{a.} The current magnetization $m(\boldsymbol{s}^{(j)})$ for individual Markov chains after $j$ iterations. Each chain illustrates a different proposal strategy with uniformly random initialization. Arrows indicate the magnetization of the ground (G), 1\textsuperscript{st}, 2\textsuperscript{nd} and 3\textsuperscript{rd} excited configurations. \textbf{b.} Convergence of the running average $\bar{m}^{(j)} = \frac{1}{j} \sum_{k=0}^j m(\boldsymbol{s}^{(k)})$ from MCMC trajectories to the true value of $\langle m \rangle_\mu$ for different proposal strategies. For each strategy, the lines and error bands show the mean and standard deviation, respectively, of $\bar{m}^{(j)}$ over 10 independent chains. The inset depicts the same chains over more iterations. We do not use a burn-in period or thinning (i.e., the running average starts at $k=0$ and includes every iteration up to $k=j$), as these practices would introduce hyperparameters that complicate the interpretation. Both panels are for the same illustrative $n=10$ instance at $T=0.1$, and use the M-H acceptance probability \eqref{eq:MH_acceptance}.
}
\label{fig:magnetization}
\end{figure} 

We used this approach to estimate $\langle m \rangle_{\mu}$ at $T=0.1$ as shown in Fig.~\ref{fig:magnetization}. At this temperature $\langle m \rangle_{\mu} \approx 0.15$, and the Boltzmann probabilities of the ground (i.e., lowest-$E$), 1\textsuperscript{st}, 2\textsuperscript{nd} and 3\textsuperscript{rd} excited configurations are approximately 43\%, 26\%, 19\% and 12\% respectively. This $T$ is therefore sufficiently high that sampling from $\mu$ is not simply an optimization problem (where $\langle m \rangle_{\mu}$ depends overwhelmingly on the ground configuration), but sufficiently low that $\langle m \rangle_{\mu}$ is mostly determined by a few low-$E$ configurations out of $2^n=1024$. Efficiently estimating $\langle m \rangle_{\mu}$ using MCMC therefore involves finding these configurations and---crucially---jumping frequently between them in proportion to their Boltzmann probabilities. The magnetization $m(\boldsymbol{s})$ for illustrative trajectories is shown in Fig.~\ref{fig:magnetization}a for the local and uniform proposal strategies, and for an experimental implementation of our quantum algorithm. While each Markov chain finds a low-$E$ configurations quickly, our quantum algorithm jumps between these configurations noticeably faster than the others. The running average estimate for $\langle m \rangle_{\mu}$ from 10 independent Markov chains of each type is shown in Fig.~\ref{fig:magnetization}b. Our quantum algorithm converges to the true value of $\langle m \rangle_{\mu}$, with no discernible bias,  substantially faster than the two classical alternatives despite experimental imperfections.

\begin{figure*}
\centering
\includegraphics[valign=t]{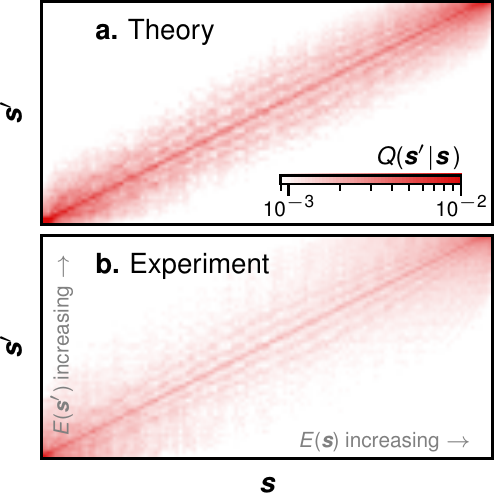}
\hfill
\includegraphics[valign=t]{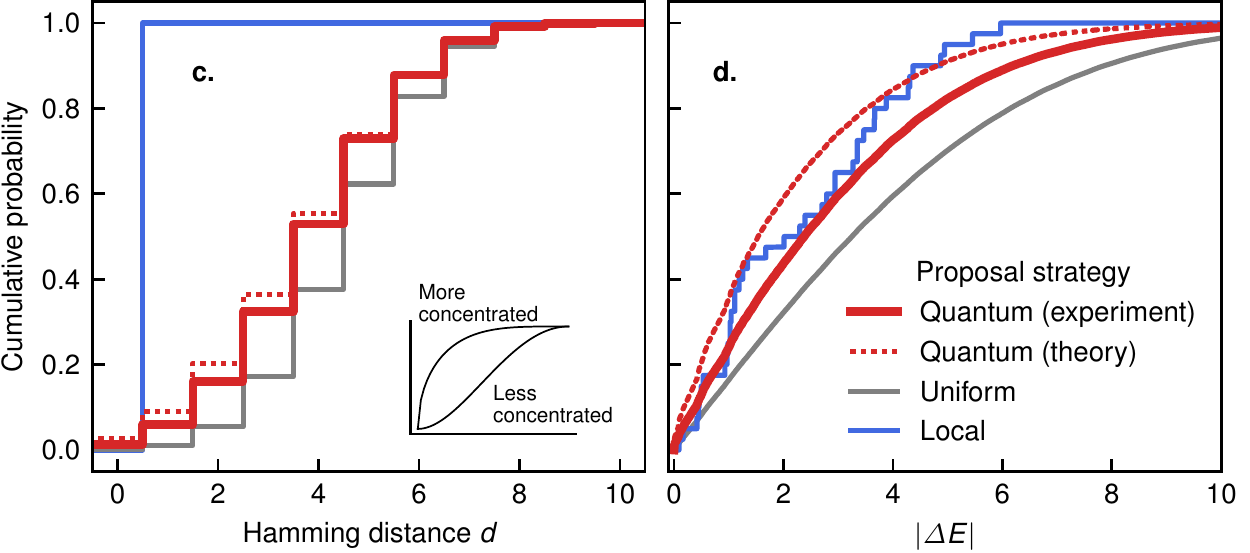}
\caption{\textbf{Quantum speedup mechanism.} \textbf{a.} The classically-simulated probabilities of $\boldsymbol{s} \rightarrow \boldsymbol{s'}$ proposals in our quantum algorithm, represented as a $2^n \times 2^n$ matrix whose columns are independent histograms. Both the initial and proposed configurations are sorted by increasing Ising energy $E$. \textbf{b.} The estimated proposal probabilities for our algorithm's experimental realization. We estimated each $Q(\boldsymbol{s'}|\,\boldsymbol{s})$ by the number of observed $\boldsymbol{s} \rightarrow \boldsymbol{s'}$ proposals normalized by the number of $\boldsymbol{s} \rightarrow \text{[anything]}$ proposals, using a total of $5.76 \times 10^7$ recorded quantum transitions. \textbf{c.} The probability distributions of Hamming distance between current ($\boldsymbol{s}$) and proposed ($\boldsymbol{s'}$) configurations, for a uniformly random current configuration. That of the experiment uses the estimated probabilities from panel b, while the rest were computed exactly. \textbf{d.} The analogous distributions for $|\Delta E| = |E(\boldsymbol{s'}) - E(\boldsymbol{s})|$ of proposed jumps. Each distribution is depicted in full detail through its cumulative distribution function, with no binning. All panels are for the same illustrative $n=10$ model instance; none depend on $T$ or on the choice of acceptance probability.
}
\label{fig:mechanism}
\end{figure*}

Finally, we examined the mechanism underlying this observed quantum speedup. Recall that the local proposal strategy typically achieves small $|\Delta E| = |E(\boldsymbol{s'}) - E(\boldsymbol{s})|$ by picking $\boldsymbol{s'}$ from the neighbors of $\boldsymbol{s}$, whereas the uniform proposal strategy typically picks $\boldsymbol{s'}$ far from $\boldsymbol{s}$ at the cost of a larger $\Delta E$, as illustrated in Fig.~\ref{fig:overview}c. Our quantum algorithm was motivated by the possibility of achieving the best of both: small $|\Delta E|$, and $\boldsymbol{s'}$ far from $\boldsymbol{s}$. This combination of features is illustrated in Fig.~\ref{fig:overview}c and borne out in Fig.~\ref{fig:mechanism}. The proposal probabilities $Q(\boldsymbol{s'} |\, \boldsymbol{s})$ arising in our algorithm for the same $n=10$ model instance are shown in Figs.~\ref{fig:mechanism}a-b for theory and experiment respectively. Both show the same effect with good qualitative agreement: starting from $\boldsymbol{s}$, our quantum algorithm mostly proposes jumps to configurations $\boldsymbol{s'}$ for which $|\Delta E|$ is small, even though $\boldsymbol{s}$ and $\boldsymbol{s'}$ may be far in Hamming distance. This effect is especially pronounced between the lowest-$E$ configurations and also between highest-$E$ ones. 

To further examine this effect we asked the following: for a uniformly random configuration $\boldsymbol{s}$, what is the probability of proposing a $\boldsymbol{s} \rightarrow \boldsymbol{s'}$ jump for which $\boldsymbol{s}$ and $\boldsymbol{s'}$ are separated by a Hamming distance $d$, or by an energy difference $|\Delta E|$? The resulting distributions are shown in Figs.~\ref{fig:mechanism}c-d respectively for the local and uniform proposal strategies, as well as for the theoretical and experimental realizations of our quantum algorithm. The Hamming distance distribution for local proposals is concentrated at $d=1$ (by definition), whereas it is much more evenly spread for both uniform proposals and for our quantum algorithm, as shown in Fig.~\ref{fig:mechanism}c. Conversely, the $\Delta E$ distribution of local proposals is more concentrated at small $\Delta E$ than that of uniform proposals. In theory, the corresponding distribution for our quantum algorithm is even more strongly concentrated at small $\Delta E$, as shown in Fig.~\ref{fig:mechanism}d. The $\Delta E$ distribution from the experimental realization of our algorithm, however, lies between those of the local and uniform proposal strategies, due to experimental imperfections.

\section*{Outlook}

Current quantum computers can sample from complicated probability distributions. We proposed and experimentally demonstrated a new quantum algorithm which leverages this ability in order to sample from the low-temperature Boltzmann distribution of classical Ising models, which is useful in many applications---not just complicated. Our algorithm uses relatively simple and shallow quantum circuits, thus enabling a quantum speedup on current hardware despite experimental imperfections. It works by alternating between quantum and classical steps on a shot-by-shot basis, unlike variational quantum algorithms, which typically run a quantum circuit many times in each step \cite{cerezo:2021}. Rather, it uses a quantum computer to propose a random bit-string, which is accepted or rejected by a classical computer. The resulting Markov chain is guaranteed to converge to the desired Boltzmann distribution, even though it may not be possible to efficiently simulate classically. In this sense our algorithm is partially heuristic, like most classical MCMC algorithms: the eventual result is theoretically guaranteed, while fast convergence is established empirically. 

Many state-of-the-art MCMC algorithms build upon simpler Markov chains in heuristically-motivated ways; for instance, by running several local M-H chains in parallel at different temperatures and occasionally swapping them \cite{swendsen:1986, houdayer:2001, zhu:2015}. Our quantum algorithm may provide a potent new ingredient for such composite algorithms in the near term. However, there remain ample opportunities for refinements and variations. For instance, a more targeted method of picking the parameters $\gamma$ and $t$ could further accelerate convergence \cite{mazzola:2021}. Indeed, $t$ did not depend on the problem size $n$ in our implementation, although in some settings it should grow with $n$ if the qubits are all to remain within each others' light cones. Moreover, different quantum processors with different connectivities, such as quantum annealers, may also be well-suited to implement our algorithm, perhaps without needing to discretize the Hamiltonian dynamics in Step~1.

Our algorithm is remarkably robust against imperfections. It achieves a speedup by proposing good jumps---but not every jump needs to be especially good for the algorithm to work well. (This is why picking $\gamma$ and $t$ at random works well: good values arise often enough.) For instance, if an error occurs in the quantum processor while a jump is being proposed, the proposal will be accepted/rejected as usual, and the Markov chain can still converge to the target distribution provided such errors do not break the $Q(\boldsymbol{s'} | \, \boldsymbol{s}) = Q(\boldsymbol{s} \,| \boldsymbol{s'})$ symmetry on average \cite{SM}. Rather than produce the wrong result, we found that such errors merely slow the convergence at low $T$ by making our algorithm more classical. Indeed, in the limit of fully depolarizing noise, our algorithm reduces to MCMC with a uniform proposal. Our simulations suggest that the quantum speedup increases with the problem size $n$. However, we also expect the quantum noise to increase with $n$, in the absence of error correction, as the number of potential errors grows. The combined effect of these competing factors at larger $n$ is currently unknown. It is interesting to note, however, that the cubic/quartic speedup we observed, should it persist at larger $n$, might give a quantum advantage on a modest fault-tolerant quantum computer despite the error correction overhead \cite{lemieux:2020, babbush:2021}.

Characterizing our algorithm at larger scales will require different methods than those employed here. For instance, a Markov chain's absolute spectral gap is a broad and unambiguous figure of merit, but it is not feasible to measure for large instances. This not an issue with our quantum algorithm in particular, but rather, with MCMC in general. Instead, a more fruitful approach may be to focus directly on how well our algorithm performs in various applications, such as in simulated annealing (for combinatorial optimization), for estimating thermal averages in many-body physics models, or for training and sampling from (classical) Boltzmann machines for machine learning applications.

\section*{Acknowledgements}

We thank Easwar Magesan and Kristan Temme for useful discussions which helped shape the project, as well as Edward Chen, Nate Earnest-Noble, Andrew Eddins and Daniel Egger for crucial help with the experiments.

\section*{Author Contributions}

D.L.\ led the theory and the experiments. G.M., R.M., M.M.\ and P.W.\ contributed to the theory; in particular, D.L.\ and G.M.\ independently proposed a variant of this algorithm which uses quantum phase estimation, described in Section V-B of \cite{SM}. J.S.K., G.M., R.M., M.M.\ and S.S.\ contributed to the design of the experiments, and S.S.\ also contributed to their implementation. D.L.\ drafted the manuscript and supplemental material; all authors contributed to revising both.

\bibliography{references_b}

\begin{thebibliography}{47}%
\makeatletter
\providecommand \@ifxundefined [1]{%
 \@ifx{#1\undefined}
}%
\providecommand \@ifnum [1]{%
 \ifnum #1\expandafter \@firstoftwo
 \else \expandafter \@secondoftwo
 \fi
}%
\providecommand \@ifx [1]{%
 \ifx #1\expandafter \@firstoftwo
 \else \expandafter \@secondoftwo
 \fi
}%
\providecommand \natexlab [1]{#1}%
\providecommand \enquote  [1]{``#1''}%
\providecommand \bibnamefont  [1]{#1}%
\providecommand \bibfnamefont [1]{#1}%
\providecommand \citenamefont [1]{#1}%
\providecommand \href@noop [0]{\@secondoftwo}%
\providecommand \href [0]{\begingroup \@sanitize@url \@href}%
\providecommand \@href[1]{\@@startlink{#1}\@@href}%
\providecommand \@@href[1]{\endgroup#1\@@endlink}%
\providecommand \@sanitize@url [0]{\catcode `\\12\catcode `\$12\catcode
  `\&12\catcode `\#12\catcode `\^12\catcode `\_12\catcode `\%12\relax}%
\providecommand \@@startlink[1]{}%
\providecommand \@@endlink[0]{}%
\providecommand \url  [0]{\begingroup\@sanitize@url \@url }%
\providecommand \@url [1]{\endgroup\@href {#1}{\urlprefix }}%
\providecommand \urlprefix  [0]{URL }%
\providecommand \Eprint [0]{\href }%
\providecommand \doibase [0]{https://doi.org/}%
\providecommand \selectlanguage [0]{\@gobble}%
\providecommand \bibinfo  [0]{\@secondoftwo}%
\providecommand \bibfield  [0]{\@secondoftwo}%
\providecommand \translation [1]{[#1]}%
\providecommand \BibitemOpen [0]{}%
\providecommand \bibitemStop [0]{}%
\providecommand \bibitemNoStop [0]{.\EOS\space}%
\providecommand \EOS [0]{\spacefactor3000\relax}%
\providecommand \BibitemShut  [1]{\csname bibitem#1\endcsname}%
\let\auto@bib@innerbib\@empty
\bibitem [{\citenamefont {Arute}\ \emph {et~al.}(2019)\citenamefont {Arute},
  \citenamefont {Arya}, \citenamefont {Babbush}, \citenamefont {Bacon},
  \citenamefont {Bardin}, \citenamefont {Barends}, \citenamefont {Biswas},
  \citenamefont {Boixo}, \citenamefont {Brandao}, \citenamefont {Buell} \emph
  {et~al.}}]{arute:2019}%
  \BibitemOpen
  \bibfield  {author} {\bibinfo {author} {\bibfnamefont {F.}~\bibnamefont
  {Arute}}, \bibinfo {author} {\bibfnamefont {K.}~\bibnamefont {Arya}},
  \bibinfo {author} {\bibfnamefont {R.}~\bibnamefont {Babbush}}, \bibinfo
  {author} {\bibfnamefont {D.}~\bibnamefont {Bacon}}, \bibinfo {author}
  {\bibfnamefont {J.~C.}\ \bibnamefont {Bardin}}, \bibinfo {author}
  {\bibfnamefont {R.}~\bibnamefont {Barends}}, \bibinfo {author} {\bibfnamefont
  {R.}~\bibnamefont {Biswas}}, \bibinfo {author} {\bibfnamefont
  {S.}~\bibnamefont {Boixo}}, \bibinfo {author} {\bibfnamefont {F.~G.}\
  \bibnamefont {Brandao}}, \bibinfo {author} {\bibfnamefont {D.~A.}\
  \bibnamefont {Buell}}, \emph {et~al.},\ }\bibfield  {title} {\bibinfo {title}
  {Quantum supremacy using a programmable superconducting processor},\ }\href
  {https://www.nature.com/articles/s41586-019-1666-5} {\bibfield  {journal}
  {\bibinfo  {journal} {Nature}\ }\textbf {\bibinfo {volume} {574}},\ \bibinfo
  {pages} {505} (\bibinfo {year} {2019})}\BibitemShut {NoStop}%
\bibitem [{\citenamefont {Wu}\ \emph {et~al.}(2021)\citenamefont {Wu} \emph
  {et~al.}}]{wu:2021}%
  \BibitemOpen
  \bibfield  {author} {\bibinfo {author} {\bibfnamefont {Y.}~\bibnamefont {Wu}}
  \emph {et~al.},\ }\bibfield  {title} {\bibinfo {title} {Strong quantum
  computational advantage using a superconducting quantum processor},\ }\href
  {https://doi.org/10.1103/PhysRevLett.127.180501} {\bibfield  {journal}
  {\bibinfo  {journal} {Phys. Rev. Lett}\ }\textbf {\bibinfo {volume} {127}},\
  \bibinfo {pages} {180501} (\bibinfo {year} {2021})}\BibitemShut {NoStop}%
\bibitem [{\citenamefont {Zhong}\ \emph {et~al.}(2021)\citenamefont {Zhong}
  \emph {et~al.}}]{zhong:2021}%
  \BibitemOpen
  \bibfield  {author} {\bibinfo {author} {\bibfnamefont {H.-S.}\ \bibnamefont
  {Zhong}} \emph {et~al.},\ }\bibfield  {title} {\bibinfo {title}
  {Phase-programmable {G}aussian boson sampling using stimulated squeezed
  light},\ }\href {https://doi.org/10.1103/PhysRevLett.127.180502} {\bibfield
  {journal} {\bibinfo  {journal} {Phys. Rev. Lett}\ }\textbf {\bibinfo {volume}
  {127}},\ \bibinfo {pages} {180502} (\bibinfo {year} {2021})}\BibitemShut
  {NoStop}%
\bibitem [{\citenamefont {Ising}(1925)}]{ising:1925}%
  \BibitemOpen
  \bibfield  {author} {\bibinfo {author} {\bibfnamefont {E.}~\bibnamefont
  {Ising}},\ }\bibfield  {title} {\bibinfo {title} {Beitrag zur {T}heorie des
  {F}erromagnetismus},\ }\href
  {https://link.springer.com/article/10.1007/BF02980577} {\bibfield  {journal}
  {\bibinfo  {journal} {Z. Phys}\ }\textbf {\bibinfo {volume} {31}},\ \bibinfo
  {pages} {253} (\bibinfo {year} {1925})}\BibitemShut {NoStop}%
\bibitem [{\citenamefont {Huang}(2008)}]{huang:2008}%
  \BibitemOpen
  \bibfield  {author} {\bibinfo {author} {\bibfnamefont {K.}~\bibnamefont
  {Huang}},\ }\href@noop {} {\emph {\bibinfo {title} {Statistical mechanics}}}\
  (\bibinfo  {publisher} {John Wiley \& Sons},\ \bibinfo {year}
  {2008})\BibitemShut {NoStop}%
\bibitem [{\citenamefont {Ackley}\ \emph {et~al.}(1985)\citenamefont {Ackley},
  \citenamefont {Hinton},\ and\ \citenamefont {Sejnowski}}]{ackley:1985}%
  \BibitemOpen
  \bibfield  {author} {\bibinfo {author} {\bibfnamefont {D.~H.}\ \bibnamefont
  {Ackley}}, \bibinfo {author} {\bibfnamefont {G.~E.}\ \bibnamefont {Hinton}},\
  and\ \bibinfo {author} {\bibfnamefont {T.~J.}\ \bibnamefont {Sejnowski}},\
  }\bibfield  {title} {\bibinfo {title} {A learning algorithm for {B}oltzmann
  machines},\ }\href
  {https://www.sciencedirect.com/science/article/abs/pii/S0364021385800124}
  {\bibfield  {journal} {\bibinfo  {journal} {Cogn. Sci}\ }\textbf {\bibinfo
  {volume} {9}},\ \bibinfo {pages} {147} (\bibinfo {year} {1985})}\BibitemShut
  {NoStop}%
\bibitem [{\citenamefont {Kirkpatrick}\ \emph {et~al.}(1983)\citenamefont
  {Kirkpatrick}, \citenamefont {Gelatt},\ and\ \citenamefont
  {Vecchi}}]{kirkpatrick:1983}%
  \BibitemOpen
  \bibfield  {author} {\bibinfo {author} {\bibfnamefont {S.}~\bibnamefont
  {Kirkpatrick}}, \bibinfo {author} {\bibfnamefont {C.~D.}\ \bibnamefont
  {Gelatt}},\ and\ \bibinfo {author} {\bibfnamefont {M.~P.}\ \bibnamefont
  {Vecchi}},\ }\bibfield  {title} {\bibinfo {title} {Optimization by simulated
  annealing},\ }\href
  {https://www.science.org/doi/10.1126/science.220.4598.671} {\bibfield
  {journal} {\bibinfo  {journal} {Science}\ }\textbf {\bibinfo {volume}
  {220}},\ \bibinfo {pages} {671} (\bibinfo {year} {1983})}\BibitemShut
  {NoStop}%
\bibitem [{\citenamefont {Lucas}(2014)}]{lucas:2014}%
  \BibitemOpen
  \bibfield  {author} {\bibinfo {author} {\bibfnamefont {A.}~\bibnamefont
  {Lucas}},\ }\bibfield  {title} {\bibinfo {title} {Ising formulations of many
  {NP} problems},\ }\href
  {https://www.frontiersin.org/article/10.3389/fphy.2014.00005} {\bibfield
  {journal} {\bibinfo  {journal} {Front. Phys}\ }\textbf {\bibinfo {volume}
  {2}} (\bibinfo {year} {2014})}\BibitemShut {NoStop}%
\bibitem [{\citenamefont {Barahona}(1982)}]{barahona:1982}%
  \BibitemOpen
  \bibfield  {author} {\bibinfo {author} {\bibfnamefont {F.}~\bibnamefont
  {Barahona}},\ }\bibfield  {title} {\bibinfo {title} {On the computational
  complexity of {I}sing spin glass models},\ }\href
  {https://iopscience.iop.org/article/10.1088/0305-4470/15/10/028} {\bibfield
  {journal} {\bibinfo  {journal} {J. Phys. A}\ }\textbf {\bibinfo {volume}
  {15}},\ \bibinfo {pages} {3241} (\bibinfo {year} {1982})}\BibitemShut
  {NoStop}%
\bibitem [{\citenamefont {Dongarra}\ and\ \citenamefont
  {Sullivan}(2000)}]{dongarra:2000}%
  \BibitemOpen
  \bibfield  {author} {\bibinfo {author} {\bibfnamefont {J.}~\bibnamefont
  {Dongarra}}\ and\ \bibinfo {author} {\bibfnamefont {F.}~\bibnamefont
  {Sullivan}},\ }\bibfield  {title} {\bibinfo {title} {Guest editors'
  introduction to the top 10 algorithms},\ }\href
  {https://www.computer.org/csdl/magazine/cs/2000/01/c1022/13rRUxBJhBm}
  {\bibfield  {journal} {\bibinfo  {journal} {Comput. Sci. Eng}\ }\textbf
  {\bibinfo {volume} {2}},\ \bibinfo {pages} {22} (\bibinfo {year}
  {2000})}\BibitemShut {NoStop}%
\bibitem [{SM()}]{SM}%
  \BibitemOpen
  \href@noop {} {}\bibinfo {note} {Supplemental Material.}\BibitemShut {Stop}%
\bibitem [{\citenamefont {Levin}\ and\ \citenamefont
  {Peres}(2017)}]{levin:2017}%
  \BibitemOpen
  \bibfield  {author} {\bibinfo {author} {\bibfnamefont {D.}~\bibnamefont
  {Levin}}\ and\ \bibinfo {author} {\bibfnamefont {Y.}~\bibnamefont {Peres}},\
  }\href@noop {} {\emph {\bibinfo {title} {Markov Chains and Mixing Times}}},\
  MBK\ (\bibinfo  {publisher} {American Mathematical Society},\ \bibinfo {year}
  {2017})\BibitemShut {NoStop}%
\bibitem [{\citenamefont {Metropolis}\ \emph {et~al.}(1953)\citenamefont
  {Metropolis}, \citenamefont {Rosenbluth}, \citenamefont {Rosenbluth},
  \citenamefont {Teller},\ and\ \citenamefont {Teller}}]{metropolis:1953}%
  \BibitemOpen
  \bibfield  {author} {\bibinfo {author} {\bibfnamefont {N.}~\bibnamefont
  {Metropolis}}, \bibinfo {author} {\bibfnamefont {A.~W.}\ \bibnamefont
  {Rosenbluth}}, \bibinfo {author} {\bibfnamefont {M.~N.}\ \bibnamefont
  {Rosenbluth}}, \bibinfo {author} {\bibfnamefont {A.~H.}\ \bibnamefont
  {Teller}},\ and\ \bibinfo {author} {\bibfnamefont {E.}~\bibnamefont
  {Teller}},\ }\bibfield  {title} {\bibinfo {title} {Equation of state
  calculations by fast computing machines},\ }\href
  {https://aip.scitation.org/doi/10.1063/1.1699114} {\bibfield  {journal}
  {\bibinfo  {journal} {J. Comp. Phys.}\ }\textbf {\bibinfo {volume} {21}},\
  \bibinfo {pages} {1087} (\bibinfo {year} {1953})}\BibitemShut {NoStop}%
\bibitem [{\citenamefont {Hastings}(1970)}]{hastings:1970}%
  \BibitemOpen
  \bibfield  {author} {\bibinfo {author} {\bibfnamefont {W.~K.}\ \bibnamefont
  {Hastings}},\ }\bibfield  {title} {\bibinfo {title} {{Monte {C}arlo sampling
  methods using {M}arkov chains and their applications}},\ }\href
  {https://doi.org/10.1093/biomet/57.1.97} {\bibfield  {journal} {\bibinfo
  {journal} {Biometrika}\ }\textbf {\bibinfo {volume} {57}},\ \bibinfo {pages}
  {97} (\bibinfo {year} {1970})}\BibitemShut {NoStop}%
\bibitem [{\citenamefont {Andrieu}\ \emph {et~al.}(2003)\citenamefont
  {Andrieu}, \citenamefont {De~Freitas}, \citenamefont {Doucet},\ and\
  \citenamefont {Jordan}}]{andrieu:2003}%
  \BibitemOpen
  \bibfield  {author} {\bibinfo {author} {\bibfnamefont {C.}~\bibnamefont
  {Andrieu}}, \bibinfo {author} {\bibfnamefont {N.}~\bibnamefont {De~Freitas}},
  \bibinfo {author} {\bibfnamefont {A.}~\bibnamefont {Doucet}},\ and\ \bibinfo
  {author} {\bibfnamefont {M.~I.}\ \bibnamefont {Jordan}},\ }\bibfield  {title}
  {\bibinfo {title} {An introduction to {MCMC} for machine learning},\ }\href
  {https://link.springer.com/article/10.1023/A:1020281327116} {\bibfield
  {journal} {\bibinfo  {journal} {Mach. Learn}\ }\textbf {\bibinfo {volume}
  {50}},\ \bibinfo {pages} {5} (\bibinfo {year} {2003})}\BibitemShut {NoStop}%
\bibitem [{\citenamefont {Swendsen}\ and\ \citenamefont
  {Wang}(1987)}]{swendsen:1987}%
  \BibitemOpen
  \bibfield  {author} {\bibinfo {author} {\bibfnamefont {R.~H.}\ \bibnamefont
  {Swendsen}}\ and\ \bibinfo {author} {\bibfnamefont {J.-S.}\ \bibnamefont
  {Wang}},\ }\bibfield  {title} {\bibinfo {title} {Nonuniversal critical
  dynamics in {M}onte {C}arlo simulations},\ }\href
  {https://doi.org/10.1103/PhysRevLett.58.86} {\bibfield  {journal} {\bibinfo
  {journal} {Phys. Rev. Lett}\ }\textbf {\bibinfo {volume} {58}},\ \bibinfo
  {pages} {86} (\bibinfo {year} {1987})}\BibitemShut {NoStop}%
\bibitem [{\citenamefont {Wolff}(1989)}]{wolff:1989}%
  \BibitemOpen
  \bibfield  {author} {\bibinfo {author} {\bibfnamefont {U.}~\bibnamefont
  {Wolff}},\ }\bibfield  {title} {\bibinfo {title} {Collective {M}onte {C}arlo
  updating for spin systems},\ }\href
  {https://doi.org/10.1103/PhysRevLett.62.361} {\bibfield  {journal} {\bibinfo
  {journal} {Phys. Rev. Lett}\ }\textbf {\bibinfo {volume} {62}},\ \bibinfo
  {pages} {361} (\bibinfo {year} {1989})}\BibitemShut {NoStop}%
\bibitem [{\citenamefont {Houdayer}(2001)}]{houdayer:2001}%
  \BibitemOpen
  \bibfield  {author} {\bibinfo {author} {\bibfnamefont {J.}~\bibnamefont
  {Houdayer}},\ }\bibfield  {title} {\bibinfo {title} {A cluster {M}onte
  {C}arlo algorithm for 2-dimensional spin glasses},\ }\href
  {https://link.springer.com/article/10.1007/PL00011151} {\bibfield  {journal}
  {\bibinfo  {journal} {Eur. Phys. J. B}\ }\textbf {\bibinfo {volume} {22}},\
  \bibinfo {pages} {479} (\bibinfo {year} {2001})}\BibitemShut {NoStop}%
\bibitem [{\citenamefont {Zhu}\ \emph {et~al.}(2015)\citenamefont {Zhu},
  \citenamefont {Ochoa},\ and\ \citenamefont {Katzgraber}}]{zhu:2015}%
  \BibitemOpen
  \bibfield  {author} {\bibinfo {author} {\bibfnamefont {Z.}~\bibnamefont
  {Zhu}}, \bibinfo {author} {\bibfnamefont {A.~J.}\ \bibnamefont {Ochoa}},\
  and\ \bibinfo {author} {\bibfnamefont {H.~G.}\ \bibnamefont {Katzgraber}},\
  }\bibfield  {title} {\bibinfo {title} {Efficient cluster algorithm for spin
  glasses in any space dimension},\ }\href
  {https://doi.org/10.1103/PhysRevLett.115.077201} {\bibfield  {journal}
  {\bibinfo  {journal} {Phys. Rev. Lett}\ }\textbf {\bibinfo {volume} {115}},\
  \bibinfo {pages} {077201} (\bibinfo {year} {2015})}\BibitemShut {NoStop}%
\bibitem [{\citenamefont {Goodfellow}\ \emph {et~al.}(2016)\citenamefont
  {Goodfellow}, \citenamefont {Bengio},\ and\ \citenamefont
  {Courville}}]{goodfellow:2016}%
  \BibitemOpen
  \bibfield  {author} {\bibinfo {author} {\bibfnamefont {I.}~\bibnamefont
  {Goodfellow}}, \bibinfo {author} {\bibfnamefont {Y.}~\bibnamefont {Bengio}},\
  and\ \bibinfo {author} {\bibfnamefont {A.}~\bibnamefont {Courville}},\
  }\href@noop {} {\emph {\bibinfo {title} {Deep Learning}}}\ (\bibinfo
  {publisher} {MIT Press},\ \bibinfo {year} {2016})\BibitemShut {NoStop}%
\bibitem [{\citenamefont {Baldwin}\ and\ \citenamefont
  {Laumann}(2018)}]{baldwin:2018}%
  \BibitemOpen
  \bibfield  {author} {\bibinfo {author} {\bibfnamefont {C.~L.}\ \bibnamefont
  {Baldwin}}\ and\ \bibinfo {author} {\bibfnamefont {C.~R.}\ \bibnamefont
  {Laumann}},\ }\bibfield  {title} {\bibinfo {title} {Quantum algorithm for
  energy matching in hard optimization problems},\ }\href
  {https://doi.org/10.1103/PhysRevB.97.224201} {\bibfield  {journal} {\bibinfo
  {journal} {Phys. Rev. B}\ }\textbf {\bibinfo {volume} {97}},\ \bibinfo
  {pages} {224201} (\bibinfo {year} {2018})}\BibitemShut {NoStop}%
\bibitem [{\citenamefont {Smelyanskiy}\ \emph {et~al.}(2020)\citenamefont
  {Smelyanskiy}, \citenamefont {Kechedzhi}, \citenamefont {Boixo},
  \citenamefont {Isakov}, \citenamefont {Neven},\ and\ \citenamefont
  {Altshuler}}]{smelyanskiy:2020}%
  \BibitemOpen
  \bibfield  {author} {\bibinfo {author} {\bibfnamefont {V.~N.}\ \bibnamefont
  {Smelyanskiy}}, \bibinfo {author} {\bibfnamefont {K.}~\bibnamefont
  {Kechedzhi}}, \bibinfo {author} {\bibfnamefont {S.}~\bibnamefont {Boixo}},
  \bibinfo {author} {\bibfnamefont {S.~V.}\ \bibnamefont {Isakov}}, \bibinfo
  {author} {\bibfnamefont {H.}~\bibnamefont {Neven}},\ and\ \bibinfo {author}
  {\bibfnamefont {B.}~\bibnamefont {Altshuler}},\ }\bibfield  {title} {\bibinfo
  {title} {Nonergodic delocalized states for efficient population transfer
  within a narrow band of the energy landscape},\ }\href
  {https://doi.org/10.1103/PhysRevX.10.011017} {\bibfield  {journal} {\bibinfo
  {journal} {Phys. Rev. X}\ }\textbf {\bibinfo {volume} {10}},\ \bibinfo
  {pages} {011017} (\bibinfo {year} {2020})}\BibitemShut {NoStop}%
\bibitem [{\citenamefont {Smelyanskiy}\ \emph {et~al.}(2019)\citenamefont
  {Smelyanskiy}, \citenamefont {Kechedzhi}, \citenamefont {Boixo},
  \citenamefont {Neven},\ and\ \citenamefont {Altshuler}}]{smelyanskiy:2019}%
  \BibitemOpen
  \bibfield  {author} {\bibinfo {author} {\bibfnamefont {V.~N.}\ \bibnamefont
  {Smelyanskiy}}, \bibinfo {author} {\bibfnamefont {K.}~\bibnamefont
  {Kechedzhi}}, \bibinfo {author} {\bibfnamefont {S.}~\bibnamefont {Boixo}},
  \bibinfo {author} {\bibfnamefont {H.}~\bibnamefont {Neven}},\ and\ \bibinfo
  {author} {\bibfnamefont {B.}~\bibnamefont {Altshuler}},\ }\bibfield  {title}
  {\bibinfo {title} {Intermittency of dynamical phases in a quantum spin
  glass},\ }\href {https://arxiv.org/abs/1907.01609} {\bibfield  {journal}
  {\bibinfo  {journal} {arXiv:1907.01609}\ } (\bibinfo {year}
  {2019})}\BibitemShut {NoStop}%
\bibitem [{\citenamefont {Lloyd}(1996)}]{lloyd:1996}%
  \BibitemOpen
  \bibfield  {author} {\bibinfo {author} {\bibfnamefont {S.}~\bibnamefont
  {Lloyd}},\ }\bibfield  {title} {\bibinfo {title} {Universal quantum
  simulators},\ }\href
  {https://www.science.org/doi/10.1126/science.273.5278.1073} {\bibfield
  {journal} {\bibinfo  {journal} {Science}\ }\textbf {\bibinfo {volume}
  {273}},\ \bibinfo {pages} {1073} (\bibinfo {year} {1996})}\BibitemShut
  {NoStop}%
\bibitem [{\citenamefont {Temme}\ \emph {et~al.}(2011)\citenamefont {Temme},
  \citenamefont {Osborne}, \citenamefont {Vollbrecht}, \citenamefont {Poulin},\
  and\ \citenamefont {Verstraete}}]{temme:2011}%
  \BibitemOpen
  \bibfield  {author} {\bibinfo {author} {\bibfnamefont {K.}~\bibnamefont
  {Temme}}, \bibinfo {author} {\bibfnamefont {T.~J.}\ \bibnamefont {Osborne}},
  \bibinfo {author} {\bibfnamefont {K.~G.}\ \bibnamefont {Vollbrecht}},
  \bibinfo {author} {\bibfnamefont {D.}~\bibnamefont {Poulin}},\ and\ \bibinfo
  {author} {\bibfnamefont {F.}~\bibnamefont {Verstraete}},\ }\bibfield  {title}
  {\bibinfo {title} {Quantum {M}etropolis sampling},\ }\href
  {https://www.nature.com/articles/nature09770} {\bibfield  {journal} {\bibinfo
   {journal} {Nature}\ }\textbf {\bibinfo {volume} {471}},\ \bibinfo {pages}
  {87} (\bibinfo {year} {2011})}\BibitemShut {NoStop}%
\bibitem [{\citenamefont {Yung}\ and\ \citenamefont
  {Aspuru-Guzik}(2012)}]{yung:2012}%
  \BibitemOpen
  \bibfield  {author} {\bibinfo {author} {\bibfnamefont {M.-H.}\ \bibnamefont
  {Yung}}\ and\ \bibinfo {author} {\bibfnamefont {A.}~\bibnamefont
  {Aspuru-Guzik}},\ }\bibfield  {title} {\bibinfo {title} {A
  quantum{\textendash}quantum {M}etropolis algorithm},\ }\href
  {https://doi.org/10.1073/pnas.1111758109} {\bibfield  {journal} {\bibinfo
  {journal} {Proc. Natl. Acad. Sci}\ }\textbf {\bibinfo {volume} {109}},\
  \bibinfo {pages} {754} (\bibinfo {year} {2012})}\BibitemShut {NoStop}%
\bibitem [{\citenamefont {Moussa}(2019)}]{moussa:2019}%
  \BibitemOpen
  \bibfield  {author} {\bibinfo {author} {\bibfnamefont {J.~E.}\ \bibnamefont
  {Moussa}},\ }\bibfield  {title} {\bibinfo {title} {Measurement-based quantum
  {M}etropolis algorithm},\ }\href {https://arxiv.org/abs/1903.01451}
  {\bibfield  {journal} {\bibinfo  {journal} {arXiv:1903.01451}\ } (\bibinfo
  {year} {2019})}\BibitemShut {NoStop}%
\bibitem [{\citenamefont {Wocjan}\ and\ \citenamefont
  {Temme}(2021)}]{wocjan:2021}%
  \BibitemOpen
  \bibfield  {author} {\bibinfo {author} {\bibfnamefont {P.}~\bibnamefont
  {Wocjan}}\ and\ \bibinfo {author} {\bibfnamefont {K.}~\bibnamefont {Temme}},\
  }\bibfield  {title} {\bibinfo {title} {Szegedy walk unitaries for quantum
  maps},\ }\href {https://arxiv.org/abs/2107.07365} {\bibfield  {journal}
  {\bibinfo  {journal} {arXiv preprint arXiv:2107.07365}\ } (\bibinfo {year}
  {2021})}\BibitemShut {NoStop}%
\bibitem [{\citenamefont {Szegedy}(2004)}]{szegedy:2004}%
  \BibitemOpen
  \bibfield  {author} {\bibinfo {author} {\bibfnamefont {M.}~\bibnamefont
  {Szegedy}},\ }\bibfield  {title} {\bibinfo {title} {Quantum speed-up of
  {M}arkov chain based algorithms},\ }in\ \href
  {https://doi.org/10.1109/FOCS.2004.53} {\emph {\bibinfo {booktitle} {45th
  Annual IEEE Symposium on Foundations of Computer Science}}}\ (\bibinfo {year}
  {2004})\ pp.\ \bibinfo {pages} {32--41}\BibitemShut {NoStop}%
\bibitem [{\citenamefont {Richter}(2007)}]{richter:2007}%
  \BibitemOpen
  \bibfield  {author} {\bibinfo {author} {\bibfnamefont {P.~C.}\ \bibnamefont
  {Richter}},\ }\bibfield  {title} {\bibinfo {title} {Quantum speedup of
  classical mixing processes},\ }\href
  {https://doi.org/10.1103/PhysRevA.76.042306} {\bibfield  {journal} {\bibinfo
  {journal} {Phys. Rev. A}\ }\textbf {\bibinfo {volume} {76}},\ \bibinfo
  {pages} {042306} (\bibinfo {year} {2007})}\BibitemShut {NoStop}%
\bibitem [{\citenamefont {Somma}\ \emph {et~al.}(2008)\citenamefont {Somma},
  \citenamefont {Boixo}, \citenamefont {Barnum},\ and\ \citenamefont
  {Knill}}]{somma:2008}%
  \BibitemOpen
  \bibfield  {author} {\bibinfo {author} {\bibfnamefont {R.~D.}\ \bibnamefont
  {Somma}}, \bibinfo {author} {\bibfnamefont {S.}~\bibnamefont {Boixo}},
  \bibinfo {author} {\bibfnamefont {H.}~\bibnamefont {Barnum}},\ and\ \bibinfo
  {author} {\bibfnamefont {E.}~\bibnamefont {Knill}},\ }\bibfield  {title}
  {\bibinfo {title} {Quantum simulations of classical annealing processes},\
  }\href {https://doi.org/10.1103/PhysRevLett.101.130504} {\bibfield  {journal}
  {\bibinfo  {journal} {Phys. Rev. Lett.}\ }\textbf {\bibinfo {volume} {101}},\
  \bibinfo {pages} {130504} (\bibinfo {year} {2008})}\BibitemShut {NoStop}%
\bibitem [{\citenamefont {Wocjan}\ and\ \citenamefont
  {Abeyesinghe}(2008)}]{wocjan:2008}%
  \BibitemOpen
  \bibfield  {author} {\bibinfo {author} {\bibfnamefont {P.}~\bibnamefont
  {Wocjan}}\ and\ \bibinfo {author} {\bibfnamefont {A.}~\bibnamefont
  {Abeyesinghe}},\ }\bibfield  {title} {\bibinfo {title} {Speedup via quantum
  sampling},\ }\href {https://doi.org/10.1103/PhysRevA.78.042336} {\bibfield
  {journal} {\bibinfo  {journal} {Phys. Rev. A}\ }\textbf {\bibinfo {volume}
  {78}},\ \bibinfo {pages} {042336} (\bibinfo {year} {2008})}\BibitemShut
  {NoStop}%
\bibitem [{\citenamefont {Harrow}\ and\ \citenamefont
  {Wei}(2020)}]{harrow:2020}%
  \BibitemOpen
  \bibfield  {author} {\bibinfo {author} {\bibfnamefont {A.~W.}\ \bibnamefont
  {Harrow}}\ and\ \bibinfo {author} {\bibfnamefont {A.~Y.}\ \bibnamefont
  {Wei}},\ }\bibfield  {title} {\bibinfo {title} {Adaptive quantum simulated
  annealing for {B}ayesian inference and estimating partition functions},\ }in\
  \href {https://epubs.siam.org/doi/pdf/10.1137/1.9781611975994.12} {\emph
  {\bibinfo {booktitle} {Proceedings of the Fourteenth Annual ACM-SIAM
  Symposium on Discrete Algorithms}}}\ (\bibinfo {organization} {SIAM},\
  \bibinfo {year} {2020})\ pp.\ \bibinfo {pages} {193--212}\BibitemShut
  {NoStop}%
\bibitem [{\citenamefont {Lemieux}\ \emph {et~al.}(2020)\citenamefont
  {Lemieux}, \citenamefont {Heim}, \citenamefont {Poulin}, \citenamefont
  {Svore},\ and\ \citenamefont {Troyer}}]{lemieux:2020}%
  \BibitemOpen
  \bibfield  {author} {\bibinfo {author} {\bibfnamefont {J.}~\bibnamefont
  {Lemieux}}, \bibinfo {author} {\bibfnamefont {B.}~\bibnamefont {Heim}},
  \bibinfo {author} {\bibfnamefont {D.}~\bibnamefont {Poulin}}, \bibinfo
  {author} {\bibfnamefont {K.}~\bibnamefont {Svore}},\ and\ \bibinfo {author}
  {\bibfnamefont {M.}~\bibnamefont {Troyer}},\ }\bibfield  {title} {\bibinfo
  {title} {Efficient quantum walk circuits for {M}etropolis-{H}astings
  algorithm},\ }\href {https://doi.org/10.22331/q-2020-06-29-287} {\bibfield
  {journal} {\bibinfo  {journal} {{Quantum}}\ }\textbf {\bibinfo {volume}
  {4}},\ \bibinfo {pages} {287} (\bibinfo {year} {2020})}\BibitemShut {NoStop}%
\bibitem [{\citenamefont {Arunachalam}\ \emph {et~al.}(2021)\citenamefont
  {Arunachalam}, \citenamefont {Havlicek}, \citenamefont {Nannicini},
  \citenamefont {Temme},\ and\ \citenamefont {Wocjan}}]{arunachalam:2021}%
  \BibitemOpen
  \bibfield  {author} {\bibinfo {author} {\bibfnamefont {S.}~\bibnamefont
  {Arunachalam}}, \bibinfo {author} {\bibfnamefont {V.}~\bibnamefont
  {Havlicek}}, \bibinfo {author} {\bibfnamefont {G.}~\bibnamefont {Nannicini}},
  \bibinfo {author} {\bibfnamefont {K.}~\bibnamefont {Temme}},\ and\ \bibinfo
  {author} {\bibfnamefont {P.}~\bibnamefont {Wocjan}},\ }\bibfield  {title}
  {\bibinfo {title} {Simpler (classical) and faster (quantum) algorithms for
  {G}ibbs partition functions},\ }in\ \href
  {https://ieeexplore.ieee.org/document/9605305/} {\emph {\bibinfo {booktitle}
  {2021 IEEE International Conference on Quantum Computing and Engineering
  (QCE)}}}\ (\bibinfo {organization} {IEEE},\ \bibinfo {year} {2021})\ pp.\
  \bibinfo {pages} {112--122}\BibitemShut {NoStop}%
\bibitem [{\citenamefont {Wild}\ \emph
  {et~al.}(2021{\natexlab{a}})\citenamefont {Wild}, \citenamefont {Sels},
  \citenamefont {Pichler}, \citenamefont {Zanoci},\ and\ \citenamefont
  {Lukin}}]{wild:2021prl}%
  \BibitemOpen
  \bibfield  {author} {\bibinfo {author} {\bibfnamefont {D.~S.}\ \bibnamefont
  {Wild}}, \bibinfo {author} {\bibfnamefont {D.}~\bibnamefont {Sels}}, \bibinfo
  {author} {\bibfnamefont {H.}~\bibnamefont {Pichler}}, \bibinfo {author}
  {\bibfnamefont {C.}~\bibnamefont {Zanoci}},\ and\ \bibinfo {author}
  {\bibfnamefont {M.~D.}\ \bibnamefont {Lukin}},\ }\bibfield  {title} {\bibinfo
  {title} {Quantum sampling algorithms for near-term devices},\ }\href
  {https://doi.org/10.1103/PhysRevLett.127.100504} {\bibfield  {journal}
  {\bibinfo  {journal} {Phys. Rev. Lett.}\ }\textbf {\bibinfo {volume} {127}},\
  \bibinfo {pages} {100504} (\bibinfo {year} {2021}{\natexlab{a}})}\BibitemShut
  {NoStop}%
\bibitem [{\citenamefont {Wild}\ \emph
  {et~al.}(2021{\natexlab{b}})\citenamefont {Wild}, \citenamefont {Sels},
  \citenamefont {Pichler}, \citenamefont {Zanoci},\ and\ \citenamefont
  {Lukin}}]{wild:2021pra}%
  \BibitemOpen
  \bibfield  {author} {\bibinfo {author} {\bibfnamefont {D.~S.}\ \bibnamefont
  {Wild}}, \bibinfo {author} {\bibfnamefont {D.}~\bibnamefont {Sels}}, \bibinfo
  {author} {\bibfnamefont {H.}~\bibnamefont {Pichler}}, \bibinfo {author}
  {\bibfnamefont {C.}~\bibnamefont {Zanoci}},\ and\ \bibinfo {author}
  {\bibfnamefont {M.~D.}\ \bibnamefont {Lukin}},\ }\bibfield  {title} {\bibinfo
  {title} {Quantum sampling algorithms, phase transitions, and computational
  complexity},\ }\href {https://doi.org/10.1103/PhysRevA.104.032602} {\bibfield
   {journal} {\bibinfo  {journal} {Phys. Rev. A}\ }\textbf {\bibinfo {volume}
  {104}},\ \bibinfo {pages} {032602} (\bibinfo {year}
  {2021}{\natexlab{b}})}\BibitemShut {NoStop}%
\bibitem [{\citenamefont {Sherrington}\ and\ \citenamefont
  {Kirkpatrick}(1975)}]{sherrington:1975}%
  \BibitemOpen
  \bibfield  {author} {\bibinfo {author} {\bibfnamefont {D.}~\bibnamefont
  {Sherrington}}\ and\ \bibinfo {author} {\bibfnamefont {S.}~\bibnamefont
  {Kirkpatrick}},\ }\bibfield  {title} {\bibinfo {title} {Solvable model of a
  spin-glass},\ }\href {https://doi.org/10.1103/PhysRevLett.35.1792} {\bibfield
   {journal} {\bibinfo  {journal} {Phys. Rev. Lett}\ }\textbf {\bibinfo
  {volume} {35}},\ \bibinfo {pages} {1792} (\bibinfo {year}
  {1975})}\BibitemShut {NoStop}%
\bibitem [{\citenamefont {Anis~Sajid}\ \emph {et~al.}(2021)\citenamefont
  {Anis~Sajid} \emph {et~al.}}]{qiskit}%
  \BibitemOpen
  \bibfield  {author} {\bibinfo {author} {\bibfnamefont {M.}~\bibnamefont
  {Anis~Sajid}} \emph {et~al.},\ }\href
  {https://doi.org/10.5281/zenodo.2573505} {\bibinfo {title} {Qiskit: An
  open-source framework for quantum computing}} (\bibinfo {year}
  {2021})\BibitemShut {NoStop}%
\bibitem [{\citenamefont {Suzuki}(1985)}]{suzuki:1985}%
  \BibitemOpen
  \bibfield  {author} {\bibinfo {author} {\bibfnamefont {M.}~\bibnamefont
  {Suzuki}},\ }\bibfield  {title} {\bibinfo {title} {Decomposition formulas of
  exponential operators and {L}ie exponentials with some applications to
  quantum mechanics and statistical physics},\ }\href
  {https://doi.org/10.1063/1.526596} {\bibfield  {journal} {\bibinfo  {journal}
  {J. Math. Phys}\ }\textbf {\bibinfo {volume} {26}},\ \bibinfo {pages} {601}
  (\bibinfo {year} {1985})}\BibitemShut {NoStop}%
\bibitem [{\citenamefont {Wallman}\ and\ \citenamefont
  {Emerson}(2016)}]{wallman:2016}%
  \BibitemOpen
  \bibfield  {author} {\bibinfo {author} {\bibfnamefont {J.~J.}\ \bibnamefont
  {Wallman}}\ and\ \bibinfo {author} {\bibfnamefont {J.}~\bibnamefont
  {Emerson}},\ }\bibfield  {title} {\bibinfo {title} {Noise tailoring for
  scalable quantum computation via randomized compiling},\ }\href
  {https://doi.org/10.1103/PhysRevA.94.052325} {\bibfield  {journal} {\bibinfo
  {journal} {Phys. Rev. A}\ }\textbf {\bibinfo {volume} {94}},\ \bibinfo
  {pages} {052325} (\bibinfo {year} {2016})}\BibitemShut {NoStop}%
\bibitem [{\citenamefont {Earnest}\ \emph {et~al.}(2021)\citenamefont
  {Earnest}, \citenamefont {Tornow},\ and\ \citenamefont
  {Egger}}]{earnest:2021}%
  \BibitemOpen
  \bibfield  {author} {\bibinfo {author} {\bibfnamefont {N.}~\bibnamefont
  {Earnest}}, \bibinfo {author} {\bibfnamefont {C.}~\bibnamefont {Tornow}},\
  and\ \bibinfo {author} {\bibfnamefont {D.~J.}\ \bibnamefont {Egger}},\
  }\bibfield  {title} {\bibinfo {title} {Pulse-efficient circuit transpilation
  for quantum applications on cross-resonance-based hardware},\ }\href
  {https://doi.org/10.1103/PhysRevResearch.3.043088} {\bibfield  {journal}
  {\bibinfo  {journal} {Phys. Rev. Research}\ }\textbf {\bibinfo {volume}
  {3}},\ \bibinfo {pages} {043088} (\bibinfo {year} {2021})}\BibitemShut
  {NoStop}%
\bibitem [{\citenamefont {Ambegaokar}\ and\ \citenamefont
  {Troyer}(2010)}]{ambegaokar:2010}%
  \BibitemOpen
  \bibfield  {author} {\bibinfo {author} {\bibfnamefont {V.}~\bibnamefont
  {Ambegaokar}}\ and\ \bibinfo {author} {\bibfnamefont {M.}~\bibnamefont
  {Troyer}},\ }\bibfield  {title} {\bibinfo {title} {Estimating errors reliably
  in {M}onte {C}arlo simulations of the {E}hrenfest model},\ }\href
  {https://doi.org/10.1119/1.3247985} {\bibfield  {journal} {\bibinfo
  {journal} {Am. J. Phys}\ }\textbf {\bibinfo {volume} {78}},\ \bibinfo {pages}
  {150} (\bibinfo {year} {2010})}\BibitemShut {NoStop}%
\bibitem [{\citenamefont {Cerezo}\ \emph {et~al.}(2021)\citenamefont {Cerezo},
  \citenamefont {Arrasmith}, \citenamefont {Babbush}, \citenamefont {Benjamin},
  \citenamefont {Endo}, \citenamefont {Fujii}, \citenamefont {McClean},
  \citenamefont {Mitarai}, \citenamefont {Yuan}, \citenamefont {Cincio} \emph
  {et~al.}}]{cerezo:2021}%
  \BibitemOpen
  \bibfield  {author} {\bibinfo {author} {\bibfnamefont {M.}~\bibnamefont
  {Cerezo}}, \bibinfo {author} {\bibfnamefont {A.}~\bibnamefont {Arrasmith}},
  \bibinfo {author} {\bibfnamefont {R.}~\bibnamefont {Babbush}}, \bibinfo
  {author} {\bibfnamefont {S.~C.}\ \bibnamefont {Benjamin}}, \bibinfo {author}
  {\bibfnamefont {S.}~\bibnamefont {Endo}}, \bibinfo {author} {\bibfnamefont
  {K.}~\bibnamefont {Fujii}}, \bibinfo {author} {\bibfnamefont {J.~R.}\
  \bibnamefont {McClean}}, \bibinfo {author} {\bibfnamefont {K.}~\bibnamefont
  {Mitarai}}, \bibinfo {author} {\bibfnamefont {X.}~\bibnamefont {Yuan}},
  \bibinfo {author} {\bibfnamefont {L.}~\bibnamefont {Cincio}}, \emph
  {et~al.},\ }\bibfield  {title} {\bibinfo {title} {Variational quantum
  algorithms},\ }\href {https://www.nature.com/articles/s42254-021-00348-9}
  {\bibfield  {journal} {\bibinfo  {journal} {Nat. Rev. Phys}\ }\textbf
  {\bibinfo {volume} {3}},\ \bibinfo {pages} {625} (\bibinfo {year}
  {2021})}\BibitemShut {NoStop}%
\bibitem [{\citenamefont {Swendsen}\ and\ \citenamefont
  {Wang}(1986)}]{swendsen:1986}%
  \BibitemOpen
  \bibfield  {author} {\bibinfo {author} {\bibfnamefont {R.~H.}\ \bibnamefont
  {Swendsen}}\ and\ \bibinfo {author} {\bibfnamefont {J.-S.}\ \bibnamefont
  {Wang}},\ }\bibfield  {title} {\bibinfo {title} {Replica {M}onte {C}arlo
  simulation of spin-glasses},\ }\href
  {https://journals.aps.org/prl/abstract/10.1103/PhysRevLett.57.2607}
  {\bibfield  {journal} {\bibinfo  {journal} {Phys. Rev. Lett}\ }\textbf
  {\bibinfo {volume} {57}},\ \bibinfo {pages} {2607} (\bibinfo {year}
  {1986})}\BibitemShut {NoStop}%
\bibitem [{\citenamefont {Mazzola}(2021)}]{mazzola:2021}%
  \BibitemOpen
  \bibfield  {author} {\bibinfo {author} {\bibfnamefont {G.}~\bibnamefont
  {Mazzola}},\ }\bibfield  {title} {\bibinfo {title} {Sampling, rates, and
  reaction currents through reverse stochastic quantization on quantum
  computers},\ }\href {https://doi.org/10.1103/PhysRevA.104.022431} {\bibfield
  {journal} {\bibinfo  {journal} {Phys. Rev. A}\ }\textbf {\bibinfo {volume}
  {104}},\ \bibinfo {pages} {022431} (\bibinfo {year} {2021})}\BibitemShut
  {NoStop}%
\bibitem [{\citenamefont {Babbush}\ \emph {et~al.}(2021)\citenamefont
  {Babbush}, \citenamefont {McClean}, \citenamefont {Newman}, \citenamefont
  {Gidney}, \citenamefont {Boixo},\ and\ \citenamefont {Neven}}]{babbush:2021}%
  \BibitemOpen
  \bibfield  {author} {\bibinfo {author} {\bibfnamefont {R.}~\bibnamefont
  {Babbush}}, \bibinfo {author} {\bibfnamefont {J.~R.}\ \bibnamefont
  {McClean}}, \bibinfo {author} {\bibfnamefont {M.}~\bibnamefont {Newman}},
  \bibinfo {author} {\bibfnamefont {C.}~\bibnamefont {Gidney}}, \bibinfo
  {author} {\bibfnamefont {S.}~\bibnamefont {Boixo}},\ and\ \bibinfo {author}
  {\bibfnamefont {H.}~\bibnamefont {Neven}},\ }\bibfield  {title} {\bibinfo
  {title} {Focus beyond quadratic speedups for error-corrected quantum
  advantage},\ }\href {https://doi.org/10.1103/PRXQuantum.2.010103} {\bibfield
  {journal} {\bibinfo  {journal} {PRX Quantum}\ }\textbf {\bibinfo {volume}
  {2}},\ \bibinfo {pages} {010103} (\bibinfo {year} {2021})}\BibitemShut
  {NoStop}%
\end{thebibliography}%


\begin{thebibliography}{55}%
\makeatletter
\providecommand \@ifxundefined [1]{%
 \@ifx{#1\undefined}
}%
\providecommand \@ifnum [1]{%
 \ifnum #1\expandafter \@firstoftwo
 \else \expandafter \@secondoftwo
 \fi
}%
\providecommand \@ifx [1]{%
 \ifx #1\expandafter \@firstoftwo
 \else \expandafter \@secondoftwo
 \fi
}%
\providecommand \natexlab [1]{#1}%
\providecommand \enquote  [1]{``#1''}%
\providecommand \bibnamefont  [1]{#1}%
\providecommand \bibfnamefont [1]{#1}%
\providecommand \citenamefont [1]{#1}%
\providecommand \href@noop [0]{\@secondoftwo}%
\providecommand \href [0]{\begingroup \@sanitize@url \@href}%
\providecommand \@href[1]{\@@startlink{#1}\@@href}%
\providecommand \@@href[1]{\endgroup#1\@@endlink}%
\providecommand \@sanitize@url [0]{\catcode `\\12\catcode `\$12\catcode
  `\&12\catcode `\#12\catcode `\^12\catcode `\_12\catcode `\%12\relax}%
\providecommand \@@startlink[1]{}%
\providecommand \@@endlink[0]{}%
\providecommand \url  [0]{\begingroup\@sanitize@url \@url }%
\providecommand \@url [1]{\endgroup\@href {#1}{\urlprefix }}%
\providecommand \urlprefix  [0]{URL }%
\providecommand \Eprint [0]{\href }%
\providecommand \doibase [0]{https://doi.org/}%
\providecommand \selectlanguage [0]{\@gobble}%
\providecommand \bibinfo  [0]{\@secondoftwo}%
\providecommand \bibfield  [0]{\@secondoftwo}%
\providecommand \translation [1]{[#1]}%
\providecommand \BibitemOpen [0]{}%
\providecommand \bibitemStop [0]{}%
\providecommand \bibitemNoStop [0]{.\EOS\space}%
\providecommand \EOS [0]{\spacefactor3000\relax}%
\providecommand \BibitemShut  [1]{\csname bibitem#1\endcsname}%
\let\auto@bib@innerbib\@empty
\bibitem [{\citenamefont {Cowles}\ and\ \citenamefont
  {Carlin}(1996)}]{cowles:1996}%
  \BibitemOpen
  \bibfield  {author} {\bibinfo {author} {\bibfnamefont {M.~K.}\ \bibnamefont
  {Cowles}}\ and\ \bibinfo {author} {\bibfnamefont {B.~P.}\ \bibnamefont
  {Carlin}},\ }\bibfield  {title} {\bibinfo {title} {Markov chain {M}onte
  {C}arlo convergence diagnostics: a comparative review},\ }\href@noop {}
  {\bibfield  {journal} {\bibinfo  {journal} {Journal of the American
  Statistical Association}\ }\textbf {\bibinfo {volume} {91}},\ \bibinfo
  {pages} {883} (\bibinfo {year} {1996})}\BibitemShut {NoStop}%
\bibitem [{\citenamefont {Levin}\ and\ \citenamefont
  {Peres}(2017)}]{levin:2017}%
  \BibitemOpen
  \bibfield  {author} {\bibinfo {author} {\bibfnamefont {D.}~\bibnamefont
  {Levin}}\ and\ \bibinfo {author} {\bibfnamefont {Y.}~\bibnamefont {Peres}},\
  }\href@noop {} {\emph {\bibinfo {title} {Markov Chains and Mixing Times}}},\
  MBK\ (\bibinfo  {publisher} {American Mathematical Society},\ \bibinfo {year}
  {2017})\BibitemShut {NoStop}%
\bibitem [{\citenamefont {Sherrington}\ and\ \citenamefont
  {Kirkpatrick}(1975)}]{sherrington:1975}%
  \BibitemOpen
  \bibfield  {author} {\bibinfo {author} {\bibfnamefont {D.}~\bibnamefont
  {Sherrington}}\ and\ \bibinfo {author} {\bibfnamefont {S.}~\bibnamefont
  {Kirkpatrick}},\ }\bibfield  {title} {\bibinfo {title} {Solvable model of a
  spin-glass},\ }\href {https://doi.org/10.1103/PhysRevLett.35.1792} {\bibfield
   {journal} {\bibinfo  {journal} {Phys. Rev. Lett}\ }\textbf {\bibinfo
  {volume} {35}},\ \bibinfo {pages} {1792} (\bibinfo {year}
  {1975})}\BibitemShut {NoStop}%
\bibitem [{\citenamefont {Callison}\ \emph {et~al.}(2019)\citenamefont
  {Callison}, \citenamefont {Chancellor}, \citenamefont {Mintert},\ and\
  \citenamefont {Kendon}}]{callison:2019}%
  \BibitemOpen
  \bibfield  {author} {\bibinfo {author} {\bibfnamefont {A.}~\bibnamefont
  {Callison}}, \bibinfo {author} {\bibfnamefont {N.}~\bibnamefont
  {Chancellor}}, \bibinfo {author} {\bibfnamefont {F.}~\bibnamefont
  {Mintert}},\ and\ \bibinfo {author} {\bibfnamefont {V.}~\bibnamefont
  {Kendon}},\ }\bibfield  {title} {\bibinfo {title} {Finding spin glass ground
  states using quantum walks},\ }\href
  {https://doi.org/10.1088/1367-2630/ab5ca2} {\bibfield  {journal} {\bibinfo
  {journal} {New Journal of Physics}\ }\textbf {\bibinfo {volume} {21}},\
  \bibinfo {pages} {123022} (\bibinfo {year} {2019})}\BibitemShut {NoStop}%
\bibitem [{\citenamefont {Hastings}(1970)}]{hastings:1970}%
  \BibitemOpen
  \bibfield  {author} {\bibinfo {author} {\bibfnamefont {W.~K.}\ \bibnamefont
  {Hastings}},\ }\bibfield  {title} {\bibinfo {title} {{Monte {C}arlo sampling
  methods using {M}arkov chains and their applications}},\ }\href
  {https://doi.org/10.1093/biomet/57.1.97} {\bibfield  {journal} {\bibinfo
  {journal} {Biometrika}\ }\textbf {\bibinfo {volume} {57}},\ \bibinfo {pages}
  {97} (\bibinfo {year} {1970})}\BibitemShut {NoStop}%
\bibitem [{\citenamefont {Chen}\ \emph {et~al.}(2002)\citenamefont {Chen},
  \citenamefont {Liu},\ and\ \citenamefont {Wang}}]{chen:2002}%
  \BibitemOpen
  \bibfield  {author} {\bibinfo {author} {\bibfnamefont {R.}~\bibnamefont
  {Chen}}, \bibinfo {author} {\bibfnamefont {J.}~\bibnamefont {Liu}},\ and\
  \bibinfo {author} {\bibfnamefont {X.}~\bibnamefont {Wang}},\ }\bibfield
  {title} {\bibinfo {title} {Convergence analyses and comparisons of {M}arkov
  chain {M}onte {C}arlo algorithms in digital communications},\ }\href
  {https://doi.org/10.1109/78.978381} {\bibfield  {journal} {\bibinfo
  {journal} {IEEE Transactions on Signal Processing}\ }\textbf {\bibinfo
  {volume} {50}},\ \bibinfo {pages} {255} (\bibinfo {year} {2002})}\BibitemShut
  {NoStop}%
\bibitem [{\citenamefont {Andrieu}\ \emph {et~al.}(2003)\citenamefont
  {Andrieu}, \citenamefont {De~Freitas}, \citenamefont {Doucet},\ and\
  \citenamefont {Jordan}}]{andrieu:2003}%
  \BibitemOpen
  \bibfield  {author} {\bibinfo {author} {\bibfnamefont {C.}~\bibnamefont
  {Andrieu}}, \bibinfo {author} {\bibfnamefont {N.}~\bibnamefont {De~Freitas}},
  \bibinfo {author} {\bibfnamefont {A.}~\bibnamefont {Doucet}},\ and\ \bibinfo
  {author} {\bibfnamefont {M.~I.}\ \bibnamefont {Jordan}},\ }\bibfield  {title}
  {\bibinfo {title} {An introduction to {MCMC} for machine learning},\ }\href
  {https://link.springer.com/article/10.1023/A:1020281327116} {\bibfield
  {journal} {\bibinfo  {journal} {Mach. Learn}\ }\textbf {\bibinfo {volume}
  {50}},\ \bibinfo {pages} {5} (\bibinfo {year} {2003})}\BibitemShut {NoStop}%
\bibitem [{\citenamefont {Neal}(1993)}]{neal:1993}%
  \BibitemOpen
  \bibfield  {author} {\bibinfo {author} {\bibfnamefont {R.~M.}\ \bibnamefont
  {Neal}},\ }\href {https://www.cs.toronto.edu/~radford/ftp/review.pdf} {\emph
  {\bibinfo {title} {Probabilistic inference using {M}arkov chain {M}onte
  {C}arlo methods}}}\ (\bibinfo  {publisher} {Department of Computer Science,
  University of Toronto Toronto, ON, Canada},\ \bibinfo {year}
  {1993})\BibitemShut {NoStop}%
\bibitem [{\citenamefont {Swendsen}\ and\ \citenamefont
  {Wang}(1987)}]{swendsen:1987}%
  \BibitemOpen
  \bibfield  {author} {\bibinfo {author} {\bibfnamefont {R.~H.}\ \bibnamefont
  {Swendsen}}\ and\ \bibinfo {author} {\bibfnamefont {J.-S.}\ \bibnamefont
  {Wang}},\ }\bibfield  {title} {\bibinfo {title} {Nonuniversal critical
  dynamics in {M}onte {C}arlo simulations},\ }\href
  {https://doi.org/10.1103/PhysRevLett.58.86} {\bibfield  {journal} {\bibinfo
  {journal} {Phys. Rev. Lett}\ }\textbf {\bibinfo {volume} {58}},\ \bibinfo
  {pages} {86} (\bibinfo {year} {1987})}\BibitemShut {NoStop}%
\bibitem [{\citenamefont {Wolff}(1989)}]{wolff:1989}%
  \BibitemOpen
  \bibfield  {author} {\bibinfo {author} {\bibfnamefont {U.}~\bibnamefont
  {Wolff}},\ }\bibfield  {title} {\bibinfo {title} {Collective {M}onte {C}arlo
  updating for spin systems},\ }\href
  {https://doi.org/10.1103/PhysRevLett.62.361} {\bibfield  {journal} {\bibinfo
  {journal} {Phys. Rev. Lett}\ }\textbf {\bibinfo {volume} {62}},\ \bibinfo
  {pages} {361} (\bibinfo {year} {1989})}\BibitemShut {NoStop}%
\bibitem [{\citenamefont {Park}\ \emph {et~al.}(2017)\citenamefont {Park},
  \citenamefont {Jang}, \citenamefont {Galanis}, \citenamefont {Shin},
  \citenamefont {Stefankovic},\ and\ \citenamefont {Vigoda}}]{park:2017}%
  \BibitemOpen
  \bibfield  {author} {\bibinfo {author} {\bibfnamefont {S.}~\bibnamefont
  {Park}}, \bibinfo {author} {\bibfnamefont {Y.}~\bibnamefont {Jang}}, \bibinfo
  {author} {\bibfnamefont {A.}~\bibnamefont {Galanis}}, \bibinfo {author}
  {\bibfnamefont {J.}~\bibnamefont {Shin}}, \bibinfo {author} {\bibfnamefont
  {D.}~\bibnamefont {Stefankovic}},\ and\ \bibinfo {author} {\bibfnamefont
  {E.}~\bibnamefont {Vigoda}},\ }\bibfield  {title} {\bibinfo {title} {{Rapid
  Mixing Swendsen-Wang Sampler for Stochastic Partitioned Attractive Models}},\
  }in\ \href {https://proceedings.mlr.press/v54/park17b.html} {\emph {\bibinfo
  {booktitle} {Proceedings of the 20th International Conference on Artificial
  Intelligence and Statistics}}},\ \bibinfo {series} {Proceedings of Machine
  Learning Research}, Vol.~\bibinfo {volume} {54},\ \bibinfo {editor} {edited
  by\ \bibinfo {editor} {\bibfnamefont {A.}~\bibnamefont {Singh}}\ and\
  \bibinfo {editor} {\bibfnamefont {J.}~\bibnamefont {Zhu}}}\ (\bibinfo
  {publisher} {PMLR},\ \bibinfo {year} {2017})\ pp.\ \bibinfo {pages}
  {440--449}\BibitemShut {NoStop}%
\bibitem [{\citenamefont {Barzegar}\ \emph {et~al.}(2018)\citenamefont
  {Barzegar}, \citenamefont {Pattison}, \citenamefont {Wang},\ and\
  \citenamefont {Katzgraber}}]{barzegar:2018}%
  \BibitemOpen
  \bibfield  {author} {\bibinfo {author} {\bibfnamefont {A.}~\bibnamefont
  {Barzegar}}, \bibinfo {author} {\bibfnamefont {C.}~\bibnamefont {Pattison}},
  \bibinfo {author} {\bibfnamefont {W.}~\bibnamefont {Wang}},\ and\ \bibinfo
  {author} {\bibfnamefont {H.~G.}\ \bibnamefont {Katzgraber}},\ }\bibfield
  {title} {\bibinfo {title} {Optimization of population annealing {M}onte
  {C}arlo for large-scale spin-glass simulations},\ }\href
  {https://doi.org/10.1103/PhysRevE.98.053308} {\bibfield  {journal} {\bibinfo
  {journal} {Phys. Rev. E}\ }\textbf {\bibinfo {volume} {98}},\ \bibinfo
  {pages} {053308} (\bibinfo {year} {2018})}\BibitemShut {NoStop}%
\bibitem [{\citenamefont {Wang}(1989)}]{wang:1989}%
  \BibitemOpen
  \bibfield  {author} {\bibinfo {author} {\bibfnamefont {J.-S.}\ \bibnamefont
  {Wang}},\ }\bibfield  {title} {\bibinfo {title} {Clusters in the
  three-dimensional {I}sing model with a magnetic field},\ }\href
  {https://doi.org/https://doi.org/10.1016/0378-4371(89)90468-8} {\bibfield
  {journal} {\bibinfo  {journal} {Physica A: Statistical Mechanics and its
  Applications}\ }\textbf {\bibinfo {volume} {161}},\ \bibinfo {pages} {249}
  (\bibinfo {year} {1989})}\BibitemShut {NoStop}%
\bibitem [{\citenamefont {{deLyra, Jorge L.}}(2006)}]{delyra:2006}%
  \BibitemOpen
  \bibfield  {author} {\bibinfo {author} {\bibnamefont {{deLyra, Jorge L.}}},\
  }\href@noop {} {\bibinfo {title} {The {W}olff algorithm with external sources
  and boundaries}},\ \bibinfo {howpublished}
  {\url{http://latt.if.usp.br/cgi-bin-delyra/cntsnd?technical-pages/twawesab/Text.pdf}}
  (\bibinfo {year} {2006})\BibitemShut {NoStop}%
\bibitem [{\citenamefont {Kent-Dobias}\ and\ \citenamefont
  {Sethna}(2018)}]{kent-dobias:2018}%
  \BibitemOpen
  \bibfield  {author} {\bibinfo {author} {\bibfnamefont {J.}~\bibnamefont
  {Kent-Dobias}}\ and\ \bibinfo {author} {\bibfnamefont {J.~P.}\ \bibnamefont
  {Sethna}},\ }\bibfield  {title} {\bibinfo {title} {Cluster representations
  and the {W}olff algorithm in arbitrary external fields},\ }\href
  {https://doi.org/10.1103/PhysRevE.98.063306} {\bibfield  {journal} {\bibinfo
  {journal} {Phys. Rev. E}\ }\textbf {\bibinfo {volume} {98}},\ \bibinfo
  {pages} {063306} (\bibinfo {year} {2018})}\BibitemShut {NoStop}%
\bibitem [{\citenamefont {Houdayer}(2001)}]{houdayer:2001}%
  \BibitemOpen
  \bibfield  {author} {\bibinfo {author} {\bibfnamefont {J.}~\bibnamefont
  {Houdayer}},\ }\bibfield  {title} {\bibinfo {title} {A cluster {M}onte
  {C}arlo algorithm for 2-dimensional spin glasses},\ }\href
  {https://link.springer.com/article/10.1007/PL00011151} {\bibfield  {journal}
  {\bibinfo  {journal} {Eur. Phys. J. B}\ }\textbf {\bibinfo {volume} {22}},\
  \bibinfo {pages} {479} (\bibinfo {year} {2001})}\BibitemShut {NoStop}%
\bibitem [{\citenamefont {He}\ \emph {et~al.}(2016)\citenamefont {He},
  \citenamefont {De~Sa}, \citenamefont {Mitliagkas},\ and\ \citenamefont
  {R\'{e}}}]{he:2016}%
  \BibitemOpen
  \bibfield  {author} {\bibinfo {author} {\bibfnamefont {B.~D.}\ \bibnamefont
  {He}}, \bibinfo {author} {\bibfnamefont {C.~M.}\ \bibnamefont {De~Sa}},
  \bibinfo {author} {\bibfnamefont {I.}~\bibnamefont {Mitliagkas}},\ and\
  \bibinfo {author} {\bibfnamefont {C.}~\bibnamefont {R\'{e}}},\ }\bibfield
  {title} {\bibinfo {title} {Scan order in {G}ibbs sampling: Models in which it
  matters and bounds on how much},\ }in\ \href
  {https://proceedings.neurips.cc/paper/2016/file/e4da3b7fbbce2345d7772b0674a318d5-Paper.pdf}
  {\emph {\bibinfo {booktitle} {Advances in Neural Information Processing
  Systems}}},\ Vol.~\bibinfo {volume} {29},\ \bibinfo {editor} {edited by\
  \bibinfo {editor} {\bibfnamefont {D.}~\bibnamefont {Lee}}, \bibinfo {editor}
  {\bibfnamefont {M.}~\bibnamefont {Sugiyama}}, \bibinfo {editor}
  {\bibfnamefont {U.}~\bibnamefont {Luxburg}}, \bibinfo {editor} {\bibfnamefont
  {I.}~\bibnamefont {Guyon}},\ and\ \bibinfo {editor} {\bibfnamefont
  {R.}~\bibnamefont {Garnett}}}\ (\bibinfo  {publisher} {Curran Associates,
  Inc.},\ \bibinfo {year} {2016})\BibitemShut {NoStop}%
\bibitem [{\citenamefont {Zhu}\ \emph {et~al.}(2015)\citenamefont {Zhu},
  \citenamefont {Ochoa},\ and\ \citenamefont {Katzgraber}}]{zhu:2015}%
  \BibitemOpen
  \bibfield  {author} {\bibinfo {author} {\bibfnamefont {Z.}~\bibnamefont
  {Zhu}}, \bibinfo {author} {\bibfnamefont {A.~J.}\ \bibnamefont {Ochoa}},\
  and\ \bibinfo {author} {\bibfnamefont {H.~G.}\ \bibnamefont {Katzgraber}},\
  }\bibfield  {title} {\bibinfo {title} {Efficient cluster algorithm for spin
  glasses in any space dimension},\ }\href
  {https://doi.org/10.1103/PhysRevLett.115.077201} {\bibfield  {journal}
  {\bibinfo  {journal} {Phys. Rev. Lett}\ }\textbf {\bibinfo {volume} {115}},\
  \bibinfo {pages} {077201} (\bibinfo {year} {2015})}\BibitemShut {NoStop}%
\bibitem [{\citenamefont {Virtanen}\ \emph {et~al.}(2020)\citenamefont
  {Virtanen} \emph {et~al.}}]{scipy}%
  \BibitemOpen
  \bibfield  {author} {\bibinfo {author} {\bibfnamefont {P.}~\bibnamefont
  {Virtanen}} \emph {et~al.},\ }\bibfield  {title} {\bibinfo {title} {{{SciPy}
  1.0: Fundamental Algorithms for Scientific Computing in Python}},\ }\href
  {https://doi.org/10.1038/s41592-019-0686-2} {\bibfield  {journal} {\bibinfo
  {journal} {Nature Methods}\ }\textbf {\bibinfo {volume} {17}},\ \bibinfo
  {pages} {261} (\bibinfo {year} {2020})}\BibitemShut {NoStop}%
\bibitem [{\citenamefont {Hsu}\ \emph {et~al.}(2015)\citenamefont {Hsu},
  \citenamefont {Kontorovich},\ and\ \citenamefont {Szepesvari}}]{hsu:2015}%
  \BibitemOpen
  \bibfield  {author} {\bibinfo {author} {\bibfnamefont {D.~J.}\ \bibnamefont
  {Hsu}}, \bibinfo {author} {\bibfnamefont {A.}~\bibnamefont {Kontorovich}},\
  and\ \bibinfo {author} {\bibfnamefont {C.}~\bibnamefont {Szepesvari}},\
  }\bibfield  {title} {\bibinfo {title} {Mixing time estimation in reversible
  {M}arkov chains from a single sample path},\ }in\ \href
  {https://proceedings.neurips.cc/paper/2015/file/7ce3284b743aefde80ffd9aec500e085-Paper.pdf}
  {\emph {\bibinfo {booktitle} {Advances in Neural Information Processing
  Systems}}},\ Vol.~\bibinfo {volume} {28},\ \bibinfo {editor} {edited by\
  \bibinfo {editor} {\bibfnamefont {C.}~\bibnamefont {Cortes}}, \bibinfo
  {editor} {\bibfnamefont {N.}~\bibnamefont {Lawrence}}, \bibinfo {editor}
  {\bibfnamefont {D.}~\bibnamefont {Lee}}, \bibinfo {editor} {\bibfnamefont
  {M.}~\bibnamefont {Sugiyama}},\ and\ \bibinfo {editor} {\bibfnamefont
  {R.}~\bibnamefont {Garnett}}}\ (\bibinfo  {publisher} {Curran Associates,
  Inc.},\ \bibinfo {year} {2015})\BibitemShut {NoStop}%
\bibitem [{\citenamefont {Layden}(2021)}]{layden:2021}%
  \BibitemOpen
  \bibfield  {author} {\bibinfo {author} {\bibfnamefont {D.}~\bibnamefont
  {Layden}},\ }\bibfield  {title} {\bibinfo {title} {First-order {T}rotter
  error from a second-order perspective},\ }\href@noop {} {\bibfield  {journal}
  {\bibinfo  {journal} {arXiv:2107.08032}\ } (\bibinfo {year}
  {2021})}\BibitemShut {NoStop}%
\bibitem [{\citenamefont {McKay}\ \emph {et~al.}(2017)\citenamefont {McKay},
  \citenamefont {Wood}, \citenamefont {Sheldon}, \citenamefont {Chow},\ and\
  \citenamefont {Gambetta}}]{mckay:2017}%
  \BibitemOpen
  \bibfield  {author} {\bibinfo {author} {\bibfnamefont {D.~C.}\ \bibnamefont
  {McKay}}, \bibinfo {author} {\bibfnamefont {C.~J.}\ \bibnamefont {Wood}},
  \bibinfo {author} {\bibfnamefont {S.}~\bibnamefont {Sheldon}}, \bibinfo
  {author} {\bibfnamefont {J.~M.}\ \bibnamefont {Chow}},\ and\ \bibinfo
  {author} {\bibfnamefont {J.~M.}\ \bibnamefont {Gambetta}},\ }\bibfield
  {title} {\bibinfo {title} {Efficient {$Z$} gates for quantum computing},\
  }\href {https://doi.org/10.1103/PhysRevA.96.022330} {\bibfield  {journal}
  {\bibinfo  {journal} {Phys. Rev. A}\ }\textbf {\bibinfo {volume} {96}},\
  \bibinfo {pages} {022330} (\bibinfo {year} {2017})}\BibitemShut {NoStop}%
\bibitem [{\citenamefont {Nielsen}\ and\ \citenamefont
  {Chuang}(2000)}]{nielsen:2000}%
  \BibitemOpen
  \bibfield  {author} {\bibinfo {author} {\bibfnamefont {M.}~\bibnamefont
  {Nielsen}}\ and\ \bibinfo {author} {\bibfnamefont {I.}~\bibnamefont
  {Chuang}},\ }\href {https://books.google.com/books?id=65FqEKQOfP8C} {\emph
  {\bibinfo {title} {Quantum Computation and Quantum Information}}},\ Cambridge
  Series on Information and the Natural Sciences\ (\bibinfo  {publisher}
  {Cambridge University Press},\ \bibinfo {year} {2000})\BibitemShut {NoStop}%
\bibitem [{\citenamefont {Sheldon}\ \emph {et~al.}(2016)\citenamefont
  {Sheldon}, \citenamefont {Magesan}, \citenamefont {Chow},\ and\ \citenamefont
  {Gambetta}}]{sheldon:2016}%
  \BibitemOpen
  \bibfield  {author} {\bibinfo {author} {\bibfnamefont {S.}~\bibnamefont
  {Sheldon}}, \bibinfo {author} {\bibfnamefont {E.}~\bibnamefont {Magesan}},
  \bibinfo {author} {\bibfnamefont {J.~M.}\ \bibnamefont {Chow}},\ and\
  \bibinfo {author} {\bibfnamefont {J.~M.}\ \bibnamefont {Gambetta}},\
  }\bibfield  {title} {\bibinfo {title} {Procedure for systematically tuning up
  cross-talk in the cross-resonance gate},\ }\href
  {https://doi.org/10.1103/PhysRevA.93.060302} {\bibfield  {journal} {\bibinfo
  {journal} {Phys. Rev. A}\ }\textbf {\bibinfo {volume} {93}},\ \bibinfo
  {pages} {060302} (\bibinfo {year} {2016})}\BibitemShut {NoStop}%
\bibitem [{\citenamefont {Sundaresan}\ \emph {et~al.}(2020)\citenamefont
  {Sundaresan}, \citenamefont {Lauer}, \citenamefont {Pritchett}, \citenamefont
  {Magesan}, \citenamefont {Jurcevic},\ and\ \citenamefont
  {Gambetta}}]{sundaresan:2020}%
  \BibitemOpen
  \bibfield  {author} {\bibinfo {author} {\bibfnamefont {N.}~\bibnamefont
  {Sundaresan}}, \bibinfo {author} {\bibfnamefont {I.}~\bibnamefont {Lauer}},
  \bibinfo {author} {\bibfnamefont {E.}~\bibnamefont {Pritchett}}, \bibinfo
  {author} {\bibfnamefont {E.}~\bibnamefont {Magesan}}, \bibinfo {author}
  {\bibfnamefont {P.}~\bibnamefont {Jurcevic}},\ and\ \bibinfo {author}
  {\bibfnamefont {J.~M.}\ \bibnamefont {Gambetta}},\ }\bibfield  {title}
  {\bibinfo {title} {Reducing unitary and spectator errors in cross resonance
  with optimized rotary echoes},\ }\href
  {https://doi.org/10.1103/PRXQuantum.1.020318} {\bibfield  {journal} {\bibinfo
   {journal} {PRX Quantum}\ }\textbf {\bibinfo {volume} {1}},\ \bibinfo {pages}
  {020318} (\bibinfo {year} {2020})}\BibitemShut {NoStop}%
\bibitem [{\citenamefont {Stenger}\ \emph {et~al.}(2021)\citenamefont
  {Stenger}, \citenamefont {Bronn}, \citenamefont {Egger},\ and\ \citenamefont
  {Pekker}}]{stenger:2021}%
  \BibitemOpen
  \bibfield  {author} {\bibinfo {author} {\bibfnamefont {J.~P.~T.}\
  \bibnamefont {Stenger}}, \bibinfo {author} {\bibfnamefont {N.~T.}\
  \bibnamefont {Bronn}}, \bibinfo {author} {\bibfnamefont {D.~J.}\ \bibnamefont
  {Egger}},\ and\ \bibinfo {author} {\bibfnamefont {D.}~\bibnamefont
  {Pekker}},\ }\bibfield  {title} {\bibinfo {title} {Simulating the dynamics of
  braiding of {M}ajorana zero modes using an {IBM} quantum computer},\ }\href
  {https://doi.org/10.1103/PhysRevResearch.3.033171} {\bibfield  {journal}
  {\bibinfo  {journal} {Phys. Rev. Research}\ }\textbf {\bibinfo {volume}
  {3}},\ \bibinfo {pages} {033171} (\bibinfo {year} {2021})}\BibitemShut
  {NoStop}%
\bibitem [{\citenamefont {Earnest}\ \emph {et~al.}(2021)\citenamefont
  {Earnest}, \citenamefont {Tornow},\ and\ \citenamefont
  {Egger}}]{earnest:2021}%
  \BibitemOpen
  \bibfield  {author} {\bibinfo {author} {\bibfnamefont {N.}~\bibnamefont
  {Earnest}}, \bibinfo {author} {\bibfnamefont {C.}~\bibnamefont {Tornow}},\
  and\ \bibinfo {author} {\bibfnamefont {D.~J.}\ \bibnamefont {Egger}},\
  }\bibfield  {title} {\bibinfo {title} {Pulse-efficient circuit transpilation
  for quantum applications on cross-resonance-based hardware},\ }\href
  {https://doi.org/10.1103/PhysRevResearch.3.043088} {\bibfield  {journal}
  {\bibinfo  {journal} {Phys. Rev. Research}\ }\textbf {\bibinfo {volume}
  {3}},\ \bibinfo {pages} {043088} (\bibinfo {year} {2021})}\BibitemShut
  {NoStop}%
\bibitem [{\citenamefont {Childs}\ \emph {et~al.}(2021)\citenamefont {Childs},
  \citenamefont {Su}, \citenamefont {Tran}, \citenamefont {Wiebe},\ and\
  \citenamefont {Zhu}}]{childs:2021}%
  \BibitemOpen
  \bibfield  {author} {\bibinfo {author} {\bibfnamefont {A.~M.}\ \bibnamefont
  {Childs}}, \bibinfo {author} {\bibfnamefont {Y.}~\bibnamefont {Su}}, \bibinfo
  {author} {\bibfnamefont {M.~C.}\ \bibnamefont {Tran}}, \bibinfo {author}
  {\bibfnamefont {N.}~\bibnamefont {Wiebe}},\ and\ \bibinfo {author}
  {\bibfnamefont {S.}~\bibnamefont {Zhu}},\ }\bibfield  {title} {\bibinfo
  {title} {Theory of {T}rotter error with commutator scaling},\ }\href
  {https://doi.org/10.1103/PhysRevX.11.011020} {\bibfield  {journal} {\bibinfo
  {journal} {Phys. Rev. X}\ }\textbf {\bibinfo {volume} {11}},\ \bibinfo
  {pages} {011020} (\bibinfo {year} {2021})}\BibitemShut {NoStop}%
\bibitem [{\citenamefont {Knee}\ and\ \citenamefont {Munro}(2015)}]{knee:2015}%
  \BibitemOpen
  \bibfield  {author} {\bibinfo {author} {\bibfnamefont {G.~C.}\ \bibnamefont
  {Knee}}\ and\ \bibinfo {author} {\bibfnamefont {W.~J.}\ \bibnamefont
  {Munro}},\ }\bibfield  {title} {\bibinfo {title} {Optimal {T}rotterization in
  universal quantum simulators under faulty control},\ }\href
  {https://doi.org/10.1103/PhysRevA.91.052327} {\bibfield  {journal} {\bibinfo
  {journal} {Phys. Rev. A}\ }\textbf {\bibinfo {volume} {91}},\ \bibinfo
  {pages} {052327} (\bibinfo {year} {2015})}\BibitemShut {NoStop}%
\bibitem [{\citenamefont {Endo}\ \emph {et~al.}(2019)\citenamefont {Endo},
  \citenamefont {Zhao}, \citenamefont {Li}, \citenamefont {Benjamin},\ and\
  \citenamefont {Yuan}}]{endo:2019}%
  \BibitemOpen
  \bibfield  {author} {\bibinfo {author} {\bibfnamefont {S.}~\bibnamefont
  {Endo}}, \bibinfo {author} {\bibfnamefont {Q.}~\bibnamefont {Zhao}}, \bibinfo
  {author} {\bibfnamefont {Y.}~\bibnamefont {Li}}, \bibinfo {author}
  {\bibfnamefont {S.}~\bibnamefont {Benjamin}},\ and\ \bibinfo {author}
  {\bibfnamefont {X.}~\bibnamefont {Yuan}},\ }\bibfield  {title} {\bibinfo
  {title} {Mitigating algorithmic errors in a {H}amiltonian simulation},\
  }\href {https://doi.org/10.1103/PhysRevA.99.012334} {\bibfield  {journal}
  {\bibinfo  {journal} {Phys. Rev. A}\ }\textbf {\bibinfo {volume} {99}},\
  \bibinfo {pages} {012334} (\bibinfo {year} {2019})}\BibitemShut {NoStop}%
\bibitem [{\citenamefont {Clinton}\ \emph {et~al.}(2021)\citenamefont
  {Clinton}, \citenamefont {Bausch},\ and\ \citenamefont
  {Cubitt}}]{clinton:2021}%
  \BibitemOpen
  \bibfield  {author} {\bibinfo {author} {\bibfnamefont {L.}~\bibnamefont
  {Clinton}}, \bibinfo {author} {\bibfnamefont {J.}~\bibnamefont {Bausch}},\
  and\ \bibinfo {author} {\bibfnamefont {T.}~\bibnamefont {Cubitt}},\
  }\bibfield  {title} {\bibinfo {title} {Hamiltonian simulation algorithms for
  near-term quantum hardware},\ }\href@noop {} {\bibfield  {journal} {\bibinfo
  {journal} {Nature communications}\ }\textbf {\bibinfo {volume} {12}},\
  \bibinfo {pages} {1} (\bibinfo {year} {2021})}\BibitemShut {NoStop}%
\bibitem [{\citenamefont {van~den Berg}\ \emph {et~al.}(2020)\citenamefont
  {van~den Berg}, \citenamefont {Minev},\ and\ \citenamefont
  {Temme}}]{vandenberg:2020}%
  \BibitemOpen
  \bibfield  {author} {\bibinfo {author} {\bibfnamefont {E.}~\bibnamefont
  {van~den Berg}}, \bibinfo {author} {\bibfnamefont {Z.~K.}\ \bibnamefont
  {Minev}},\ and\ \bibinfo {author} {\bibfnamefont {K.}~\bibnamefont {Temme}},\
  }\bibfield  {title} {\bibinfo {title} {Model-free readout-error mitigation
  for quantum expectation values},\ }\href@noop {} {\bibfield  {journal}
  {\bibinfo  {journal} {arXiv:2012.09738}\ } (\bibinfo {year}
  {2020})}\BibitemShut {NoStop}%
\bibitem [{\citenamefont {Knill}(2004)}]{knill:2004}%
  \BibitemOpen
  \bibfield  {author} {\bibinfo {author} {\bibfnamefont {E.}~\bibnamefont
  {Knill}},\ }\bibfield  {title} {\bibinfo {title} {Fault-tolerant postselected
  quantum computation: Threshold analysis},\ }\href@noop {} {\bibfield
  {journal} {\bibinfo  {journal} {quant-ph/0404104}\ } (\bibinfo {year}
  {2004})}\BibitemShut {NoStop}%
\bibitem [{\citenamefont {Wallman}\ and\ \citenamefont
  {Emerson}(2016)}]{wallman:2016}%
  \BibitemOpen
  \bibfield  {author} {\bibinfo {author} {\bibfnamefont {J.~J.}\ \bibnamefont
  {Wallman}}\ and\ \bibinfo {author} {\bibfnamefont {J.}~\bibnamefont
  {Emerson}},\ }\bibfield  {title} {\bibinfo {title} {Noise tailoring for
  scalable quantum computation via randomized compiling},\ }\href
  {https://doi.org/10.1103/PhysRevA.94.052325} {\bibfield  {journal} {\bibinfo
  {journal} {Phys. Rev. A}\ }\textbf {\bibinfo {volume} {94}},\ \bibinfo
  {pages} {052325} (\bibinfo {year} {2016})}\BibitemShut {NoStop}%
\bibitem [{\citenamefont {Kim}\ \emph {et~al.}(2021)\citenamefont {Kim},
  \citenamefont {Wood}, \citenamefont {Yoder}, \citenamefont {Merkel},
  \citenamefont {Gambetta}, \citenamefont {Temme},\ and\ \citenamefont
  {Kandala}}]{kim:2021}%
  \BibitemOpen
  \bibfield  {author} {\bibinfo {author} {\bibfnamefont {Y.}~\bibnamefont
  {Kim}}, \bibinfo {author} {\bibfnamefont {C.~J.}\ \bibnamefont {Wood}},
  \bibinfo {author} {\bibfnamefont {T.~J.}\ \bibnamefont {Yoder}}, \bibinfo
  {author} {\bibfnamefont {S.~T.}\ \bibnamefont {Merkel}}, \bibinfo {author}
  {\bibfnamefont {J.~M.}\ \bibnamefont {Gambetta}}, \bibinfo {author}
  {\bibfnamefont {K.}~\bibnamefont {Temme}},\ and\ \bibinfo {author}
  {\bibfnamefont {A.}~\bibnamefont {Kandala}},\ }\bibfield  {title} {\bibinfo
  {title} {Scalable error mitigation for noisy quantum circuits produces
  competitive expectation values},\ }\href@noop {} {\bibfield  {journal}
  {\bibinfo  {journal} {arXiv:2108.09197}\ } (\bibinfo {year}
  {2021})}\BibitemShut {NoStop}%
\bibitem [{\citenamefont {Cross}\ \emph {et~al.}(2021)\citenamefont {Cross},
  \citenamefont {Javadi-Abhari}, \citenamefont {Alexander}, \citenamefont
  {de~Beaudrap}, \citenamefont {Bishop}, \citenamefont {Heidel}, \citenamefont
  {Ryan}, \citenamefont {Smolin}, \citenamefont {Gambetta},\ and\ \citenamefont
  {Johnson}}]{cross:2021}%
  \BibitemOpen
  \bibfield  {author} {\bibinfo {author} {\bibfnamefont {A.~W.}\ \bibnamefont
  {Cross}}, \bibinfo {author} {\bibfnamefont {A.}~\bibnamefont
  {Javadi-Abhari}}, \bibinfo {author} {\bibfnamefont {T.}~\bibnamefont
  {Alexander}}, \bibinfo {author} {\bibfnamefont {N.}~\bibnamefont
  {de~Beaudrap}}, \bibinfo {author} {\bibfnamefont {L.~S.}\ \bibnamefont
  {Bishop}}, \bibinfo {author} {\bibfnamefont {S.}~\bibnamefont {Heidel}},
  \bibinfo {author} {\bibfnamefont {C.~A.}\ \bibnamefont {Ryan}}, \bibinfo
  {author} {\bibfnamefont {J.}~\bibnamefont {Smolin}}, \bibinfo {author}
  {\bibfnamefont {J.~M.}\ \bibnamefont {Gambetta}},\ and\ \bibinfo {author}
  {\bibfnamefont {B.~R.}\ \bibnamefont {Johnson}},\ }\bibfield  {title}
  {\bibinfo {title} {{OpenQASM} 3: A broader and deeper quantum assembly
  language},\ }\href@noop {} {\bibfield  {journal} {\bibinfo  {journal}
  {arXiv:2104.14722}\ } (\bibinfo {year} {2021})}\BibitemShut {NoStop}%
\bibitem [{\citenamefont {C\'orcoles}\ \emph {et~al.}(2021)\citenamefont
  {C\'orcoles}, \citenamefont {Takita}, \citenamefont {Inoue}, \citenamefont
  {Lekuch}, \citenamefont {Minev}, \citenamefont {Chow},\ and\ \citenamefont
  {Gambetta}}]{corcoles:2021}%
  \BibitemOpen
  \bibfield  {author} {\bibinfo {author} {\bibfnamefont {A.~D.}\ \bibnamefont
  {C\'orcoles}}, \bibinfo {author} {\bibfnamefont {M.}~\bibnamefont {Takita}},
  \bibinfo {author} {\bibfnamefont {K.}~\bibnamefont {Inoue}}, \bibinfo
  {author} {\bibfnamefont {S.}~\bibnamefont {Lekuch}}, \bibinfo {author}
  {\bibfnamefont {Z.~K.}\ \bibnamefont {Minev}}, \bibinfo {author}
  {\bibfnamefont {J.~M.}\ \bibnamefont {Chow}},\ and\ \bibinfo {author}
  {\bibfnamefont {J.~M.}\ \bibnamefont {Gambetta}},\ }\bibfield  {title}
  {\bibinfo {title} {Exploiting dynamic quantum circuits in a quantum algorithm
  with superconducting qubits},\ }\href
  {https://doi.org/10.1103/PhysRevLett.127.100501} {\bibfield  {journal}
  {\bibinfo  {journal} {Phys. Rev. Lett.}\ }\textbf {\bibinfo {volume} {127}},\
  \bibinfo {pages} {100501} (\bibinfo {year} {2021})}\BibitemShut {NoStop}%
\bibitem [{\citenamefont {Bowker}(1948)}]{bowker:1948}%
  \BibitemOpen
  \bibfield  {author} {\bibinfo {author} {\bibfnamefont {A.~H.}\ \bibnamefont
  {Bowker}},\ }\bibfield  {title} {\bibinfo {title} {A test for symmetry in
  contingency tables},\ }\href@noop {} {\bibfield  {journal} {\bibinfo
  {journal} {Journal of the American Statistical Association}\ }\textbf
  {\bibinfo {volume} {43}},\ \bibinfo {pages} {572} (\bibinfo {year}
  {1948})}\BibitemShut {NoStop}%
\bibitem [{\citenamefont {McNemar}(1947)}]{mcnemar:1947}%
  \BibitemOpen
  \bibfield  {author} {\bibinfo {author} {\bibfnamefont {Q.}~\bibnamefont
  {McNemar}},\ }\bibfield  {title} {\bibinfo {title} {Note on the sampling
  error of the difference between correlated proportions or percentages},\
  }\href@noop {} {\bibfield  {journal} {\bibinfo  {journal} {Psychometrika}\
  }\textbf {\bibinfo {volume} {12}},\ \bibinfo {pages} {153} (\bibinfo {year}
  {1947})}\BibitemShut {NoStop}%
\bibitem [{\citenamefont {Krampe}\ and\ \citenamefont
  {Kuhnt}(2007)}]{krampe:2007}%
  \BibitemOpen
  \bibfield  {author} {\bibinfo {author} {\bibfnamefont {A.}~\bibnamefont
  {Krampe}}\ and\ \bibinfo {author} {\bibfnamefont {S.}~\bibnamefont {Kuhnt}},\
  }\bibfield  {title} {\bibinfo {title} {Bowker's test for symmetry and
  modifications within the algebraic framework},\ }\href@noop {} {\bibfield
  {journal} {\bibinfo  {journal} {Computational statistics \& data analysis}\
  }\textbf {\bibinfo {volume} {51}},\ \bibinfo {pages} {4124} (\bibinfo {year}
  {2007})}\BibitemShut {NoStop}%
\bibitem [{\citenamefont {Cho}\ and\ \citenamefont {Meyer}(2001)}]{cho:2001}%
  \BibitemOpen
  \bibfield  {author} {\bibinfo {author} {\bibfnamefont {G.~E.}\ \bibnamefont
  {Cho}}\ and\ \bibinfo {author} {\bibfnamefont {C.~D.}\ \bibnamefont
  {Meyer}},\ }\bibfield  {title} {\bibinfo {title} {Comparison of perturbation
  bounds for the stationary distribution of a {M}arkov chain},\ }\href@noop {}
  {\bibfield  {journal} {\bibinfo  {journal} {Linear Algebra and its
  Applications}\ }\textbf {\bibinfo {volume} {335}},\ \bibinfo {pages} {137}
  (\bibinfo {year} {2001})}\BibitemShut {NoStop}%
\bibitem [{\citenamefont {{StataCorp}}()}]{stata}%
  \BibitemOpen
  \bibfield  {author} {\bibinfo {author} {\bibnamefont {{StataCorp}}},\
  }\href@noop {} {\bibinfo {title} {Stata manual: symmetry}},\ \bibinfo
  {howpublished}
  {\url{https://www.stata.com/manuals/rsymmetry.pdf}}\BibitemShut {NoStop}%
\bibitem [{\citenamefont {Gilchrist}\ \emph {et~al.}(2005)\citenamefont
  {Gilchrist}, \citenamefont {Langford},\ and\ \citenamefont
  {Nielsen}}]{gilchrist:2005}%
  \BibitemOpen
  \bibfield  {author} {\bibinfo {author} {\bibfnamefont {A.}~\bibnamefont
  {Gilchrist}}, \bibinfo {author} {\bibfnamefont {N.~K.}\ \bibnamefont
  {Langford}},\ and\ \bibinfo {author} {\bibfnamefont {M.~A.}\ \bibnamefont
  {Nielsen}},\ }\bibfield  {title} {\bibinfo {title} {Distance measures to
  compare real and ideal quantum processes},\ }\href
  {https://doi.org/10.1103/PhysRevA.71.062310} {\bibfield  {journal} {\bibinfo
  {journal} {Phys. Rev. A}\ }\textbf {\bibinfo {volume} {71}},\ \bibinfo
  {pages} {062310} (\bibinfo {year} {2005})}\BibitemShut {NoStop}%
\bibitem [{\citenamefont {Sakurai}\ and\ \citenamefont
  {Napolitano}(2011)}]{sakurai:2011}%
  \BibitemOpen
  \bibfield  {author} {\bibinfo {author} {\bibfnamefont {J.~J.}\ \bibnamefont
  {Sakurai}}\ and\ \bibinfo {author} {\bibfnamefont {J.}~\bibnamefont
  {Napolitano}},\ }\href {https://cds.cern.ch/record/1341875} {\emph {\bibinfo
  {title} {{Modern quantum mechanics; 2nd ed.}}}}\ (\bibinfo  {publisher}
  {Addison-Wesley},\ \bibinfo {address} {San Francisco, CA},\ \bibinfo {year}
  {2011})\BibitemShut {NoStop}%
\bibitem [{\citenamefont {Farhi}\ and\ \citenamefont
  {Gutmann}(1998)}]{farhi:1998}%
  \BibitemOpen
  \bibfield  {author} {\bibinfo {author} {\bibfnamefont {E.}~\bibnamefont
  {Farhi}}\ and\ \bibinfo {author} {\bibfnamefont {S.}~\bibnamefont
  {Gutmann}},\ }\bibfield  {title} {\bibinfo {title} {Analog analogue of a
  digital quantum computation},\ }\href
  {https://doi.org/10.1103/PhysRevA.57.2403} {\bibfield  {journal} {\bibinfo
  {journal} {Phys. Rev. A}\ }\textbf {\bibinfo {volume} {57}},\ \bibinfo
  {pages} {2403} (\bibinfo {year} {1998})}\BibitemShut {NoStop}%
\bibitem [{\citenamefont {Blanes}\ \emph {et~al.}(2009)\citenamefont {Blanes},
  \citenamefont {Casas}, \citenamefont {Oteo},\ and\ \citenamefont
  {Ros}}]{blanes:2009}%
  \BibitemOpen
  \bibfield  {author} {\bibinfo {author} {\bibfnamefont {S.}~\bibnamefont
  {Blanes}}, \bibinfo {author} {\bibfnamefont {F.}~\bibnamefont {Casas}},
  \bibinfo {author} {\bibfnamefont {J.}~\bibnamefont {Oteo}},\ and\ \bibinfo
  {author} {\bibfnamefont {J.}~\bibnamefont {Ros}},\ }\bibfield  {title}
  {\bibinfo {title} {The {M}agnus expansion and some of its applications},\
  }\href {https://doi.org/https://doi.org/10.1016/j.physrep.2008.11.001}
  {\bibfield  {journal} {\bibinfo  {journal} {Physics Reports}\ }\textbf
  {\bibinfo {volume} {470}},\ \bibinfo {pages} {151} (\bibinfo {year}
  {2009})}\BibitemShut {NoStop}%
\bibitem [{\citenamefont {Zueco}\ \emph {et~al.}(2009)\citenamefont {Zueco},
  \citenamefont {Reuther}, \citenamefont {Kohler},\ and\ \citenamefont
  {H\"anggi}}]{zueco:2009}%
  \BibitemOpen
  \bibfield  {author} {\bibinfo {author} {\bibfnamefont {D.}~\bibnamefont
  {Zueco}}, \bibinfo {author} {\bibfnamefont {G.~M.}\ \bibnamefont {Reuther}},
  \bibinfo {author} {\bibfnamefont {S.}~\bibnamefont {Kohler}},\ and\ \bibinfo
  {author} {\bibfnamefont {P.}~\bibnamefont {H\"anggi}},\ }\bibfield  {title}
  {\bibinfo {title} {Qubit-oscillator dynamics in the dispersive regime:
  Analytical theory beyond the rotating-wave approximation},\ }\href
  {https://doi.org/10.1103/PhysRevA.80.033846} {\bibfield  {journal} {\bibinfo
  {journal} {Phys. Rev. A}\ }\textbf {\bibinfo {volume} {80}},\ \bibinfo
  {pages} {033846} (\bibinfo {year} {2009})}\BibitemShut {NoStop}%
\bibitem [{\citenamefont {Childs}\ \emph {et~al.}(2003)\citenamefont {Childs},
  \citenamefont {Cleve}, \citenamefont {Deotto}, \citenamefont {Farhi},
  \citenamefont {Gutmann},\ and\ \citenamefont {Spielman}}]{childs:2003}%
  \BibitemOpen
  \bibfield  {author} {\bibinfo {author} {\bibfnamefont {A.~M.}\ \bibnamefont
  {Childs}}, \bibinfo {author} {\bibfnamefont {R.}~\bibnamefont {Cleve}},
  \bibinfo {author} {\bibfnamefont {E.}~\bibnamefont {Deotto}}, \bibinfo
  {author} {\bibfnamefont {E.}~\bibnamefont {Farhi}}, \bibinfo {author}
  {\bibfnamefont {S.}~\bibnamefont {Gutmann}},\ and\ \bibinfo {author}
  {\bibfnamefont {D.~A.}\ \bibnamefont {Spielman}},\ }\bibfield  {title}
  {\bibinfo {title} {Exponential algorithmic speedup by a quantum walk},\ }in\
  \href {https://doi.org/10.1145/780542.780552} {\emph {\bibinfo {booktitle}
  {Proceedings of the Thirty-Fifth Annual ACM Symposium on Theory of
  Computing}}},\ \bibinfo {series and number} {STOC '03}\ (\bibinfo
  {publisher} {Association for Computing Machinery},\ \bibinfo {address} {New
  York, NY, USA},\ \bibinfo {year} {2003})\ p.\ \bibinfo {pages}
  {59–68}\BibitemShut {NoStop}%
\bibitem [{\citenamefont {Kechedzhi}\ \emph {et~al.}(2018)\citenamefont
  {Kechedzhi}, \citenamefont {Smelyanskiy}, \citenamefont {McClean},
  \citenamefont {Denchev}, \citenamefont {Mohseni}, \citenamefont {Isakov},
  \citenamefont {Boixo}, \citenamefont {Altshuler},\ and\ \citenamefont
  {Neven}}]{kechedzhi:2018}%
  \BibitemOpen
  \bibfield  {author} {\bibinfo {author} {\bibfnamefont {K.}~\bibnamefont
  {Kechedzhi}}, \bibinfo {author} {\bibfnamefont {V.}~\bibnamefont
  {Smelyanskiy}}, \bibinfo {author} {\bibfnamefont {J.~R.}\ \bibnamefont
  {McClean}}, \bibinfo {author} {\bibfnamefont {V.~S.}\ \bibnamefont
  {Denchev}}, \bibinfo {author} {\bibfnamefont {M.}~\bibnamefont {Mohseni}},
  \bibinfo {author} {\bibfnamefont {S.}~\bibnamefont {Isakov}}, \bibinfo
  {author} {\bibfnamefont {S.}~\bibnamefont {Boixo}}, \bibinfo {author}
  {\bibfnamefont {B.}~\bibnamefont {Altshuler}},\ and\ \bibinfo {author}
  {\bibfnamefont {H.}~\bibnamefont {Neven}},\ }\bibfield  {title} {\bibinfo
  {title} {{Efficient Population Transfer via Non-Ergodic Extended States in
  Quantum Spin Glass}},\ }in\ \href {https://doi.org/10.4230/LIPIcs.TQC.2018.9}
  {\emph {\bibinfo {booktitle} {13th Conference on the Theory of Quantum
  Computation, Communication and Cryptography (TQC 2018)}}},\ \bibinfo {series}
  {Leibniz International Proceedings in Informatics (LIPIcs)}, Vol.\ \bibinfo
  {volume} {111},\ \bibinfo {editor} {edited by\ \bibinfo {editor}
  {\bibfnamefont {S.}~\bibnamefont {Jeffery}}}\ (\bibinfo  {publisher} {Schloss
  Dagstuhl--Leibniz-Zentrum fuer Informatik},\ \bibinfo {address} {Dagstuhl,
  Germany},\ \bibinfo {year} {2018})\ pp.\ \bibinfo {pages}
  {9:1--9:16}\BibitemShut {NoStop}%
\bibitem [{\citenamefont {Smelyanskiy}\ \emph {et~al.}(2020)\citenamefont
  {Smelyanskiy}, \citenamefont {Kechedzhi}, \citenamefont {Boixo},
  \citenamefont {Isakov}, \citenamefont {Neven},\ and\ \citenamefont
  {Altshuler}}]{smelyanskiy:2020}%
  \BibitemOpen
  \bibfield  {author} {\bibinfo {author} {\bibfnamefont {V.~N.}\ \bibnamefont
  {Smelyanskiy}}, \bibinfo {author} {\bibfnamefont {K.}~\bibnamefont
  {Kechedzhi}}, \bibinfo {author} {\bibfnamefont {S.}~\bibnamefont {Boixo}},
  \bibinfo {author} {\bibfnamefont {S.~V.}\ \bibnamefont {Isakov}}, \bibinfo
  {author} {\bibfnamefont {H.}~\bibnamefont {Neven}},\ and\ \bibinfo {author}
  {\bibfnamefont {B.}~\bibnamefont {Altshuler}},\ }\bibfield  {title} {\bibinfo
  {title} {Nonergodic delocalized states for efficient population transfer
  within a narrow band of the energy landscape},\ }\href
  {https://doi.org/10.1103/PhysRevX.10.011017} {\bibfield  {journal} {\bibinfo
  {journal} {Phys. Rev. X}\ }\textbf {\bibinfo {volume} {10}},\ \bibinfo
  {pages} {011017} (\bibinfo {year} {2020})}\BibitemShut {NoStop}%
\bibitem [{\citenamefont {Smelyanskiy}\ \emph {et~al.}(2019)\citenamefont
  {Smelyanskiy}, \citenamefont {Kechedzhi}, \citenamefont {Boixo},
  \citenamefont {Neven},\ and\ \citenamefont {Altshuler}}]{smelyanskiy:2019}%
  \BibitemOpen
  \bibfield  {author} {\bibinfo {author} {\bibfnamefont {V.~N.}\ \bibnamefont
  {Smelyanskiy}}, \bibinfo {author} {\bibfnamefont {K.}~\bibnamefont
  {Kechedzhi}}, \bibinfo {author} {\bibfnamefont {S.}~\bibnamefont {Boixo}},
  \bibinfo {author} {\bibfnamefont {H.}~\bibnamefont {Neven}},\ and\ \bibinfo
  {author} {\bibfnamefont {B.}~\bibnamefont {Altshuler}},\ }\bibfield  {title}
  {\bibinfo {title} {Intermittency of dynamical phases in a quantum spin
  glass},\ }\href {https://arxiv.org/abs/1907.01609} {\bibfield  {journal}
  {\bibinfo  {journal} {arXiv:1907.01609}\ } (\bibinfo {year}
  {2019})}\BibitemShut {NoStop}%
\bibitem [{\citenamefont {Baldwin}\ and\ \citenamefont
  {Laumann}(2018)}]{baldwin:2018}%
  \BibitemOpen
  \bibfield  {author} {\bibinfo {author} {\bibfnamefont {C.~L.}\ \bibnamefont
  {Baldwin}}\ and\ \bibinfo {author} {\bibfnamefont {C.~R.}\ \bibnamefont
  {Laumann}},\ }\bibfield  {title} {\bibinfo {title} {Quantum algorithm for
  energy matching in hard optimization problems},\ }\href
  {https://doi.org/10.1103/PhysRevB.97.224201} {\bibfield  {journal} {\bibinfo
  {journal} {Phys. Rev. B}\ }\textbf {\bibinfo {volume} {97}},\ \bibinfo
  {pages} {224201} (\bibinfo {year} {2018})}\BibitemShut {NoStop}%
\bibitem [{\citenamefont {Crosson}\ and\ \citenamefont
  {Lidar}(2021)}]{crosson:2021}%
  \BibitemOpen
  \bibfield  {author} {\bibinfo {author} {\bibfnamefont {E.}~\bibnamefont
  {Crosson}}\ and\ \bibinfo {author} {\bibfnamefont {D.}~\bibnamefont
  {Lidar}},\ }\bibfield  {title} {\bibinfo {title} {Prospects for quantum
  enhancement with diabatic quantum annealing},\ }\href@noop {} {\bibfield
  {journal} {\bibinfo  {journal} {Nature Reviews Physics}\ ,\ \bibinfo {pages}
  {1}} (\bibinfo {year} {2021})}\BibitemShut {NoStop}%
\bibitem [{\citenamefont {Sels}\ and\ \citenamefont
  {Polkovnikov}(2017)}]{sels:2017}%
  \BibitemOpen
  \bibfield  {author} {\bibinfo {author} {\bibfnamefont {D.}~\bibnamefont
  {Sels}}\ and\ \bibinfo {author} {\bibfnamefont {A.}~\bibnamefont
  {Polkovnikov}},\ }\bibfield  {title} {\bibinfo {title} {Minimizing
  irreversible losses in quantum systems by local counterdiabatic driving},\
  }\href {https://doi.org/10.1073/pnas.1619826114} {\bibfield  {journal}
  {\bibinfo  {journal} {Proceedings of the National Academy of Sciences}\
  }\textbf {\bibinfo {volume} {114}},\ \bibinfo {pages} {E3909} (\bibinfo
  {year} {2017})}\BibitemShut {NoStop}%
\bibitem [{\citenamefont {Kolodrubetz}\ \emph {et~al.}(2017)\citenamefont
  {Kolodrubetz}, \citenamefont {Sels}, \citenamefont {Mehta},\ and\
  \citenamefont {Polkovnikov}}]{kolodrubetz:2017}%
  \BibitemOpen
  \bibfield  {author} {\bibinfo {author} {\bibfnamefont {M.}~\bibnamefont
  {Kolodrubetz}}, \bibinfo {author} {\bibfnamefont {D.}~\bibnamefont {Sels}},
  \bibinfo {author} {\bibfnamefont {P.}~\bibnamefont {Mehta}},\ and\ \bibinfo
  {author} {\bibfnamefont {A.}~\bibnamefont {Polkovnikov}},\ }\bibfield
  {title} {\bibinfo {title} {Geometry and non-adiabatic response in quantum and
  classical systems},\ }\href
  {https://doi.org/https://doi.org/10.1016/j.physrep.2017.07.001} {\bibfield
  {journal} {\bibinfo  {journal} {Physics Reports}\ }\textbf {\bibinfo {volume}
  {697}},\ \bibinfo {pages} {1} (\bibinfo {year} {2017})}\BibitemShut {NoStop}%
\end{thebibliography}%

\end{document}


\renewcommand{\theequation}{S\arabic{equation}}
\renewcommand{\thefigure}{S\arabic{figure}}
\renewcommand{\thetable}{S\arabic{table}}
\renewcommand{\thealgocf}{S\arabic{algocf}}

\title{Supplementary information for:\\
``Quantum-enhanced Markov chain Monte Carlo''}

\author{David Layden}
\email{david.layden@ibm.com}
\affiliation{IBM Quantum, Almaden Research Center, San Jose, California 95120, USA}

\author{Guglielmo Mazzola}
\affiliation{IBM Quantum, IBM Research -- Zurich, 8803 R\"uschlikon, Switzerland}

\author{Ryan V. Mishmash}
\affiliation{IBM Quantum, Almaden Research Center, San Jose, California 95120, USA}

\author{Mario Motta} 
\affiliation{IBM Quantum, Almaden Research Center, San Jose, California 95120, USA}

\author{Pawel Wocjan}
\affiliation{IBM Quantum, T. J. Watson Research Center, Yorktown Heights, NY 10598, USA}

\author{Jin-Sung Kim}
\altaffiliation[Current affiliation: ]{NVIDIA}
\affiliation{IBM Quantum, Almaden Research Center, San Jose, California 95120, USA}

\author{Sarah Sheldon}
\affiliation{IBM Quantum, Almaden Research Center, San Jose, California 95120, USA}

\maketitle
\tableofcontents

\setlength{\parindent}{0em} 
\setlength{\parskip}{1em} 

\section{Notation}

We will use $\boldsymbol{P}$ to denote a $2^n \times 2^n$ transition matrix formed from transition probabilities $\{ P(\boldsymbol{s'} | \, \boldsymbol{s}) \}$ in lexicographic order, and likewise $\boldsymbol{Q}$ for a $2^n \times 2^n$ matrix of proposal probabilities $\{ Q(\boldsymbol{s'} | \, \boldsymbol{s}) \}$. We will also use $\vec{\mu} \in \mathbb{R}^{2^n}$ for the vector of probabilities $\{ \mu(\boldsymbol{s}) \}$ with the same ordering. To simplify the notation, we will index both with integers in $[0, 2^n-1]$ whose binary representation encodes a spin configuration, e.g.:
\begin{center}
\begin{tabular}{c|c}
Integer & Spin configuration \\
\hline
$j=0$ & $\boldsymbol{s} = (1, \dots, 1, 1)$\\ 
$j=1$ & $\boldsymbol{s} = (1, \dots, 1, -1)$\\ 
$j=2$ & $\boldsymbol{s} = (1, \dots, -1, 1)$\\ 
$\vdots$ & $\vdots$ \\ 
$j=2^n-1$ \vspace{1ex} & \vspace{1ex} $\boldsymbol{s} = (-1, \dots, -1, -1)$
\end{tabular}
\end{center}
We will refer to a spin configuration by its vector ($\boldsymbol{s} \in \{1, -1\}^n$) or integer ($j \in [0, 2^n-1]$) representation interchangeably. Moreover, to be consistent with the convention used in quantum mechanics, we will treat $\vec{\mu}$ as a column vector, and $\boldsymbol{P}$ and $\boldsymbol{Q}$ as left-stochastic matrices, meaning their columns sum to 1.

\section{Algorithm details}

Our full algorithm is shown in Algorithm~\ref{alg:M-H}. This version uses the Metropolis-Hastings (M-H) acceptance probability from Eq.~\eqref{eq:MH_acceptance} of the main text on line \ref{algline:acceptance}, although any alternative satisfying detailed balance would also work. In Section \ref{secS:glauber}, for instance, we consider the Gibbs sampler acceptance probability, but find little difference in performance. Gauging MCMC convergence can be difficult in practice, and is usually done heuristically \cite{cowles:1996}. This is a property of MCMC in general, regardless of whether moves are proposed by classical or quantum means. For the sake of clarity then, Algorithm~\ref{alg:M-H} assumes that the user already has a convergence diagnostic in mind that is well-suited for the application at hand. We will occasionally write the Hamiltonian $H$ from Eq.~\eqref{eq:H} of the main text as $H(\gamma)$ to emphasize its dependence on the parameter $\gamma \in [0,1]$.

\begin{algorithm}[H]
\SetStartEndCondition{ }{}{}
\SetKwProg{Fn}{def}{\string:}{}\SetKwFunction{Range}{range}
\SetKw{KwTo}{in}\SetKwFor{For}{for}{\string:}{}
\SetKwIF{If}{ElseIf}{Else}{if}{:}{elif}{else:}{}
\SetKwFor{While}{while}{:}{fintq}

$\boldsymbol{s}$ = initial spin configuration (often chosen uniformly at random)\\
\BlankLine
 \While{not converged}{
 \BlankLine
  \textbf{Propose jump (quantum step 1)}\\
  $\gamma$ = random.uniform(0.25, 0.6) \label{algline:gamma_sample}\\
  $t$ = random.uniform(2, 20) \label{algline:t_sample}\\
  $\ket{\psi} = \exp[-i H(\gamma) \, t] \ket{\boldsymbol{s}}$ on quantum device\\
  $\boldsymbol{s'}$ = result of measuring $\ket{\psi}$ in computational basis\\
  \BlankLine
  \textbf{Accept/reject jump (classical step 2)}\\
  $A = \min(1, \,  e^{[E(\boldsymbol{s})-E(\boldsymbol{s'})]/T})$ \label{algline:acceptance} \\
  \If{$A \ge \mathrm{random.uniform}(0, 1)$}{$\boldsymbol{s} = \boldsymbol{s'}$}
}
 \caption{Quantum-enhanced Markov chain Monte Carlo}
 \label{alg:M-H}
\end{algorithm}

As discussed in the main text, we have not sought to optimize the free parameters $\gamma$ and $t$ in our quantum proposal mechanism, as the optimal values seem to have a complicated dependence on the model instance. Rather, we sampled them randomly in each MCMC iteration according to $\gamma \sim \text{uniform}([0.25, 0.6])$ and $t \sim \text{uniform}([2, 20])$. We settled on these \textit{ad hoc} distributions through trial and error on a small number of fully-connected random model instances, and found them to be robust across a wide range of instances for various $n$ and model connectivities. For a given instance, certain values of $(\gamma, t)$ might lead to particularly good MCMC moves (e.g., between distant local minima) while others do not. Picking $(\gamma, t)$ at random ensures that these good values (or nearby ones) arise regularly, even if we don't know what they are \textit{a priori}. Accordingly, some fraction of the moves will be particularly good, which proves sufficient for a quantum enhancement in the absolute spectral gap $\delta$. This randomized strategy works reasonably well, although more sophisticated strategies are a promising area of future research. 

\subsection{Irreducibility and aperiodicity}
\label{secS:ergodic}
We begin with some basic definitions and results concerning Markov chains. A Markov chain with transition matrix $\boldsymbol{P}$ is said to be irreducible if for all $j,k \in [0, 2^n-1]$ there exists an integer $m>0$ (possibly depending on $j$ and $k$) such that $(\boldsymbol{P}^m)_{jk} > 0$. Let
\begin{equation}
\mathcal{M}(j) = \{ m \ge 1 \, | \,  (\boldsymbol{P}^m)_{jj} > 0 \},
\label{eqS:periodic}
\end{equation}
then the period of configuration $j$ is defined to be the greatest common divisor of $\mathcal{M}(j)$. A Markov chain is called aperiodic if all configurations $j$ have period 1, otherwise it is called periodic \cite{levin:2017} (Section 1.3). A distribution $\vec{\mu}$ is said to be stationary for a Markov chain if $\boldsymbol{P} \vec{\mu} = \vec{\mu}$. If $\vec{\mu}$ satisfies the detailed balance condition for a Markov chain, Eq.~\eqref{eq:detailed_balance} of the main text, then $\vec{\mu}$ is a stationary distribution of the chain \cite{levin:2017} (Proposition 1.19). An irreducible Markov chain has a unique stationary distribution $\vec{\mu}$ \cite{levin:2017} (Corollary 1.17). Finally, if a Markov chain is irreducible and aperiodic with stationary distribution $\vec{\mu}$ then any initial distribution of the chain will converge to $\vec{\mu}$ \cite{levin:2017} (Theorem 4.9). Putting this all together, an irreducible and aperiodic Markov chain that satisfies detailed balance with respect to $\vec{\mu}$ will converge to $\vec{\mu}$, as stated in the main text. This a powerful result, as it guarantees convergence to $\vec{\mu}$ for all initial states, under mild conditions, even if $\boldsymbol{P}$ is too large to diagonalize (or store in memory). 

Detailed balance is satisfied by design in our algorithm through the use of an appropriate acceptance probability, e.g., Eq.~\eqref{eq:MH_acceptance} of the main text. (In that equation we assumed for simplicity that $Q(\boldsymbol{s'}|\boldsymbol{s}) \neq 0$ for an $\boldsymbol{s}\rightarrow \boldsymbol{s'}$ proposal. If $Q(\boldsymbol{s'}|\boldsymbol{s})=0$ then the value of $A(\boldsymbol{s'}|\boldsymbol{s})$ is irrelevant since there will never be an opportunity to accept or reject a $\boldsymbol{s}\rightarrow \boldsymbol{s'}$ proposal to begin with.) What remains to be shown is that our algorithm produces an irreducible and aperiodic Markov chain.

Suppose that $Q(\boldsymbol{s'}| \boldsymbol{s})>0$ for all configurations $\boldsymbol{s}$ and $\boldsymbol{s'}$. Since $\mu(\boldsymbol{s})>0$ for all $\boldsymbol{s}$ at nonzero temperature, it follows from Eq.~\eqref{eq:MH_acceptance} of the main text that $A(\boldsymbol{s'}| \boldsymbol{s})>0$ and therefore $P(\boldsymbol{s'}| \boldsymbol{s})>0$ for all $\boldsymbol{s}$ and $\boldsymbol{s'}$. The corresponding Markov chain is therefore irreducible (with $m=1$) by definition. Similarly, since $Q(\boldsymbol{s}| \boldsymbol{s})>0$ by assumption and $A(\boldsymbol{s}| \boldsymbol{s})=1$ for all $\boldsymbol{s}$, $P(\boldsymbol{s}| \boldsymbol{s})>0$ so every state $\boldsymbol{s}$ has period 1 and the chain is aperiodic. In summary, a sufficient condition for the chain to be irreducible and aperiodic is for all the proposal probabilities $Q(\boldsymbol{s'}| \boldsymbol{s})$ to be nonzero. Formally, this could be guaranteed by proposing moves as follows: for some small $\epsilon>0$, propose the next configuration $\boldsymbol{s'}$ uniformly at random (i.e., use the uniform proposal) with probability $\epsilon$, otherwise propose $\boldsymbol{s'}$ using a quantum computer as in our algorithm. In practice, however, we found no need for this formality. Absent some special symmetry, no $\boldsymbol{s} \rightarrow \boldsymbol{s'}$ quantum transition should be completely forbidden. Considering also experimental noise, $Q(\boldsymbol{s'}|\boldsymbol{s})>0$ for all $\boldsymbol{s}$ and $\boldsymbol{s'}$ is all-but formally assured without having to incorporate the uniform proposal as described above. The resulting Markov chain therefore converges to $\vec{\mu}$ on a combination of physical and mathematical grounds.

\section{Numerical details}

This section describes the numerics underlying Figs.~\ref{fig:bulk_numerics}, \ref{fig:n=10_gap} and \ref{fig:mechanism} of the main text (not including the experimental data in Figs.~\ref{fig:n=10_gap} and \ref{fig:mechanism}, which is treated in Section~\ref{secS:experiment}), and also presents a range of complementary numerical results.

\subsection{Implementation}

We analyzed the spectral gap $\delta$ for the local, uniform, quantum and mismatched quantum proposals as follows. For a given model instance and temperature $T$, we explicitly computed every proposal probability $Q(\boldsymbol{s'}|\boldsymbol{s})$ and in turn each transition probability $P(\boldsymbol{s'}|\boldsymbol{s})$ (i.e., we did not approximate either through sampling). We then formed $\{P(\boldsymbol{s'}|\boldsymbol{s})\}$ into a $2^n \times 2^n$ transition matrix $\boldsymbol{P}$, which we numerically diagonalized to find $\delta$. We repeated this procedure for each $n$, $T$, and model instance to form Fig.~\ref{fig:bulk_numerics} of the main text. We applied the same steps for a single model instance in Fig.~\ref{fig:n=10_gap} of the main text. In Fig.~\ref{fig:mechanism} of the main text we only computed $Q(\boldsymbol{s'}|\boldsymbol{s})$, which does not depend on $T$. For the mismatched quantum proposal, we picked the ``mismatched'' parameters $\tilde{J}_{jk}$ and $\tilde{h}_j$ from the same (standard normal) distribution as the actual parameters $J_{jk}$ and $h_k$ which define the model instance. For each model instance we picked new mismatched parameters. This approach gives meaningful results over an ensemble of model instances, but not for individual instances, since there could be significant variation in the ``mismatched'' $\delta$ depending on which $\tilde{J}_{jk}$ and $\tilde{h}_j$ are drawn. We therefore did not consider the mismatched quantum proposal in Figs.~\ref{fig:n=10_gap} and \ref{fig:mechanism}, nor in similar analyses below.

For our quantum algorithm, we simulated noiseless dynamics generated by $H$. Viewed as a quantum channel $\mathcal{C}$ acting on an initial state $\rho = \ket{\boldsymbol{s}}\!\bra{\boldsymbol{s}}$, the quantum proposal mechanism described in Algorithm~\ref{alg:M-H} is
\begin{equation}
\mathcal{C}(\rho)
=
\frac{1}{(\gamma_\text{max}-\gamma_\text{min})(t_\text{max}-t_\text{min})}
\int_{\gamma_\text{min}}^{\gamma_\text{max}} d\gamma
\int_{t_\text{min}}^{t_\text{max}} dt 
\; \;
U(\gamma, t) \, \rho \, U(\gamma,t)^\dagger,
\label{eqS:channel}
\end{equation}
where $U(\gamma, t)=e^{-iH(\gamma)t}$, $[\gamma_\text{min}, \gamma_\text{max}]=[0.25, 0.6]$, and $[t_\text{min}, t_\text{max}]=[2,20]$. In our numerics, however, we implemented a slightly modified version of Algorithm~\ref{alg:M-H} for simplicity, corresponding to a different channel $\mathcal{C'} \approx \mathcal{C}$, where $\gamma$ is drawn from a discrete, rather than continuous, uniform distribution. The issue is that the integral over $t$ in Eq.~\eqref{eqS:channel} is easy to evaluate analytically, but we had to resort to numerical integration for the one over $\gamma$, which we approximated by a midpoint Riemann sum with 20 subintervals of equal size. (Increasing the number of subintervals made little difference to the results, suggesting a good approximation to Eq.~\eqref{eqS:channel}.) This amounts to sampling 
\begin{equation}
\gamma \sim \text{uniform}\left( \left\{0.25+ \frac{\Delta \gamma}{2},\;
0.25+ \frac{3 \Delta \gamma}{2} , \;
0.25+ \frac{5 \Delta \gamma}{2} , \;
\dots, \;
0.6 - \frac{\Delta \gamma}{2} 
\right\} \right)
\label{eqS:gamma_dist}
\end{equation}
for each MCMC step (Algorithm~\ref{alg:M-H} line~\ref{algline:gamma_sample}), with a subinterval width of $\Delta \gamma=0.0175$, rather than $\gamma \sim \text{uniform}([0.25, 0.6])$. Algorithm~\ref{alg:M-H} uses the latter distribution for conceptual simplicity, but in practice the choice seems to make little difference.

\subsection{Random instances and connectivity}
\label{secS:inst_and_conn}

To create Fig.~\ref{fig:bulk_numerics} of the main text and similar figures below, we generated random model instances by drawing coefficients $J_{jk}$ and $h_j$ that are independent and identically distributed (IID) from standard normal distributions (i.e., with zero mean and unit variance). Including non-zero fields $\{h_j\}$ of comparable size to the couplings serves to complicated the sampling problem: at low $T$, the spins seek to align with the local fields, but also to satisfy all the couplings. It is not typically possible to do both, and neither objective dominates. In effect, an $n$-spin system with non-zero fields $\{h_j\}$ is equivalent to an $(n+1)$-spin system without fields, where the original spins $j \in \{1, \dots, n \}$ are instead also coupled with strength $J_{n+1,j}=h_j$ to an additional spin that is held fixed in the +1 state (sometimes called a ``ghost spin''). For fully-connected models this makes little difference: a fully-connected $n$-spin model with fields amounts to a fully-connected $(n+1)$-spin model without fields. The difference is more pronounced when the connectivity is sparser, like in the instances we implemented experimentally. For example, an $n$-spin model with 1D nearest-neighbor connectivity and non-zero fields amounts to a field-less $(n+1)$-spin model which does not form a 1D lattice since each spin couples to its neighbors, as well as to a common spin.

Drawing $J_{jk}$ and $h_j$ from independent standard normal distributions gives an ensemble of model instances similar to the well-known Sherrington-Kirkpatrick (S-K) model \cite{sherrington:1975}, but with two main differences: (i) the original S-K model does not include the random local fields discussed above, and (ii) it draws $\{J_{jk}\}$ from normal distributions with standard deviation $1/\sqrt{n}$ rather than 1. (Or equivalently, it draws $\{J_{jk}\}$ from standard normal distributions but includes a $1/\sqrt{n}$ prefactor in the energy $E(\boldsymbol{s})$ that is not present in Eq.~\eqref{eq:E_function} of the main text.) Our variation of the S-K model is also common, see, e.g., Ref.~\cite{callison:2019}. We have chosen to use a distribution that does not depend on $n$ for two main reasons. First, the S-K $\;1/\sqrt{n}$ scaling is used frequently in the physics literature because it ensures that the $n\rightarrow \infty$ thermodynamic limit is well-behaved. However, coefficients arising in other applications (e.g., in Boltzmann machines) need not exhibit such finely-tuned scaling. The second reason has to do with connectivity. Interpreting $(J_{jk})$ as the adjacency matrix of a graph whose vertices are spins and whose edges are couplings, picking all $\{J_{jk}\}_{j>k}$ at random---without fixing any to be zero---gives a fully-connected graph (almost surely). This is the case we focused on in our numerics. Other graphs of lower degree (where some of the $J_{jk}$ are set to zero) are also of interest, however, such as the ones used in our experiments, where the spins form a 1D chain with nearest-neighbor connectivity. It is not always clear how the S-K $\;1/\sqrt{n}$ scaling should be generalized for models with different connectivity. We therefore chose to avoid such scaling altogether. However, it would be trivial to introduce this $1/\sqrt{n}$ scaling into our results by simply rescaling the temperature as $T \rightarrow T/\sqrt{n}$.

Note that the $1/\sqrt{n}$ scaling of the original S-K model arises naturally in the quantum part of our algorithm through the $\alpha = \|H_\text{mix}\|_\textsc{f} /\|H_\text{prob}\|_\textsc{f}$ scaling factor in Eq.~\eqref{eq:H} of the main text. To see how, suppose there are $N-n$ couplings $\{J_{jk}\}$ that are not fixed to 0, while the rest are IID as  $J_{jk}, h_j \sim \mathcal{N}(0,\sigma^2)$ for some standard deviation $\sigma$. Let $\vec{x} \in \mathbb{R}^N$ be the vector formed by stacking the fields $\{h_j\}$ and the non-zero couplings, which follows a multivariate normal distribution $p(\vec{x}) = (2 \pi \sigma^2)^{-N/2} \exp(-\|\vec{x} \|^2/2 \sigma^2)$. Integrating this probability density over $\vec{x} \in \mathbb{R}^N$ in $N$-dimensional spherical coordinates gives
\begin{equation}
1
=
\int d^N x \; p(\vec{x}) 
= (2 \pi \sigma^2)^{-N/2}
\int_0^\infty dr \, e^{-r^2/2\sigma^2} r^{N-1} \int d\Omega
=
\frac{\Gamma(N/2)}{2 \pi^{N/2}}  \int d\Omega,
\end{equation}
where $\Gamma$ denotes the gamma function, $r=\|\vec{x}\|$, and the differential solid angle $d\Omega$ is integrated over the full surface of an $N$-dimensional sphere. The average scaling factor can then be found by integrating $\alpha$ and solving for $\int d\Omega$ in the previous equation:
\begin{equation}
\langle \alpha \rangle_{J,h}
=
\sqrt{n} \int d^Nx \;
\| \vec{x} \|^{-1} p(\vec{x}) \
=
\frac{\sqrt{n}}{(2\pi\sigma^2)^{N/2}} \int_0^\infty dr e^{-r^2/2\sigma^2} r^{N-2} \int d\Omega
=
\frac{\sqrt{n}}{\sigma \sqrt{2}} \, \Gamma \left(\frac{N-1}{2} \right) \Big/ \Gamma \left( \frac{N}{2} \right).
\end{equation}
For a fully-connected model (depicted in Fig.~\ref{fig:overview}a of the main text) $N=n + \binom{n}{2}$, so asymptotically
\begin{equation}
\langle \alpha \rangle_{J,h}
= \frac{\sqrt{2}}{\sigma \sqrt{n}} + O\left( \sigma^{-1} n^{-3/2} \right).
\end{equation}
Setting $\sigma=1$ as in our numerics, $H_\text{prob}$, and in turn $E(\boldsymbol{s})$, gets scaled by a factor of $\sqrt{2/n}$ on average in Eq.~\eqref{eq:H}. Had we used the S-K convention of $\sigma=1/\sqrt{n}$, $H_\text{prob}$ would only get scaled be $\sqrt{2}$, producing the same result. In other words, for the quantum part of our algorithm it doesn't matter whether we use the conventional $1/\sqrt{n}$ S-K scaling since it arises naturally in Eq.~\eqref{eq:H} regardless. 

For models with different connectivity the story is different. A 1D chain of $n$ spins with nearest-neighbor couplings (depicted in Fig.~\ref{fig:overview}b of the main text), as in our experiments, has $N=2n-1$ so
\begin{equation}
\langle \alpha \rangle_{J,h}
=
\frac{1}{\sigma \sqrt{2}} + O\left (\sigma^{-1} n^{-1} \right),
\end{equation}
meaning the scaling factor in Eq.~\eqref{eq:H} tends towards $1/\sqrt{2}$ on average for large $n$ when $\sigma=1$. We give numerical results for such 1D model instances in the next section.

\subsection{Supplemental data}

\subsubsection{1D model instances} 

\begin{figure}[h]
\includegraphics{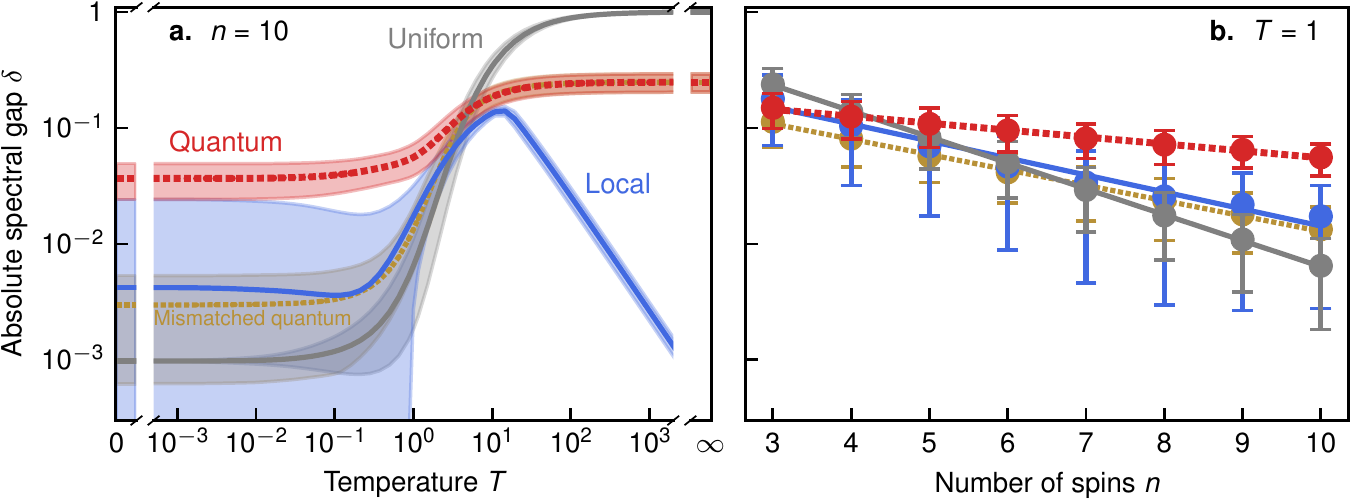}
\hfill
\raisebox{8.5em}{
\begin{tabular}{ c c }
\multicolumn{2}{c}{
\textbf{c.} $\boldsymbol{\;\langle \delta \rangle \propto 2^{-kn}}$ \textbf{fits} } \\
\hline
\color{tab:red} Quantum & $\color{tab:red}k=0.198(5)$\\
\color{bronze} Mismatched & \multirow{2}{*}{$\color{bronze} k=0.440(9)$} \\[-1ex]
\color{bronze} quantum & \\
\color{royalblue} Local & $ \color{royalblue} k=0.49(3)$ \\ 
\color{mpl_gray} Uniform & $\color{mpl_gray}  k=0.743(4)$\\
\hline 
\multicolumn{2}{c}{
Quantum enhancement}\\[-1ex]
\multicolumn{2}{c}{
factor in $k$: 2.5(2)}

\end{tabular}
}
\caption{\textbf{Average-case convergence rate simulations for 1D model instances.} Analogous data to that shown in Fig.~\ref{fig:bulk_numerics} of the main text, but for 1D model instances rather than fully-connected ones.  All strategies were simulated classically. Lines/markers show the average $\delta$ over 500 random 1D Ising model instances for each $n$; error bands/bars show the standard deviation in $\delta$ over these instances. Dotted lines are for visibility. \textbf{a.} The slow-down of each strategy at low $T$. The local proposal strategy performs better on average here than in Fig.~\ref{fig:bulk_numerics}a due to the larger fraction of easy instances. The change in connectivity has only a minor effect on $\delta$ for the other strategies. \textbf{b.} Problem size dependence, with least squares exponential fits to the average $\delta$, weighted by the standard error of the mean. \textbf{c.} The resulting fit parameters and the average quantum enhancement exponent, which is the ratio of $k$ for the quantum algorithm and the smallest $k$ among classical proposal strategy (the local strategy, here). Uncertainties are from the fit covariance matrices.}
\label{figS:bulk_numerics_line}
\end{figure}

In the main text we describe our analysis of $\delta$ over many random fully-connected model instances, summarized in Fig.~\ref{fig:bulk_numerics}. We also performed a similar analysis for 1D instances (with open boundaries), which we present here. We picked $J_{j+1,j}$ and $h_j$ IID from standard normal distributions, $\mathcal{N}(0,1)$, and fixed all other $J_{jk}$ to zero. Fig.~\ref{figS:bulk_numerics_line} shows the analogous results to Fig.~\ref{fig:bulk_numerics} of the main text but for 500 random instances per $n$ with 1D nearest-neighbor connectivity. The mismatched quantum proposal also uses $\tilde{J}_{jk}$ with 1D connectivity; that is, $\tilde{J}_{j+1,j}, \tilde{h}_j \sim \mathcal{N}(0,1)$ all IID, and all other $\tilde{J}_{jk}=0$. For clarity, all other settings are identical to those used in Fig.~\ref{fig:bulk_numerics}. The average $\delta$ (which we denote $\expval{\delta}$) vs. $T$ curve for our quantum algorithm in Fig.~\ref{figS:bulk_numerics_line}a remains very similar to that in Fig.~\ref{fig:bulk_numerics}a for fully-connected instances. The main difference is in $\expval{\delta}$ for the local proposal, which is much larger on 1D instances than on fully-connected ones. This is expected: sparser connectivity reduces the potential for frustration, thus increasing the fraction of easy instances. The scaling advantage of $\expval{\delta}$ versus $n$ at $T=1$ from the main text persists in Fig.~\ref{figS:bulk_numerics_line}b, though it is less pronounced due to enhanced performance of the local proposal. Note, finally, that care must be taken when comparing such plots across different model connectivities, since the typical scale of $E(\boldsymbol{s})$ depends on the number of nonzero $J_{jk}$ coefficients. This effectively rescales the temperature in a way that depends on $n$, thus introducing some ambiguity when comparing results for different connectivities at a fixed $T$.

\subsubsection{Different $n$ and $T$}

\begin{figure}[h!]
\centering
\includegraphics[width=0.32\textwidth]{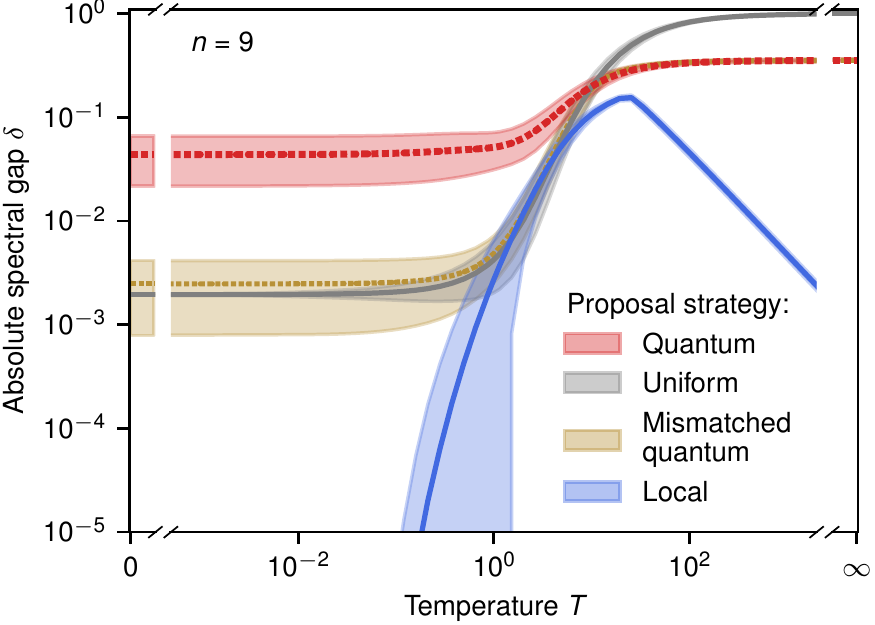} \hfill
\includegraphics[width=0.32\textwidth]{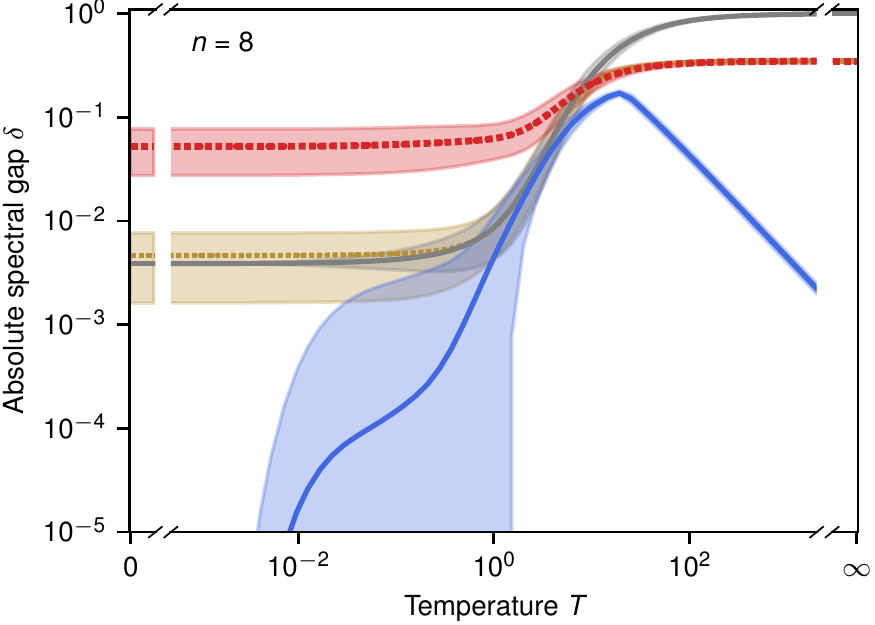} \hfill
\includegraphics[width=0.32\textwidth]{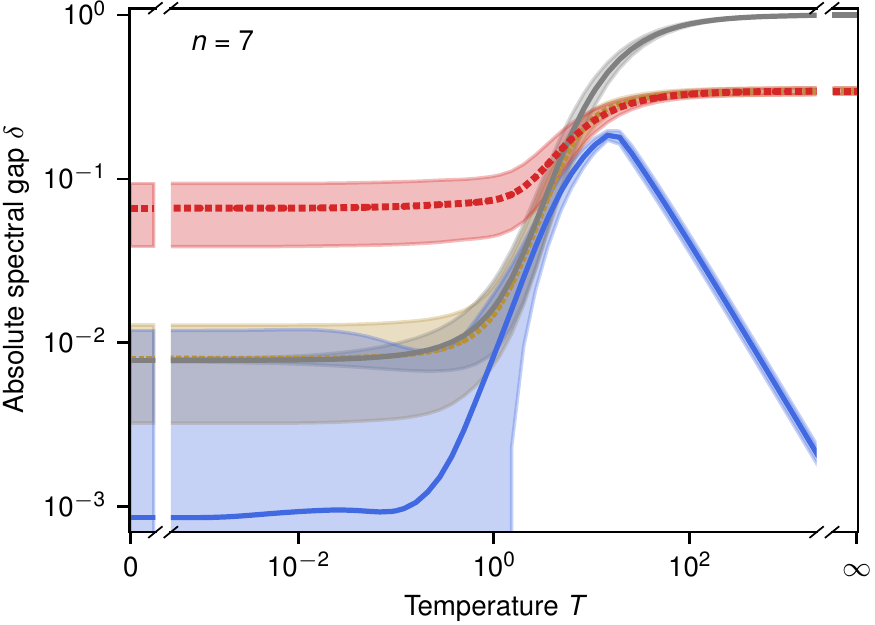}\\
\includegraphics[width=0.32\textwidth]{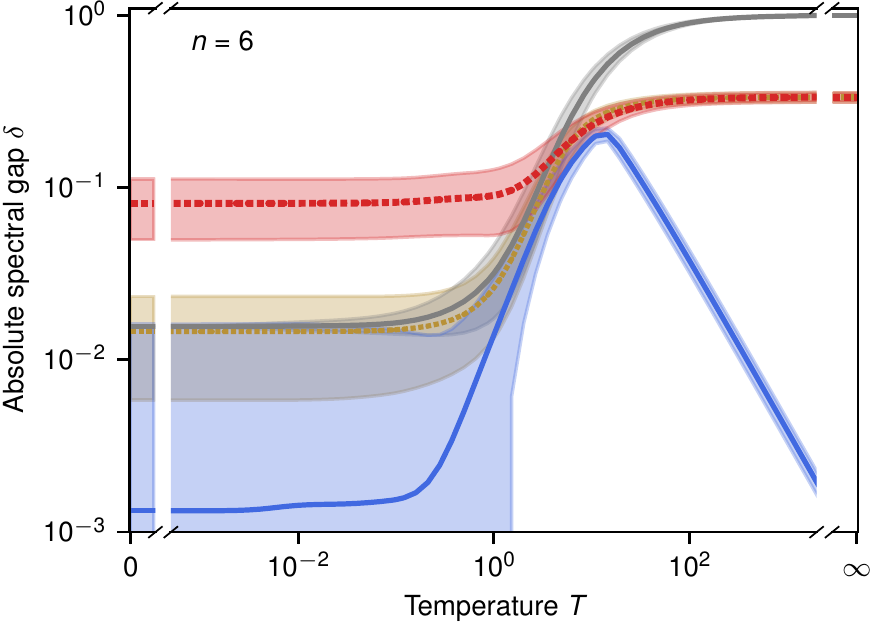} \hfill
\includegraphics[width=0.32\textwidth]{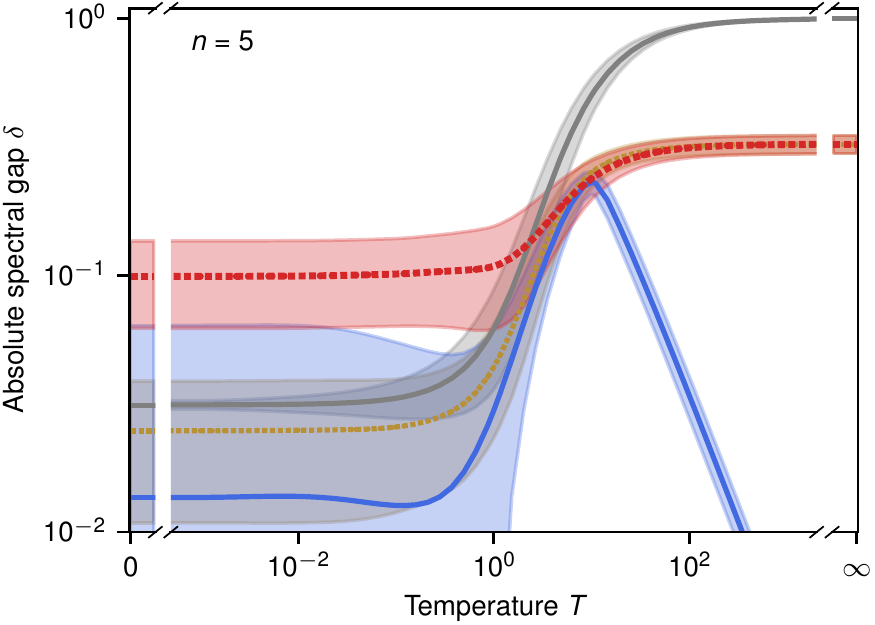} \hfill
\includegraphics[width=0.32\textwidth]{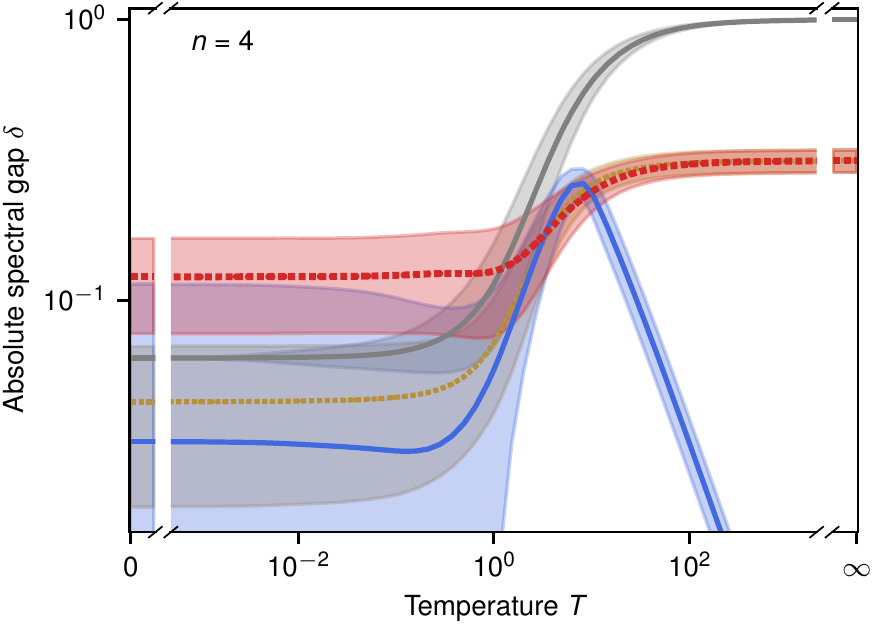}
\caption{\textbf{Average-case M-H convergence rate simulations for different $\boldsymbol{n}$.} Analogous plots to Fig.~\ref{fig:bulk_numerics}a from the main text, but for different values of $n$. All strategies were simulated classically. Lines show the average $\delta$ over 500 random fully-connected Ising model instances for each $n$; error bands show the standard deviation in $\delta$ over these instances. Dotted lines are for visibility. The same model instances were used to make Fig.~\ref{fig:bulk_numerics} of the main text.}
\label{figS:delta_vs_T_MH}
\end{figure}

\begin{figure}[h!]
\centering
\includegraphics[width=0.32\textwidth]{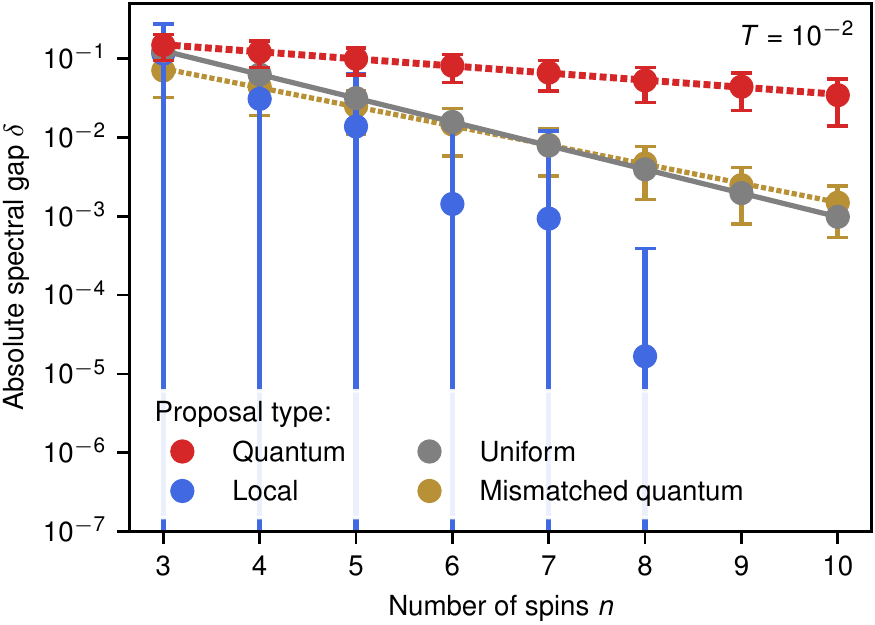} \hfill
\includegraphics[width=0.32\textwidth]{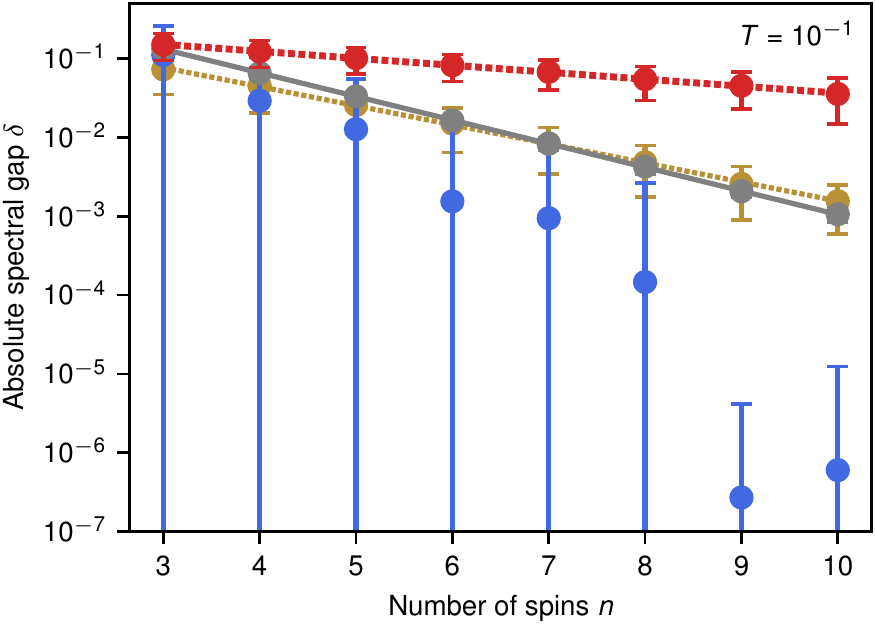} \hfill
\includegraphics[width=0.32\textwidth]{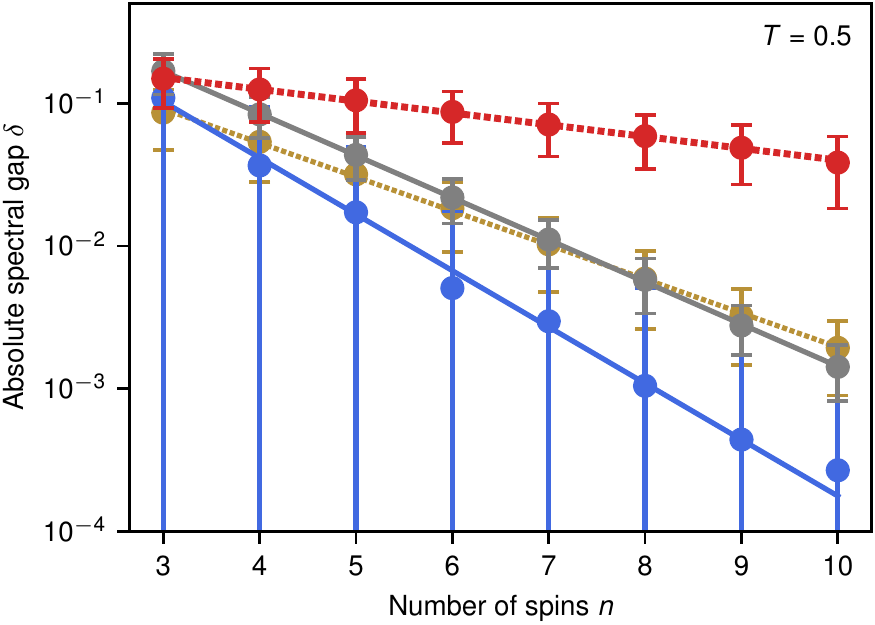}\\ 
\includegraphics[width=0.32\textwidth]{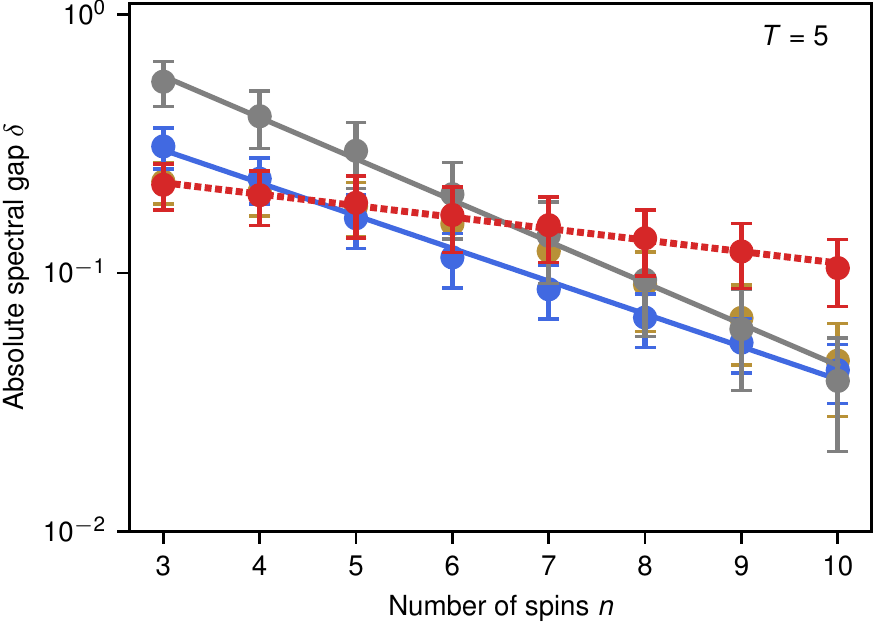} \hfill
\includegraphics[width=0.32\textwidth]{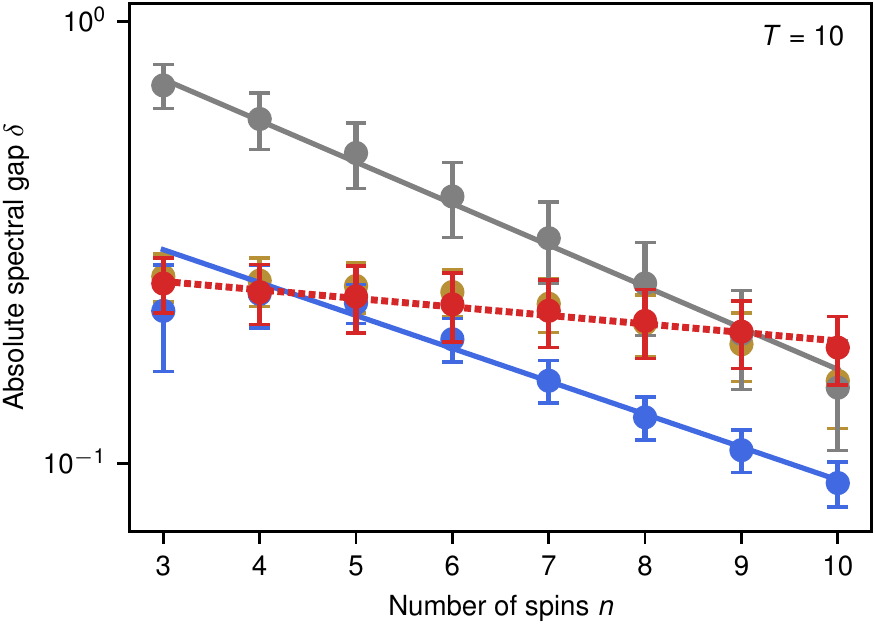} \hfill
\includegraphics[width=0.32\textwidth]{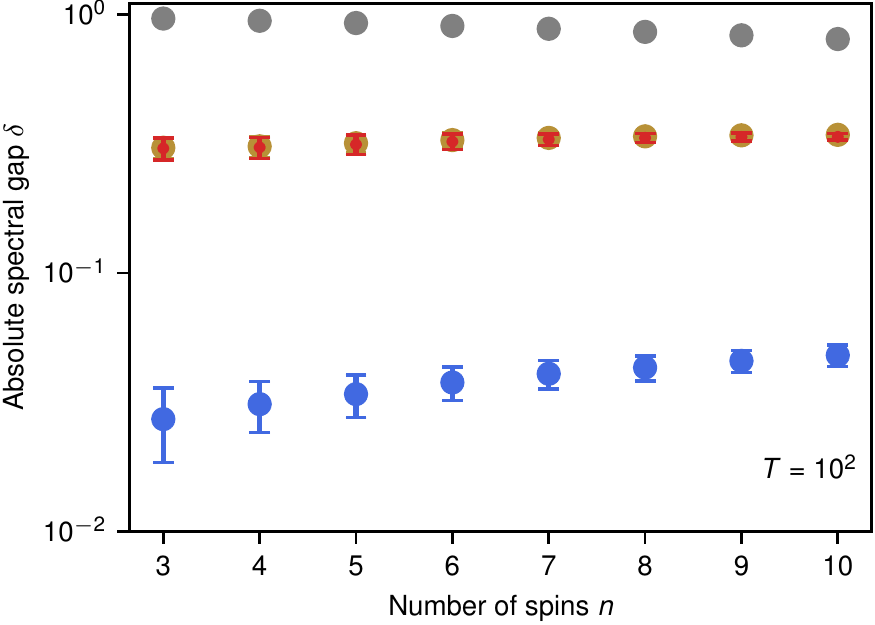}
\caption{\textbf{Average-case M-H convergence rate simulations for different $\boldsymbol{T}$.} Analogous plots to Fig.~\ref{fig:bulk_numerics}b from the main text, but for different values of $T$. All strategies were simulated classically. Markers show the average $\delta$ over 500 random fully-connected Ising model instances for each $n$; error bars show the standard deviation in $\delta$ over these instances. Lines show least squares exponential fits $\expval{\delta} \propto 2^{-kn}$, weighted by the standard error of the mean, wherever such fits are reasonably good. The resulting values of $k$ are given in Table~\ref{tabS:fits_MH}. Dotted lines are for visibility. The same model instances were used to make Fig.~\ref{fig:bulk_numerics} of the main text.}
\label{figS:delta_vs_n_MH}
\end{figure}

We now return to fully-connected model instances, and further analyze the numerical data underlying Fig.~\ref{fig:bulk_numerics} of the main text. Two parameters, $n$ and $T$, control the average problem difficulty in our numerics. Fig.~\ref{fig:bulk_numerics} in the main text shows two representative slices through $n$-$T$ parameter space. Figs.~\ref{figS:delta_vs_T_MH} and \ref{figS:delta_vs_n_MH} show the same quantities for other values of $n$ and $T$. Each panel of  Fig.~\ref{figS:delta_vs_T_MH} is qualitatively similar to Fig.~\ref{fig:bulk_numerics}a of the main text, but they show a growing separation in $\delta$ with increasing $n$ at low $T$ between our quantum algorithm and the purely classical Markov chains. Similarly, the panels of Fig.~\ref{figS:delta_vs_n_MH} show the onset of an apparent scaling advantage of $\delta$ with $n$, illustrated in Fig.~\ref{fig:bulk_numerics}b of the main text, which gets increasingly pronounced as $T$ decreases. The exponential fit parameters (as in Fig.~\ref{fig:bulk_numerics}c of the main text) for Fig.~\ref{figS:delta_vs_n_MH} are given in Table~\ref{tabS:fits_MH}.

\newpage

\begin{table}[h!]
\setlength{\tabcolsep}{8pt}
\begin{center}
\begin{tabular}{| c | c c c c c c|}
\hline
Proposal type & $T=10^{-2}$ &  $T=10^{-1}$ &  $T=0.5$ & $T=5$ & $T=10$ & $T=10^2$\\
\hline
Quantum & $k=0.299(3)$ & $k=0.293(2)$ & $k=0.275(4)$ & $k=0.147(6)$ & $k=0.064(4)$ & -\\
Mismatched & \multirow{2}{*}{$k=0.806(9)$} & \multirow{2}{*}{$k=0.803(8)$} & \multirow{2}{*}{$k=0.792(9)$} & \multirow{2}{*}{-} &  \multirow{2}{*}{-}&  \multirow{2}{*}{-} \\[-1ex]
quantum & & & & & & \\
Local & - & - & $k=1.31(4)$&  $\boldsymbol{k=0.42(1)}$ & $\boldsymbol{k=0.25(2)}$  & -\\ 
Uniform & $\boldsymbol{ k=0.9999(2)}$ & $\boldsymbol{k=0.995(3)}$ & $\boldsymbol{k=0.981(4)}$ & $k=0.53(2)$ & $k=0.31(1)$ & -\\
\hline 
Quantum & \multirow{2}{*}{3.35(3)} & \multirow{2}{*}{3.39(3)} & \multirow{2}{*}{3.57(6)} & \multirow{2}{*}{2.8(1)} & \multirow{2}{*}{3.9(4)} & \multirow{2}{*}{-} \\[-1ex] 
enhancement & &&&&&\\
\hline
\end{tabular}
\end{center}
\caption{\textbf{M-H $\boldsymbol{\expval{\delta}}$ vs.\ $\boldsymbol{n}$ fits on fully-connected instances.} The exponential fit parameters $k$ from $\expval{\delta} \propto 2^{-kn}$ in Fig.~\ref{figS:delta_vs_n_MH} for different proposal strategies, wherever such fits are reasonably good. The resulting average quantum enhancement exponent for each $T$, which is the ratio of $k$ for the quantum algorithm and the smallest $k$ among classical proposal strategies (shown in bold), is given in the bottom row. Uncertainties are from the fit covariance matrices.}
\label{tabS:fits_MH}
\end{table}

\newpage

\subsubsection{Gibbs sampler acceptance probability}
\label{secS:glauber}

The results presented above and in the main text use the Metropolis-Hastings (M-H) acceptance probability from Eq.~\eqref{eq:MH_acceptance}, which we will denote here as $A_\textsc{mh}$. As stated in the main text, however, there is a continuum of possible acceptance probabilities that satisfy detailed balance \cite{hastings:1970}. Yet, to the best of our knowledge, the only two that seem to be used frequently in practice are $A_\textsc{mh}$ and 
\begin{equation}
A_\textsc{g}(\boldsymbol{s'}|\boldsymbol{s}) = \left[ 
1+
\left( \frac{\mu(\boldsymbol{s'})}{\mu(\boldsymbol{s})}
\frac{Q(\boldsymbol{s}\,|\boldsymbol{s'})}{Q(\boldsymbol{s'}|\,\boldsymbol{s})} \right)^{-1}
\right]^{-1},
\label{eqS:gibbs_acceptance}
\end{equation}
which is variously named after Gibbs (hence the \textsc{g} subscript), Glauber, Boltzmann and Barker. The Markov chain resulting from $A_\textsc{g}$ with a local proposal is called a Gibbs sampler; we will therefore refer to Eq.~\eqref{eqS:gibbs_acceptance} as the Gibbs sampler acceptance probability. (The Gibbs sampler algorithm is sometimes expressed in a way that combines the proposal and accept/reject steps into a single step \cite{chen:2002, andrieu:2003}. In the context of the Ising model, however, this ``rejection-free'' version is equivalent to proposing local jumps and accepting them with probability $A_\textsc{g}$; see \cite{neal:1993} \S4.4.) Neither $A_\textsc{mh}$ nor $A_\textsc{g}$ is strictly better than the other. $A_\textsc{mh}$ is always larger than $A_\textsc{g}$, so it tends to produce slightly faster convergence at low $T$, where frequent rejections are the main bottleneck. (The difference is typically small, however, since the two approach each other as $T\rightarrow 0$.) We focused on M-H because we are primarily interested in the hard, low-$T$ regime. However, whereas $A_\textsc{mh} \rightarrow 1$ for all moves as $T\rightarrow \infty$, $A_\textsc{g} \rightarrow 1/2$ in the same limit. This can actually accelerate convergence compared to M-H, notably for local proposals at high $T$. The reason is that the M-H transition matrix for a local proposal at high $T$, $\boldsymbol{P}_\textsc{mh}$, is nearly periodic and has an eigenvalue near $-1$. Using the Gibbs sampler remedies this issue (cf. Eq.~\eqref{eqS:periodic}) and prevents the spectral gap from vanishing as $T\rightarrow \infty$. 

\begin{figure}[ht!]
\centering
\includegraphics[width=0.32\textwidth]{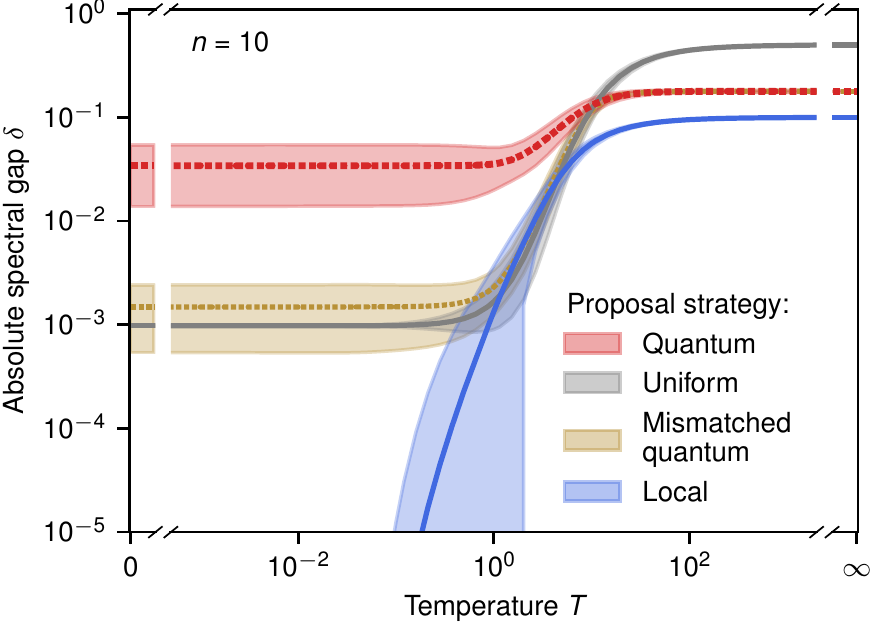} \hfill
\includegraphics[width=0.32\textwidth]{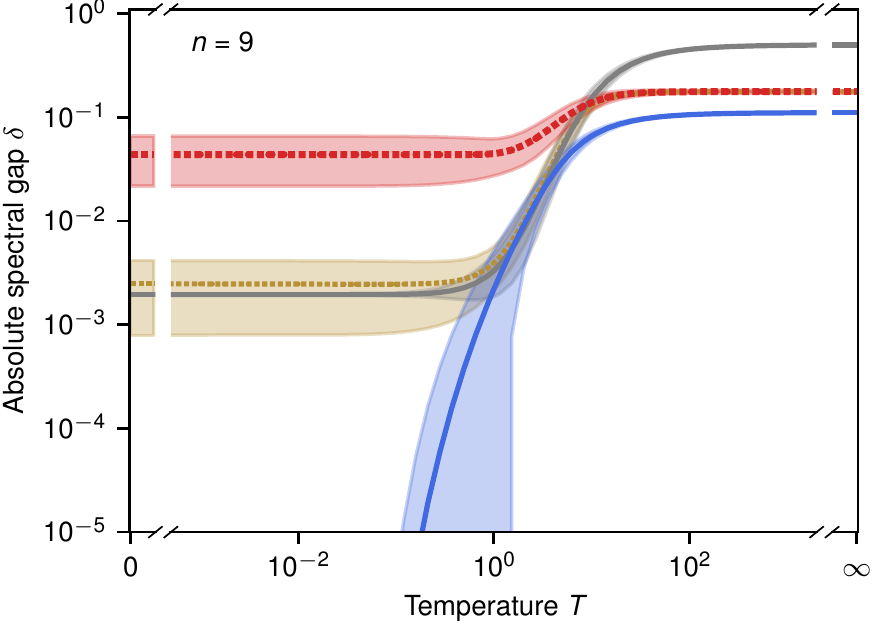} \hfill
\includegraphics[width=0.32\textwidth]{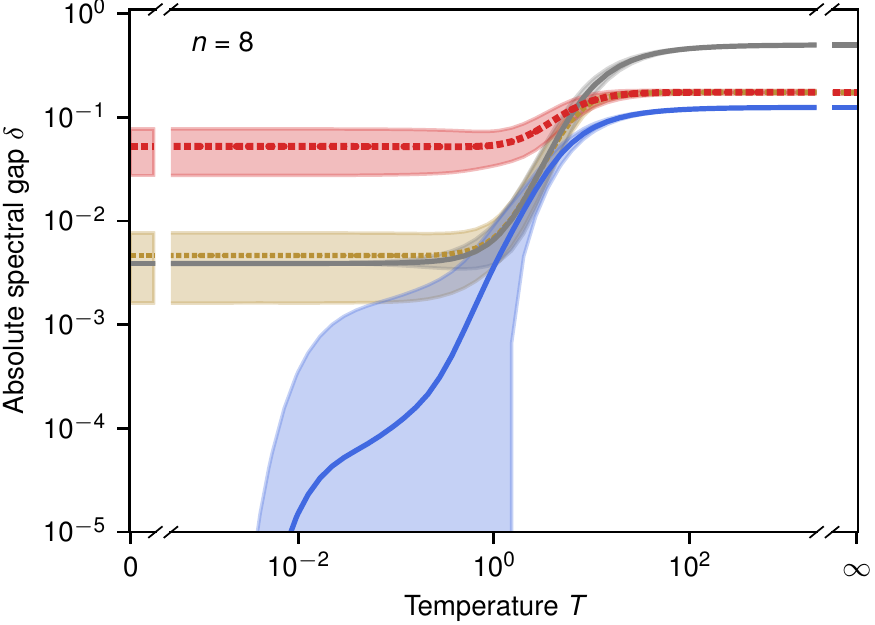}\\
\includegraphics[width=0.32\textwidth]{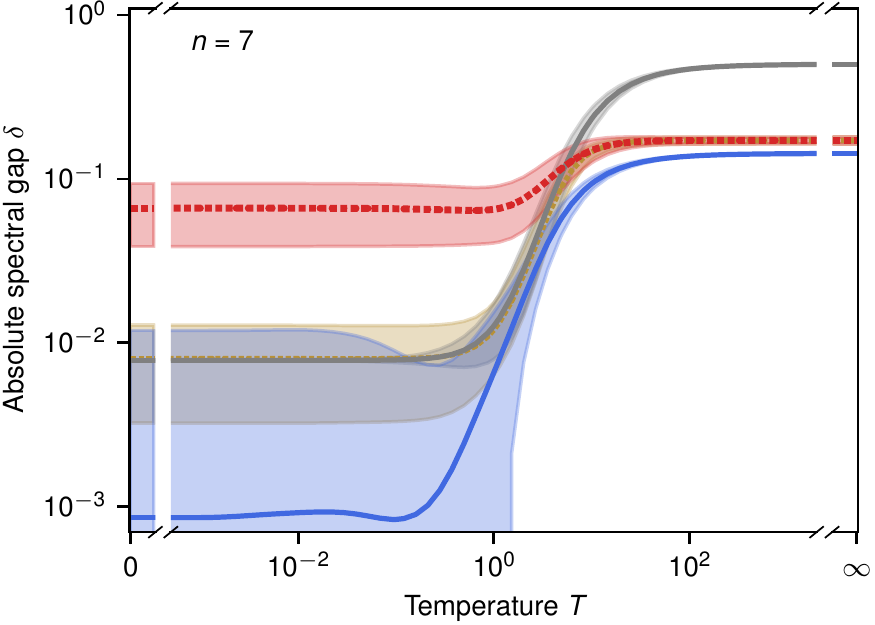} \hfill
\includegraphics[width=0.32\textwidth]{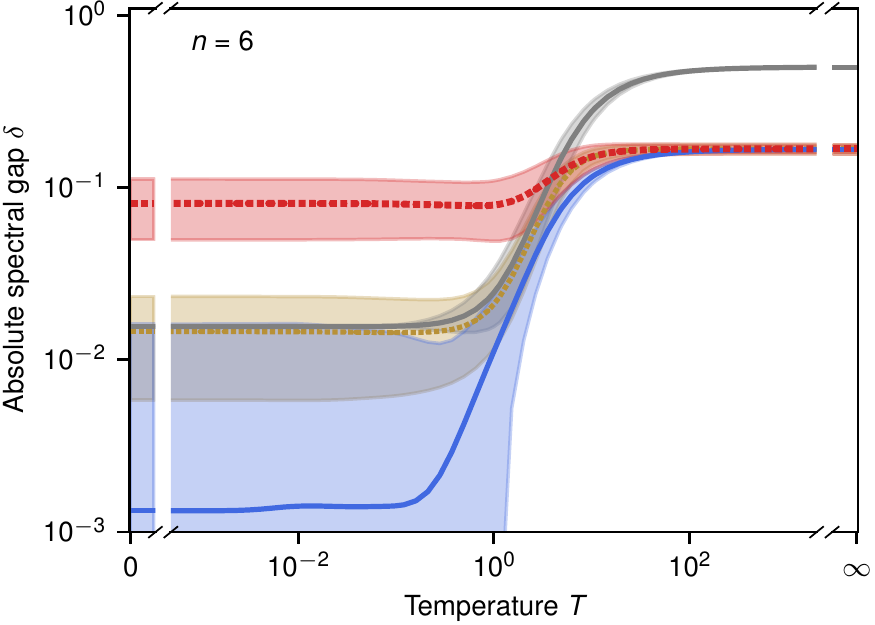} \hfill
\includegraphics[width=0.32\textwidth]{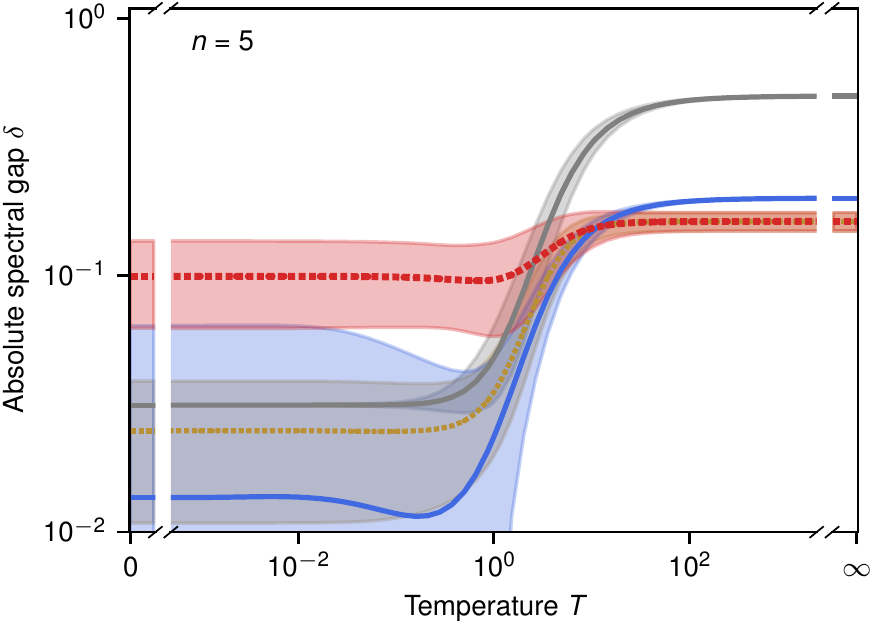}\\
\includegraphics[width=0.32\textwidth]{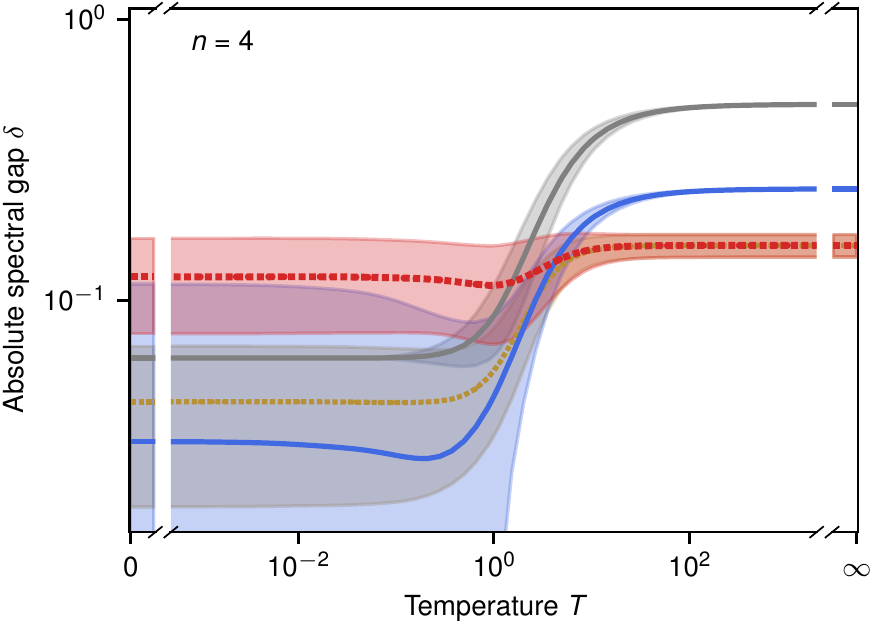}
\caption{\textbf{Average-case Gibbs sampler convergence rate simulations for different $\boldsymbol{n}$.} Analogous plots to Fig.~\ref{figS:delta_vs_T_MH} using the Gibbs sampler acceptance probability $A_\textsc{g}$ from Eq.~\eqref{eqS:gibbs_acceptance}, rather than the Metropolis-Hastings one from Eq.~\eqref{eq:MH_acceptance} of the main text. All strategies were simulated classically. Lines show the average $\delta$ over 500 random fully-connected Ising model instances for each $n$; error bands show the standard deviation in $\delta$ over these instances. Dotted lines are for visibility. The same model instances were used to make Fig.~\ref{fig:bulk_numerics} of the main text.}
\label{figS:delta_vs_T_glauber}
\end{figure}

\begin{figure}[ht!]
\centering
\includegraphics[width=0.32\textwidth]{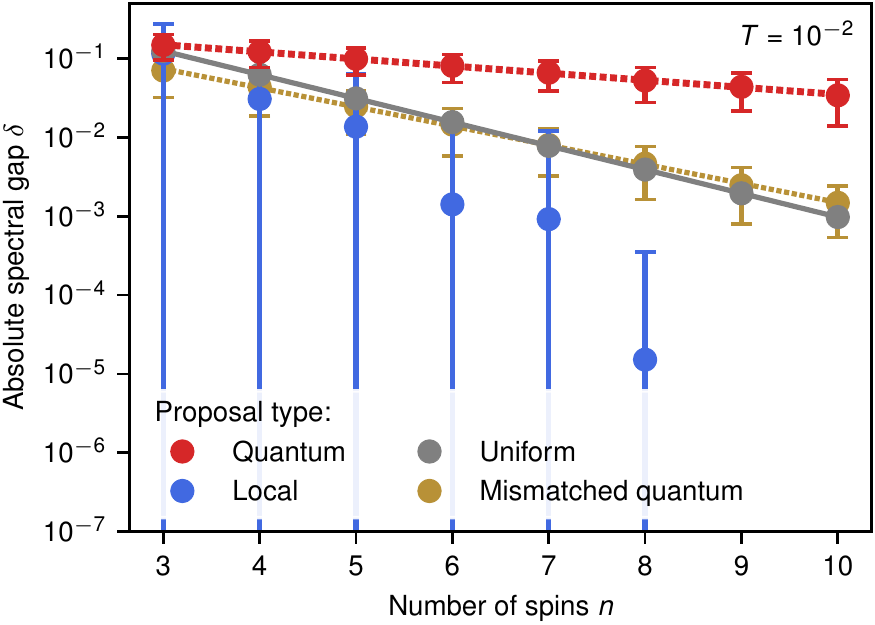} \hfill
\includegraphics[width=0.32\textwidth]{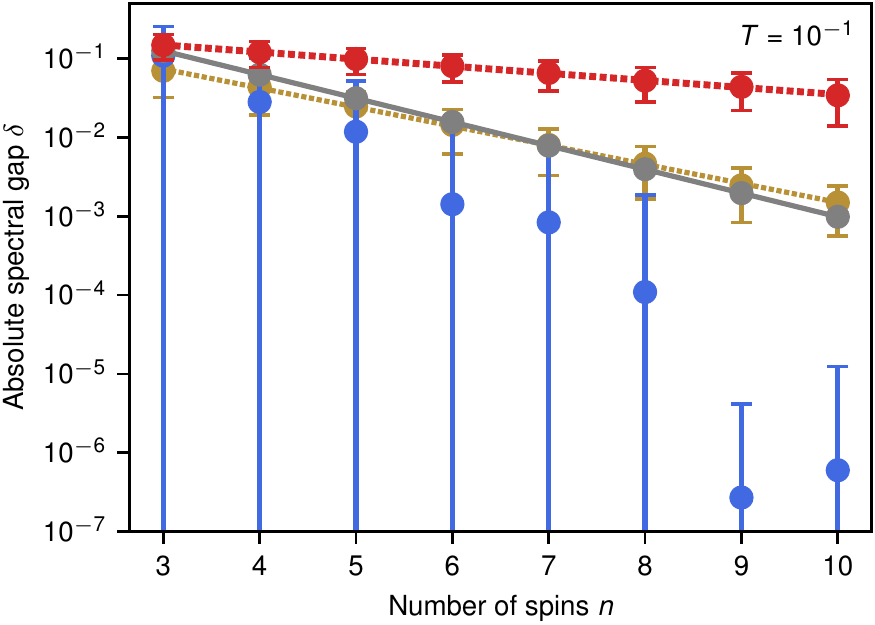} \hfill
\includegraphics[width=0.32\textwidth]{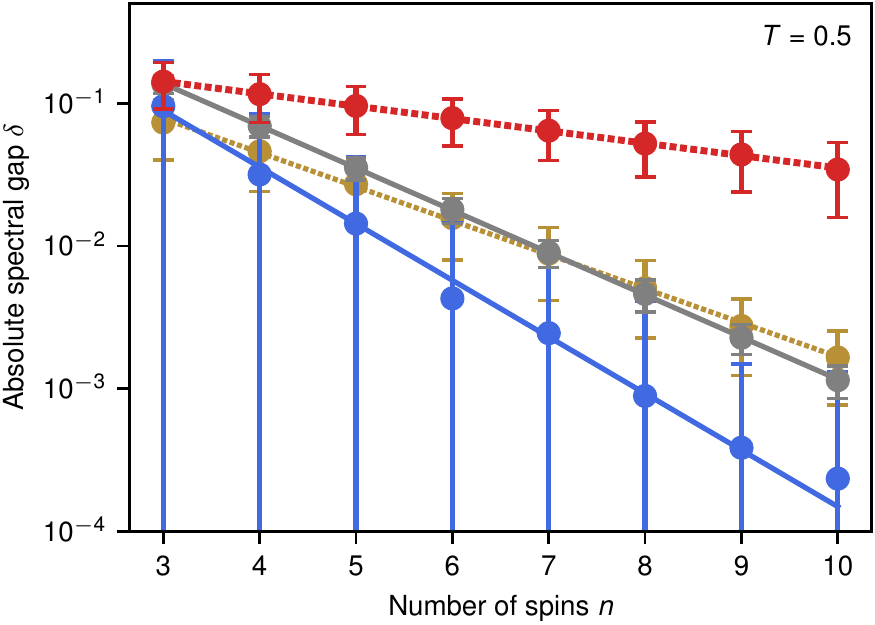}\\
\includegraphics[width=0.32\textwidth]{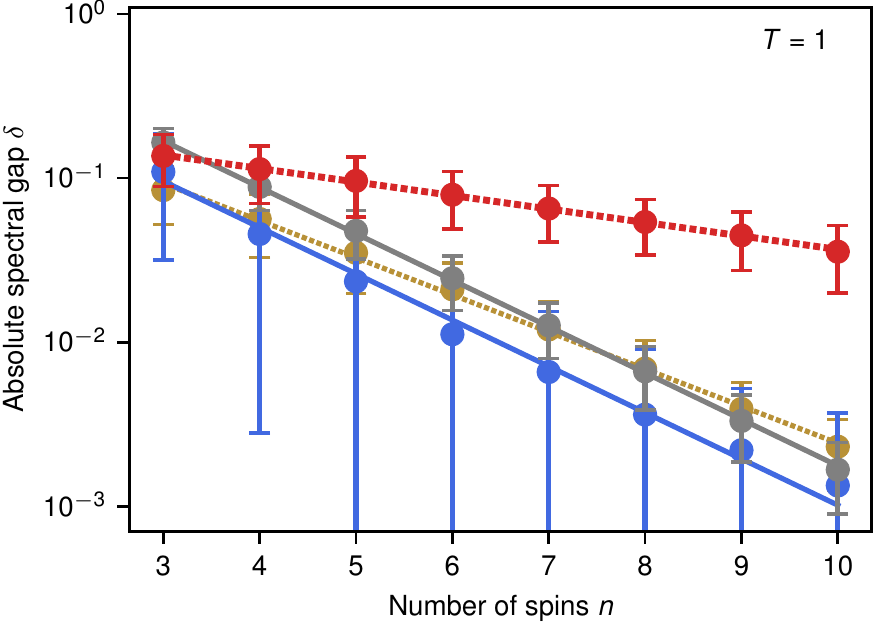} \hfill
\includegraphics[width=0.32\textwidth]{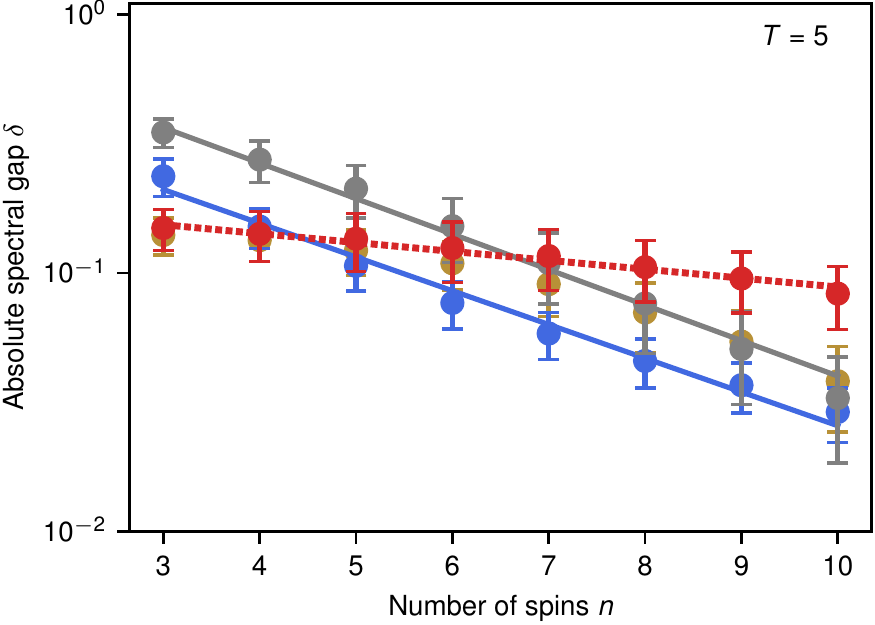} \hfill
\includegraphics[width=0.32\textwidth]{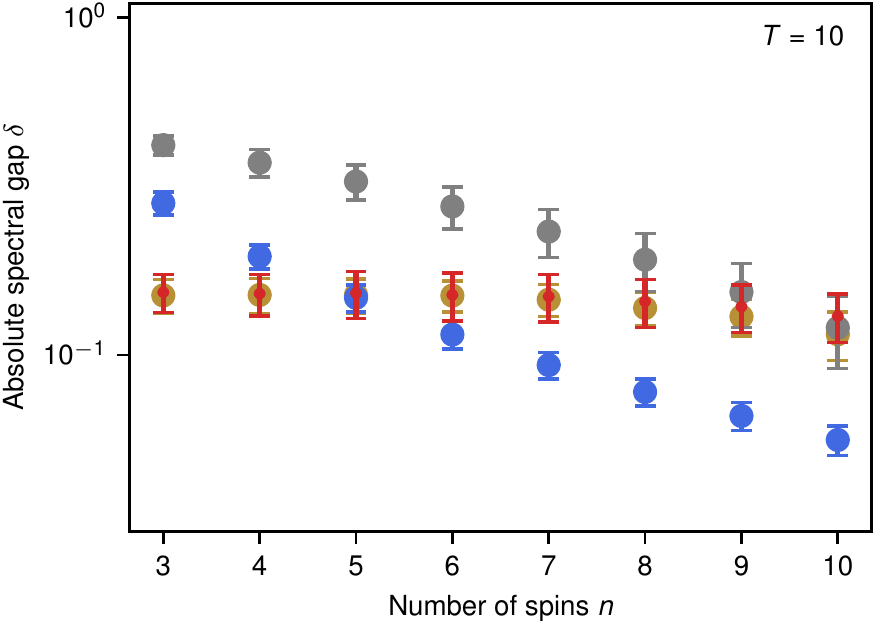}\\
\includegraphics[width=0.32\textwidth]{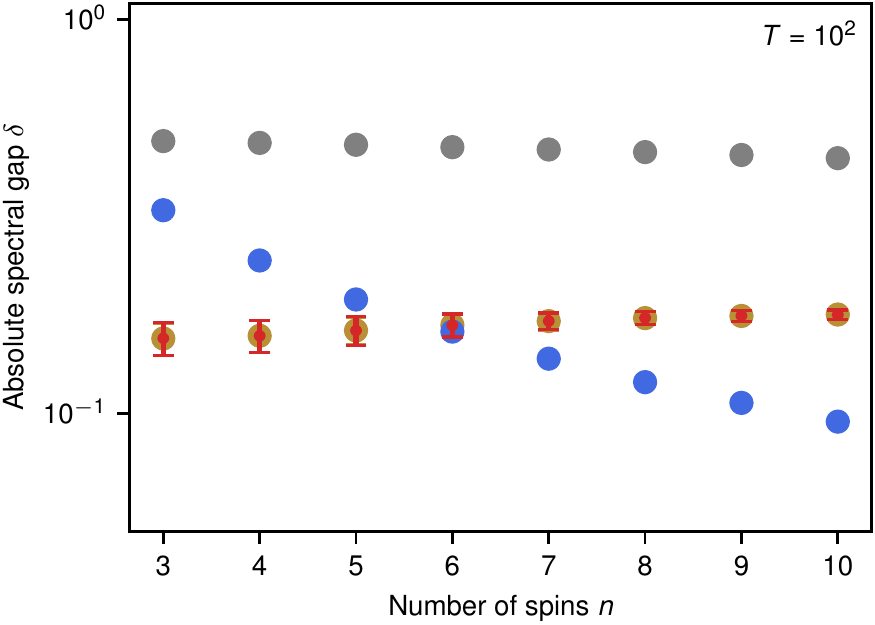}
\caption{\textbf{Average-case Gibbs sampler convergence rate simulations for different $\boldsymbol{T}$.} Analogous plots to Fig.~\ref{figS:delta_vs_n_MH} using the Gibbs sampler acceptance probability $A_\textsc{g}$ from Eq.~\eqref{eqS:gibbs_acceptance}, rather than the Metropolis-Hastings one from Eq.~\eqref{eq:MH_acceptance} of the main text. All strategies were simulated classically. Markers show the average $\delta$ over 500 random fully-connected Ising model instances for each $n$; error bars show the standard deviation in $\delta$ over these instances. Lines show least squares exponential fits $\expval{\delta} \propto 2^{-kn}$, weighted by the standard error of the mean, wherever such fits are reasonably good. The resulting values of $k$ are given in Table~\ref{tabS:fits_glauber}. Dotted lines are for visibility. The same model instances were used to make Fig.~\ref{fig:bulk_numerics} of the main text.}
\label{figS:delta_vs_n_glauber}
\end{figure}

While we used $A_\textsc{mh}$ in Algorithm~\ref{alg:M-H}, we could just as well have used $A_\textsc{g}$ (or any other acceptance probability with similar dependence on $\mu$ and $Q$ satisfying detailed balance) \cite{hastings:1970}. We found this choice to have little impact on $\delta$ in the low-$T$ regime of interest, with $A_\textsc{g}$ typically producing marginally slower convergence than $A_\textsc{mh}$, as shown in Figs.~\ref{figS:delta_vs_T_glauber} and \ref{figS:delta_vs_n_glauber}, and Table~\ref{tabS:fits_glauber}. Crucially, the arguments in the main text and in Section~\ref{secS:ergodic} about irreducibility, aperiodicity, and the computational efficiency of evaluating $A_\textsc{mh}$ hold straightforwardly for $A_\textsc{g}$ too.

\clearpage

\vfill

\begin{table}[hb!]
\setlength{\tabcolsep}{8pt}
\begin{center}
\begin{tabular}{| c | c c c c c c|}
\hline
Proposal type & $T=10^{-2}$ &  $T=10^{-1}$ &  $T=0.5$ & $T=1$ & $T=5$ & $T=10$\\
\hline
Quantum & $k=0.299(3)$ & $k=0.298(3)$ & $k=0.288(3)$ & $k=0.273(4)$ & $k=0.114(8)$ & -\\
Mismatched & \multirow{2}{*}{$k=0.806(9)$} & \multirow{2}{*}{$k=0.803(9)$} & \multirow{2}{*}{$k=0.79(1)$} & \multirow{2}{*}{$k=0.75(1)$} &  \multirow{2}{*}{-}&  \multirow{2}{*}{-} \\[-1ex]
quantum & & & & & & \\
Local & - & - & $k=1.32(4)$&  $\boldsymbol{k=0.94(4)}$ & $\boldsymbol{k=0.43(2)}$  & -\\ 
Uniform & $\boldsymbol{ k=1}$ & $\boldsymbol{k=0.9989(4)}$ & $\boldsymbol{k=0.983(2)}$ & $k=0.940(7)$ & $k=0.46(2)$ & -\\
\hline 
Quantum & \multirow{2}{*}{3.34(3)} & \multirow{2}{*}{3.35(3)} & \multirow{2}{*}{3.42(3)} & \multirow{2}{*}{3.4(1)} & \multirow{2}{*}{3.8(3)} & \multirow{2}{*}{-} \\[-1ex] 
enhancement & &&&&&\\
\hline
\end{tabular}
\end{center}
\caption{\textbf{Gibbs sampler $\boldsymbol{\expval{\delta}}$ vs.\ $\boldsymbol{n}$ fits on fully-connected instances.} The exponential fit parameters $k$ from $\expval{\delta} \propto 2^{-kn}$ in Fig.~\ref{figS:delta_vs_n_glauber} for different proposal strategies, wherever such fits are reasonably good. The resulting average quantum enhancement
exponent for each $T$, which is the ratio of $k$ for the quantum algorithm and the smallest $k$ among classical proposal strategies (shown in bold), is given in the bottom row. The parameters here are analogous to those in Table~\ref{tabS:fits_MH}, but for the Gibbs sampler acceptance probability $A_\textsc{g}$ from Eq.~\eqref{eqS:gibbs_acceptance} rather than the Metropolis-Hastings one from Eq.~\eqref{eq:MH_acceptance} of the main text.}
\label{tabS:fits_glauber}
\end{table}

\newpage
\subsubsection{Lazy chains}
\label{secS:lazy_chains}
Section~\ref{secS:glauber} highlighted the possibility that $\delta$ can be small because the MCMC transition matrix $\boldsymbol{P}$ has an eigenvalue close to -1. (This effect occurs for the local proposal at high $T$ in Figs.~\ref{fig:bulk_numerics}a and \ref{fig:n=10_gap} of the main text.) While this indeed reflects slow convergence, it can be remedied in a trivial way by using a lazy Markov chain. Suppose $\boldsymbol{P}$ is the transition matrix for a Markov chain; the lazy counterpart of this chain has transition matrix $\boldsymbol{P}_\text{lazy}=\frac{1}{2} (\boldsymbol{P}+\boldsymbol{I})$. To realize the lazy chain, at each step one either jumps according to $\boldsymbol{P}$ or stays at the current configuration, each with $50\%$ probability. Perhaps surprisingly, this lazy chain can converge faster than the regular chain. If $\boldsymbol{P}$ has eigenvalues $\{\lambda\}$, then $\boldsymbol{P}_\text{lazy}$ has eigenvalues $\{\frac{\lambda+1}{2}\}$. So if an eigenvalue of $\lambda \approx -1$ limits $\delta$ in the former, it typically does not in the latter, resulting in faster convergence.

\begin{figure}[h]
\centering
\includegraphics[width=0.32\textwidth]{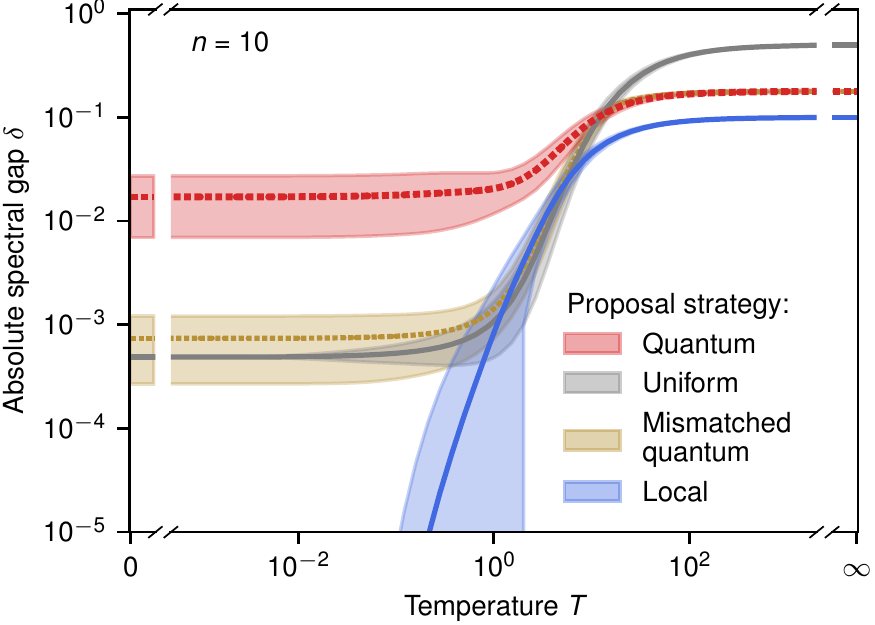} \hfill
\includegraphics[width=0.32\textwidth]{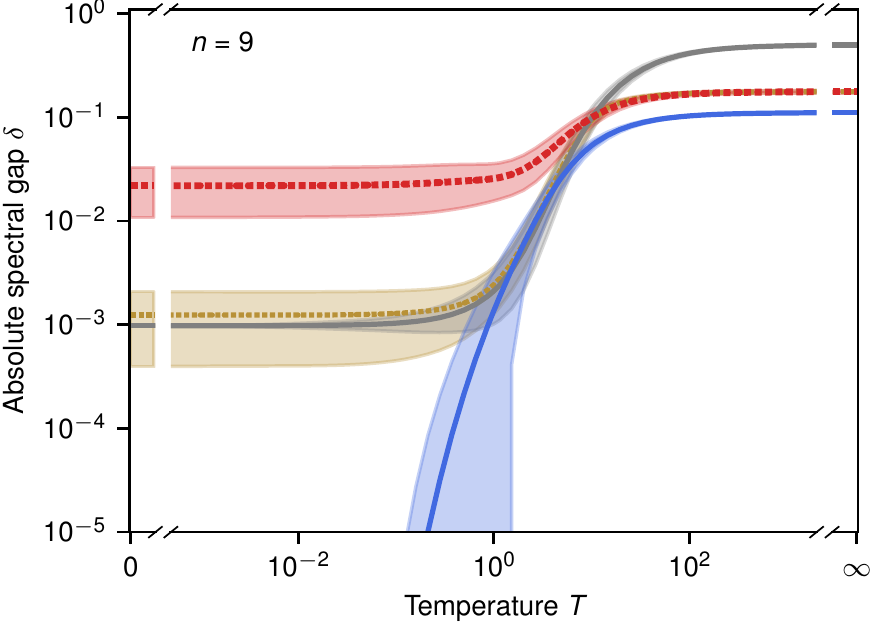} \hfill
\includegraphics[width=0.32\textwidth]{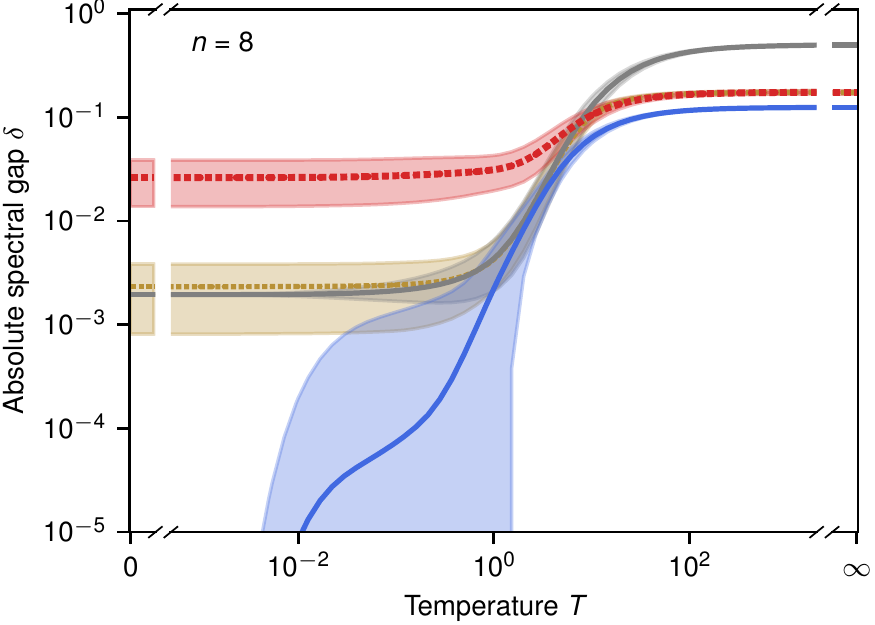}\\
\includegraphics[width=0.32\textwidth]{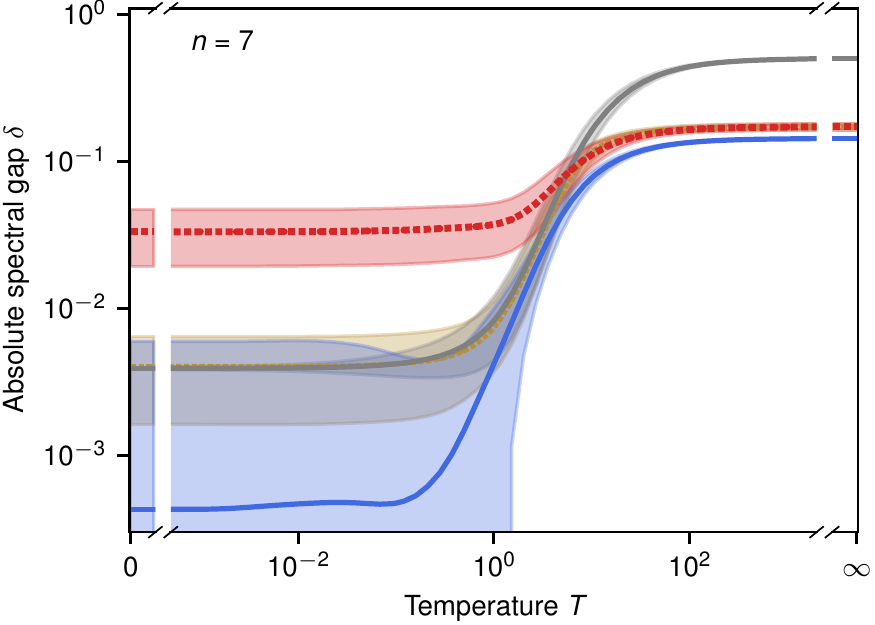} \hfill
\includegraphics[width=0.32\textwidth]{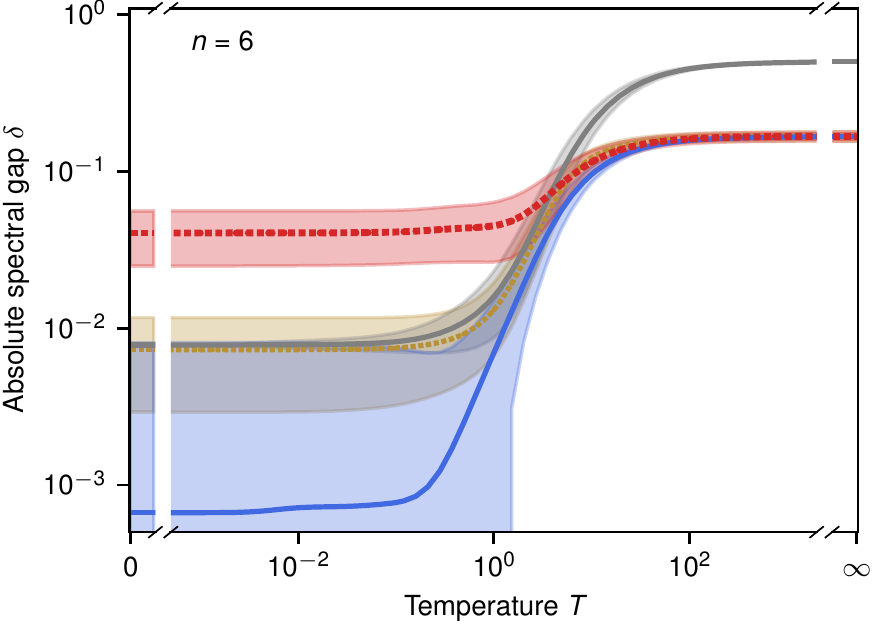} \hfill
\includegraphics[width=0.32\textwidth]{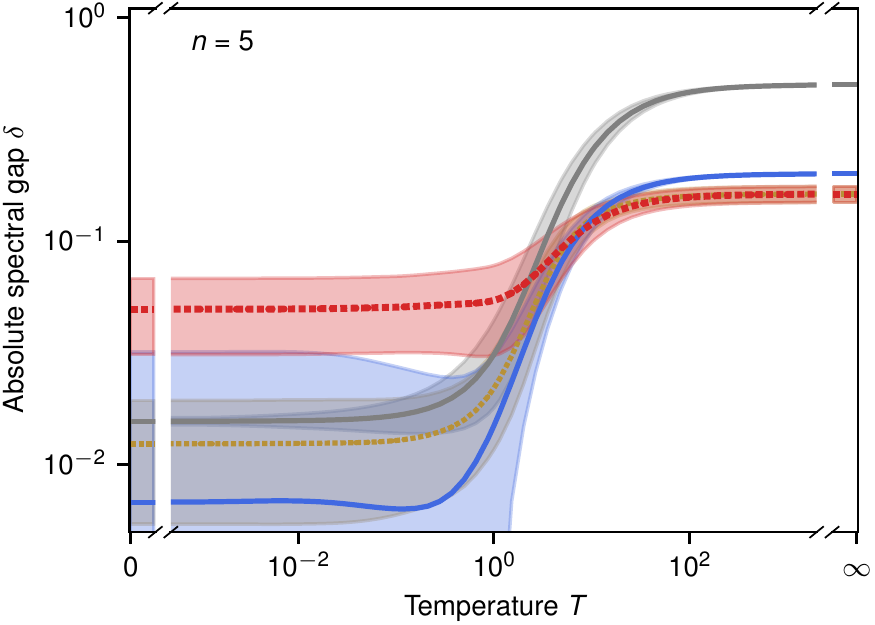}\\
\includegraphics[width=0.32\textwidth]{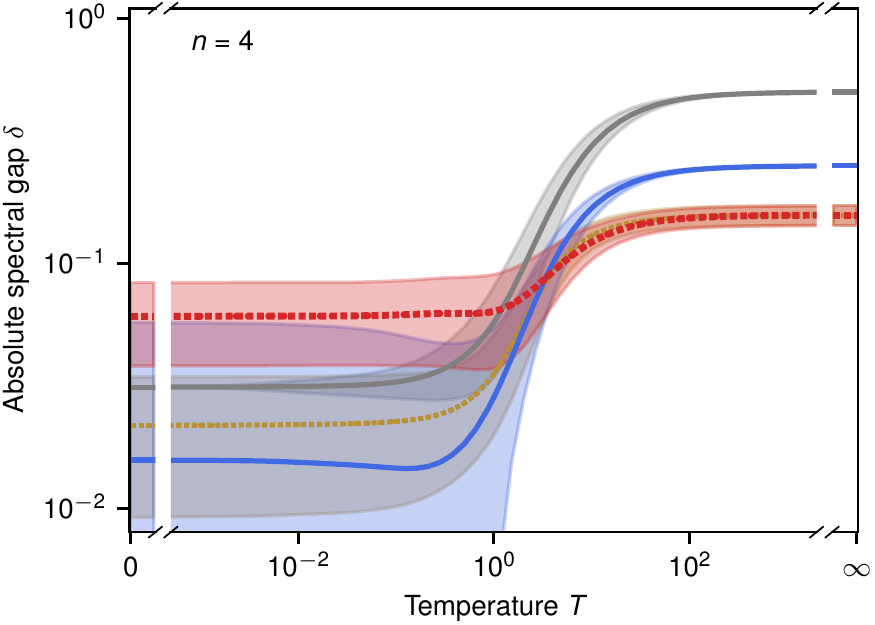}
\caption{\textbf{Average-case lazy M-H convergence rate simulations for different $\boldsymbol{n}$.} Analogous plots to Fig.~\ref{figS:delta_vs_T_MH} using lazy Metropolis-Hastings chains. All strategies were simulated classically. Lines show the average $\delta$ over 500 random fully-connected Ising model instances for each $n$; error bands show the standard deviation in $\delta$ over these instances. Dotted lines are for visibility. The same model instances were used to make Fig.~\ref{fig:bulk_numerics} of the main text.}
\label{figS:delta_vs_T_lazy}
\end{figure}

\begin{figure}[h]
\centering
\includegraphics[width=0.32\textwidth]{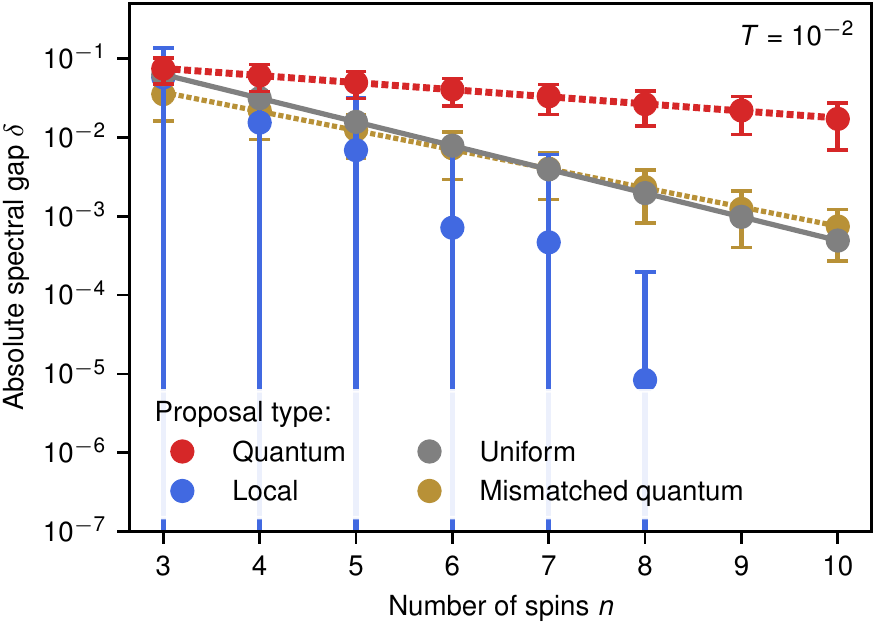} \hfill
\includegraphics[width=0.32\textwidth]{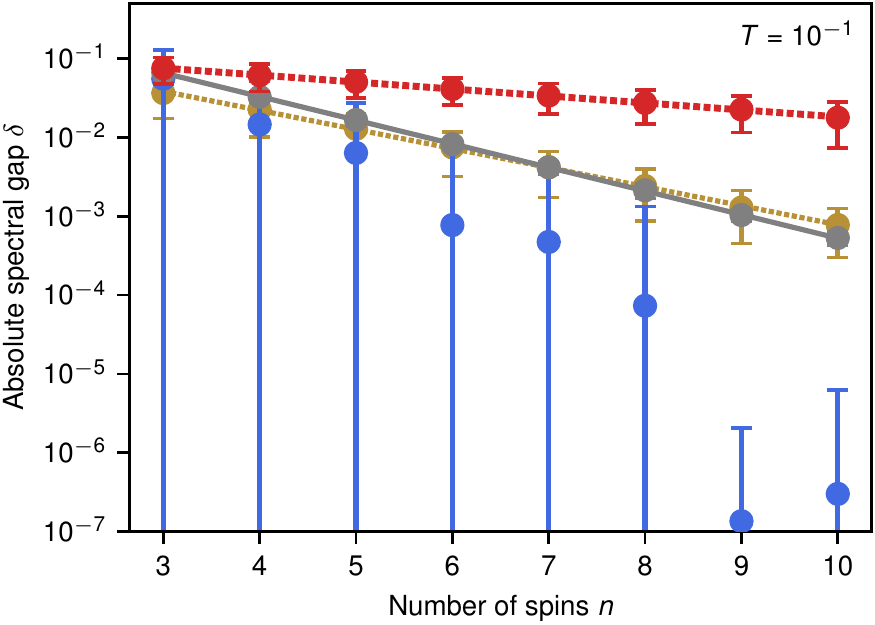} \hfill
\includegraphics[width=0.32\textwidth]{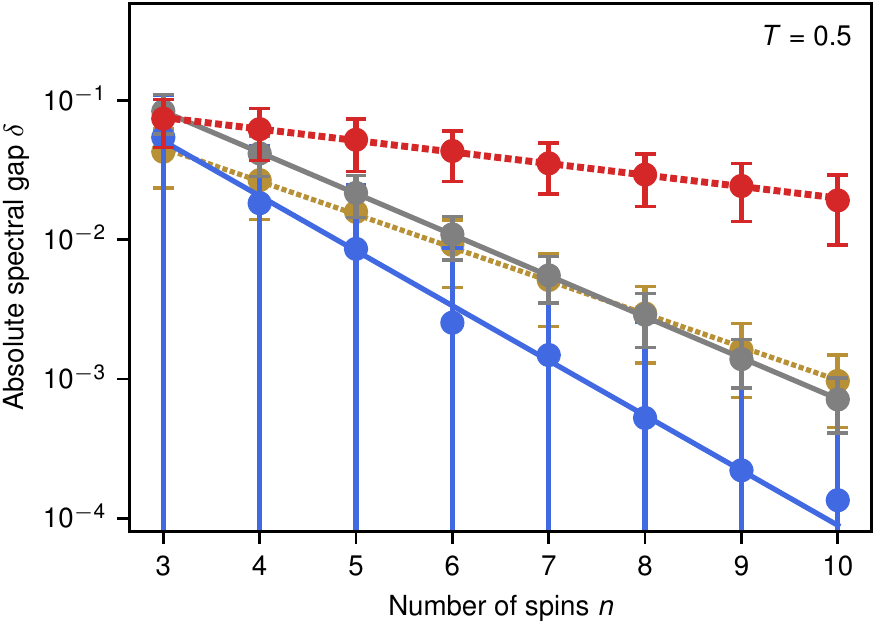}\\
\includegraphics[width=0.32\textwidth]{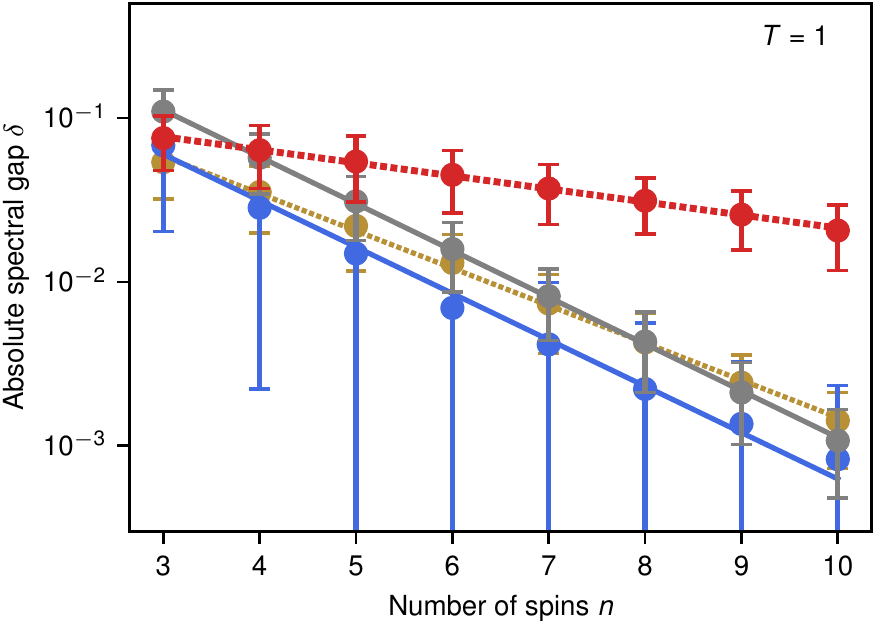} \hfill
\includegraphics[width=0.32\textwidth]{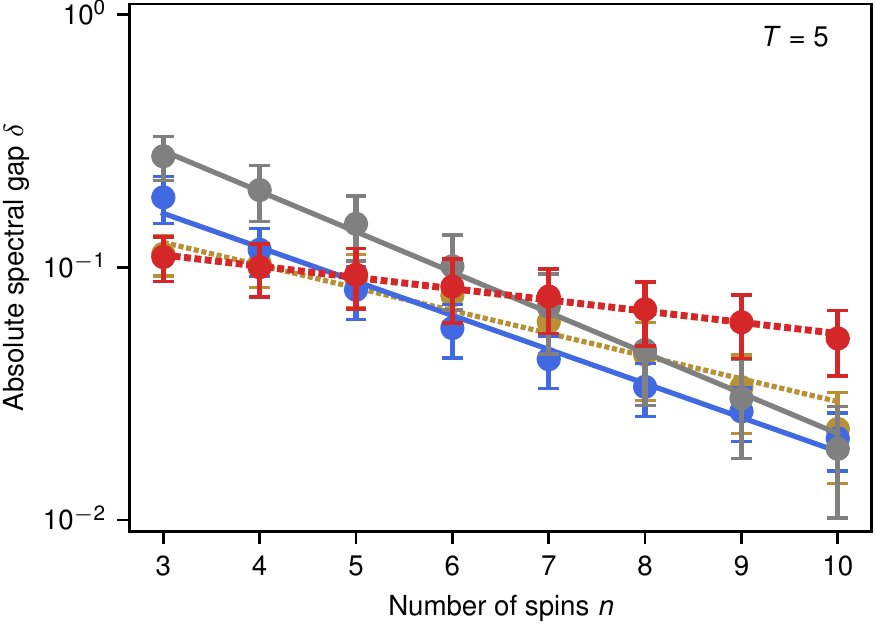} \hfill
\includegraphics[width=0.32\textwidth]{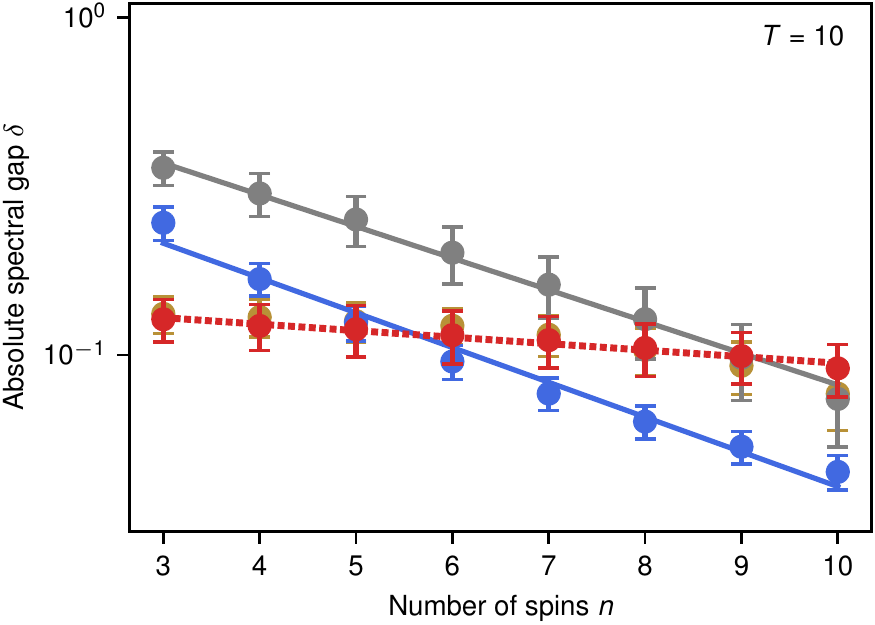}\\
\includegraphics[width=0.32\textwidth]{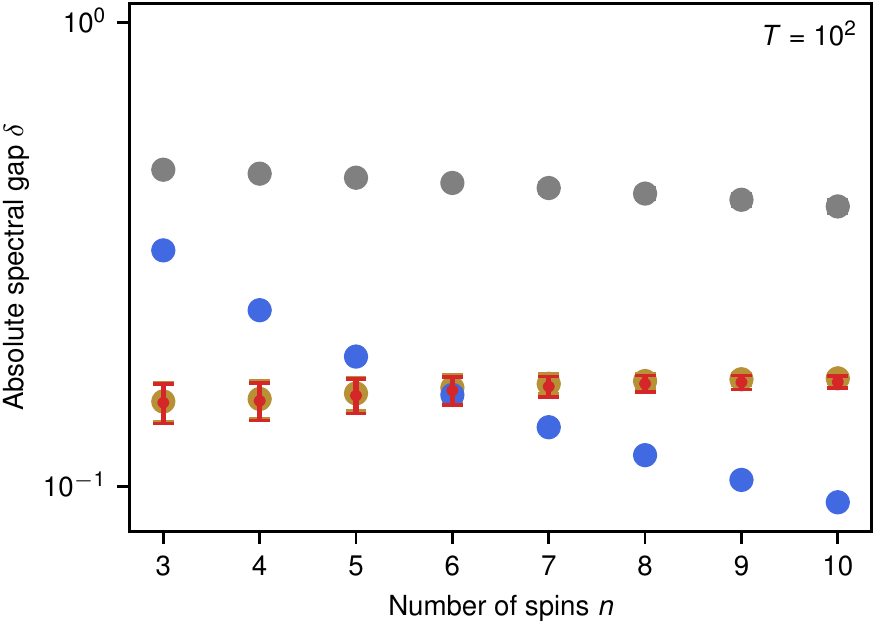}
\caption{\textbf{Average-case lazy M-H convergence rate simulations for different $\boldsymbol{T}$.} Analogous plots to Fig.~\ref{figS:delta_vs_n_MH} using lazy Metropolis-Hastings chains. All strategies were simulated classically. Markers show the average $\delta$ over 500 random fully-connected Ising model instances for each $n$; error bars show the standard deviation in $\delta$ over these instances. Lines show least squares exponential fits $\expval{\delta} \propto 2^{-kn}$, weighted by the standard error of the mean, wherever such fits are reasonably good. The resulting values of $k$ are given in Table~\ref{tabS:fits_lazy}. Dotted lines are for visibility. The same model instances were used to make Fig.~\ref{fig:bulk_numerics} of the main text.}
\label{figS:delta_vs_n_lazy}
\end{figure}

Given the trivial nature of the speedup that can be obtained through a lazy chain, it is essential to check that the quantum enhancement in $\delta$ we observed is not due to a similar effect, owing to an eigenvalue near $-1$. Indeed, it is not, as shown in Figs.~\ref{figS:delta_vs_T_lazy} and \ref{figS:delta_vs_n_lazy}, and Table~\ref{tabS:fits_lazy}. In fact, the only setting where lazy chains enhance $\delta$ is for the local proposal at high $T$. Otherwise, they only serve to slightly reduce $\delta$.

\begin{table}[hb!]
\setlength{\tabcolsep}{8pt}
\begin{center}
\begin{tabular}{| c | c c c c c c|}
\hline
Proposal type & $T=10^{-2}$ &  $T=10^{-1}$ &  $T=0.5$ & $T=1$ & $T=5$ & $T=10$ \\
\hline
Quantum & $k=0.299(3)$ & $k=0.293(2)$ & $k=0.275(4)$ & $k=0.264(4)$ & $k=0.147(6)$ & $k=0.064(4)$\\
Mismatched & \multirow{2}{*}{$k=0.806(9)$} & \multirow{2}{*}{$k=0.803(8)$} & \multirow{2}{*}{$k=0.792(9)$} &
\multirow{2}{*}{$k=0.76(1)$} &
\multirow{2}{*}{$k=0.30(3)$} &  \multirow{2}{*}{-}  \\[-1ex]
quantum & & & & & & \\
Local & - & - & $k=1.31(4)$& $\boldsymbol{k=0.94(4)}$ &  $\boldsymbol{k=0.45(2)}$ & $k=0.34(2)$  \\ 
Uniform & $\boldsymbol{ k=0.9999(2)}$ & $\boldsymbol{k=0.995(3)}$ & $\boldsymbol{k=0.981(4)}$ & $k=0.948(7)$ & $k=0.53(2)$ & $\boldsymbol{k=0.31(1)}$ \\
\hline 
Quantum & \multirow{2}{*}{3.35(3)} & \multirow{2}{*}{3.39(3)} & \multirow{2}{*}{3.57(6)} & \multirow{2}{*}{3.6(1)} & \multirow{2}{*}{3.0(2)} & \multirow{2}{*}{4.9(4)}  \\[-1ex] 
enhancement & &&&&&\\
\hline
\end{tabular}
\end{center}
\caption{\textbf{Lazy M-H $\boldsymbol{\expval{\delta}}$ vs.\ $\boldsymbol{n}$ fits on fully-connected instances.} The exponential fit parameters $k$ from $\expval{\delta} \propto 2^{-kn}$ in Fig.~\ref{figS:delta_vs_n_lazy} for different proposal strategies, wherever such fits are reasonably good. The resulting average quantum enhancement exponent for each $T$, which is the ratio of $k$ for the quantum algorithm and the smallest $k$ among classical proposal strategies (shown in bold), is given in the bottom row. The parameters here are analogous to those in Table~\ref{tabS:fits_MH}, but for lazy Metropolis-Hastings chains.}
\label{tabS:fits_lazy}
\end{table}

\subsection{Cluster Algorithms}
\label{secS:clusters}

In Section~\ref{secS:experiment} we show additional experimental data similar to Fig.~\ref{fig:n=10_gap} of the main text. (Namely, Figs.~\ref{figS:gap_n=10}, \ref{figS:gap_n=9} and \ref{figS:gap_n=8}.) For completeness, we also computed $\delta$ for five common MCMC cluster algorithms (which are the most common ones, to the best of our knowledge) in these supplementary plots. One would not expect them to converge particularly fast, since none of them are intended for fully-connected or 1D spin glasses. Indeed we found their spectral gaps to be comparable with those of the local and uniform proposals, despite the underlying algorithms being substantially more complicated, both conceptually and practically.

\subsubsection{Swendsen-Wang and Wolff clusters}
\label{secS:swwolff}

The Swendsen-Wang (S-W) and Wolff MCMC algorithms work by grouping spins into clusters based on the model instance, the current configuration $\boldsymbol{s}$, and the temperature $T$ \cite{swendsen:1987, wolff:1989}. (See Refs.~\cite{park:2017, barzegar:2018} for spin glass implementations.) In the S-W algorithm multiple clusters can then be flipped (flipping a cluster means $s_j \mapsto -s_j$ for each spin $j$ in the cluster), whereas in the Wolff algorithm only a single one can. We are aware of two main variants (which are in fact distinct algorithms) for both the S-W and Wolff algorithms, which differ in how the fields $\{h_j\}$ are handled. The first uses an additional ``ghost spin,'' whose state is held fixed, to turn the fields into couplings as discussed in Section~\ref{secS:inst_and_conn} \cite{wang:1989}. The other uses an additional accept/reject step that depends only on $\{h_j\}$ and $T$, unlike the more common accept/reject probabilities discussed in the main text \cite{delyra:2006, park:2017, kent-dobias:2018}. These two versions of the S-W and Wolff algorithms account for four of the five cluster algorithms considered here. We are not aware of any closed-form expression for the transition probabilities $P(\boldsymbol{s'} | \boldsymbol{s})$ for any of these four algorithms. Rather, we estimated each $P(\boldsymbol{s'} | \boldsymbol{s})$ by 
\begin{equation}
\hat{P}(\boldsymbol{s'} | \boldsymbol{s})
=
\frac{\text{number of } \boldsymbol{s} \rightarrow \boldsymbol{s'} \text{ transitions observed}}
{\text{number of } \boldsymbol{s} \rightarrow \text{[anything] transitions observed}},
\end{equation}
where initial states $\boldsymbol{s}$ were chosen uniformly at random in each repetition. Finally, we formed $\{\hat{P}(\boldsymbol{s'} | \boldsymbol{s}) \}$ into a $2^n \times 2^n$ stochastic matrix $\hat{\boldsymbol{P}}$, which we diagonalized to yield an estimator $\hat{\delta}$ for the true absolute spectral gap $\delta$ as described in the main text. Finally, we repeatedly resampled the data, where each data point consists of an observed $\boldsymbol{s} \rightarrow \boldsymbol{s'}$ transition, to compute bootstrap confidence intervals (CIs) for $\delta$. We picked the initial states $\boldsymbol{s}$ uniformly at random when gathering data (rather than iterating over $\boldsymbol{s} \in \{-1,1\}^n$ in some fixed order) so that the data points would be IID, which simplifies the computation of bootstrap CIs.

The resulting absolute spectral gaps are shown in Figs.~\ref{figS:gap_n=10}, \ref{figS:gap_n=9} and \ref{figS:gap_n=8} for illustrative model instances, together with experimental data for our quantum algorithm. We did not repeat this analysis  on large ensembles of instances as in Fig.~\ref{fig:bulk_numerics} of the main text, however, for three main reasons. First, the analysis is computationally intensive, especially since the cluster formation probabilities depend on $T$, so this sampling procedure must be repeated not just for each instance, but also for each $T$. Second, this approach inevitably entails some uncertainty in $\delta$, which is hard to control in an automated way, and makes it complicated to average $\delta$ over many model instances. Indeed, estimating $\delta$ confidence intervals for a single instance through bootstrapping (which must be done separately for each value of $T$ considered) is, in itself, quite computationally intensive. Finally, these algorithms are not designed for spin glasses. The mediocre absolute spectral gaps we observed as a result on illustrative model instances did not justify the considerable computational cost required to analyze them over many such instances. 

Note finally that the S-W and Wolff algorithms (even the versions with an accept/reject step) do not fit into the simple framework described in the main text wherein a jump $\boldsymbol{s} \rightarrow \boldsymbol{s'}$ is proposed with some $T$-independent probability $Q(\boldsymbol{s'}|\boldsymbol{s})$ and then accepted with some $T$-dependent probability $A(\boldsymbol{s'}|\boldsymbol{s})$, since the cluster formation process depends on $T$. This extra temperature dependence prevents us from showing these cluster algorithms in Fig.~\ref{fig:mechanism} of the main text, and the like, as there is no obvious quantity for these algorithms admitting a fair comparison to $Q(\boldsymbol{s'}|\boldsymbol{s})$ for the local, uniform and quantum proposals.

\subsubsection{Houdayer clusters}

The other MCMC cluster algorithm we considered is that of Houdayer \cite{houdayer:2001}. Analyzing its performance poses two main difficulties: First, finding $\delta$ can be computationally intensive. Second, the algorithm uses more resources than those considered in the main text, making it difficult to achieve a fair comparison. For one, Houdayer's algorithm uses $R \ge 2$ independent replicas of an Ising model instance simultaneously at the same $T$. (Here, ``replicas'' refer to multiple model instances with the same couplings $\{J_{jk}\}$ and fields $\{h_j\}$ but in independent spin configurations, i.e., the simulation cell consists of $nR$ total spins.) In each iteration, the replicas are paired together and their states are compared in such a way as to form clusters, which are then flipped. To achieve the fairest possible comparison with the other algorithms we considered, all of which use a single replica, we focus on the simplest case of $R=2$ replicas.

Other issues hindering fair comparison arise immediately. For instance, unlike in the S-W and Wolff algorithms, this cluster flipping procedure alone does not guarantee convergence. Rather, it is meant to be paired with $n$ local Metropolis-Hastings jumps on each cluster. (In fact, in each replica, each spin $j \in [1,n]$ is meant to be flipped sequentially, with an accept/reject step after each. This is different than the local proposal described in the main text, where a random spin is picked each time, and does not constitute a Markov chain in the same way. However, the ultimate effect is likely very similar to performing $n$ local Metropolis-Hastings jumps on each replica \cite{he:2016}, so we will treat the two as equivalent.) This means that every Houdayer iteration, as described so far, comprises $nR+1$ jumps. To allow a fair comparison, we instead consider a variant of the algorithm wherein each iteration performs a local Metropolis-Hastings jump on all replicas with probability $n/(n+1)$, or a cluster flip with probability $1/(n+1)$. This way, the average behavior is that of the standard Houdayer algorithm, but only a single operation per cluster is performed in every iteration.

Finally, these steps are often wrapped in a parallel tempering scheme, i.e., they are performed for many different values of $T$ in parallel, and the replicas at different temperatures are made to interact periodically. However, as discussed in the main text, our quantum algorithm could also be wrapped in similar ways, although such generalizations are beyond the scope of this initial work. To allow a fair comparison, then, we use only two replicas at a single temperature $T$. (Note that this restriction to a single $T$ precludes comparison with another cluster algorithm called isoenergetic cluster moves \cite{zhu:2015}.)

The MCMC transition matrix $\boldsymbol{P}$ for this implementation of the Houdayer algorithm can be constructed exactly, without resorting to sampling of the sort described in Section~\ref{secS:swwolff}. However, even for the simplest case of $R=2$ replicas, $\boldsymbol{P}$ is $2^{2n} \times 2^{2n}$, so for an $n=10$ instance it has $2^{40} \approx 10^{12}$ matrix elements. Fortunately it is sparse for the model instances we considered, so $\delta$ can be found using an iterative eigenvalue algorithm. Still, doing so up to $n=10$ is computationally expensive, and difficult to automate. Moreover, the resulting $\delta$ follows that of the local proposal closely or exactly. (This was expected, as the algorithm is designed for 2D models, not 1D or fully-connected ones.) So as with the S-W and Wolff algorithms, we analyzed the Houdayer algorithm only on illustrative model instances, rather than for large ensembles of random instances.

We found $\delta$ using the \texttt{sparse.linalg.eigs} function from the SciPy library \cite{scipy}, which finds the largest eigenvalues in absolute value of a matrix. For $\boldsymbol{P}$ these are $\lambda_1 = 1$ and $\lambda_2$, where the latter sets the absolute spectral gap as $\delta = 1-|\lambda_2| > 0$. Unfortunately, the underlying eigenvalue algorithm can be slow or inaccurate when applied directly to $\boldsymbol{P}$ with $|\lambda_2| \approx 1$, as the near-degeneracy makes it hard to resolve these eigenvalues from each other. We instead preconditioned the problem by first constructing a new linear operator related to $\boldsymbol{P}$ but without this near-degeneracy, allowing us to find $\delta$ faster and more reliably. Suppose $\boldsymbol{P}$ is the transition matrix for a Markov chain satisfying detailed balance with a unique stationary distribution $\vec{p} = \boldsymbol{P} \vec{p}$ whose entries are all positive. Then the matrix
\begin{equation}
\boldsymbol{L} = \text{diag}(\vec{p}^{\,\,\, -1/2})
\, \boldsymbol{P} \,
\text{diag}(\vec{p}^{\,\, 1/2})
= \boldsymbol{L}^T,
\end{equation}
where $\vec{p}^{\,\,\, \pm 1/2}$ is defined element-wise, is symmetric \cite{hsu:2015}. Moreover, since $\boldsymbol{L}$ and $\boldsymbol{P}$ are similar matrices, they have the same eigenvalues. Because $\boldsymbol{L}$ is symmetric, however, its eigenvectors corresponding to different eigenvalues are orthogonal. In particular, $\sqrt{\vec{p}}$ is the $\lambda=1$ eigenvector of $\boldsymbol{L}$ satisfying $\| \sqrt{\vec{p}} \, \|=1$ since $\sum_j p_j=1$. Subtracting the orthogonal projector $\sqrt{\vec{p}} \, \sqrt{\vec{p}}^{\,T}$ onto this eigenspace from $\boldsymbol{L}$ gives the matrix 
\begin{align}
\tilde{\boldsymbol{L}}
&=
\boldsymbol{L}
- \sqrt{\vec{p}} \, \sqrt{\vec{p}}^{\,T}\\ 
&= 
\text{diag}(\vec{p}^{\,\,\, -1/2})
\, \big[\boldsymbol{P} - \boldsymbol{M} \big]\,
\text{diag}(\vec{p}^{\,\, 1/2}), \nonumber
\end{align}
where $\boldsymbol{M}=(\vec{p} \, | \cdots | \, \vec{p} \,)$ has the elements of $\vec{p}$ in each column. $\boldsymbol{L}$ and $\tilde{\boldsymbol{L}}$ have all the same eigenvectors, associated with all the same eigenvalues except for one: $\sqrt{\vec{p}}$ is an eigenvector of $\boldsymbol{L}$ with eigenvalue 1 but an eigenvector of $\tilde{\boldsymbol{L}}$ with eigenvalue 0. Finally, since $\tilde{\boldsymbol{L}}$ and $\boldsymbol{P} - \boldsymbol{M}$ are similar matrices, they have the same spectrum. This means that the largest eigenvalue of $\boldsymbol{P} - \boldsymbol{M}$, in absolute value, is $\lambda_2$ rather than 1. Applying an iterative eigenvalue algorithm to $\boldsymbol{P} - \boldsymbol{M}$ rather than $\boldsymbol{P}$ therefore avoids the issue of near-degeneracy due to $|\lambda_2| \approx 1$. And while $\boldsymbol{M}$ is a dense $2^{2n} \times 2^{2n}$ matrix, $\boldsymbol{M} \vec{x} = \vec{p} \, \sum_j x_j$ for any vector $\vec{x}$. One can therefore apply \texttt{sparse.linalg.eigs} to the linear operator $\mathcal{P}$ defined by the action
\begin{equation}
\mathcal{P}(\vec{x})
=
\boldsymbol{P} \vec{x} - \vec{p} \, \sum_j x_j,
\end{equation}
without having to store $\boldsymbol{M}$ in memory. We used this method to compute $\delta$ as a function of $T$ for the Houdayer algorithm, for which $\boldsymbol{P}$ satisfies detailed balance by construction and the stationary distribution is $\vec{p} = \vec{\mu} \otimes \vec{\mu}$. We constructed $\vec{p}$ by first finding $\vec{\mu}$ by diagonalizing the transition matrix of a single-replica Metropolis-Hastings chain. The results are shown in Figs.~\ref{figS:gap_n=10}, \ref{figS:gap_n=9} and \ref{figS:gap_n=8} for illustrative model instances, together with S-W and Wolff gaps, and experimental data for our quantum algorithm. Despite their relative complexity, these cluster algorithms perform comparably to M-H with a local or uniform proposal, and converge especially slowly in the low-$T$ regime of interest.

\section{Experimental details}
\label{secS:experiment}

We performed all of the experiments on \textit{ibmq\_mumbai}, a quantum processor with 27 fixed-frequency superconducting transmon qubits forming a heavy hexagonal lattice. In this section we describe the quantum circuits that were implemented, how the resulting data was analyzed, and finally, we show additional experimental data similar to Figs.~\ref{fig:n=10_gap}-\ref{fig:mechanism} of the main text but for other illustrative model instances.

\subsection{Quantum circuits} 

\subsubsection{Trotter circuits and gates}

To simplify the notation, let 
\begin{equation}
H_1 = -(1-\gamma)\alpha \sum_{j=1}^n h_j Z_j + \gamma  \sum_{j=1}^n X_j
\end{equation}
and
\begin{equation}
H_2 = -(1-\gamma) \, \alpha \sum_{j>k=1}^n J_{jk} Z_j Z_k
\end{equation}
so that $H$ from Eq.~\eqref{eq:H} of the main text can be written as $H = H_1 + H_2$, where $H_1$ and $H_2$ contain all the 1- and 2-body terms respectively. For any value of $\gamma$ and $t$, we approximated the dynamics $U=e^{-iHt}$ by a 2\textsuperscript{nd}-order product formula (PF)
\begin{equation}
V = \Big( e^{-i H_2 \Delta t/2} \, e^{-i H_1 \Delta t} \, e^{-i H_2 \Delta t/2} \Big)^r,
\label{eqS:PF2}
\end{equation}
where $\Delta t = t/r$ for an integer number of timesteps $r$. The unitary $e^{-i H_1 \Delta t}$ can be realized through single-qubit gates applied in parallel, while $e^{-i H_2 \Delta t/2}$ is implemented through two layers of 2-qubit gates, as detailed below. In theory, one could equally well exchange $H_1 \leftrightarrow H_2$ in Eq.~\eqref{eqS:PF2}. However, the order we used requires fewer 2-qubit gates than the alternative. This is because our algorithm only involves initial states and measurements in the computational basis, so the first and last $e^{-i H_2 \Delta t/2}$ layers have no impact on the measurement statistics, and therefore need not be implemented at all. Rather, one can implement
\begin{align}
\label{eqS:V_tilde}
\tilde{V} &= \Big( e^{-i H_1 \Delta t} \, e^{-i H_2 \Delta t/2} \Big) \, 
\Big( e^{-i H_2 \Delta t/2} \, e^{-i H_1 \Delta t} \, e^{-i H_2 \Delta t/2} \Big)^{r-2} \,
\Big( e^{-i H_2 \Delta t/2} \, e^{-i H_1 \Delta t} \Big) \\
&= 
e^{-i H_1 \Delta t} \Big( e^{-i H_2 \Delta t} e^{-i H_1 \Delta t} \Big)^{r-1} \nonumber
\end{align}
in place of $V$, therefore realizing a 2\textsuperscript{nd}-order $r$-step PF (or equivalently, a 1\textsuperscript{st}-order $r$-step PF \cite{layden:2021}) using only $r-1$ applications of the entangling Trotter step $e^{-i H_2 \Delta t}$.

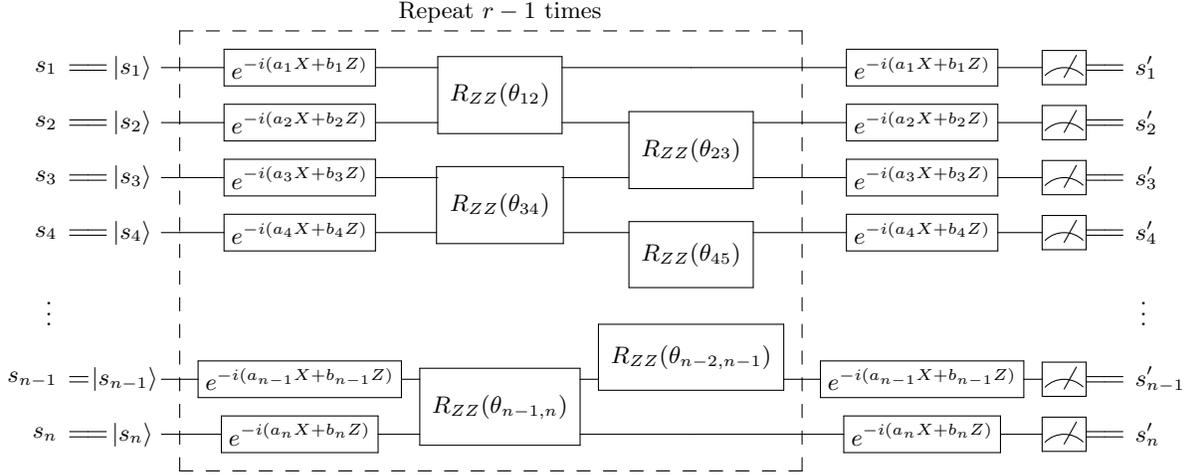
\begin{figure}[h]
\[
\Qcircuit @C=0.75em @R=0.75em {
 & & & & & & & & \mbox{Repeat $r-1$ times} & & & & & & & \\ 
 & & & & & & & & & & & & & & & \\ 
 \lstick{s_1}  & \cw & \cw & \hspace{1.5ex}\ket{s_1} &&& \qw & \gate{e^{-i (a_1 X + b_1 Z)}} & \multigate{1}{R_{ZZ}(\theta_{12})} & \qw & \qw & \gate{e^{-i (a_1 X + b_1 Z)}} & \meter & \cw & \rstick{s_1'} \cw \\
 \lstick{s_2}  & \cw & \cw & \hspace{1.5ex}\ket{s_2} &&& \qw & \gate{e^{-i (a_2 X + b_2 Z)}} & \ghost{R_{ZZ}(\theta_{12})} & \multigate{1}{R_{ZZ}(\theta_{23})} & \qw &  \gate{e^{-i (a_2 X + b_2 Z)}} & \meter & \cw & \rstick{s_2'} \cw \\
 \lstick{s_3}  & \cw & \cw & \hspace{1.5ex}\ket{s_3} &&& \qw & \gate{e^{-i (a_3 X + b_3 Z)}} & \multigate{1}{R_{ZZ}(\theta_{34})} & \ghost{R_{ZZ}(\theta_{23})} & \qw & \gate{e^{-i (a_3 X + b_3 Z)}} & \meter & \cw & \rstick{s_3'} \cw \\
 \lstick{s_4} & \cw & \cw & \hspace{1.5ex}\ket{s_4} &&& \qw & \gate{e^{-i (a_4 X + b_4 Z)}} & \ghost{R_{ZZ}(\theta_23)} & \multigate{2}{R_{ZZ}(\theta_{45})} & \qw & \gate{e^{-i (a_4 X + b_4 Z)}} & \meter & \cw & \rstick{s_4'} \cw \\
 & & & & & & & & & & & & & & \\ 
 & & & & & & & & & & & & & & \\ 
 \lstick{\vdots \hspace{0.5ex}} & & & & & & & & & & & & & & \rstick{\hspace{0.5ex}\vdots} \\
 & & & & & & & & & & & & & & \\
 & & & & & & & & & & & & & & \\ 
 \lstick{s_{n-1}}  & \cw & & \hspace{0ex}\ket{s_{n-1}} &&& \qw & \gate{e^{-i (a_{n-1} X + b_{n-1} Z)}} & \multigate{1}{R_{ZZ}(\theta_{n-1,n})} & \multigate{-2}{R_{ZZ}(\theta_{n-2,n-1})} & \qw & \gate{e^{-i (a_{n-1} X + b_{n-1} Z)}} & \meter & \cw & \rstick{s_{n-1}'} \cw \\
 \lstick{s_n}  & \cw & \cw & \hspace{1.5ex} \ket{s_n} &&& \qw & \gate{e^{-i (a_n X + b_n Z)}} & \ghost{R_{ZZ}(\theta_{n-1,n})} & \qw & \qw & \gate{e^{-i (a_n X + b_n Z)}} & \meter & \cw & \rstick{s_n'} \cw
\gategroup{2}{7}{14}{11}{0em}{--}   \\ 
 & & & & & & & & & & & & & 
}
\]
\caption{\textbf{Trotter circuit.} A $n$-qubit circuit approximating $e^{-iHt}$ through an $r$-step 2\textsuperscript{nd} order product formula. It applies the unitary $\tilde{V}$ from Eq.~\eqref{eqS:V_tilde} to an initial computational state $\ket{\boldsymbol{s}}$, then performs a measurement in the computational basis which yields an $n$-bit string $\boldsymbol{s'}$.}
\label{figS:trotter_circ}
\end{figure}

The resulting Trotter circuit is shown in Fig.~\ref{figS:trotter_circ}. Each $e^{-i H_1 \Delta t}$ step can be implemented by applying $\exp[-i(a_j X_j + b_j Z_j)]$ for $a_j = \gamma \Delta t$ and $b_j = -(1-\gamma) \alpha h_j \Delta t$ on all qubits $j$ in parallel. We realized these 1-qubit unitaries using Qiskit's ``$U$'' gates, which are implemented using two X90 pulses along with virtual $Z$ rotations \cite{mckay:2017}. Each $e^{-i H_2 \Delta t}$ step can be decomposed in terms of composite $R_{ZZ}(\theta) = e^{-i \theta/2 \, Z \otimes Z}$ operations, with rotation angles of $\theta_{jk} = -2J_{jk} (1-\gamma) \alpha \Delta t$ between qubits $j$ and $k$. For this initial demonstration, we implemented our algorithm exclusively for 1D problem instances in order to avoid SWAP gates, and so that each $e^{-i H_2 \Delta t}$ step could be realized through only two layers of parallel $R_{ZZ}$ rotations, therefore reducing the circuit depth for fixed $r$.

\begin{figure}[h]
\subfloat[\label{figS:Rzx_wrapping_a}]{
\Qcircuit @C=1em @R=1em {
&  \multigate{1}{R_{ZX}(\theta)} & \qw\\
& \ghost{R_{ZX}(\theta)} & \qw 
}}
\hspace{3em}
\subfloat[\label{figS:Rzx_wrapping_b}]{
\Qcircuit @C=1em @R=1em {
&  \multigate{1}{R_{ZX}(\theta \pm 2k \pi)} & \qw\\
& \ghost{R_{ZX}(\theta \pm 2k \pi)} & \qw 
}}
\hspace{3em}
\subfloat[\label{figS:Rzx_wrapping_c}]{
\Qcircuit @C=1em @R=1em {
& \multigate{1}{R_{ZX}(\theta \pm \pi)} & \gate{Z} & \qw\\
&\ghost{R_{ZX}(\theta \pm \pi)} & \gate{X} & \qw \\
}}
\caption{\textbf{Wrapping $\boldsymbol{R_{ZX}(\theta)}$ angles.} Equivalent circuits for arbitrary angles $\theta$ and integers $k$.}
\label{figS:Rzx_wrapping}
\end{figure}
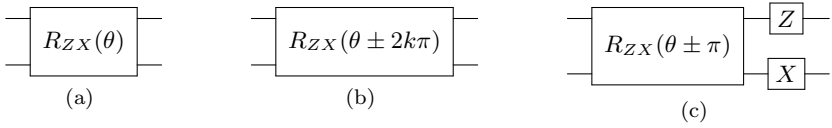

A common way to realize $R_{ZZ}(\theta)$ rotations is through the double-CNOT decomposition in Fig.~\ref{figS:Rzz_approaches_b}, which uses two fully-entangling CNOT gates and a single-qubit rotation $R_Z(\theta) = e^{-i \theta/2 \, Z}$ \cite{nielsen:2000}. CNOTs on \textit{ibmq\_mumbai} are implemented through an echoed cross-resonance (CR) gate which (ideally) performs an $R_{ZX}(\theta) = e^{-i \theta Z \otimes X /2}$ rotation with $\theta=\pi/2$ using two flat-top CR pulses and rotary tones together with an echo sequence \cite{sheldon:2016, sundaresan:2020}. We instead realized $R_{ZZ}(\theta)$ rotations through the pulse-efficient method of Refs.~\cite{stenger:2021, earnest:2021}, shown in Fig.~\ref{figS:Rzz_approaches_c} and available through Qiskit, which scales the area of the CNOT CR and rotary pulses to implement echoed $R_{ZX}(\theta)$ gates for arbitrary angles $\theta$. This approach was thoroughly characterized in \cite{earnest:2021}, where it led to an error reduction of up to 50\% compared to the double-CNOT decomposition due to its substantially shorter pulse schedule. Crucially, since it scales the pulse parameters from calibrated CNOT gates to realize arbitrary angles $\theta$, this pulse-efficient approach requires no additional calibration. Arbitrary angles $\theta$ can be folded into the interval $[-\pi, \pi]$ by adding/subtracting integer multiples of $2\pi$ since $R_{ZX}(\theta + \phi) = R_{ZX}(\theta) R_{ZX}(\phi)$ and $R_{ZX}(\pm 2\pi) = -I$, as shown in Fig.~\ref{figS:Rzx_wrapping_b}. They can then be folded further into $[-\pi/2, \pi/2]$ by adding/subtracting $\pi$ and introducing additional 1-qubit gates since $R_{ZX}(\pm \pi) = \mp i Z \otimes X$, as shown in Fig.~\ref{figS:Rzx_wrapping_c}. Using this approach, we implemented every $R_{ZZ}(\theta)$ rotation in each $e^{-i H_2 \Delta t}$ Trotter step using a pulse-efficient $R_{ZX}$ rotation with an angle in $[-\pi/2, \pi/2]$; that is, with a fraction of a CNOT rather than two CNOTs, together with single-qubit gates. For each $R_{ZX}$ gate, we picked the control ($Z$) and target ($X$) qubits based on the direction that gave the fastest CNOT.

\begin{figure}[h]
\subfloat[\label{figS:Rzz_approaches_a}]{
\Qcircuit @C=1em @R=1em {
&  \multigate{1}{R_{ZZ}(\theta)} & \qw\\
& \ghost{R_{ZZ}(\theta)} & \qw 
}}
\hspace{3em}
\subfloat[\label{figS:Rzz_approaches_b}]{
\Qcircuit @C=1em @R=1em {
& \ctrl{1} & \qw & \ctrl{1} & \qw \\
& \targ & \gate{R_Z(\theta)} & \targ & \qw
}}
\hspace{3em}
\subfloat[\label{figS:Rzz_approaches_c}]{
\Qcircuit @C=1em @R=1em {
& \qw & \multigate{1}{R_{ZX}(\theta)} & \qw & \qw\\
& \gate{\mathrm{H}} & \ghost{R_{ZX}(\theta)} & \gate{\mathrm{H}} & \qw 
}}
\caption{\textbf{Ways to implement $\boldsymbol{R_{ZZ}(\theta)}$.} The target operation $R_{ZZ}(\theta)$ in panel (a) is often implemented using the equivalent circuit in (b), which we call the double-CNOT decomposition. We instead implemented it through the equivalent circuit in (c), where H denotes a Hadamard gate (not the Hamiltonian $H$).}
\label{figS:Rzz_approaches}
\end{figure}
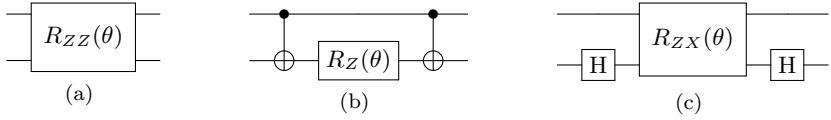

On an ideal quantum computer, the Trotter timestep duration $\Delta t$ could be chosen to ensure that the final state is within some desired error tolerance of the target state given by the Schr\"odinger equation \cite{childs:2021}. On an actual, noisy quantum computer, however, one can only hope to pick $\Delta t$ (or equivalently, the number of timesteps $r$) minimizing the aggregate error from discretizing the target dynamics (i.e., Trotter error) and from gate errors \cite{knee:2015, endo:2019, clinton:2021}. When $r$ is small (and $\Delta t$ is large) the former type of error dominates, while when $r$ is large (and $\Delta t$ is small) the latter dominates. The optimal choice of $\Delta t$ could therefore depend on $H$, $t$, and on the nature of the experimental gate errors. We used a Trotter timestep of $\Delta t=0.8$ for all experiments in this work. We found this value to give a reasonably good trade-off between gate and Trotter errors for the implementation described above, independent of $t$ (as predicted in Ref.~\cite{knee:2015}), $\gamma$, and the model instance. Moreover, it is sufficiently large that all $R_{ZX}$ gates could be implemented by shortening the flat-top portion of CR pulses, rather than scaling their amplitude \cite{earnest:2021}. 

\subsubsection{Twirling}

We sought not only to minimize experimental errors, but also to ensure that any remaining errors satisfied the symmetry requirement $Q(\boldsymbol{s'}|\,\boldsymbol{s}) = Q(\boldsymbol{s}\, | \boldsymbol{s'})$ for our algorithm. (We emphasize that quantum errors do not necessarily bias our MCMC algorithm, i.e., cause it to converge to the wrong distribution. Rather, it is asymmetry in the quantum errors that can have this effect.) We found that implementing bare Trotter circuits like that in Fig.~\ref{figS:trotter_circ} clearly breaks this symmetry, which could bias our algorithm. We attribute this to biased errors in the two-qubit gates and the readout, which we expect to be the noisiest components of our experiments. By randomizing the errors in both we were able to restore the necessary symmetry.

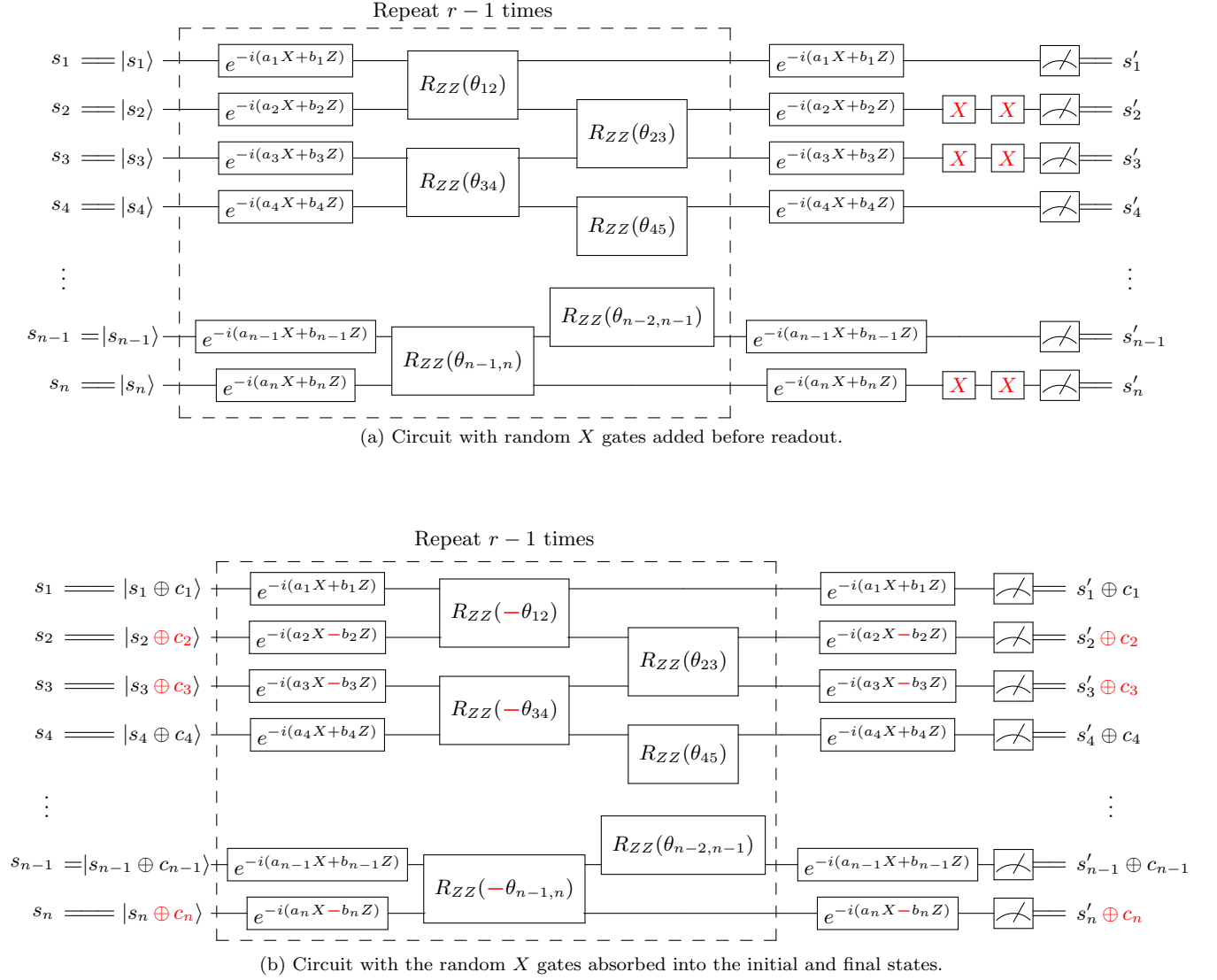
\begin{figure}[h]
\centering
\subfloat[Circuit with random $X$ gates added before readout.\label{figS:SPAM_twirl_a}]{
\Qcircuit @C=0.75em @R=0.75em {
 & & & & & & & & \mbox{Repeat $r-1$ times} & & & & & & & &&\\ 
 & & & & & & & & & & & & & & & &&\\ 
 \lstick{s_1}  & \cw & \cw & \hspace{1.5ex}\ket{s_1} &&& \qw & \gate{e^{-i (a_1 X + b_1 Z)}} & \multigate{1}{R_{ZZ}(\theta_{12})} & \qw & \qw & \gate{e^{-i (a_1 X + b_1 Z)}} & \qw & \qw & \meter & \cw & \rstick{s_1'} \cw \\
 \lstick{s_2}  & \cw & \cw & \hspace{1.5ex}\ket{s_2} &&& \qw & \gate{e^{-i (a_2 X + b_2 Z)}} & \ghost{R_{ZZ}(\theta_{12})} & \multigate{1}{R_{ZZ}(\theta_{23})} & \qw &  \gate{e^{-i (a_2 X + b_2 Z)}} & \gate{\red{X}} & \gate{\red{X}} & \meter & \cw & \rstick{s_2'} \cw \\
 \lstick{s_3}  & \cw & \cw & \hspace{1.5ex}\ket{s_3} &&& \qw & \gate{e^{-i (a_3 X + b_3 Z)}} & \multigate{1}{R_{ZZ}(\theta_{34})} & \ghost{R_{ZZ}(\theta_{23})} & \qw & \gate{e^{-i (a_3 X + b_3 Z)}} & \gate{\red{X}} & \gate{\red{X}}  & \meter & \cw & \rstick{s_3'} \cw \\
 \lstick{s_4} & \cw & \cw & \hspace{1.5ex}\ket{s_4} &&& \qw & \gate{e^{-i (a_4 X + b_4 Z)}} & \ghost{R_{ZZ}(\theta_23)} & \multigate{2}{R_{ZZ}(\theta_{45})} & \qw & \gate{e^{-i (a_4 X + b_4 Z)}} & \qw & \qw & \meter & \cw & \rstick{s_4'} \cw \\
 & & & & & & & & & & & & & & & & \\ 
 & & & & & & & & & & & & & & & &\\ 
 \lstick{\vdots \hspace{0.5ex}} & & & & & & & & & & & & & & & & \rstick{\hspace{0.5ex}\vdots} \\
 & & & & & & & & & & & & & & & &\\
 & & & & & & & & & & & & & & & &\\ 
 \lstick{s_{n-1}}  & \cw & & \hspace{0ex}\ket{s_{n-1}} &&& \qw & \gate{e^{-i (a_{n-1} X + b_{n-1} Z)}} & \multigate{1}{R_{ZZ}(\theta_{n-1,n})} & \multigate{-2}{R_{ZZ}(\theta_{n-2,n-1})} & \qw & \gate{e^{-i (a_{n-1} X + b_{n-1} Z)}} & \qw & \qw & \meter & \cw & \rstick{s_{n-1}'} \cw \\
 \lstick{s_n}  & \cw & \cw & \hspace{1.5ex} \ket{s_n} &&& \qw & \gate{e^{-i (a_n X + b_n Z)}} & \ghost{R_{ZZ}(\theta_{n-1,n})} & \qw & \qw & \gate{e^{-i (a_n X + b_n Z)}} & \gate{\red{X}} & \gate{\red{X}} & \meter & \cw & \rstick{s_n'} \cw
\gategroup{2}{7}{14}{11}{0em}{--}   \\ 
 & & & & & & & & & & & & & 
}}
\vspace{3em}
\subfloat[Circuit with the random $X$ gates absorbed into the initial and final states.\label{figS:SPAM_twirl_b}]{
\hspace{-4em}
\Qcircuit @C=0.75em @R=0.75em {
 & && & && & & & & & \mbox{Repeat $r-1$ times} & & & & & & \\ 
 & && & && & & & & & & & & & & & \\ 
 \lstick{s_1}  & \cw & \cw & \cw & && \ket{s_1 \oplus c_1} &&&& \gate{e^{-i (a_1 X + b_1 Z)}} & \multigate{1}{R_{ZZ}(\red{\boldsymbol{-}} \theta_{12})} & \qw & \qw & \gate{e^{-i (a_1 X + b_1 Z)}} & \meter & \cw & \rstick{s_1' \oplus c_1} \cw \\
 \lstick{s_2}  & \cw & \cw & \cw &&& \ket{s_2 \,\red{\oplus}\, \red{c_2}} &&&& \gate{e^{-i (a_2 X \red{\boldsymbol{-}} b_2 Z)}} & \ghost{R_{ZZ}(-\theta_{12})} & \multigate{1}{R_{ZZ}(\theta_{23})} & \qw &  \gate{e^{-i (a_2 X \red{\boldsymbol{-}} b_2 Z)}} & \meter & \cw & \rstick{s_2' \, \red{\oplus} \, \red{c_2}} \cw \\
 \lstick{s_3}  & \cw & \cw & \cw &&& \ket{s_3 \, \red{\oplus}\, \red{c_3}} &&&& \gate{e^{-i (a_3 X \red{\boldsymbol{-}} b_3 Z)}} & \multigate{1}{R_{ZZ}(\red{\boldsymbol{-}} \theta_{34})} & \ghost{R_{ZZ}(\theta_{23})} & \qw & \gate{e^{-i (a_3 X \red{\boldsymbol{-}} b_3 Z)}} & \meter & \cw & \rstick{s_3' \, \red{\oplus} \, \red{c_3}} \cw \\
 \lstick{s_4} & \cw & \cw & \cw &&& \ket{s_4 \oplus c_4} &&&& \gate{e^{-i (a_4 X + b_4 Z)}} & \ghost{R_{ZZ}(-\theta_34)} & \multigate{2}{R_{ZZ}(\theta_{45})} & \qw & \gate{e^{-i (a_4 X + b_4 Z)}} & \meter & \cw & \rstick{s_4' \oplus c_4} \cw \\
 & && & & & & & & & & & & &&& & \\ 
 & && & & & & & & & & & & &&& & \\ 
 \lstick{\vdots \hspace{0.5ex}} && &&& & & & & & & & & & & & & \rstick{\hspace{3.5ex}\vdots} \\
 & & && & & & & & & & & && && & \\
 & & && & & & & & & & & & && && \\ 
 \lstick{s_{n-1}}  & \cw &&&&& \ket{s_{n-1}\oplus c_{n-1}} \hspace{3.5ex} &&&& \gate{e^{-i (a_{n-1} X + b_{n-1} Z)}} & \multigate{1}{R_{ZZ}(\red{\boldsymbol{-}} \theta_{n-1,n})} & \multigate{-2}{R_{ZZ}(\theta_{n-2,n-1})} & \qw & \gate{e^{-i (a_{n-1} X + b_{n-1} Z)}} & \meter & \cw & \rstick{s_{n-1}' \oplus c_{n-1}} \cw \\
 \lstick{s_n}  & \cw & \cw & \cw &&& \ket{s_n \, \red{\oplus} \, \red{c_n}} &&&& \gate{e^{-i (a_n X \red{\boldsymbol{-}} b_n Z)}} & \ghost{R_{ZZ}(-\theta_{n-1,n})} & \qw & \qw & \gate{e^{-i (a_n X \red{\boldsymbol{-}} b_n Z)}} & \meter & \cw & \rstick{s_n' \, \red{\oplus} \, \red{c_n} } \cw
\gategroup{2}{10}{14}{14}{-0.5em}{--}   \\ 
 & & & & & && & & & & & & & &&
}
}
\caption{\textbf{SPAM twirling.} The top circuit (a) is equivalent to the bare Trotter circuit in Fig.~\ref{figS:trotter_circ}, but has pairs of $X$ gates added to random qubits (qubits 2, 3 and $n$ in this example, shown in red). To produce the bottom circuit (b), one $X$ gate from each pair is commuted through $\tilde{V}$ and into the initial state, while the other is moved past the readout. The impact is highlighted in red. The bottom circuit produces identical statistics to the top one in theory, provided one prepares the initial state $\ket{\boldsymbol{s} \oplus \boldsymbol{c}}$ for a current MCMC configuration $\boldsymbol{s}$ and XORs the measured string (denoted $\boldsymbol{s'} \oplus \boldsymbol{c}$) with $\boldsymbol{c}$ to recover $\boldsymbol{s'} = \boldsymbol{c} \oplus (\boldsymbol{s'} \oplus \boldsymbol{c})$. In this example $\boldsymbol{c}=(1,-1,-1,1,\dots, 1, -1)$.}
\label{figS:SPAM_twirl}
\end{figure}

To randomize the readout errors, we picked a subset of qubits uniformly at random and---in principle---added two adjacent $X$ gates (which is equivalent to the identity gate) to each of those qubits before the readout, as shown in Fig.~\ref{figS:SPAM_twirl_a}. For each selected qubit, we commuted one $X$ gate through the measurement, transforming it into a classical NOT gate on the output bit. This removes any preferred direction in the readout errors on each qubit (e.g., if a $1\rightarrow 0$ error was more likely to occur than a $0\rightarrow 1$ error). Rather than physically implementing the remaining $X$ gates, we chose to commute them through the Trotter unitaries and into the initial state, in the spirit of \cite{vandenberg:2020}, as shown in Fig.~\ref{figS:SPAM_twirl_b}. This leaves the circuit's structure unchanged, but maps $Z_j \mapsto -Z_j$ for all selected qubits $j$. The net effect is to randomize both the state preparation and measurement (SPAM) errors---accordingly, we refer to this procedure as \textit{SPAM twirling}. We will describe which qubits get selected by an $n$-bit \textit{key} $\boldsymbol{c}$. In keeping with our convention of $\boldsymbol{s} \in \{-1,1\}^n$ rather than $\{0,1\}^n$ for the Ising model, we take $\boldsymbol{c} \in \{-1,1\}^n$ and $c_j=-1$ to mean qubit $j$ is selected ($c_j=1$ otherwise). We use $\boldsymbol{s} \oplus \boldsymbol{c}$ to denote bitwise XOR, which in our notation is simply $(s_1 c_1, \dots, s_n c_n)$. The above procedure of adding random $X$ gates is logically equivalent to:
\begin{enumerate}
    \item picking a random key $\boldsymbol{c} \in \text{uniform}(\{-1,1\}^n)$,
    \item preparing an initial state $\ket{\boldsymbol{s} \oplus \boldsymbol{c}}$ in place of $\ket{\boldsymbol{s}}$
    \item implementing a Trotter circuit with $Z_j \mapsto -Z_j$ for all $j$ where $c_j=-1$, or equivalently, $J_{jk} \mapsto J_{jk} c_j c_k$ and $h_j \mapsto h_j c_j$,
    \item applying $\boldsymbol{c} \, \oplus \, \cdot$ to the measured bit string (i.e., XOR-ing it with $\boldsymbol{c}$) to recover the intended output $\boldsymbol{s'}$.
\end{enumerate}
This is the procedure we implemented in all of the experiments presented here. On an ideal quantum computer, it would yield identical measurement statistics (for $\boldsymbol{s'}$, as a function of $\boldsymbol{s}$) to the bare circuit in Fig.~\ref{figS:trotter_circ} for any key $\boldsymbol{c}$. (Note though that Figs.~\ref{figS:trotter_circ} and \ref{figS:SPAM_twirl_b} do not implement the same unitary; rather, they produce equivalent quantum transition probabilities after post-processing the input and output strings. For brevity, we will nonetheless call the circuits equivalent.) On a noisy quantum computer, however, some readout errors (e.g., $T_1$ errors during measurement) may be more likely than others. But the effect of these more-likely errors on $\boldsymbol{s'}$ depends on which key $\boldsymbol{c}$ is used. Using many random keys (together with the 2-qubit gate twirling described below) removed any detectable asymmetry from our experimental results, as detailed in Section~\ref{secS:supplemental_data}.

\begin{figure}[h]
\subfloat[\label{figS:gate_twirl_a}]{
\Qcircuit @C=1em @R=1em {
& & \push{\rule{0em}{0.8em}} & & \\
&  \multigate{1}{R_{ZX}(\theta)} & \qw\\
& \ghost{R_{ZX}(\theta)} & \qw \\
}}
\hspace{3em}
\subfloat[\label{figS:gate_twirl_b}]{
\Qcircuit @C=1em @R=1em {
& & [ZX,P_1 P_2] = 0 & &\\
& \gate{P_1} & \multigate{1}{R_{ZX}(\theta)} & \gate{P_1} & \qw\\
& \gate{P_2} & \ghost{R_{ZX}(\theta)} & \gate{P_2} & \qw 
}}
\hspace{3em}
\subfloat[\label{figS:gate_twirl_c}]{
\Qcircuit @C=1em @R=1em {
& & \{ ZX,P_1 P_2 \} = 0 & &\\
& \gate{P_1} & \multigate{1}{R_{ZX}(-\theta)} & \gate{P_1} & \qw\\
& \gate{P_2} &\ghost{R_{ZX}(-\theta)} & \gate{P_2} & \qw
}}
\caption{\textbf{Gate twirling.} Rather than implement the target gate $R_{ZX}(\theta)$ in panel (a) directly, we drew Paulis $P_1$ and $P_2$ uniformly at random, independently. Depending on whether $P_1 \otimes P_2$ commuted or anti-commuted with $Z \otimes X$, we implemented the equivalent circuits (b) or (c) respectively.}
\label{figS:gate_twirl}
\end{figure}
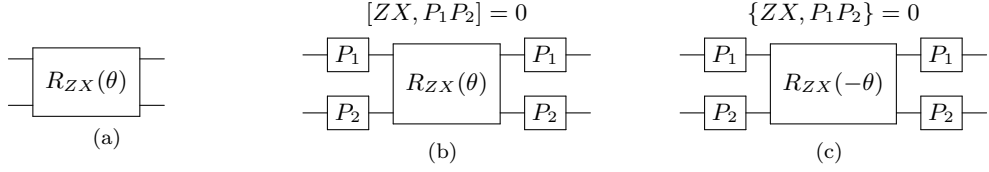

We also randomized every $R_{ZX}(\theta)$ gate independently as shown in Fig.~\ref{figS:gate_twirl}, inspired by Refs.~\cite{knill:2004, wallman:2016, kim:2021}. That is, for each occurrence we picked IID random single-qubit Paulis $P_1, P_2 \sim \text{uniform}(\{I,X,Y,Z\})$ to add before $R_{ZX}$, as in Fig.~\ref{figS:gate_twirl_b}. We then sought to undo the effect of these additional Paulis after the $R_{ZX}$ gate. However, because $R_{ZX}(\theta)$ is not generally a Clifford gate, it is not always possible to find Paulis (nor single-qubit gates, more generally) $P_3$ and $P_4$ such that $R_{ZX}(\theta) = (P_3 \otimes P_4) R_{ZX}(\theta) (P_1 \otimes P_2)$. Rather, if $[P_1 \otimes P_2, Z \otimes X]=0$ we undid these Paulis by re-applying them after $R_{ZX}$ (i.e., we took $P_3=P_1$ and $P_4=P_2$), as in Fig.~\ref{figS:gate_twirl_b}. Otherwise, if $\{ P_1 \otimes P_2, Z \otimes X \}=0$, we did the same thing but with $\theta \mapsto -\theta$, as shown in Fig.~\ref{figS:gate_twirl_c}. Finally, we consolidated all adjacent 1-qubit gates on each qubit into single gates before executing the circuit. We refer to this procedure, informally, as \textit{gate twirling}. (Informally, because the two-qubit gate we apply depends on the choice of $P_1$ and $P_2$, so the process may or may not be considered ``twirling,'' depending on one's definition.) Like SPAM twirling, the motivation for this procedure is to remove any preferred direction from the noise on average---here in the $R_{ZX}(\theta)$ gates rather than the readout. Unlike SPAM twirling, however, gate twirling does not change the unitary described by the circuit. We used both SPAM and gate twirling in all of the experiments presented here.

\subsubsection{Overall structure of experiments}
\label{secS:exp_structure}
All of the experiments in this work consisted of repeatedly preparing a computational state $\ket{\boldsymbol{s}}$, implementing a twirled Trotter circuit, and counting the measured computational states $\ket{\boldsymbol{s'}}$. Combining such counts for different circuits gave estimates for all the proposal probabilities $Q(\boldsymbol{s'}|\, \boldsymbol{s})$, as shown in Fig.~\ref{fig:mechanism}b of the main text and similar figures in Section~\ref{secS:supplemental_data}, which we then used to estimate $P(\boldsymbol{s'} |\boldsymbol{s})$ and $\delta$. We also used the same data to extract MCMC chains, as shown in Fig.~\ref{fig:magnetization} of the main text and the like in Section \ref{secS:supplemental_data}, using a resampling procedure described in Section~\ref{secS:data_analysis}. In principle, we could have collected such data by iterating over all $2^n$ initial states $\ket{\boldsymbol{s}}$ for each circuit and recorded measurement counts for each. In practice, however, it was more experimentally convenient to prepare initial states uniformly at random by preparing a $\ket{+}^{\otimes n} = 2^{-n/2} \sum_{\boldsymbol{s}} \ket{\boldsymbol{s}}$ state, collapsing it to a random $\ket{\boldsymbol{s}}$ through a first measurement, then applying a Trotter circuit followed by a second measurement, as shown in Fig.~\ref{figS:state_prep}a. This approach greatly reduces the time spent loading circuits into the classical control hardware, but increases the number of shots required to observe a fixed number of quantum transitions. The time required to initialize the classical control hardware versus that required for each additional shot makes this a favorable trade-off.

\begin{figure}[h]
\centering
\subfloat[]{
\Qcircuit @C=1em @R=1em {
\lstick{\ket{0}^{\otimes n}} & {/} \qw & \gate{\text{H}^{\otimes{n}} } & \meter \cwx[1] & \gate{\text{Trotter circuit}} & \meter & \rstick{\boldsymbol{s'}} \cw \\
& & & & \cw & \cw & \rstick{\boldsymbol{s}} \cw
}}
\vspace{1em}

\subfloat[]{
\Qcircuit @C=1em @R=1em {
\lstick{\ket{0}^{\otimes n}} & {/} \qw & \gate{\text{H}^{\otimes{n}} } & \meter \cwx[1] & \gate{\text{Gate-twirled Trotter circuit for key } \boldsymbol{c}} & \meter & \rstick{\boldsymbol{s'} \oplus \boldsymbol{c} } \cw \\
& & & & \cw & \cw & \rstick{\boldsymbol{s} \,\, \oplus \boldsymbol{c}} \cw
}}
\caption{\textbf{Overall structure of experiments.} The top circuit (a) represents the logical structure of the experiments, where ``Trotter circuit'' refers to the unitary in Fig.~\ref{figS:trotter_circ} and H denotes a Hadamard gate (not the Hamiltonian $H$). Running this circuit repeatedly samples quantum transitions $\ket{\boldsymbol{s}} \rightarrow \ket{\boldsymbol{s'}}$ according to $\text{Pr}(\boldsymbol{s})=2^{-n}$ and $\text{Pr}(\boldsymbol{s'} | \, \boldsymbol{s}) = \big| \bra{\boldsymbol{s'}} V \ket{\boldsymbol{s}} \big|^2$. The bottom circuit (b) is the one we implemented experimentally. It is logically equivalent to (a), but uses SPAM twirling with a random key $\boldsymbol{c}$ as in Fig.~\ref{figS:SPAM_twirl_b}, and gate twirling on all the resulting $R_{ZX}$ gates independently, as in Fig.~\ref{figS:gate_twirl}. We then XOR-ed both classical registers with $\boldsymbol{c}$ to recover the intended $\boldsymbol{s}$ and $\boldsymbol{s'}$.}
\label{figS:state_prep}
\end{figure}

To further reduce the time required to collect data, we considered a close variant of Algorithm~\ref{alg:M-H} in which both $\gamma$ and $t$ are picked from discrete, rather than continuous, uniform distributions in lines~\ref{algline:gamma_sample} and \ref{algline:t_sample}. For $\gamma$, we discretized the interval $[0.25, 0.6]$ into 10 subintervals (half the number used for numerics, cf.~Eq.~\eqref{eqS:gamma_dist}) and considered $\gamma \in \Gamma$ for
\begin{equation}
\Gamma = \left\{0.25+ \frac{\Delta \gamma}{2},\;
0.25+ \frac{3 \Delta \gamma}{2} , \;
0.25+ \frac{5 \Delta \gamma}{2} , \;
\dots, \;
0.6 - \frac{\Delta \gamma}{2} 
\right\}
\end{equation}
using a subinterval width of $\gamma = 0.035$. Similarly, rather than use $t \in [2,20]$ as in our numerics, we considered only integer values of the Trotter timestep $\Delta t=0.8$, namely $t\in \mathcal{T}$ for
\begin{equation}
\mathcal{T} = \{r \Delta t \;|\; 2 \le r \le 25\}
=
\{ 1.6, 2.4, 3.2, \dots, 19.2, 20\},
\end{equation}
where $r$ indicates the number of Trotter steps used for each $t$. For every $\gamma \in \Gamma$ and $t \in \mathcal{T}$, we implemented $N_\text{twirl}$ distinct (but equivalent) twirled Trotter circuits with keys $\boldsymbol{c}$ and gate twirls picked uniformly at random. We implemented each such circuit experimentally as shown in Fig.~\ref{figS:state_prep}b, for $N_\text{shots}$ shots each. We recorded the number of observed $\ket{\boldsymbol{s}} \rightarrow \ket{\boldsymbol{s'}}$ quantum transitions between all computational states $\ket{\boldsymbol{s}}$ and $\ket{\boldsymbol{s'}}$ separately for each circuit, after accounting for the key $\boldsymbol{c}$. We then iterated this procedure over all $\gamma \in \Gamma$ and $t \in \mathcal{T}$. We therefore ran a total of $N_\text{circ} \equiv |\Gamma| \times |\mathcal{T}| \times N_\text{twirl} = 10 \times 24 \times N_\text{twirl}$ different circuits per experiment, and observed a total of $N_\text{circ} \times N_\text{shots}$ quantum transitions. We picked $N_\text{twirl}$ and $N_\text{shots}$ differently based on the number of qubits $n$, as given in Section~\ref{secS:supplemental_data}.

\subsection{Data analysis}
\label{secS:data_analysis}

We stored the data from each experiment in a 3-dimensional array $C$ with dimensions $2^n \times 2^n \times N_\text{circ}$, where the first, second and third dimensions encode the final states, initial states, and the different circuits respectively. That is, for integers $k,j \in [0,2^n-1]$ encoding the initial and final states $\ket{\boldsymbol{s}}=\ket{k}$ and $\ket{\boldsymbol{s'}} = \ket{j}$ respectively, and for the $\ell^\text{th}$ Trotter circuit out of $N_\text{circ}$:
\begin{equation}
C[j,k,\ell] = \text{number of } \ket{k} \rightarrow \ket{j} \text{ quantum transitions observed for the } \ell^\text{th} \text{ Trotter circuit}.
\end{equation}
We estimated the experimental proposal probabilities $Q(j|k)$ by combining the counts from all circuits; that is, using the estimator
\begin{equation}
\hat{Q}(j|k) = \frac{
\sum_{\ell = 1}^{N_\text{circ}} C[j,k,\ell]
}{ 
\sum_{j=0}^{2^n-1} \sum_{\ell = 1}^{N_\text{circ}} C[j,k,\ell]
}.
\label{eqS:Q_hat}
\end{equation}
The resulting matrix $\hat{\boldsymbol{Q}}$ is shown explicitly in Fig.~\ref{fig:mechanism}b and was used to construct the cumulative histograms for our experiment in Fig.~\ref{fig:mechanism}c-d, all for the $n=10$ instance in the main text. The same approach was used for the corresponding figures in Section~\ref{secS:supplemental_data}. We then used $\hat{\boldsymbol{Q}}$ to construct an estimator $\hat{P}(j|k)$ for $P(j|k)$ as
\begin{equation}
\hat{P}(j|k)
=
\begin{cases}
A(j|k) \, \hat{Q}(j|k) & j \neq k \\ 
1- \sum_{\substack{m=1 \\ m \neq j}}^{2^n-1} \hat{P}(j|m) & j = k
\end{cases},
\label{eqS:P_hat}
\end{equation}
where 
\begin{equation}
A(j|k) = \text{min} \left( 1, \frac{\mu(j)}{\mu(k)} \right)
\end{equation}
is the Metropolis-Hastings acceptance probability from Eq.~\eqref{eq:MH_acceptance} of the main text for a symmetric proposal strategy. Finally, denoting the eigenvalues of the corresponding matrix $\hat{\boldsymbol{P}}$ (which is left-stochastic, by construction) as $\hat{\lambda}_1, \hat{\lambda}_2, \dots, \hat{\lambda}_{2^n}$ where $1=|\hat{\lambda}_1| \ge |\hat{\lambda}_2| \ge  \cdots \ge  |\hat{\lambda}_{2^n}|$, we estimated the absolute spectral gap $\delta$ by $\hat{\delta} = 1 - |\hat{\lambda}_2|$. This point estimate, as a function of the temperature $T$, is shown as a solid red line in Fig.~\ref{fig:n=10_gap} of the main text. The same approach was used for the corresponding figures in Section~\ref{secS:supplemental_data}. 

\subsubsection{MCMC trajectories}
\label{secS:trajectories}

We did not experimentally implement our MCMC algorithm ``online'' by alternating between quantum and classical steps in real time. Doing so would have been impractically slow due to limitations in the classical infrastructure around our quantum processor. The dynamic circuit capabilities required for an online implementation are under development \cite{cross:2021}, and some components have been demonstrated in Ref.~\cite{corcoles:2021}. Rather, we generated the MCMC trajectories in Fig.~\ref{fig:magnetization} of the main text, and the corresponding figures in Section~\ref{secS:supplemental_data}, by running our algorithm ``offline.'' That is, we first formed a cache---namely, the array $C$ described above---of observed quantum transitions by executing circuits as described above and depicted in Fig.~\ref{figS:state_prep}b. We then iterated through the steps of our algorithm, but rather than getting proposed jumps $\boldsymbol{s} \rightarrow \boldsymbol{s'}$ from the quantum processor in real time, we drew them at random without replacement from the cache. This means that we extracted MCMC trajectories by subsampling the same data used to estimate $Q$, $P$ and $\delta$ above.

In effect, for every circuit $\ell$ and every state $k$ we formed a list $L_{\ell k}$ of experimentally observed $\ket{k} \rightarrow \ket{j}$ transitions repeated by multiplicity. E.g., for circuit $\ell=2$ and state $k=1$, if we observed $\ket{k} \rightarrow \ket{0}$ three times, $\ket{k} \rightarrow \ket{4}$ once, $\ket{k} \rightarrow \ket{5}$ twice, and no other $\ket{k} \rightarrow$ [anything] transitions, the corresponding list would be $L_{\ell=2, k=1} = [0,0,0,4,5,5]$ or some permutation thereof. (The order does not matter.) We then ran our algorithm offline by picking an initial state $k$ uniformly at random, a random circuit index $\ell \sim \text{uniform}(\{1,\dots,N_\text{circ}\})$ IID in each iteration (with replacement), and then drawing a proposed jump $j \sim  \text{uniform}(L_{\ell, k})$ without replacement. For instance, if we drew $\ell=2$ and $k=1$, we would then pick $j$ uniformly from $L_{\ell=2, k=1} = [0,0,0,4,5,5]$ (i.e., with Pr$(j=0)=1/2$, Pr$(j=4)=1/6$, Pr$(j=5)=1/3$). If we drew $j=0$, we would remove it from the list so that $L_{\ell=2, k=1} = [0,0,4,5,5]$. We then accepted or rejected the $k\rightarrow j$ jump as usual, i.e., used $k=j$ to start the next iteration if accepted, otherwise we used the same $k$ again. We repeated this process until we hit an empty list $L_{\ell k}$, meaning we ran out of data (even though not all lists were necessarily empty). We call this process \textit{Markov chain subsampling}. In Fig.~\ref{fig:magnetization} of the main text and corresponding ones in Section~\ref{secS:supplemental_data}, we generated MCMC trajectories of 1000 iterations each in this way. We did not reset the cache for each trajectory, i.e., we did not re-use any data. This means that the chains are statistically independent from each other.

Note that this subsampling procedure is \textit{not} equivalent to sampling $j$ from the estimated distributions $[\hat{Q}(j|k)]_{j=0}^{2^n-1}$ or $[\hat{P}(j|k)]_{j=0}^{2^n-1}$ in each iteration. These alternative approaches would only approximately mimic our algorithm. Our approach, on the other hand, involves post-processing but no approximations.

\subsubsection{Symmetry testing} 
\label{secS:symmetry}

Our quantum algorithm assumes the symmetry $Q(\boldsymbol{s'}|\,\boldsymbol{s}) = Q(\boldsymbol{s}\,|\boldsymbol{s'})$ for all computational states $\boldsymbol{s}$ and $\boldsymbol{s'}$ so that the classical accept/reject step can be done efficiently. This symmetry is satisfied in theory by our quantum circuits, and we expect the twirling techniques described above to suppress asymmetry which might arise from experimental imperfections. We use a statistical hypothesis test called the Bowker test to check whether our experimental data is consistent with such symmetry. Even if $\boldsymbol{Q}$ were symmetric, one should not expect the number of experimentally observed $\ket{\boldsymbol{s}} \rightarrow \ket{\boldsymbol{s'}}$ quantum transitions to exactly equal that of $\ket{\boldsymbol{s'}} \rightarrow \ket{\boldsymbol{s}}$ transitions for all $\boldsymbol{s}$ and $\boldsymbol{s'}$, just as one should not expect a fair coin to turn up ``heads'' exactly half the time. Rather, some asymmetry in the experimental counts is inevitable even if the underlying process is symmetric, due to statistical fluctuations. It is therefore appropriate to use a statistical test to quantify how asymmetric the counts are, and whether that degree of asymmetry is consistent with the underlying $Q$ being symmetric.

The Bowker test \cite{bowker:1948}, which generalizes the well-known McNemar test \cite{mcnemar:1947}, serves this purpose. Consider $N$ independent data points, where each datum has two attributes, both of which are labelled $1,\dots, d$. (For instance, a datum could be $(d-2, 3)$.) Form a $d \times d$ matrix $M = (m_{jk})_{j,k=1}^d$ where $m_{jk}$ is the number of data points whose first and second attributes equal $j$ and $k$ respectively (so $\sum_{jk}m_{jk}=N$). The Bowker test aims to determine whether the distribution from which the data was drawn is symmetric; that is, whether $\text{Pr}[(j,k)]=\text{Pr}[(k,j)]$ for all attributes $j,k \in \{1,\dots,d\}$. It takes this symmetry as the null hypothesis, and constructs the test statistic
\begin{equation}
\chi^2 = \sum_{j>k=1}^d \frac{(m_{jk} - m_{kj})^2}{m_{jk} + m_{kj}}.
\label{eqS:chi2}
\end{equation}
Intuitively, $\chi^2$ is a clear measure of asymmetry. It comprises a sum of non-negative terms quantifying the discrepancy in counts between each pair of bins $m_{jk}$ and $m_{kj}$. If the bins contain the same number of counts ($m_{jk} = m_{kj}$) they add nothing to the overall statistic. If they are unequal but contain few data points ($m_{jk}$ and $m_{kj}$ are small) they contribute only slightly to $\chi^2$ since the discrepancy could easily be due to shot noise. But if they are unequal and contain many data points, they contribute strongly to $\chi^2$ since this constitutes strong evidence that $\text{Pr}[(j,k)] \neq \text{Pr}[(k,j)]$. Therefore, a small value of $\chi^2$ suggests symmetry while a large value suggests asymmetry in the underlying distribution (where the meaning of ``small'' and ``large'' depends on the number of terms $\binom{d}{2}$ in the sum). More formally, under the null hypothesis, $\chi^2$ follows a chi-squared distribution asymptotically with $\binom{d}{2} = d(d-1)/2$ degrees of freedom \cite{bowker:1948, krampe:2007}. Let $F$ be the cumulative distribution function for such a chi-squared distribution, then the $p$-value 
\begin{equation}
p_\text{sym} = 1 - F(\chi^2)
\end{equation}
answers the following question: If the underlying distribution were symmetric (null hypothesis), what is the probability of getting data that is as asymmetric, or more asymmetric, than my data $M$? We take $p_\text{sym} \le 1\%$ as the threshold for significance throughout (for consistency with the $99\%$ confidence intervals used throughout), meaning if $p_\text{sym} \le 0.01$ we reject the null hypothesis and conclude that the underlying distribution is asymmetric. Otherwise, the null hypothesis holds, meaning the data is consistent with symmetry.

In our experiments each datum is an observed $\ket{\boldsymbol{s}} \rightarrow \ket{\boldsymbol{s'}}$ quantum transition, $k$ and $j$ represent the initial and final states respectively, and $d=2^n$. When $N_\text{circ}>1$ (which is the case in all of the experiments presented here), Bowker's test cannot be applied directly to the full set of counts (i.e., we cannot simply take $m_{jk} = \sum_{\ell=1}^{N_\text{circ}} C[j,k,\ell]$) because the data points are not IID, as required by the test. The issue is that we recorded quantum transitions for different quantum circuits, but we iterated through these circuits sequentially rather than picking them at random for each shot. This latter approach would have yielded IID data as required by Bowker's test, but would have been experimentally impractical. As a workaround, we subsample our data to mimic this latter approach, apply the Bowker test to the resulting subset of data, and repeat until the full data set has been tested. 

More precisely, we initialized $M$ to all zeros and used the same initial step as in Markov chain subsampling: for every circuit $\ell$ and every state $k$ we effectively formed a list $L_{\ell k}$ of experimentally observed $\ket{k} \rightarrow \ket{j}$ transitions repeated by multiplicity. We then picked a circuit $\ell \sim \text{uniform}(\{1, \dots, N_\text{circ}\})$ and an initial state $k \sim \text{uniform}(\{0, \dots, 2^n-1\})$ uniformly at random, then drew a proposed jump $j \sim \text{uniform}(L_{\ell, k})$. We removed an occurrence of $j$ from $L_{\ell, k}$ and increased $m_{jk}$ by 1. We repeated this process until we hit an empty $L_{\ell, k}$. Unlike in Markov chain subsampling, we picked $k$ uniformly at random in each step, rather than accepting/rejecting the previous $j$ to form a Markov chain. Since this process produces a random subset of IID data, we call it \textit{IID subsampling}. We then applied Bowker's test to the resulting $M$ and computed $p_\text{sym}$. Finally, we repeated these steps many times (starting over with the full data set each time) until every datum had been used in at least one Bowker test with overwhelming probability. In effect, we tested for symmetry in the full data set by testing for it in many random subsets which, together, covered the full set. There is some ambiguity in how to combine the $p$-values resulting from the different random subsets, especially since they are correlated (so for a fixed data set they need not be uniformly distributed, even under the null hypothesis). However, the observed $p$-values, given in Section~\ref{secS:supplemental_data}, were so large that they clearly did not show any statistically significant asymmetry, regardless of how they are combined.

Note that exact $Q(\boldsymbol{s'}|\,\boldsymbol{s}) = Q(\boldsymbol{s}\,|\boldsymbol{s'})$ symmetry is impossible in a complex engineered system like a quantum computer, or a classical computer for that matter. Given enough data, it will always be possible to resolve some slight asymmetry. However, this is less of a practical issue and more of a philosophical one. Practically, if we assume the ratio $Q(\boldsymbol{s}\,|\boldsymbol{s'}) / Q(\boldsymbol{s'}|\,\boldsymbol{s}) = 1$ in the M-H acceptance probability (Eq.~\eqref{eq:MH_acceptance} of the main text) but the true value is slightly different, the resulting Markov chain may converge to a slightly different distribution than intended \cite{cho:2001}, which may slightly bias thermal average estimates like in Eq.~\eqref{eq:avg_magnetization} of the main text. We could not resolve any such biases in our experiments. More broadly, we propose the following rule of thumb: if the (potential) asymmetry in $Q$ is too small to detect from the data at hand, it is probably too small to make much difference in your MCMC results. This heuristic cuts both ways of course, and suggests that long Markov chains are more sensitive to such asymmetry. We raise this issue to clarify the following point: we do not claim that any of our experiments realized a $\boldsymbol{Q}$ that was \textit{exactly} symmetric, only that it was symmetric enough for practical purposes.

Finally, there are two main ways to handle empty pairs of bins ($m_{jk}=m_{kj}=0$), to our knowledge. In both, the corresponding $0/0$ terms in Eq.~\eqref{eqS:chi2} are set to zero, since they give no evidence of asymmetry. (A different way to motivate this is to substitute $m_{jk}$ with its expectation value $N \, \text{Pr}[(j,k)]$ in Eq.~\eqref{eqS:chi2}, and likewise for $m_{kj}$. The corresponding term in $\chi^2$ tends towards zero as $N$ decreases.) The traditional Bowker test uses $k=\binom{2^n}{2}$ chi-squared degrees of freedom regardless of the observed data; however, a variant (used in Stata statistics software \cite{stata}, for instance) accounts for empty bin pairs by using $k=\binom{2^n}{2} - N_\text{empty}$ instead, where $N_\text{empty}$ is the number of empty pairs. We call this variant the modified Bowker test. We give $p$-values for both variants in Section~\ref{secS:supplemental_data}; when there are empty bin pairs the traditional Bowker test tends to be more conservative (i.e., to give larger $p$-values) than the modified test.

\subsubsection{Bootstrapping} 
\label{secS:bootstrapping}

In Fig.~\ref{fig:n=10_gap} of the main text and similar figures in Section~\ref{secS:supplemental_data}, we computed $\delta$ confidence intervals (CIs) for our experiments through bootstrapping. However, we faced a similar issue as in Section~\ref{secS:symmetry}, in that bootstrapping assumes IID data, whereas our experiments naturally produced correlated data. We therefore employed IID subsampling to compute CIs. That is, for every $T$ we generated an IID subsample $M$ of the full dataset as described Section~\ref{secS:symmetry}. We then used the \texttt{stats.bootstrap} function from the SciPy library \cite{scipy} with the ``basic'' (i.e., reverse percentile) setting to resample this $M$ (rather than the full dataset) 200 times to estimate the 99\% confidence interval for $\delta$ at this $T$. (We picked the most stringent bootstrap settings possible within a reasonable computational cost. The 99\% confidence level was chosen to match the 1\% significant threshold used in Section~\ref{secS:symmetry}.) We estimated $Q(j|k)$ from $M$ using 
\begin{equation}
\hat{Q}(j|k) = \frac{m_{jk}}{\sum_{j=0}^{2^n-1} m_{jk} },
\end{equation}
then estimated $\boldsymbol{P}$ and $\delta$ following Eq.~\eqref{eqS:P_hat}, and used this latter estimator for bootstrapping. We repeated this process for every $T$, generating a new subsample $M$ for each temperature. This means that while we used the full dataset to form the point estimate of $\delta$ at each $T$, i.e., the solid line in Fig.~\ref{fig:n=10_gap} and the like, we used different random subsets of the data to compute the CIs at each $T$. This subsampling, together with the stochastic nature of bootstrapping, is why the estimated $\delta$ forms a smooth curve while the CIs fluctuate more noticeably. This approach also means that the computed CIs are almost certainly pessimistic (i.e., overly broad), since they do not use all the data. Finally, note that the CIs at adjacent temperatures all agree reasonably well (in that there are no large, erratic fluctuations), despite using different random subsets of data. This self-consistency lends credence to the bootstrap settings described above, and to this approach to estimating $\delta$ CIs more broadly.

Note that the numerical data used to estimate $\delta$ for the S-W and Wolff cluster algorithms in Figs.~\ref{figS:gap_n=10}, \ref{figS:gap_n=9} and \ref{figS:gap_n=8} was IID by design, so we could bootstrap it directly, with no need to subsample it first.

\subsection{Supplemental data}
\label{secS:supplemental_data}

In this section we give some supplementary figures of merit for the $n=10$ experiment presented in the main text, as well as data for two other experiments on illustrative model instances with $n=8$ and $n=9$ spins.

\newpage
\subsubsection{$n=10$ instance}

\begin{figure}[h]
\centering 
\includegraphics[width=0.4\textwidth]{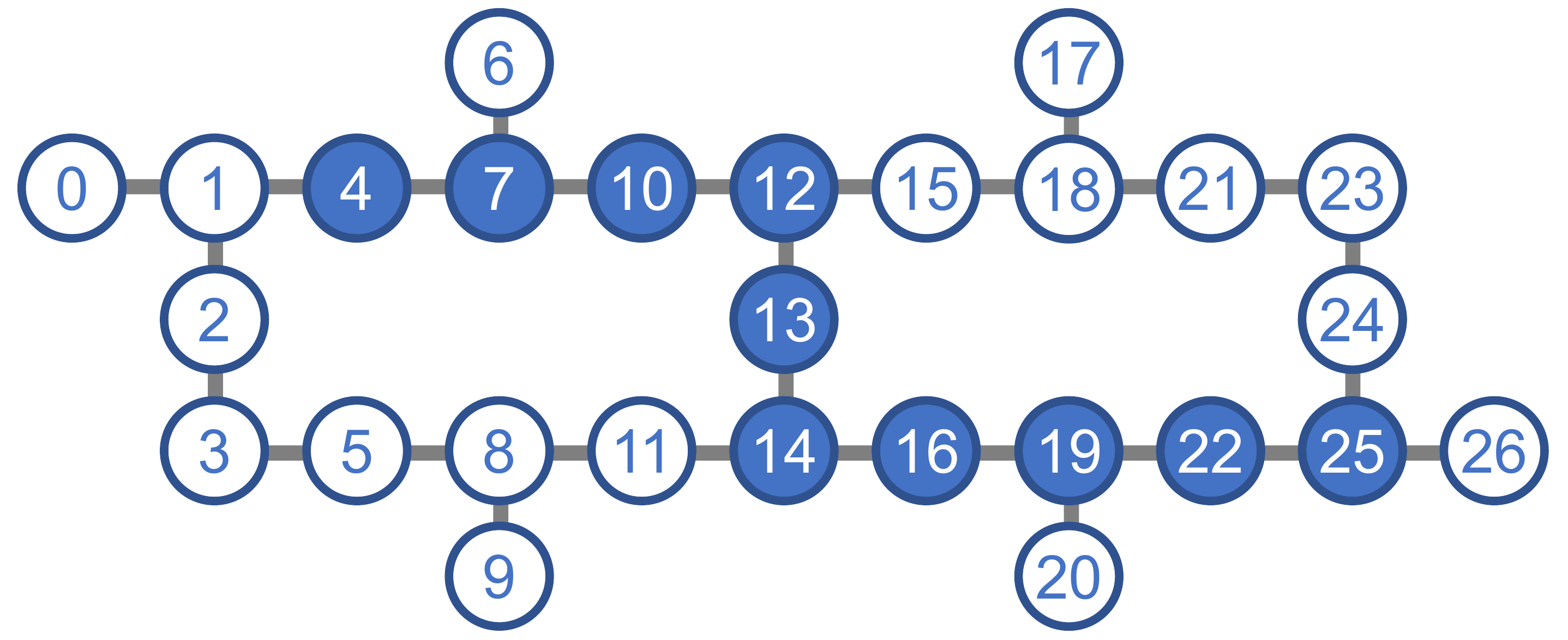}
\hspace{5em}
\raisebox{4.5em}{
\setlength{\tabcolsep}{12pt}
\begin{tabular}{ c c }
CNOT error (mean) & 0.66\% \\
CNOT error (worst) & 0.98\% \\
Readout assignment error (mean) & 2.06\%  \\ 
Readout assignment error (worst) & 4.94\% \\
\end{tabular}
}
\caption{\textbf{Device information for $\boldsymbol{n=10}$ experiment}. Left: the 1D chain of qubits on \textit{ibmq\_mumbai} used for this experiment, where Ising model spins $1, \dots, 10$ were mapped to qubits $4,\dots,25$ respectively. Right: representative figures of merit for this chain of qubits at the time of the experiment.} 
\label{figS:n=10_qubits}
\end{figure}

The qubits used for the $n=10$ experiment presented in the main text, together with representative figures of merit from the last calibration before the experiment, are shown in Fig.~\ref{figS:n=10_qubits}. The problem instance is defined by the couplings and fields

\begin{align}
(J_{j+1,j})_{j=1}^{n-1}
&= \big( -0.99121054, \; 0.84436089, \; -0.83043895,  \; 0.95766024, \; -1.02814718, \; 1.24969204, \nonumber \\
& \qquad \qquad \qquad \qquad 0.81649925, \; -0.92147578,  \; 1.11394418 \big) \\
(h_j)_{j=1}^n
&=
\big( 0.47921821,  \;0.10207621, \; -0.4780673 ,\; -0.39407213, \; 0.15239487,\; 0.44938277, \nonumber \\
& \qquad \qquad \qquad 0.91715616, \; 0.73303354, \; 0.40145444,\; 0.55915183
\big), \nonumber
\end{align}
with all other $J_{jk}$ equal to zero.

\begin{figure}[h]
\centering
\setlength{\tabcolsep}{12pt}
\raisebox{9em}{
\begin{tabular}{c c} 
$N_\text{twirl}$ & 6\\
$N_\text{shots}$ & $4 \times 10^4$\\ 
 Total shots & $5.76 \times 10^7$ \\[2ex]
TV error & 0.206 \\ 
$\expval{p_\text{sym}}$ traditional & 36.7\%  \\
$\expval{p_\text{sym}}$ modified & 25.1\% 
\end{tabular}
}
\hspace{3em}
\includegraphics{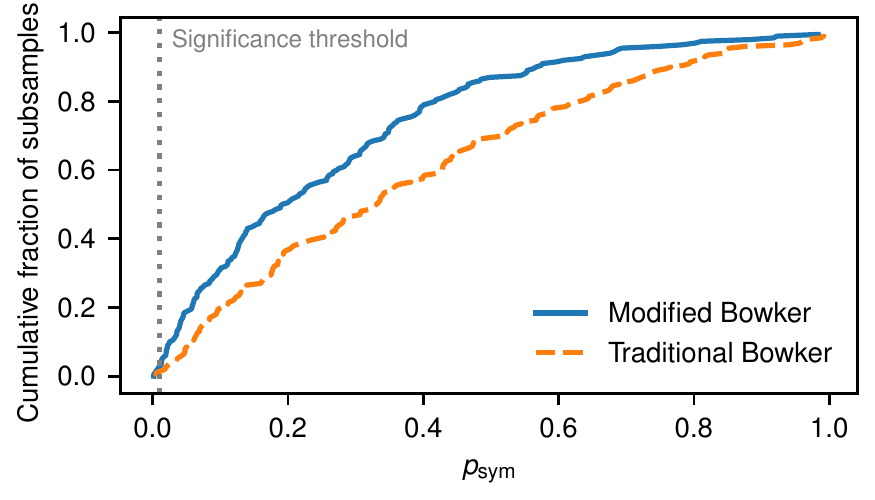}
\caption{\textbf{Parameters and figures of merit for $\boldsymbol{n=10}$ experiment.} Left: experimental parameters (top) and the resulting figures of merit (bottom). Right: Cumulative histograms showing the distribution of $p$-values from Bowker tests run on random subsets of the full dataset. The $1\%$ significance threshold used throughout is shown for comparison.}
\label{figS:n=10_FoM}
\end{figure}

Let $\boldsymbol{Q}_\text{th}$ be the $2^n \times 2^n$ matrix of proposal probabilities for our algorithm obtained through numerical simulation, and let $\hat{\boldsymbol{Q}}_\text{exp}$ be that of experimentally estimated probabilities defined by Eq.~\eqref{eqS:Q_hat}. The total variation (TV) distance, defined as
\begin{equation}
\|\vec{p} - \vec{q} \|_\textsc{tv} = \frac{1}{2} \sum_{i=0}^{2^n-1} \big|p_i - q_i \big| = \frac{1}{2} \|\vec{p} - \vec{q} \, \|_1
\end{equation}
for distributions $\vec{p}, \vec{q} \in \mathbb{R}^{2^n}$, is the most common measure of distance between distributions in MCMC. Accordingly, we use 
\begin{equation}
\text{TV error}
=
\big \| \boldsymbol{Q}_\text{th} - \hat{\boldsymbol{Q}}_\text{exp} \big \|_\text{TV,avg}
=
\frac{1}{2^{n+1}} 
\sum_{i,j=0}^{2^n-1} 
\Big| 
(\boldsymbol{Q}_\text{th})_{ij} - (\hat{\boldsymbol{Q}}_\text{exp})_{ij}
\Big|
\end{equation}
as our measure of experimental error; namely, the average TV distance between all columns of $\boldsymbol{Q}_\text{th}$ and $\hat{\boldsymbol{Q}}_\text{exp}$ (each of which is a probability distribution). The extreme values of 0 and 1 represent the smallest and largest possible error, respectively. Since our algorithm only uses initial states and measurements in the computational basis, the experimental error is completely captured by comparing the quantum transition probabilities in $\boldsymbol{Q}_\text{th}$ and $\hat{\boldsymbol{Q}}_\text{exp}$. This makes TV error a better figure of merit in this context than more typical measures of distance between quantum channels (e.g., process fidelity, diamond distance etc.) encompassing all possible initial states and measurement bases \cite{gilchrist:2005}, most of which are \textit{a priori} irrelevant here. The TV error for this experiment, together with other experimental parameters defined in Section~\ref{secS:exp_structure}, is given in the left panel of Fig.~\ref{figS:n=10_FoM}. For comparison, the average TV distances between $\hat{\boldsymbol{Q}}_\text{exp}$ and the uniform proposal, the local proposal, and the identity matrix $I$ are larger: 0.259, 0.953 and 0.987 respectively.

Fig.~\ref{figS:n=10_FoM} also gives $p$-values from the traditional and modified Bowker tests, averaged over 200 IID subsamples as described in Section~\ref{secS:symmetry}. Both are well above the 1\% significance threshold, and are therefore consistent with the experimental quantum proposal mechanism being symmetric. Typically, about 15\% of the full dataset makes it into each IID subsample used for the Bowker tests. There were a total of $5.76 \times 10^7$ shots, so the expected number of data points that did not make it into any subsample is approximately $5.76 \times 10^7 \times (1-0.15)^{200} \sim  10^{-7} \ll 1$. The 200 subsamples used to test for symmetry therefore cover the full dataset. The integrated histogram of $p$-values for both Bowker test variants over these random subsamples are shown explicitly in the right panel of Fig.~\ref{figS:n=10_FoM}. Neither is sharply concentrated on $p_\text{sym} \lesssim 1\%$, which constitutes further evidence of symmetry.

\begin{figure}[h!]
\centering
\includegraphics[valign=c]{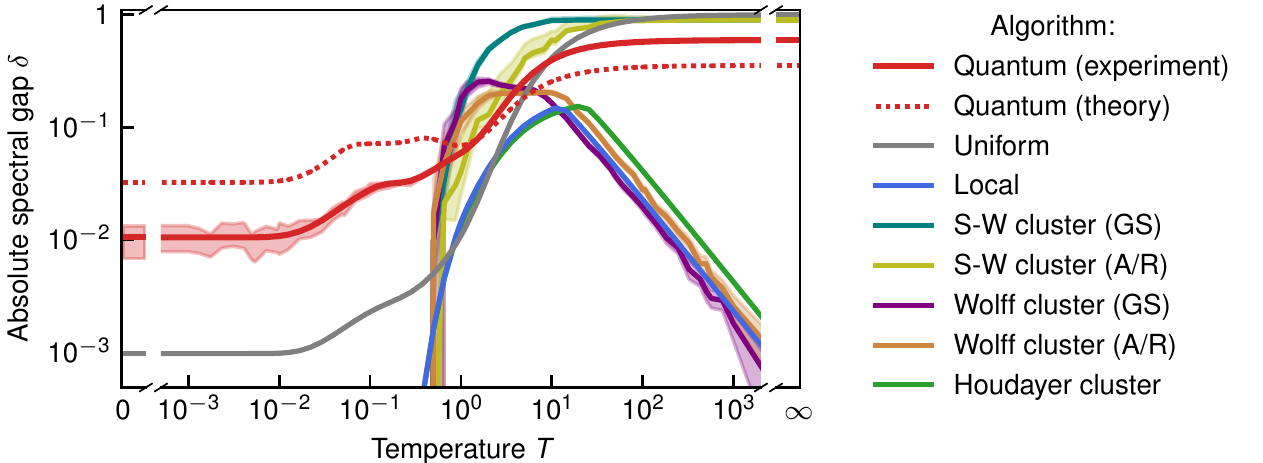}
\caption{\textbf{Convergence rates for $\boldsymbol{n=10}$ instance.} The absolute spectral gap $\delta$ for various MCMC algorithms on the same $n=10$ model instance. The quantum, local and uniform proposal strategies were combined with the Metropolis-Hastings acceptance probability \eqref{eq:MH_acceptance} of the main text. The data shown for these is the same as in Fig.~\ref{fig:n=10_gap} of the main text. It is re-plotted here for comparison with the five cluster algorithms discussed in Section~\ref{secS:clusters}. ``GS'' and ``A/R'' denote ``ghost spin'' and ``accept/reject'' versions respectively of the Swendsen-Wang (S-W) and Wolff algorithms. Error bands, where present, denote 99\% confidence intervals found through basic (i.e., reverse percentile) bootstrapping using 200 resamples independently for each $T$. For the experimental realization of our quantum algorithm we bootstrapped random subsets of the full dataset, which comprises $5.76\times 10^7$ shots, as described in Sec.~\ref{secS:bootstrapping}. For the S-W and Wolff algorithms there was no need for such subsampling. Rather, we formed point estimates and confidence intervals for their absolute spectral gaps at different temperatures by generating $10^5$ IID cluster moves separately for each $T$.}
\label{figS:gap_n=10}
\end{figure}

\begin{figure}[h]
\centering
\subfloat[Lazy M-H]{%
\raisebox{0.5ex}{%
\includegraphics[width=0.59\textwidth, height=1.5in]{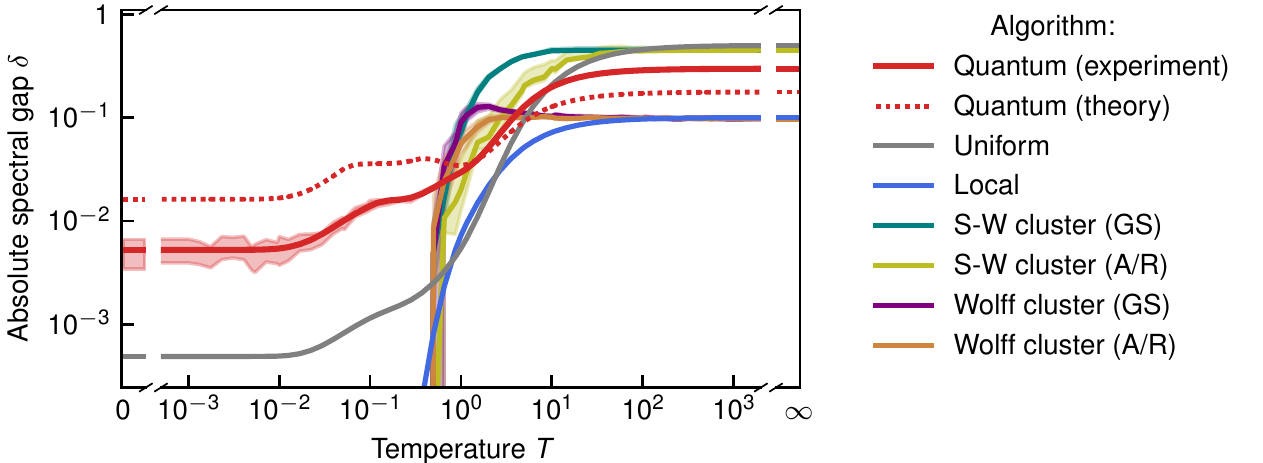}}
}
\hfill
\subfloat[Gibbs sampler]{%
\includegraphics[width=0.38\textwidth, height=1.53in]{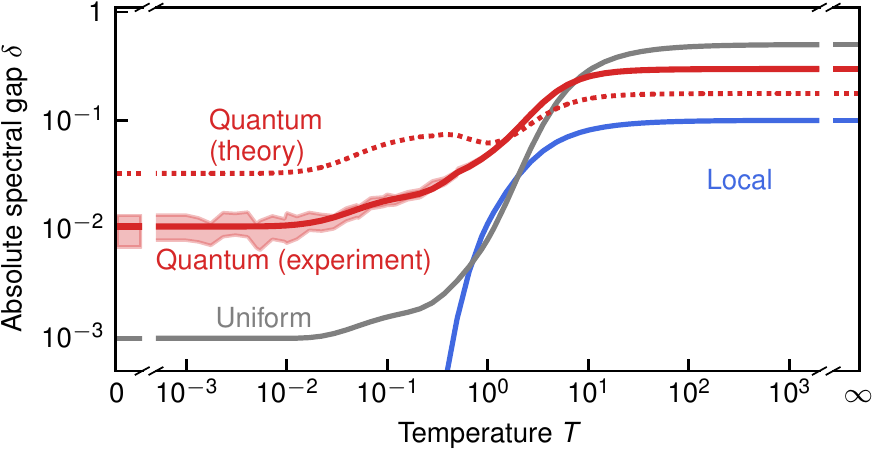}
}
\caption{\textbf{Convergence rates for $\boldsymbol{n=10}$ instance---variants.} The absolute spectral gap $\delta$ for various MCMC algorithms on the same $n=10$ model instance. (a) Lazy versions of the Markov chains considered in Fig.~\ref{fig:n=10_gap} of the main text and Fig.~\ref{figS:gap_n=10}. The Houdayer cluster algorithm is not shown since there is some ambiguity in how a lazy version should be defined, and because its $\delta$ at low $T$ is set by a positive eigenvalue, meaning lazy variants should offer no speedup anyways. ``GS'' and ``A/R'' denote ``ghost spin'' and ``accept/reject'' versions respectively of the Swendsen-Wang (S-W) and Wolff algorithms. (b) Absolute spectral gaps based on the Gibbs sampler acceptance probability in Eq.~\eqref{eqS:gibbs_acceptance}, rather than the M-H probability in Eq.~\eqref{eq:MH_acceptance} of the main text. Error bands in both panels, where present, denote 99\% bootstrap confidence intervals constructed identically to those in Fig.~\ref{figS:gap_n=10}.} 
\label{figS:gap_n=10_lazyglauber}
\end{figure}

\begin{figure}[h]
\centering
\includegraphics[width=2.5in]{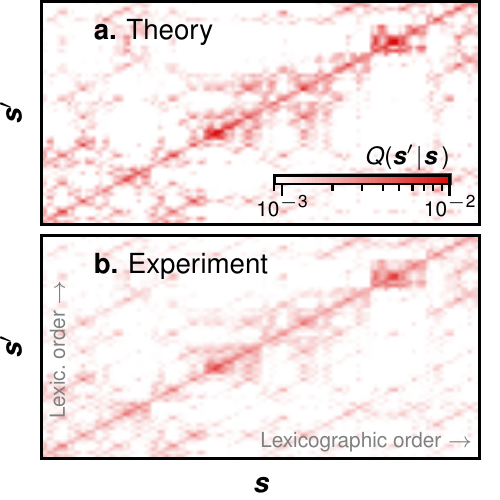}
\caption{The same data shown in Figs.~\ref{fig:mechanism}a-b of the main text, but with $\boldsymbol{s}$ and $\boldsymbol{s'}$ sorted in lexicographic order, rather than by increasing Ising energy $E$.}
\label{figS:Q_lex_n=10}
\end{figure}

Figs.~\ref{figS:gap_n=10}, \ref{figS:gap_n=10_lazyglauber} and \ref{figS:Q_lex_n=10} show the same experimental data as Figs.~\ref{fig:n=10_gap} and \ref{fig:mechanism}a-b of the main text, but in different ways. Specifically, Fig.~\ref{figS:gap_n=10} shows the same results as Fig.~\ref{fig:n=10_gap}, but also plots the spectral gaps for the five MCMC cluster algorithms introduced in Sec~\ref{secS:clusters}. As discussed in the main text, they are substantially more complicated than the other classical MCMC algorithms considered, but offer no significant $\delta$ improvement in the low-$T$ regime of interest, which is why they are relegated to here. For completeness, Fig.~\ref{figS:gap_n=10_lazyglauber} similarly shows the absolute spectral for two M-H variants discussed in Sections~\ref{secS:glauber} and \ref{secS:lazy_chains}: lazy M-H and the Gibbs sampler acceptance probability. The quantum enhancement in $\delta$ at low $T$ is nearly identical in both variants. Finally, Fig.~\ref{figS:Q_lex_n=10} shows the same proposal probabilities $Q(\boldsymbol{s'}|\,\boldsymbol{s})$ as Fig.~\ref{fig:mechanism}a-b, but with the spin configurations sorted lexicographically, rather than by increasing $E$.

\clearpage
\subsubsection{$n=9$ instance}

\begin{figure}[h]
\centering 
\includegraphics[width=0.4\textwidth]{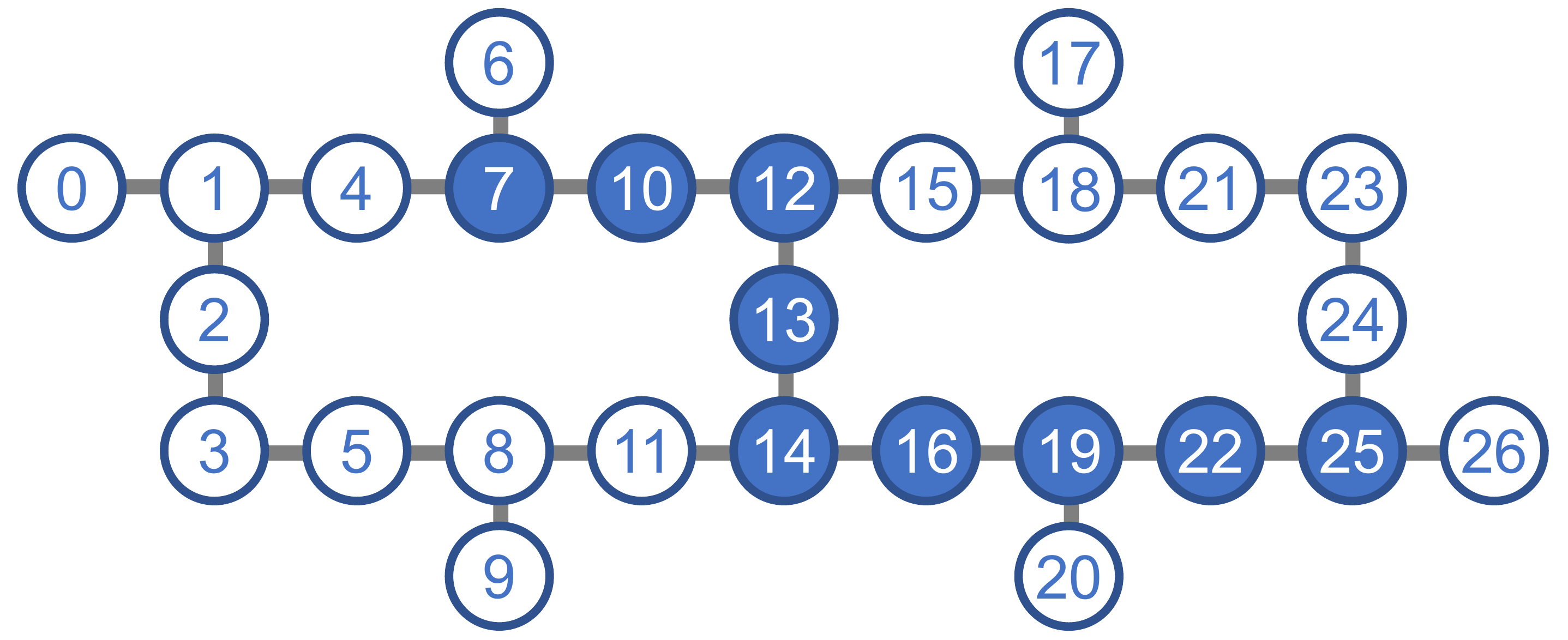}
\hspace{5em}
\raisebox{4.5em}{
\setlength{\tabcolsep}{12pt}
\begin{tabular}{ c c }
CNOT error (mean) & 0.71\% \\
CNOT error (worst) & 0.88\% \\
Readout assignment error (mean) & 1.97\%  \\ 
Readout assignment error (worst) & 4.53\% \\
\end{tabular}
}
\caption{\textbf{Device information for $\boldsymbol{n=9}$ experiment}. Left: the 1D chain of qubits on \textit{ibmq\_mumbai} used for this experiment, where Ising model spins $1, \dots, 9$ were mapped to qubits $7,\dots,25$ respectively. Right: representative figures of merit for this chain of qubits at the time of the experiment.}
\label{figS:n=9_qubits}
\end{figure}

We implemented our algorithm for another illustrative model instance with $n=9$ spins. We present the experimental results here, which are analogous to those for the $n=10$ instance from the main text. This $n=9$ instance is defined by the couplings and fields
\begin{align}
(J_{j+1,j})_{j=1}^{n-1}
&= \big( 0.89504271, \; 0.85382636, \; 1.01494031,\; 1.04243257,\; 1.19556769, \; 1.03985729\nonumber\\
&       \qquad \qquad \qquad \qquad \qquad  1.15258874, \; 1.17542484 \big) \\
(h_j)_{j=1}^n
&=
\big( -0.41017914, \; 0.69312341, \; 0.61866226, \; 0.45567707, \; -0.32062001,\;  0.31684772 \nonumber \\ 
& \qquad \qquad \qquad -0.62064839, \; 0.04672033, \; -0.48091904
\big), \nonumber
\end{align}
with all other $J_{jk}$ equal to zero. Its four lowest-$E$ configurations are all local minima, and the lowest three are nearly degenerate: at $T=0.1$ they have Boltzmann probabilities of approximately 37\%, 36\% and 27\% for the ground, first- and second-excited configurations respectively. The qubits used in this experiment, together with representative figures of merit from the last calibration before the experiment, are shown in Fig.~\ref{figS:n=9_qubits}.

\begin{figure}[h]
\centering
\setlength{\tabcolsep}{12pt}
\raisebox{9em}{
\begin{tabular}{c c} 
$N_\text{twirl}$ & 4\\
$N_\text{shots}$ & $3 \times 10^4$\\ 
 Total shots & $2.88 \times 10^7$ \\[2ex]
TV error & 0.189 \\ 
$\expval{p_\text{sym}}$ traditional & 12.7\%  \\
$\expval{p_\text{sym}}$ modified & 12.7\% 
\end{tabular}
}
\hspace{3em}
\includegraphics{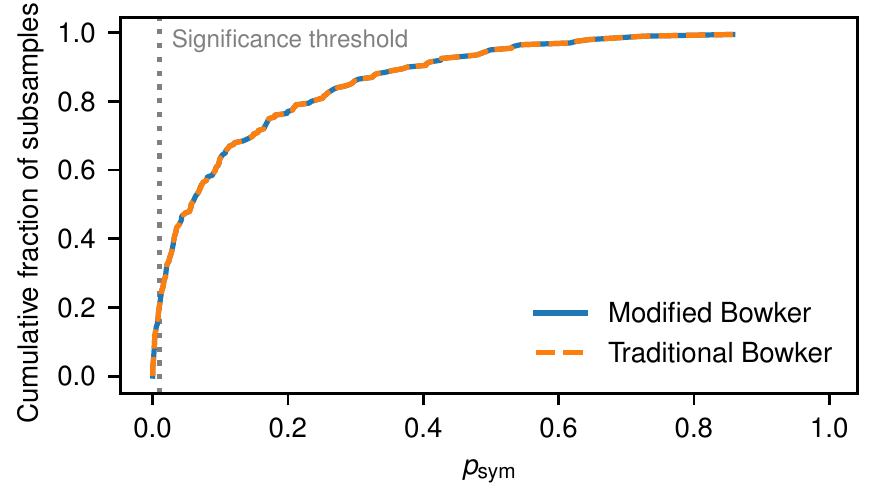}
\caption{\textbf{Parameters and figures of merit for $\boldsymbol{n=9}$ experiment.} Left: experimental parameters (top) and the resulting figures of merit (bottom). Right: Cumulative histograms showing the distribution of $p$-values from Bowker tests run on random subsets of the full dataset. The $1\%$ significance threshold used throughout is shown for comparison.}
\label{figS:n=9_FoM}
\end{figure}

The TV error for this experiment, together with other experimental parameters defined in Section~\ref{secS:exp_structure}, is given in the left panel of Fig.~\ref{figS:n=9_FoM}. For comparison, the average TV distances between $\hat{\boldsymbol{Q}}_\text{exp}$ for this instance and the uniform proposal, the local proposal, and the identity matrix $I$ are larger: 0.233, 0.935 and 0.980 respectively. Fig.~\ref{figS:n=9_FoM} also gives $p$-values from the traditional and modified Bowker tests, averaged over 200 IID subsamples as described in Section~\ref{secS:symmetry}. Both are well above the 1\% significance threshold, and are therefore consistent with the experimental quantum proposal mechanism being symmetric. Typically, about 30\% of the full dataset makes it into each IID subsample used for the Bowker tests. There were a total of $2.88 \times 10^7$ shots, so the expected number of data points that did not make it into any subsample is approximately $2.88 \times 10^7 \times (1-0.3)^{200} \sim  10^{-24} \ll 1$. The 200 subsamples used to test for symmetry therefore cover the full dataset. The integrated histogram of $p$-values for both Bowker test variants over these random subsamples are shown explicitly in the right panel of Fig.~\ref{figS:n=9_FoM}. The two agree almost exactly since the data is denser in this experiment than in the $n=10$ one (i.e., the ratio of total shots to the number of possible transitions is larger), so there are few empty pairs of bins. The distributions are more sharply concentrated around small $p_\text{sym}$ than the corresponding $n=10$ ones in Fig.~\ref{figS:n=10_FoM}, which we attribute to having used a smaller number of random twirls $N_\text{twirl}$ per $\gamma$ and $t$ here. Nevertheless, their means are both well above the 1\% significance threshold for asymmetry.

\begin{figure}[h]
\centering
\includegraphics[valign=c]{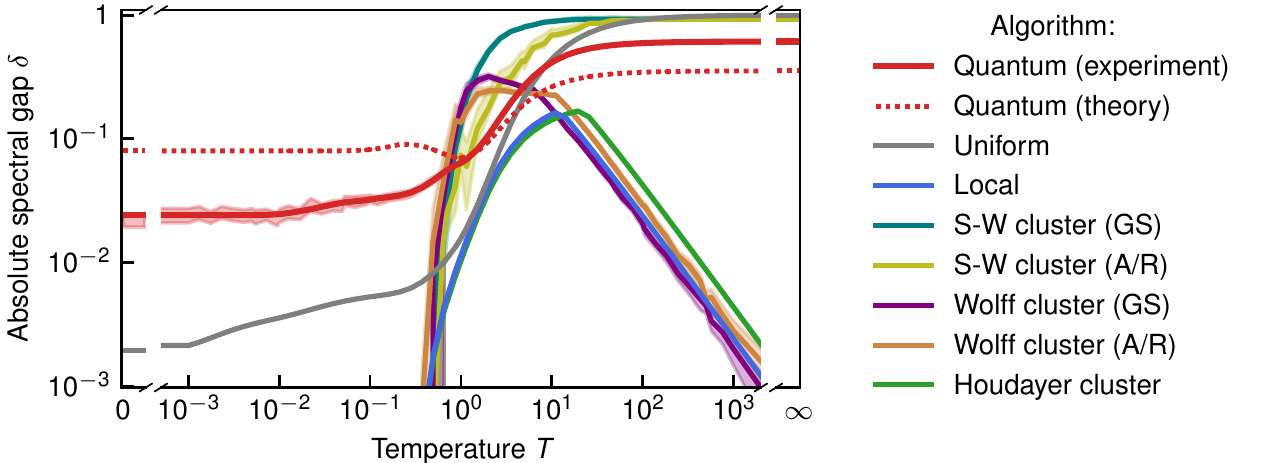}
\caption{\textbf{Convergence rates for $\boldsymbol{n=9}$ instance.} The absolute spectral gap $\delta$ for various MCMC algorithms on the same $n=9$ model instance. This plot was constructed in the same way as Fig.~\ref{figS:gap_n=10}. The quantum, local and uniform proposal strategies were combined with the Metropolis-Hastings acceptance probability \eqref{eq:MH_acceptance} of the main text, and the cluster algorithms are discussed in Section~\ref{secS:clusters}. ``GS'' and ``A/R'' denote ``ghost spin'' and ``accept/reject'' versions respectively of the Swendsen-Wang (S-W) and Wolff algorithms. Error bands, where present, denote 99\% confidence intervals found through basic (i.e., reverse percentile) bootstrapping using 200 resamples independently for each $T$. For the experimental realization of our quantum algorithm we bootstrapped random subsets of the full dataset, which comprises $2.88 \times 10^7$ shots, as described in Sec.~\ref{secS:bootstrapping}. For the S-W and Wolff algorithms there was no need for such subsampling. Rather, we formed point estimates and confidence intervals for their absolute spectral gaps at different temperatures by generating $10^5$ IID cluster moves separately for each $T$.}
\label{figS:gap_n=9}
\end{figure}

\begin{figure}[h]
\centering
\subfloat[Lazy M-H]{%
\raisebox{0.5ex}{%
\includegraphics[width=0.59\textwidth, height=1.5in]{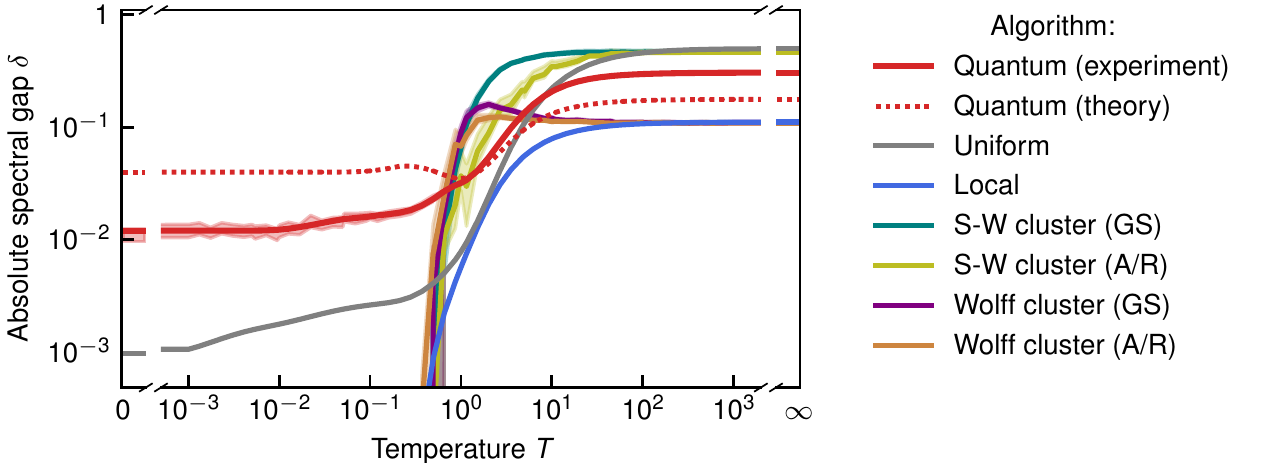}}
}
\hfill
\subfloat[Gibbs sampler]{%
\includegraphics[width=0.38\textwidth, height=1.53in]{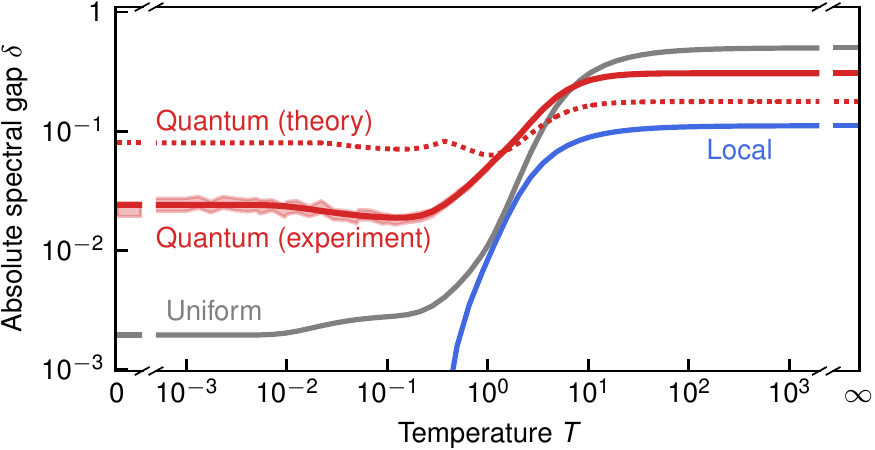}
}
\caption{\textbf{Convergence rates for $\boldsymbol{n=9}$ instance---variants.} The absolute spectral gap $\delta$ for various MCMC algorithms on the same $n=9$ model instance. (a) Lazy versions of the Markov chains considered in Fig.~\ref{figS:gap_n=9}. The Houdayer cluster algorithm is not shown since there is some ambiguity in how a lazy version should be defined, and because its $\delta$ at low $T$ is set by a positive eigenvalue, meaning lazy variants should offer no speedup anyways. ``GS'' and ``A/R'' denote ``ghost spin'' and ``accept/reject'' versions respectively of the Swendsen-Wang (S-W) and Wolff algorithms. (b) Absolute spectral gaps based on the Gibbs sampler acceptance probability in Eq.~\eqref{eqS:gibbs_acceptance}, rather than the M-H probability in Eq.~\eqref{eq:MH_acceptance} of the main text. Error bands in both panels, where present, denote 99\% bootstrap confidence intervals constructed identically to those in Fig.~\ref{figS:gap_n=9}.}
\label{figS:gap_n=9_lazyglauber}
\end{figure}

\begin{figure}[h]
\centering
\includegraphics[valign=c]{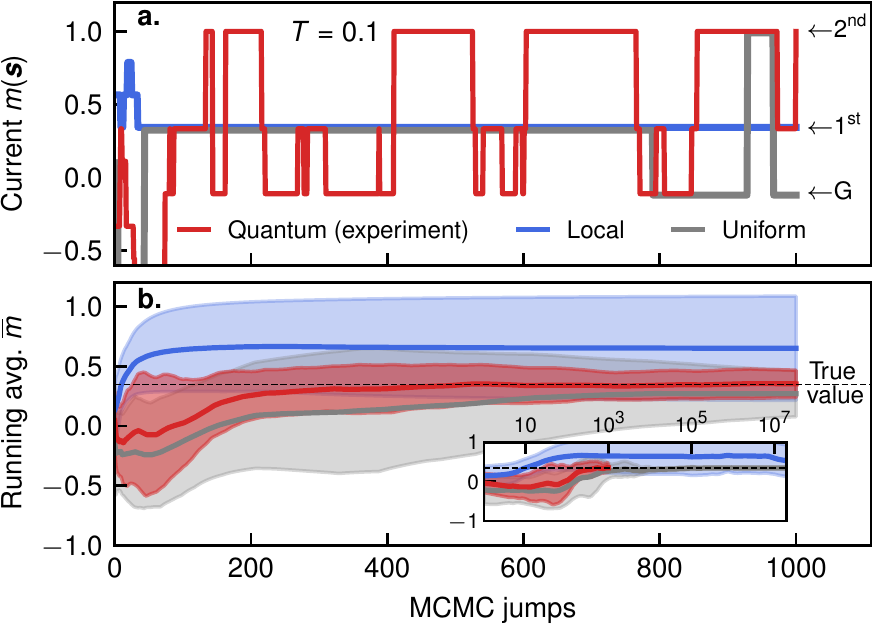}
\caption{\textbf{Magnetization estimate experiment with $\boldsymbol{n=9}$.} Analogous data to Fig.~\ref{fig:magnetization} of the main text but for the $n=9$ instance, collected as described in Sec.~\ref{secS:trajectories}. \textbf{a.} The current magnetization $m(\boldsymbol{s}^{(j)})$ for individual Markov chains after $j$ iterations. Each chain illustrates a different proposal strategy with uniformly random initialization. Arrows indicate the magnetization of the ground (G), 1\textsuperscript{st} and 2\textsuperscript{nd} excited configurations.
\textbf{b.} Convergence of the running average $\bar{m}^{(j)} = \frac{1}{j} \sum_{k=0}^j m(\boldsymbol{s}^{(k)})$ from MCMC trajectories to the true value of $\langle m \rangle_\mu$ for different proposal strategies. For each strategy, the lines and error bands show the mean and standard deviation, respectively, of $\bar{m}^{(j)}$ over 10 independent chains. The inset depicts the same chains over more iterations. We do not use a burn-in period or thinning (i.e., the running average starts at $k=0$ and includes every iteration up to $k=j$), as these practices would introduce hyperparameters that complicate the interpretation. Both panels are for $T=0.1$ and use the Metropolis-Hastings acceptance probability in Eq.~\eqref{eq:MH_acceptance} of the main text.}
\label{figS:trajectories_n=9}
\end{figure}

\begin{figure}[h]
\centering
\includegraphics[valign=t]{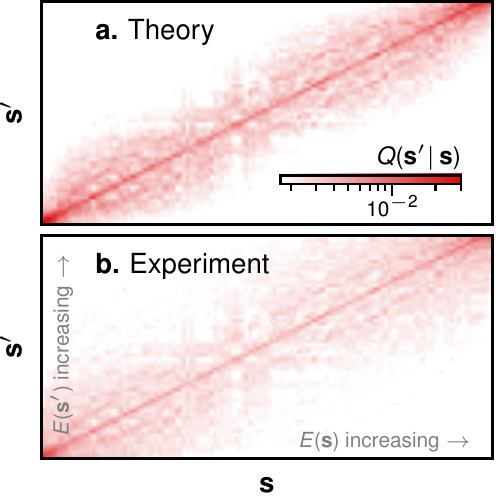}
\hfill
\includegraphics[valign=t]{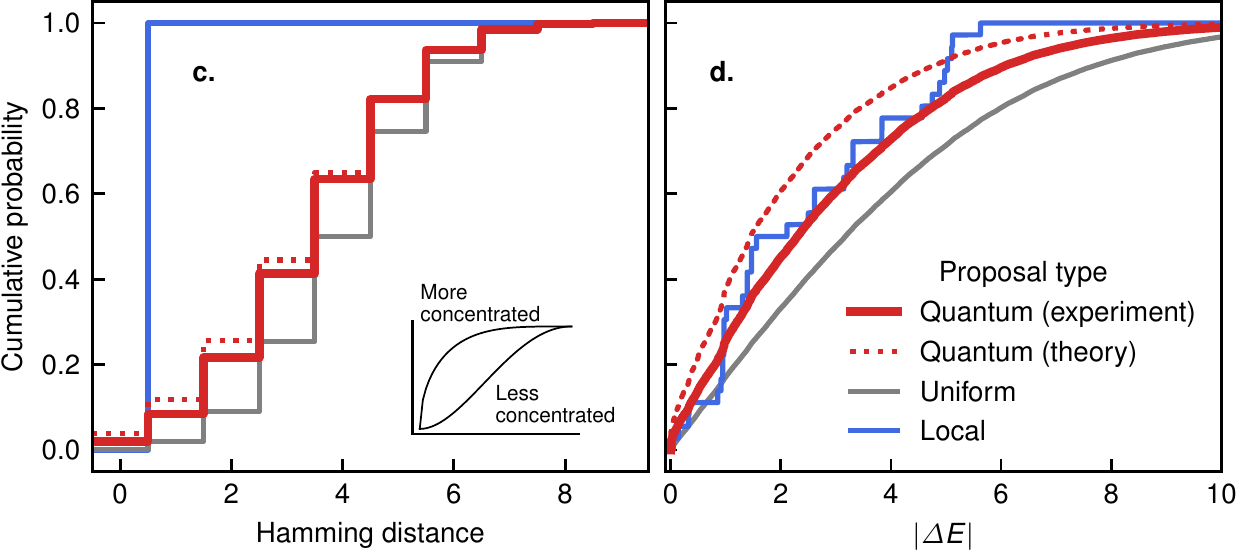}
\caption{\textbf{Quantum speedup mechanism for $\boldsymbol{n=9}$.} This figure shows data analogous to Fig.~\ref{fig:mechanism} of the main text, but for the $n=9$ instance. \textbf{a.} The classically-simulated probabilities of $\boldsymbol{s} \rightarrow \boldsymbol{s'}$ proposals in our quantum algorithm, represented as a $2^n \times 2^n$ matrix whose columns are independent histograms. Both the initial and proposed configurations are sorted by increasing Ising energy $E$. The same data is shown with lexicographic ordering in Fig.~\ref{figS:Q_lex_n=9}. \textbf{b.} The estimated proposal probabilities for our algorithm's experimental realization. We estimated each $Q(\boldsymbol{s'}|\,\boldsymbol{s})$ as described in Sec.~\ref{secS:data_analysis}. \textbf{c.} The probability distributions of Hamming distance between current ($\boldsymbol{s}$) and proposed ($\boldsymbol{s'}$) configurations, for a uniformly random current configuration. That of the experiment uses the estimated probabilities from panel b, while the rest were computed exactly. \textbf{d.} The analogous distributions for $|\Delta E| = |E(\boldsymbol{s'}) - E(\boldsymbol{s})|$ of proposed jumps. Each distribution is depicted in full detail through its cumulative distribution function, with no binning. None of the panels depend on $T$ or on the choice of acceptance probability.}
\label{figS:mechanism_n=9}
\end{figure}

\begin{figure}[h]
\centering
\includegraphics[width=2.5in]{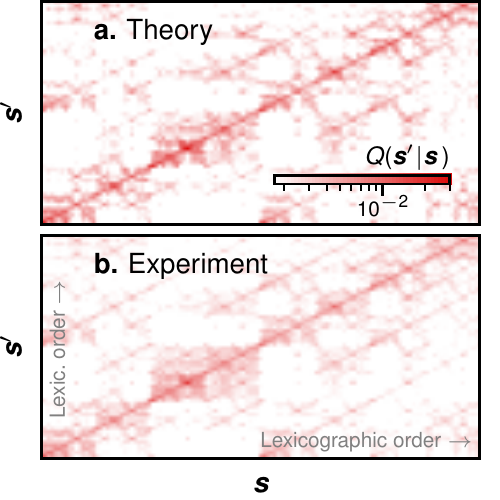}
\caption{The same data shown in Figs.~\ref{figS:mechanism_n=9}a-b, but with $\boldsymbol{s}$ and $\boldsymbol{s'}$ sorted in lexicographic order, rather than by increasing Ising energy $E$.}
\label{figS:Q_lex_n=9}
\end{figure}

Figs.~\ref{figS:gap_n=9}--\ref{figS:mechanism_n=9} show results analogous to Figs.~\ref{fig:n=10_gap}--\ref{fig:mechanism} from the main text for this experiment. Specifically, the inferred absolute spectral gap $\delta$ for M-H, as a function of temperature, is shown in Fig.~\ref{figS:gap_n=9}. As in the $n=10$ experiment, there is a significant quantum enhancement in $\delta$ at low $T$. For completeness, Fig.~\ref{figS:gap_n=9_lazyglauber} similarly shows the absolute spectral gap for two M-H variants discussed in Sections~\ref{secS:glauber} and \ref{secS:lazy_chains}: lazy M-H and the Gibbs sampler acceptance probability. The quantum enhancement in $\delta$ at low $T$ is nearly identical in both variants. Fig.~\ref{figS:trajectories_n=9}a shows the magnetization $m(\boldsymbol{s})$ for illustrative MCMC trajectories at $T=0.1$. Our quantum algorithm jumps between the lowest-$E$ configurations noticeably more often than the classical alternatives. Accordingly, the running average estimate for $\expval{m}_\mu$ converges more quickly to the true value of $\expval{m}_\mu \approx 0.35$ in Fig.~\ref{figS:trajectories_n=9}b. Fig.~\ref{figS:mechanism_n=9} shows the distribution of jumps for our quantum algorithm, and reveals the same enhancement mechanism discussed in the main text for $n=10$. Finally, Fig.~\ref{figS:Q_lex_n=9} shows the same proposal probabilities $Q(\boldsymbol{s'}|\,\boldsymbol{s})$ as Figs.~\ref{figS:mechanism_n=9}a-b, but with the spin configurations sorted lexicographically, rather than by increasing $E$.

\clearpage
\subsubsection{$n=8$ instance}

\begin{figure}[h]
\centering 
\includegraphics[width=0.4\textwidth]{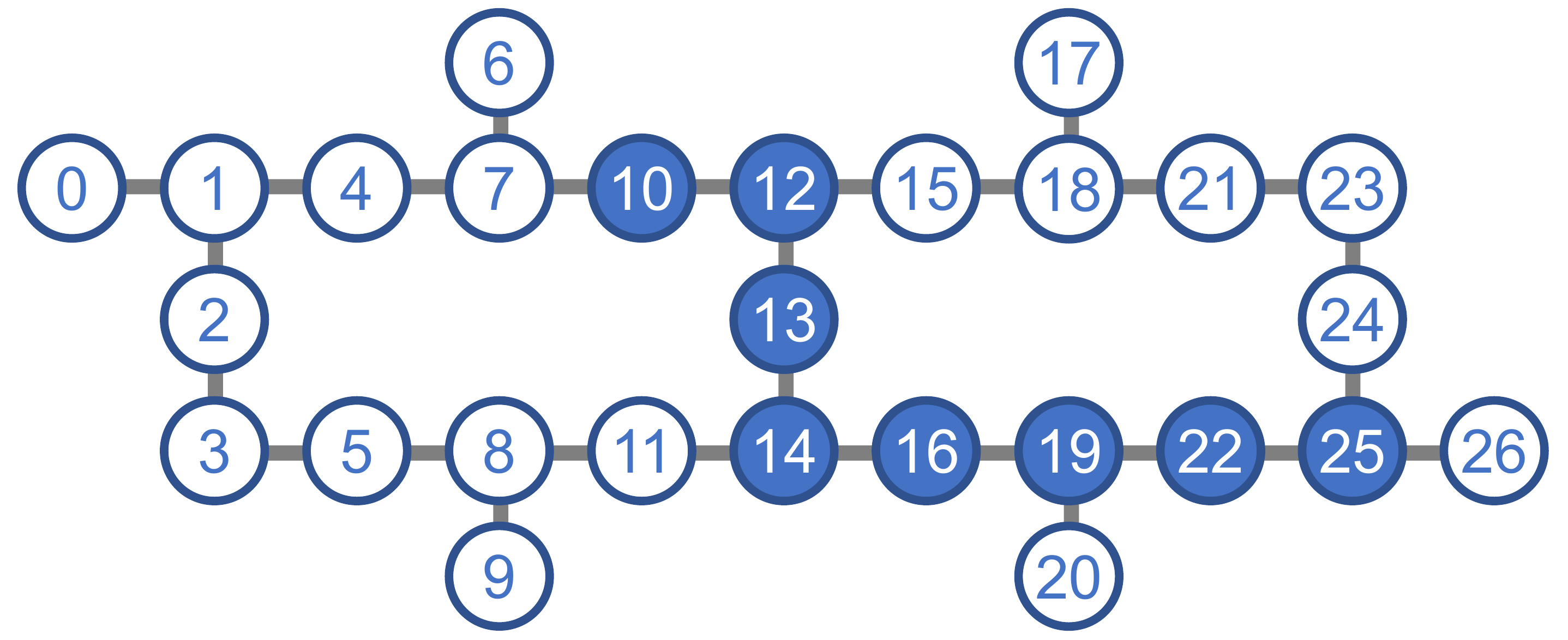}
\hspace{5em}
\raisebox{4.5em}{
\setlength{\tabcolsep}{12pt}
\begin{tabular}{ c c }
CNOT error (mean) & 0.70\% \\
CNOT error (worst) & 0.88\% \\
Readout assignment error (mean) & 2.05\%  \\ 
Readout assignment error (worst) & 4.53\% \\
\end{tabular}
}
\caption{\textbf{Device information for $\boldsymbol{n=8}$ experiment}. Left: the 1D chain of qubits on \textit{ibmq\_mumbai} used for this experiment, where Ising model spins $1, \dots, 8$ were mapped to qubits $10,\dots,25$ respectively. Right: representative figures of merit for this chain of qubits at the time of the experiment.}
\label{figS:n=8_qubits}
\end{figure}

We implemented our algorithm for another illustrative model instance with $n=8$ spins. We present the experimental results here, which are analogous to those for the $n=10$ instance from the main text. This $n=8$ instance is defined by the couplings and fields
\begin{gather}
(J_{j+1,j})_{j=1}^{n-1}
= \big( -0.94496973, \; 1.01674697, \; 1.05072852,\;  1.07515862,  \; 0.90289512, \;0.98594583,  \; 0.9361144\big) \\
(h_j)_{j=1}^n
=
\big( 0.53727449, \; -0.24671696, \; 0.4930372 , \; -0.92341807, \;  0.75710839, \; -0.65808939, \; 0.30686208, \;  0.68948975
\big), \nonumber
\end{gather}
with all other $J_{jk}$ equal to zero. Its four lowest-$E$ configurations are all local minima, and the lowest three are nearly degenerate: at $T=0.1$ they have Boltzmann probabilities of approximately 39\%, 32\% and 26\% for the ground, first- and second-excited configurations respectively. The qubits used in this experiment, together with representative figures of merit from the last calibration before the experiment, are shown in Fig.~\ref{figS:n=8_qubits}.

\begin{figure}[h]
\centering
\setlength{\tabcolsep}{12pt}
\raisebox{9em}{
\begin{tabular}{c c} 
$N_\text{twirl}$ & 4\\
$N_\text{shots}$ & $10^4$\\ 
 Total shots & $9.6 \times 10^6$ \\[2ex]
TV error & 0.177 \\ 
$\expval{p_\text{sym}}$ traditional & 41.4\%  \\
$\expval{p_\text{sym}}$ modified & 41.4\% 
\end{tabular}
}
\hspace{3em}
\includegraphics{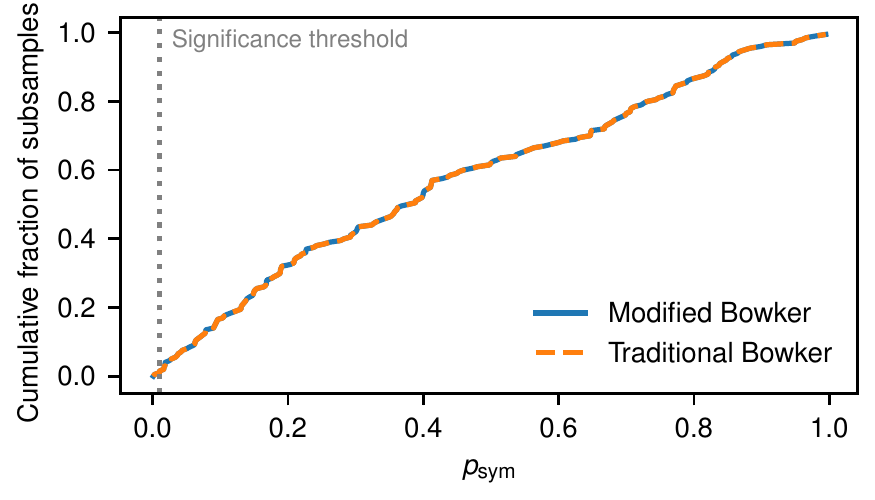}
\caption{\textbf{Parameters and figures of merit for $\boldsymbol{n=8}$ experiment.} Left: experimental parameters (top) and the resulting figures of merit (bottom). Right: Cumulative histograms showing the distribution of $p$-values from Bowker tests run on random subsets of the full dataset. The $1\%$ significance threshold used throughout is shown for comparison.}
\label{figS:n=8_FoM}
\end{figure}

The TV error for this experiment, together with other experimental parameters defined in Section~\ref{secS:exp_structure}, is given in the left panel of Fig.~\ref{figS:n=8_FoM}. For comparison, the average TV distances between $\hat{\boldsymbol{Q}}_\text{exp}$ for this instance and the uniform proposal, the local proposal, and the identity matrix $I$ are larger: 0.238, 0.911 and 0.971 respectively. Fig.~\ref{figS:n=8_FoM} also gives $p$-values from the traditional and modified Bowker tests, averaged over 200 IID subsamples as described in Section~\ref{secS:symmetry}. Both are well above the 1\% significance threshold, and are therefore consistent with the experimental quantum proposal mechanism being symmetric. Typically, about 20\% of the full dataset makes it into each IID subsample used for the Bowker tests. There were a total of $9.6 \times 10^6$ shots, so the expected number of data points that did not make it into any subsample is approximately $9.6 \times 10^6 \times (1-0.2)^{200} \sim  10^{-13} \ll 1$. The 200 subsamples used to test for symmetry therefore cover the full dataset. The integrated histogram of $p$-values for both Bowker test variants over these random subsamples are shown explicitly in the right panel of Fig.~\ref{figS:n=8_FoM}. The two agree almost exactly since the data is denser in this experiment than in the $n=10$ one (i.e., the ratio of total shots to the number of possible transitions is larger), so there are few empty pairs of bins. Neither is sharply concentrated on $p_\text{sym} \lesssim 1\%$, which constitutes further evidence of symmetry.

\begin{figure}[h]
\centering
\includegraphics[valign=c]{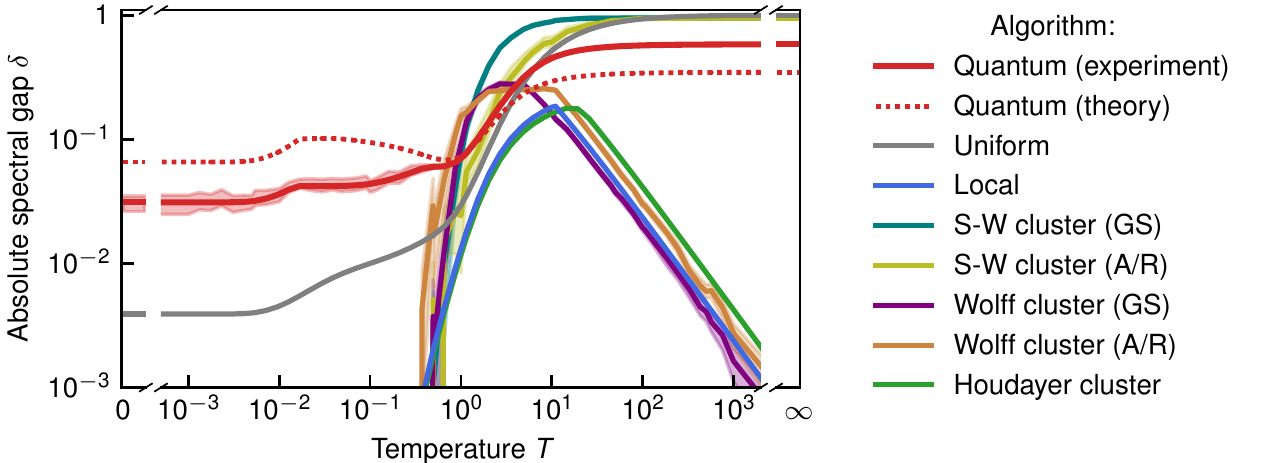}
\caption{\textbf{Convergence rates for $\boldsymbol{n=8}$ instance.} The absolute spectral gap $\delta$ for various MCMC algorithms on the same $n=8$ model instance. This plot was constructed in the same way as Fig.~\ref{figS:gap_n=10}. The quantum, local and uniform proposal strategies were combined with the Metropolis-Hastings acceptance probability \eqref{eq:MH_acceptance} of the main text, and the cluster algorithms are discussed in Section~\ref{secS:clusters}. ``GS'' and ``A/R'' denote ``ghost spin'' and ``accept/reject'' versions respectively of the Swendsen-Wang (S-W) and Wolff algorithms. Error bands, where present, denote 99\% confidence intervals found through basic (i.e., reverse percentile) bootstrapping using 200 resamples independently for each $T$. For the experimental realization of our quantum algorithm we bootstrapped random subsets of the full dataset, which comprises $9.6 \times 10^6$ shots, as described in Sec.~\ref{secS:bootstrapping}. For the S-W and Wolff algorithms there was no need for such subsampling. Rather, we formed point estimates and confidence intervals for their absolute spectral gaps at different temperatures by generating $10^5$ IID cluster moves separately for each $T$.}
\label{figS:gap_n=8}
\end{figure}

\begin{figure}[h]
\centering
\subfloat[Lazy M-H]{%
\raisebox{0.5ex}{%
\includegraphics[width=0.59\textwidth, height=1.5in]{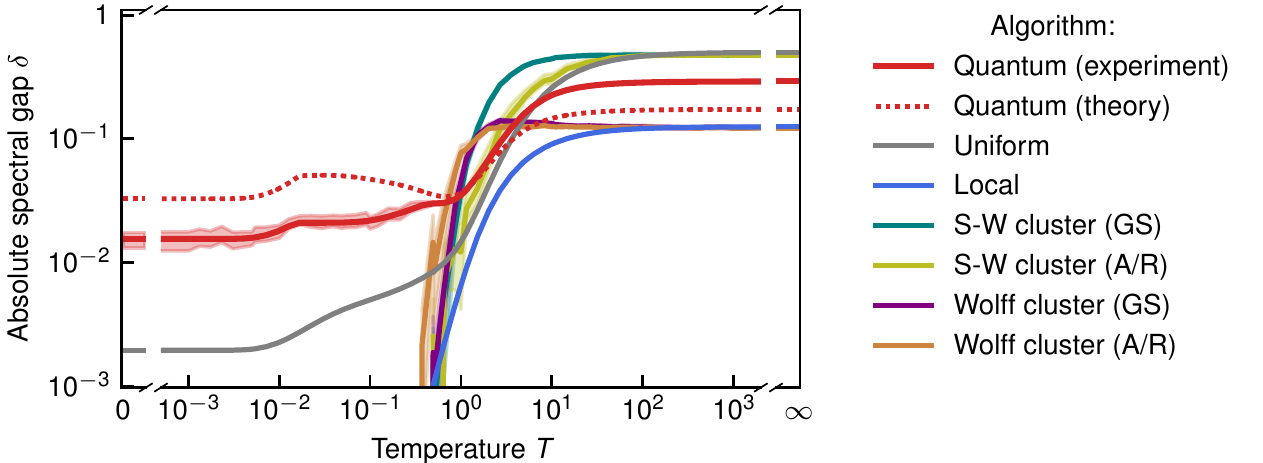}}
}
\hfill
\subfloat[Gibbs sampler]{%
\includegraphics[width=0.38\textwidth, height=1.53in]{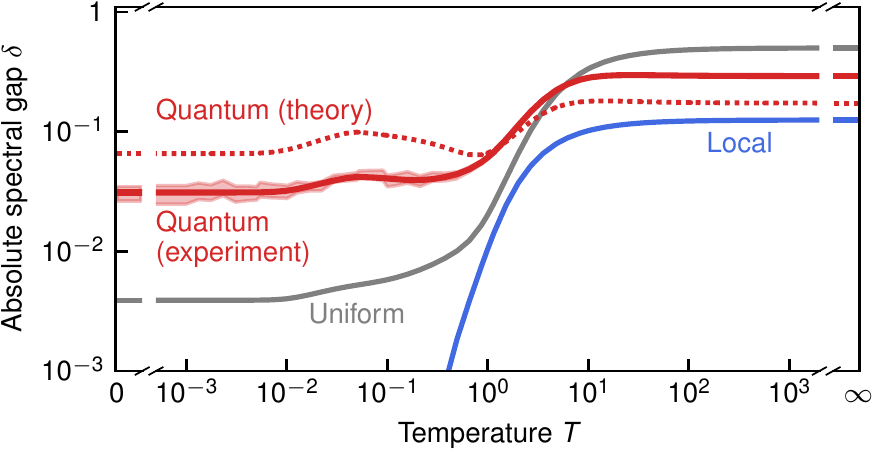}
}
\caption{\textbf{Convergence rates for $\boldsymbol{n=8}$ instance---variants.} The absolute spectral gap $\delta$ for various MCMC algorithms on the same $n=8$ model instance. (a) Lazy versions of the Markov chains considered in Fig.~\ref{figS:gap_n=8}. The Houdayer cluster algorithm is not shown since there is some ambiguity in how a lazy version should be defined, and because its $\delta$ at low $T$ is set by a positive eigenvalue, meaning lazy variants should offer no speedup anyways. ``GS'' and ``A/R'' denote ``ghost spin'' and ``accept/reject'' versions respectively of the Swendsen-Wang (S-W) and Wolff algorithms. (b) Absolute spectral gaps based on the Gibbs sampler acceptance probability in Eq.~\eqref{eqS:gibbs_acceptance}, rather than the M-H probability in Eq.~\eqref{eq:MH_acceptance} of the main text. Error bands in both panels, where present, denote 99\% bootstrap confidence intervals constructed identically to those in Fig.~\ref{figS:gap_n=8}.}
\label{figS:gap_n=8_lazyglauber}
\end{figure}

\begin{figure}[h]
\centering
\includegraphics[valign=c]{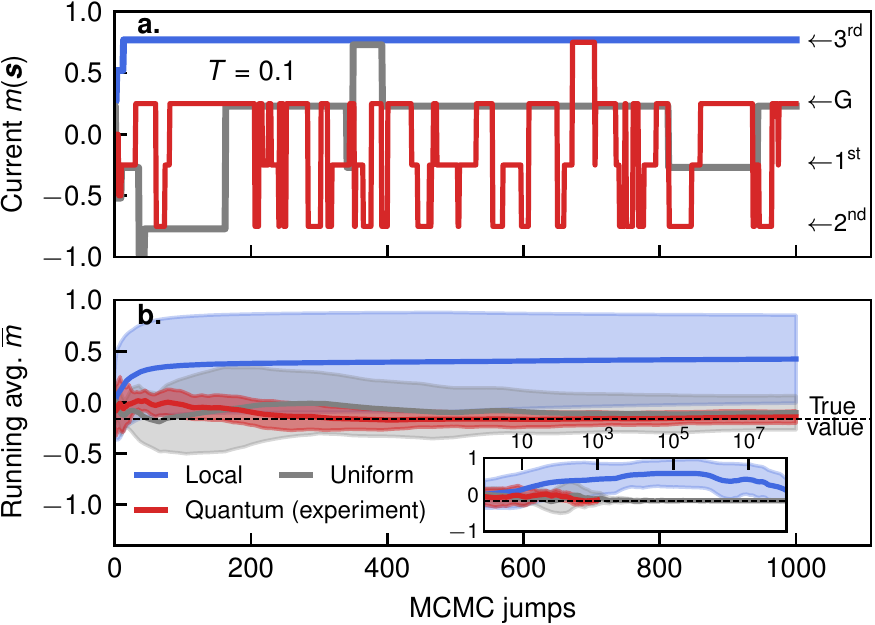}
\caption{\textbf{Magnetization estimate experiment with $\boldsymbol{n=8}$.} Analogous data to Fig.~\ref{fig:magnetization} of the main text but for the $n=8$ instance, collected as described in Sec.~\ref{secS:trajectories}. \textbf{a.} The current magnetization $m(\boldsymbol{s}^{(j)})$ for individual Markov chains after $j$ iterations. Each chain illustrates a different proposal strategy with uniformly random initialization. Arrows indicate the magnetization of the ground (G), 1\textsuperscript{st}, 2\textsuperscript{nd} and 3\textsuperscript{rd} excited configurations.
\textbf{b.} Convergence of the running average $\bar{m}^{(j)} = \frac{1}{j} \sum_{k=0}^j m(\boldsymbol{s}^{(k)})$ from MCMC trajectories to the true value of $\langle m \rangle_\mu$ for different proposal strategies. For each strategy, the lines and error bands show the mean and standard deviation, respectively, of $\bar{m}^{(j)}$ over 10 independent chains. The inset depicts the same chains over more iterations. We do not use a burn-in period or thinning (i.e., the running average starts at $k=0$ and includes every iteration up to $k=j$), as these practices would introduce hyperparameters that complicate the interpretation. Both panels are for $T=0.1$ and use the Metropolis-Hastings acceptance probability in Eq.~\eqref{eq:MH_acceptance} of the main text.}
\label{figS:trajectories_n=8}
\end{figure}

\begin{figure}[h]
\centering
\includegraphics[valign=t]{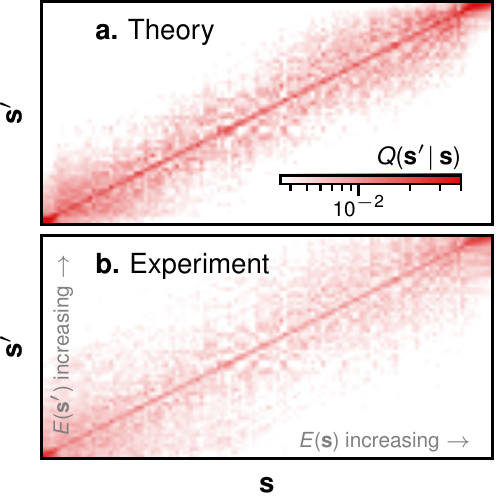}
\hfill
\includegraphics[valign=t]{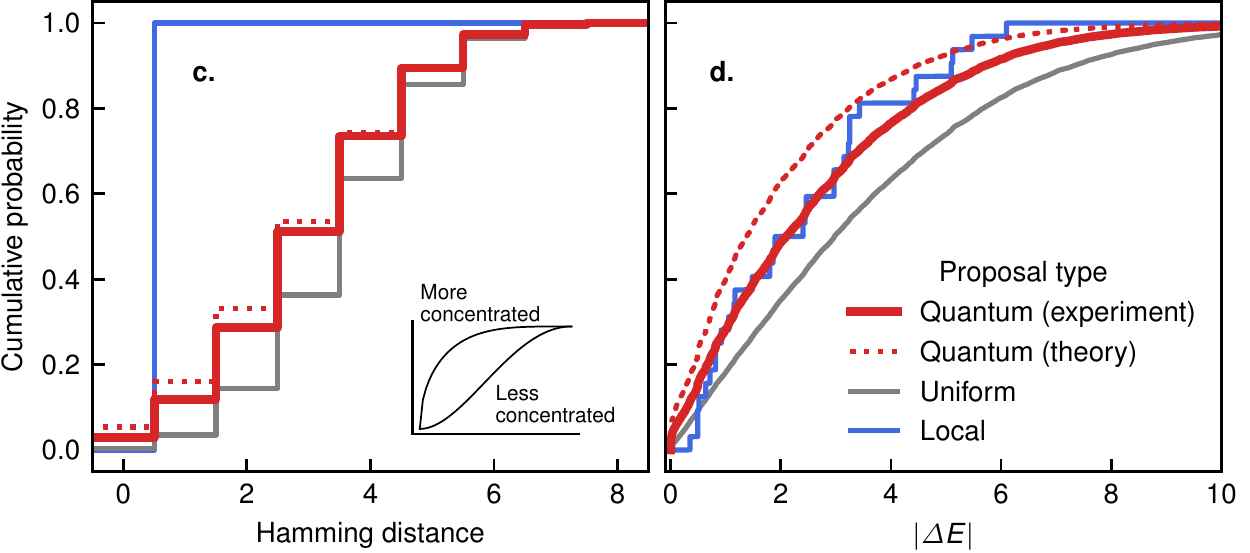}
\caption{\textbf{Quantum speedup mechanism for $\boldsymbol{n=8}$.} This figure shows data analogous to Fig.~\ref{fig:mechanism} of the main text, but for the $n=8$ instance. \textbf{a.} The classically-simulated probabilities of $\boldsymbol{s} \rightarrow \boldsymbol{s'}$ proposals in our quantum algorithm, represented as a $2^n \times 2^n$ matrix whose columns are independent histograms. Both the initial and proposed configurations are sorted by increasing Ising energy $E$. The same data is shown with lexicographic ordering in Fig.~\ref{figS:Q_lex_n=8}. \textbf{b.} The estimated proposal probabilities for our algorithm's experimental realization. We estimated each $Q(\boldsymbol{s'}|\,\boldsymbol{s})$ as described in Sec.~\ref{secS:data_analysis}. \textbf{c.} The probability distributions of Hamming distance between current ($\boldsymbol{s}$) and proposed ($\boldsymbol{s'}$) configurations, for a uniformly random current configuration. That of the experiment uses the estimated probabilities from panel b, while the rest were computed exactly. \textbf{d.} The analogous distributions for $|\Delta E| = |E(\boldsymbol{s'}) - E(\boldsymbol{s})|$ of proposed jumps. Each distribution is depicted in full detail through its cumulative distribution function, with no binning. None of the panels depend on $T$ or on the choice of acceptance probability.}
\label{figS:mechanism_n=8}
\end{figure}

\begin{figure}[h]
\centering
\includegraphics[width=2.5in]{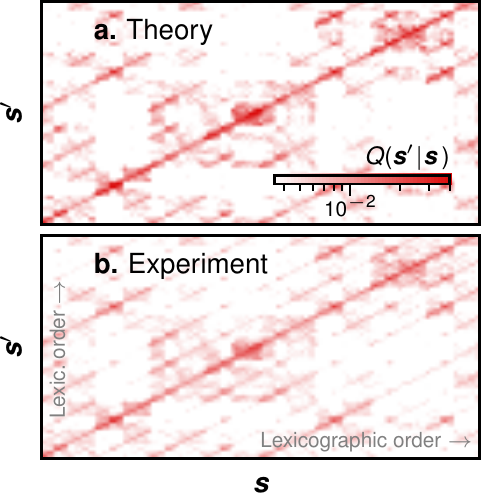}
\caption{The same data shown in Figs.~\ref{figS:mechanism_n=8}a-b, but with $\boldsymbol{s}$ and $\boldsymbol{s'}$ sorted in lexicographic order, rather than by increasing Ising energy $E$.}
\label{figS:Q_lex_n=8}
\end{figure}

Figs.~\ref{figS:gap_n=8}--\ref{figS:Q_lex_n=8} show results analogous to Figs.~\ref{fig:n=10_gap}--\ref{fig:mechanism} from the main text for this experiment. Specifically, the inferred absolute spectral gap $\delta$ for M-H, as a function of temperature, is shown in Fig.~\ref{figS:gap_n=8}. As in the $n=9$ and $10$ experiments, there is a significant quantum enhancement in $\delta$ at low $T$. For completeness, Fig.~\ref{figS:gap_n=8_lazyglauber} similarly shows the absolute spectral gap for two M-H variants discussed in Sections~\ref{secS:glauber} and \ref{secS:lazy_chains}: lazy M-H and the Gibbs sampler acceptance probability. The quantum enhancement in $\delta$ at low $T$ is nearly identical in both variants. Fig.~\ref{figS:trajectories_n=8}a shows the magnetization $m(\boldsymbol{s})$ for illustrative MCMC trajectories at $T=0.1$. Our quantum algorithm jumps between the lowest-$E$ configurations noticeably more often than the classical alternatives. Accordingly, the running average estimate for $\expval{m}_\mu$ converges more quickly to the true value of $\expval{m}_\mu \approx -0.16$ in Fig.~\ref{figS:trajectories_n=8}b. Fig.~\ref{figS:mechanism_n=8} shows the distribution of jumps for our quantum algorithm, and reveals the same enhancement mechanism discussed in the main text for $n=10$. Finally, Fig.~\ref{figS:Q_lex_n=8} shows the same proposal probabilities $Q(\boldsymbol{s'}|\,\boldsymbol{s})$ as Figs.~\ref{figS:mechanism_n=8}a-b, but with the spin configurations sorted lexicographically, rather than by increasing $E$.

\clearpage
\section{Quantum proposal strategies: motivation and alternatives}

In this section we discuss the motivation for our quantum proposal strategy as well as possible alternatives.

\subsection{Perturbative regime}
\label{secS:perturbation_theory}

We begin with the pertubative calculation to which the main text refers. Consider $H$ as given by Eqs.~\eqref{eq:H} and \eqref{eq:H_prob} of the main text, where $H_\text{mix}$ is some as-yet unspecified Hamiltonian with $\bra{k} H_\text{mix} \ket{j} \in \mathbb{R}$ for all computational basis states $\ket{j}$ and $\ket{k}$. To emphasize the dependence on $\gamma$ we will sometimes write $H = H(\gamma)$. We will compute the $\ket{j} \rightarrow \ket{k}$ quantum transition probability perturbatively to leading order in $\gamma$, assuming for now that $\gamma$ is sufficiently small for perturbation theory to hold.

It is convenient to move to the interaction picture defined by $\alpha H_\text{prob}$, where the Hamiltonian becomes
\begin{equation}
\tilde{H}(\gamma, t) = e^{i t \alpha  H_\text{prob}} \big[ H(\gamma) - \alpha H_\text{prob} \big] e^{-i t \alpha H_\text{prob}}
=
\gamma \left( 
e^{i t \alpha H_\text{prob}} H_\text{mix} \, e^{-i t \alpha H_\text{prob}} - \alpha H_\text{prob}
\right).
\end{equation}
Let $U(t) = e^{-i H(\gamma) t}$ be the propagator from time $0$ to $t$ in the Schr\"odinger picture, and $\tilde{U}(t)$ be that in the interaction picture (satisfying $i \tilde{U}' = \tilde{H} \tilde{U}$ and $\tilde{U}(0) = I$). Since $H_\text{prob}$ is diagonal in the computational basis, the quantum transition probabilities are simply the absolute-squared matrix elements of the propagator in either picture:
\begin{equation} 
\text{Pr}\Big( \ket{j} \rightarrow \ket{k} \Big) = 
\big|\bra{k} \tilde{U}(t) \ket{j} \big |^2 = \big|\bra{k} U(t) \ket{j} \big|^2.
\label{eqS:Q_transition_probs}
\end{equation}
We can expand $\tilde{U}(t)$ perturbatively in $\gamma$ through a Dyson series \cite{sakurai:2011}:
\begin{equation}
\tilde{U}(t) = I - i \int_0^t \tilde{H}(\gamma, t') \, dt' + O(\gamma^2).
\end{equation}
To leading order in $\gamma$, its matrix elements are
\begin{equation}
\bra{k} \tilde{U} (t) \ket{j}
=
\begin{cases}
\gamma \int_0^t e^{i t' \alpha \Delta E} dt' \bra{k} H_\text{mix} \ket{j} + O(\gamma^2) & j \neq k\\ 
1 - O(\gamma) & j = k,
\end{cases}
\end{equation}
where $\Delta E = E(k) - E(j)$ is the difference in classical energies, defined in Eq.~\eqref{eq:E_function} of the main text, between spin configurations $j$ and $k$. Using Eq.~\eqref{eqS:Q_transition_probs} we find
\begin{equation}
\text{Pr}\Big( \ket{j} \rightarrow \ket{k} \Big) = 
\begin{cases}
\gamma^2 t^2 \, \text{sinc}^2 \left( \frac{t \alpha \Delta E}{2} \right) 
\big| \bra{k} H_\text{mix} \ket{j} \big|^2 + O(\gamma^3) & j \neq k\\ 
1- O(\gamma^2) & j = k.
\end{cases}
\label{eqS:transition_dyson}
\end{equation}
The sinc-squared factor above acts as a bandpass filter for $\Delta E$, since $\text{sinc}^2(x)$ is of order unity within the passband $x \in [-\pi, \pi]$ but nearly 0 otherwise. Suppose that $H_\text{mix}$ is dense, e.g., $\big| \bra{k} H_\text{mix} \ket{j} \big| = 1$ as invoked in early formulations of Grover's algorithm using quantum walks \cite{farhi:1998}. Then writing $t=2\pi/\alpha \epsilon$ for some energy scale $\epsilon$, we see that nontrivial $\ket{j} \rightarrow \ket{k}$ quantum transitions occur predominantly when $|\Delta E| \lesssim \epsilon$. In other words, they occur primarily between configurations $j$ and $k$ whose classical energies $E(j) \approx E(k)$ are close on the scale set by $\epsilon$, even if they are far in Hamming distance.

The same conclusion can be reached using a Magnus series rather than a Dyson series \cite{blanes:2009}. That is, we can define an effective Hamiltonian $\tilde{H}_\text{eff}$ term-by-term in powers of $\gamma$ such that $\tilde{U}(t) = \exp(-i t \tilde{H}_\text{eff})$ (with no time ordering). To leading order in $\gamma$, it has the form $\tilde{H}_\text{eff} = \frac{1}{t} \int_0^t \tilde{H}(\gamma, t') \, dt' + O(\gamma^2)$ and matrix elements
\begin{equation}
\bra{k} \tilde{H}_\text{eff} \ket{j} = 
\frac{\gamma}{t} \int_0^t e^{i t' \alpha \Delta E} dt' \bra{k} H_\text{mix}
\ket{j} + O(\gamma^2)
\end{equation}
in the computational basis. Invoking the rotating wave approximation, the fast-rotating elements (with $\Delta E \gtrsim 2\pi/\alpha t$) average out and nearly vanish, but the slow ones do not \cite{zueco:2009}. The effective Hamiltonian therefore primarily couples states $\ket{j}$ and $\ket{k}$ for which $E(j) \approx E(k)$, even if $j$ and $k$ are far in Hamming distance, causing $\ket{j} \rightarrow \ket{k}$ transitions between such states to dominate.

Both of these methods assume $t$ to be sufficiently small; however, the same phenomenon also occurs at long times. We return to the Schr\"odinger picture for this calculation. Suppose one prepares a state $\ket{j}$, evolves it under $H(\gamma)$ for a time $t$ and measures in the computational basis. For a random $t\sim \text{uniform}([0, t'])$, the probability of measuring $\ket{k}$ approaches 
\begin{equation}
\text{Pr}\Big( \ket{j} \rightarrow \ket{k} \Big) 
= 
\sum_{\ell=0}^{2^n-1} \big | 
\braket{k}{\lambda_\ell} \! \braket{\lambda_\ell}{j}
\big|^2
\label{eqS:infinite_t_transition}
\end{equation}
as $t' \rightarrow \infty$ \cite{childs:2003}, where $\{\ket{\lambda_\ell}\}_{\ell=0}^{2^n-1}$ are the eigenvectors of $H(\gamma)$, indexed such that $\lim_{\gamma \rightarrow 0} \ket{\lambda_j} = \ket{j}$. For simplicity, we assume that $H(\gamma)$ has a non-degenerate spectrum. Expanding $\ket{\lambda_\ell}$ in powers of $\gamma$ through time-independent, non-degenerate perturbation theory gives
\begin{equation}
\braket{k}{\lambda_\ell} = 
\begin{cases}
\frac{\gamma}{\alpha} \frac{\bra{k} (H_\text{mix} - \alpha H_\text{prob}) \ket{\ell}}{E_\ell - E_k} + O(\gamma^2) & k \neq \ell \\ 
1 - O(\gamma) & k = \ell.
\end{cases}
\end{equation}
Combining this with Eq.~\eqref{eqS:infinite_t_transition}, we get
\begin{equation}
\text{Pr}\Big( \ket{j} \rightarrow \ket{k} \Big) 
= 
\begin{cases}
2 \left( \frac{\gamma}{\alpha \Delta E} \right)^2 
\big| \bra{k} H_\text{mix} \ket{j} \big|^2
+ O(\gamma^3) & j \neq k \\ 
1 - O(\gamma^2) & j = k.
\end{cases}
\end{equation}
This equation exhibits the same $\Delta E^{-2}$ scaling as the sinc-squared term of Eq.~\eqref{eqS:transition_dyson}, leading to the same conclusion: non-trivial transitions occur primarily between states with similar classical energies, even if they are far in Hamming distance. This last calculation, however, highlights the role of $H$'s eigenvectors in driving quantum transitions between such states. Namely, these transitions occur because $H$ has eigenvectors (specifically, $\ket{\lambda_j}$ and $\ket{\lambda_k}$) that overlap substantially with both $\ket{j}$ and $\ket{k}$ when $E(j) \approx E(k)$, but that are nearly orthogonal to other computational states $\ket{\ell}$ for which $E(j) \not \approx E(\ell)$. 

A similar, albeit much more complicated, effect can also occur outside the perturbative regime for $\gamma$. Certain quantum spin glass Hamiltonians have been shown to possess eigenvectors with components concentrated on (potentially distant) low-$E$ configurations forming local minima of similar energies \cite{kechedzhi:2018, smelyanskiy:2020, smelyanskiy:2019}. These are the spin configurations that often cause bottlenecks in MCMC at low temperatures $T$, as discussed in the main text. Evolution under such Hamiltonians can therefore produce relatively frequent quantum transitions between these configurations, analogous to those in the previous, perturbative calculation. There are important advantages to using a large $\gamma$ beyond the scope of perturbation theory, however: (i) $H_\text{mix}$ can be sparse, which is easier to realize experimentally. That is, quantum transitions can occur with high probability between states $\ket{j}$ and $\ket{k}$ even if $\bra{k} H_\text{mix} \ket{j}=0$. (ii) Trivial $\ket{j} \rightarrow \ket{j}$ ``transitions'' need not dominate, as they do in the perturbative analysis above. (iii) The resulting dynamics can be hard to simulate classically. However, it is not clear \textit{a priori} how the results of Refs.~\cite{kechedzhi:2018, smelyanskiy:2020, smelyanskiy:2019} translate into MCMC performance. For instance, the spin glasses studied in these works are idealizations of those we consider. We therefore view the perturbative calculations above, and Refs.~\cite{kechedzhi:2018, baldwin:2018, smelyanskiy:2020, smelyanskiy:2019}, as providing motivation for our algorithm---which we then evaluate empirically---rather than any formal guarantee of performance.

\subsection{Quantum phase estimation}
\label{secS:QPE}

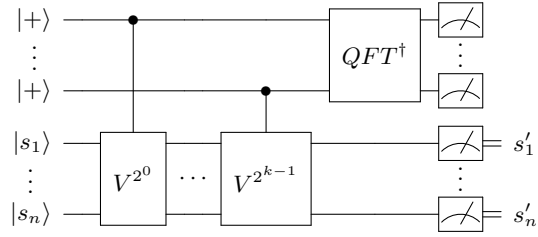
\begin{figure}[h]
\[
\Qcircuit @C=0.75em @R=0.75em {
\lstick{\ket{+}} & \qw & \ctrl{3} & \qw & \qw & \qw & \multigate{2}{QFT^\dagger} & \meter\\
\lstick{\raisebox{1ex}{\vdots \;\;}}  & & &  & & & & \raisebox{1.5ex}{\vdots} \\
\lstick{\ket{+}} & \qw & \qw & \qw & \qw & \ctrl{1} & \ghost{QFT^\dagger} & \meter \\ 
\lstick{\ket{s_1}} & \qw &  \multigate{2}{V^{2^0}} & \qw & \qw & \multigate{2}{V^{2^{k-1}}} & \qw & \meter & \rstick{s'_1} \cw \\
\lstick{\raisebox{1ex}{\vdots\;\;\,}} & & & \;\;\, \cdots & & & & \raisebox{1.5ex}{\vdots} \\
\lstick{\ket{s_n}} & \qw & \ghost{V^{2^0}} & \qw  & \qw & \ghost{V^{2^{k-1}}} & \qw & \meter & \rstick{s'_n} \cw \\ 
}
\]
\caption{An alternative proposal mechanism based on quantum phase estimation. Here $\ket{+}=\frac{1}{\sqrt{2}} (\ket{0}+\ket{1})$, $V=e^{-i H \tau}$ for some desired timescale $\tau$, and $QFT$ is the quantum Fourier transform. The circuit comprises two quantum registers: the top one consists of $k$ qubits initialized in $\ket{+}^{\otimes k}$, while the bottom one consists of $n$ qubits initialized in the computational state $\ket{\boldsymbol{s}}$.}
\label{figS:QPE}
\end{figure}

The final calculation of the Section \ref{secS:perturbation_theory} involved time evolution $e^{-i H t}$ for a uniformly random $t$. This dephases an initial state $\ket{j}$ in the eigenbasis of $H$. One could therefore realize the same effect by preparing an initial state $\ket{j}$, measuring it in the eigenbasis of $H$, then measuring the resulting state in the computational basis. While it is not typically feasible to measure $H$ directly, the conventional way to approximate such a measurement is through quantum phase estimation (QPE), as shown in Fig.~\ref{figS:QPE}. The measurement result from the top register encodes a $k$-bit approximation to an eigenvalue $\lambda_\ell$ of $H$, while the bottom register---just prior to measurement---is approximately in the corresponding eigenstate $\ket{\lambda_\ell}$. The measurement result $\boldsymbol{s'}$ from the bottom register gives the proposed move. It is easy to show that this realizes a symmetric proposal $Q(\boldsymbol{s'}|\boldsymbol{s}) = Q(\boldsymbol{s}|\boldsymbol{s'})$.

While using QPE to propose moves as in Fig.~\ref{figS:QPE} may be of theoretical interest, we see little practical appeal in this approach at the moment. The issue is that the measurement result from the top register is not currently used. This means that one could forego the quantum Fourier transform without affecting the bottom register. In fact, there is no reason at present to use the top register at all. Instead, one could achieve the same effect on the bottom register by picking a random $m \sim \text{uniform} ( \{ 2^0, 2^1, \dots, 2^{k-1} \} )$, applying $V^m$ (not conditioned on any quantum state) to $\ket{\boldsymbol{s}}$, then measuring in the computational basis. This latter scheme, in turn, is equivalent to propagating $\ket{\boldsymbol{s}}$ by $e^{-iHt}$ for a random $t \sim \text{uniform} (\{ \tau 2^0, \tau 2^1, \dots, \tau 2^{k-1} \} )$. In effect then, the only difference between this QPE method and the random-$t$ method in Algorithm~\ref{alg:M-H} is the set of possible values from which $t$ is drawn.

\subsection{Reverse annealing}
\label{secS:annealing}

We have focused so far on time-independent Hamiltonians $H$ given by Eqs.~\eqref{eq:H} and \eqref{eq:H_prob} of the main text, where $\bra{k} H_\text{mix} \ket{j} \in \mathbb{R}$ for all computational states $\ket{j}$ and $\ket{k}$. This ensures that $H$ is real and symmetric (not just Hermitian), so it has real eigenvalues $\{\lambda_\ell\}_{\ell=0}^{2^n-1} \subset \mathbb{R}$ and a set of real orthonormal eigenvectors $\{\ket{\lambda_\ell} \}_{\ell=0}^{2^n-1} \subset \mathbb{R}^{2^n}$ which give the spectral decomposition
\begin{equation}
H = \sum_{\ell=0}^{2^n-1} \lambda_\ell \, \ket{\lambda_\ell} \! \bra{\lambda_\ell}.
\end{equation}
Since we have chosen $\ket{\lambda_\ell}$ to be real, $\ket{\lambda_\ell} \! \bra{\lambda_\ell}^T = \ket{\lambda_\ell} \! \bra{\lambda_\ell}$, so $U =e^{-iHt}$ is symmetric (though not typically Hermitian):
\begin{equation}
U^T
=
\sum_{\ell=0}^{2^n-1} e^{i \lambda_\ell t} \, \ket{\lambda_\ell} \! \bra{\lambda_\ell}^T
=
U,
\end{equation}
therefore satisfying Eq.~\eqref{eq:symmetry} of the main text.

Time-dependent Hamiltonians can also satisfy this symmetry constraint. Consider more generally a Hamiltonian $H(t)$ and its corresponding propagator $U(t, t')$ from $t$ to $t'$ satisfying $i \partial_{t'} U(t, t') = H(t') U(t, t')$ and $U(t, t)=I$ for all $t \le t'$ in $[0, \tau]$, where $\tau$ is the total evolution time. A sufficient condition for $U(0, \tau)$ to satisfy Eq.~\eqref{eq:symmetry} of the main text is that $H(t)^T = H(t)$ and $H(t)=H(\tau-t)$ for all $t \in [0,\tau]$. (We assume that appropriate continuity in $H(t)$ is assured on physical grounds.)

\begin{quote}
\textit{Proof.} For $0 \le t \le \tau/2$ we claim that $U(t, \tau-t) = U(\tau/2, \tau-t) \, U(t, \tau/2)$ is symmetric. Since 
\begin{equation}
\big[ U(\tau/2, \tau-t) \, U(t, \tau/2) \big]^T
=
U(t, \tau/2)^T \, U(\tau/2, \tau-t)^T,
\end{equation}
it suffices to show that $U(\tau/2, \tau-t)^T$ equals $U(t, \tau/2)$. We will do so by showing that both satisfy the same first-order differential equation with the same initial condition. First:
\begin{equation}
\begin{split}
i \frac{\partial}{\partial t} U(\tau/2, \tau-t)^T
&=
-\big[ 
H(\tau-t)\,  U(\tau/2, \tau-t)
\big]^T\\ 
&= -U(\tau/2, \tau-t)^T \, H(t).
\label{eqS:DE1}
\end{split}
\end{equation}
Next, differentiating both sides of $U(0, \tau/2) = U(t, \tau/2) \, U(0, t)$ with respect to $t$ gives
\begin{align}
0 &= \left[ 
\frac{\partial}{\partial t} U(t, \tau/2)
\right] U(0, t) + 
U(t, \tau/2)  \; \frac{\partial}{\partial t} U(0, t)
\end{align}
and therefore
\begin{equation}
\begin{split}
\frac{\partial}{\partial t} U(t, \tau/2)
&= 
- U(t, \tau/2) \left[ 
\frac{\partial}{\partial t} U(0, t)
\right] 
U(0, t)^\dagger\\
&= 
i U(t, \tau/2)\, H(t).
\label{eqS:DE2}
\end{split}
\end{equation}
Since $U(\tau/2, \tau-t)^T$ and $U(t, \tau/2)$, viewed as functions of $t$, satisfy the same differential equation in Eqs.~\eqref{eqS:DE1} and \eqref{eqS:DE2}, and both equal $I$ when $t=\tau/2$, they are equal for all $t \in [0, \tau/2]$. Therefore
\begin{align}
U(t, \tau-t)^T
=
U(t, \tau/2)^T \, U(\tau/2, \tau-t)^T
=
U(\tau/2, \tau-t) \, U(t, \tau/2)
=
U(t, \tau-t).
\end{align}
Setting $t=0$ gives $U(0, \tau)^T = U(0, \tau)$. $\square$
\end{quote}

A particularly interesting way to propose MCMC moves that are close in energy but potentially far in Hamming distance is through reverse quantum annealing \cite{crosson:2021}. That is, by evolving under a Hamiltonian of the form
\begin{equation}
H(t) = \big[1-f(t) \big] \alpha H_\text{prob} + f(t) H_\text{mix}
\end{equation}
where $f: [0,\tau] \rightarrow [0,1]$ is an even function about $t=\tau/2$ that starts at $f(0)=0$, gradually ramps up to some maximum value $f(\tau/2)>0$ and then ramps back down symmetrically to $f(\tau)=0$. $H_\text{mix}$ could be proportional to $\sum_{j=1}^n X_j$ or some other easily-realizable, real matrix. In the limit of $\tau \rightarrow \infty$, the final state would be the same as the initial state due to the adiabatic theorem. For finite $\tau$, however, Landau–Zener transitions can occur at avoided crossings. These transitions are typically harmful in adiabatic quantum computing, but in this context they are beneficial: when a small number of them occur, the measured configuration $k$ should be close in energy to the initial one $j$, but generically far in Hamming distance.

One way to see this is by moving to a rotating frame defined by the eigenvectors of $H(t)$. Let $\{\ket{\lambda_j(t)} \}_{j=0}^{2^n-1}$ be the eigenvectors of $H(t)$, as in Section~\ref{secS:perturbation_theory}, with corresponding eigenvalues $\{ \lambda_j(t) \}_{j=0}^{2^n-1}$, and let
\begin{equation}
V(t) = \sum_{j=0}^{2^n-1} \ket{\lambda_j(t)} \! \bra{j}
\end{equation}
be the unitary that changes between the ``lab frame'' (Schr\"odinger picture) and a rotating frame (interaction picture). That is, a state $\ket{\psi}$ in the former frame becomes $V^\dagger \ket{\psi}$ in the latter, and an observable $O$ becomes $V^\dagger O V$. (We take this as the mathematical definition of the rotating frame.) Since $V(0)=V(\tau)=I$, we are free to compute transition probabilities in the rotating frame, without ever having to explicitly convert back to the lab frame. The rotating frame dynamics are generated by the Hamiltonian $\tilde{H}(t)$, given by:
\begin{equation}
i \frac{d}{dt} V^\dagger \ket{\psi} 
=
\underbrace{\big( V^\dagger H(t) V + i \dot{V}^\dagger V \big)}_{\tilde{H}(t)} V^\dagger \ket{\psi}.
\end{equation}
While $V^\dagger H(t) V = \sum_{j=0}^{2^n-1} \lambda_j(t) \ket{j} \! \bra{j}$ is diagonal, $\tilde{H}(t)$ is generally not, due to the $i \dot{V}^\dagger V$ term. This latter term arises because the eigenbasis of $H(t)$ changes with time, unlike in Section~\ref{secS:perturbation_theory}. It scales inversely with $\tau$, and is typically highly non-local \cite{sels:2017, kolodrubetz:2017}. In this rotating frame, we clearly see that an initial computational state $\ket{j}$ will always produce a final state $\ket{j}$ (up to a global phase) in the $\tau \rightarrow \infty$ limit, as per the adiabatic theorem. When $\tau$ is finite, however, $i \dot{V}^\dagger V$ will generate Landau-Zener transitions between instantaneous energy eigenstates. Notice that $\tilde{H}(t)$ here is analogous to $H(\gamma)$ in Section~\ref{secS:perturbation_theory}, with the diagonal part $V^\dagger H(t) V$ playing the role of $H_\text{prob}$ and $i \dot{V}^\dagger V$ that of $H_\text{mix}$. Invoking the arguments from Section~\ref{secS:perturbation_theory}, we expect this reverse annealing scheme to generate transitions between configurations that are close in energy but potentially far in Hamming distance.

\bibliography{references_b}

\makeatletter\@input{auxfile.tex}\makeatother